\documentclass[referee]{aa}
\usepackage{amsmath}
\usepackage{txfonts}
\usepackage{braket}
\usepackage{amssymb}
\usepackage{enumerate}
\usepackage{graphicx}
\usepackage{subcaption}

\DeclareMathOperator{\atan}{atan}
\begin{document}
\title{Characterizing maser polarization: effects of saturation, anisotropic pumping and hyperfine structure}
\titlerunning{Maser Polarization}
\author{Boy Lankhaar
        \and
        Wouter Vlemmings
        }
\institute{Department of Space, Earth and Environment, Chalmers University of Technology, Onsala Space Observatory, 439 92 Onsala, Sweden \\
\email{boy.lankhaar@chalmers.se}}

\date{Received ... ; accepted ...}

\abstract
   {The polarization of masers contains information on the magnetic field strength and direction of the regions they occur in. Many maser polarization observations have been performed over the last 30 years. However, versatile maser polarization models that can aide in the interpretation of these observations are not available.}
    {We aim to develop a program suite that can compute the polarization by a magnetic field of any non-paramagnetic maser specie at arbitrarily high maser saturation. Furthermore, we aim to investigate the polarization of masers by non-Zeeman polarizing effects. We aim to present a general interpretive structure for maser polarization observations.}
    {We expand existing maser polarization theories of non-paramagnetic molecules and incorporate these in a numerical modeling program suite.}
    {We present a modeling program that CHAracterizes Maser Polarization (CHAMP) that can examine the polarization of masers of arbitrarily high maser saturation and high angular momentum. Also, hyperfine multiplicity of the maser-transition can be incorporated. The user is able to investigate non-Zeeman polarizing mechanisms such as anisotropic pumping and polarized incident seed radiation. We present an analysis of the polarization of $v=1$ SiO masers and the $22$ GHz water maser. We comment on the underlying polarization mechanisms, and also investigate non-Zeeman effects.}
    {We identify the regimes where different polarizing mechanisms will be dominant and present the polarization characteristics of the SiO and water masers. From the results of our calculations, we identify markers to recognize alternative polarization mechanisms. We show that comparing randomly generated linear vs.~circular polarization ($p_L-p_V$) scatter-plots at fixed magnetic field strength to the observationally obtained $p_L-p_V$ scatter can be a promising method to ascertain the average magnetic field strength of a large number of masers.}

\maketitle
\section{Introduction}
Observation of the polarized emission from masers is an established method to obtain information on the magnetic field in the maser region. Linear polarization reveals the (projected) magnetic field direction and circular polarization reveals information on the magnetic field strength. Maser polarization observations have been performed for OH \citep[e.g.][]{baudry:98, fish:06}, H$_2$O \citep[e.g.][]{vlemmings:06a}, SiO \citep[e.g.][]{kemball:97,kemball:09,herpin:06} and methanol \citep[e.g.][]{vlemmings:08,vlemmings:11a,lankhaar:18}. Such observations have indicated, among other things, an ordered magnetic field around asymptotic giant branch (AGB) stars, such as TX Cam \citep{kemball:97}, a magnetically collimated jet from an evolved star \citep{vlemmings:06b}, the first extragalactic Zeeman-effect detection in (ultra)luminous infrared galaxies \citep{robishaw:08} and the magnetically regulated infall of mass on a massive protostellar disk \citep{vlemmings:10}.

The analysis of maser-polarization observations is often based on the theories of \citet{goldreich:73} (GKK73), that are derived analytically for masers under the limiting conditions of (i) strong saturation, where the rate of stimulated emission, $R$, is significantly higher than the isotropic decay rate, $\Gamma$ ($R \gg \Gamma$), (ii) (very) strong magnetic fields, where the magnetic precession rate $g\Omega$ is significantly higher than $R$ ($g\Omega \gg R$) and (iii) high thermal widths, where the thermal broadening, $\Delta \omega$ in frequency units, is significantly higher than $g\Omega$ ($\Delta \omega \gg g\Omega$). In fact, these requirements are seldom fulfilled and one needs to invoke numerical approaches to ascertain the maser polarization characteristics at intermediary conditions. Well known numerical approaches to characterizing maser-polarization have been presented by \citet{deguchi:90} (D\&W90) and \citet{gray:95} (G\&F95). The latter models are aimed at the polarization of masers arising from paramagnetic molecules like OH, but can be generalized to masers from a non-paramagnetic species \citep{gray:12}. Even though the G\&F95 and the D\&W90 models have been shown to be isomorphic \citep{gray:03}, they have made different assumptions in their formulation. For instance, the direct time-dependence of the population ($\rho_{aa}$ and $\rho_{bb}$), coupling elements ($\rho_{ab}$) and the electric field elements have been integrated out in the D\&W90 models. This was shown by \citet{trung:09} to have no impact on the simulation results. However, reversely, the G\&F95 models do not take into account the off-diagonal elements of the state-populations ($\rho_{aa'}, a \neq a'$). Especially in the regions where magnetic field interactions become comparable to the rates of stimulated emission ($g\Omega \sim R$), or when accounting for non-Zeeman effects as anisotropic pumping of the maser or partially polarized incident radiation, this approximation is not valid.

The D\&W90 models have been applied in a number of incarnations:
\begin{enumerate}[i)]
  \item In \citet{nedoluha:90} (N\&W90), the maser polarization model of D\&W90 is applied for one frequency. Circular polarization can not be computed. Linear polarization can be computed in both the Stokes-$U$ and -$Q$ parameters. It is possible to introduce anisotropic pumping. Only one hyperfine sub-transition can be accounted for. N\&W90 report that simulations can be made at up to $J=3-2$ transitions. Convergence issues arise for higher angular momentum transitions.
  \item In \citet{nedoluha:92} (N\&W92), the maser polarization model of D\&W90 is applied under the limiting condition of $g\Omega \gg R$. Therefore, the Stokes-$U$ component of the radiation can be ignored, and only diagonal elements of the density-population matrices need to be regarded. It is in this respect, that this variant of the D\&W90 models is similar to the G\&F95 models. Circular polarization can be computed, as well as linear polarization, but the polarization angle can only be $\chi = 0^o$ or $\chi = 90^o$. Multiple hyperfine sub-transition can be included.
  \item In \citet{nedoluha:94} (N\&W94), we find the most extensive variant of the D\&W90 models. In the N\&W94 models, one accounts for off-diagonal elements in the population-densities, the Stokes-$U$ component of the radiation field, and multiple frequency bins along the maser-line. Anisotropic pumping can be introduced, but multiple hyperfine sub-transitions cannot be included. Because of the computational costs of this approach, N\&W94 give only results for the $J=1-0$ transition.
\end{enumerate}
However, only the qualitative results of these approaches are available.

In this paper, we present a program that we call CHAMP (CHAracterizing Maser Polarization) that simulates the propagation of maser radiation through a medium permeated by a magnetic field. The user is able to use the three approaches of N\&W90, N\&W92 and  N\&W94. We have reproduced these models, and made two significant improvements: (i) the transition of arbitrary angular momentum can be simulated, and (ii) we have expanded the N\&W94 formalism to include multiple (and high $F$) hyperfine transitions. These improvements are vital when analyzing the polarization of high-frequency masers that have become more relevant in the era of ALMA and its full polarization capabilities \citep[see, e.g.][]{perez:13}. The source code of CHAMP and a number of standard input files are available on GitHub at \url{https://github.com/blankhaar/CHAMP}.

As a way of outlining the capabilities of CHAMP, we will perform a range of simulations of the non-paramagnetic maser species SiO and H$_2$O, and comment on their relation to simplified methods of analysis performed in the past. We will focus on a range of SiO $v=1$, $J-(J-1)$ maser transitions, and the $22$ GHz water maser transition. We will present simulations of non-Zeeman polarizing effects, like anisotropy in the maser pumping and polarized seed radiation. We leave maser polarization simulations and analysis of methanol masers, including its complex hyperfine structure \citep{lankhaar:16}, for a later publication. It is possible to investigate the polarization of any non-paramagnetic maser (e.g.~formaldehyde) with CHAMP. The paramagnetic OH masers can also be investigated with these models, but this would be an unnecessary complication of the maser polarization theory, because simplifications from the complete spectral decoupling of the magnetic sub-transitions are not utilized.

This paper is built-up as follows. In § $2$, we will recall the theory of maser-radiation put forth by GKK73 and D\&W90, and expand their work by considering multiple hyperfine-transitions within a certain rotational maser line. In § $3$, we will present the three numerical approaches, based on N\&W90, N\&W92 and  N\&W94, to solve the polarized maser-propagation simulations. We will dedicate extra attention in this section to the improvements made that dealt with previous convergence issues. In § $4$ we will apply our models to simulate the polarization of SiO and water masers. In § $5$, the results will be evaluated by outlining distinguishable polarizing mechanisms, along with an evaluation of some of the existing maser polarization literature. We will conclude with a summary of the results in § $6$.   

\section{Theory}
\subsection{Maser polarization by a magnetic field}
The theory presented here is based on GKK73, and the extension for numerical modeling by \citep{western:84, deguchi:90, nedoluha:94}. We extend these formalisms by considering multiple hyperfine-transitions that lie close to each other in frequency. Often, it is a rotational transition that is masing, and we consider only the states relevant---the hyperfine manifold and magnetic substates---to this transition. Interactions of these states with other molecular states (collisionally or radiatively) are absorbed into the phenomenological pumping and decay term. Because the maser-molecules are permeated by a magnetic field, the degeneracy of the magnetic substates is lifted. The consequential spectral decoupling of the photon-transitions with different helicity will cause a polarization of the radiation. This means that we have to consider all the magnetic substates of the maser-transitions, as well as all the modes of polarization in the radiation.

Up to this point, we have not mentioned the analytical maser polarization theory by Elitzur \citep[see][]{elitzur:91, elitzur:93, elitzur:95, elitzur:98}. In this elegant formalism, the no-divergence requirement of the electric component of the radiation field is shown to put constraints on the phases of the propagated electric field polarizations. These phase-relations yield the polarization solutions of GKK73, but general to any degree of saturation. This result is very different from the other theories of maser polarization, including the one we present here. The idealized presuppositions of the Elitzur models, such as the equal populations of magnetic substates throughout propagation, are however not reproduced by the D\&W90 (and the CHAMP) models---while the radiation field is always subject to the constraint of no-divergence. We work within the D\&W90 formalism, because of mutual confirmation between G\&F95 and D\&W90 on multiple levels of analysis: the isomorphism of the theories of D\&W90 and the G\&F95 models \citep{gray:03, trung:09}, their reproduction of the earlier derived theories of GKK73 and their strong resemblance to the non-maser radiative transfer models of \citet{degl:06}.

Setting up the theory of maser radiation propagation can be divided in two parts. On the one hand, we will present a model on the occupation of the molecular state-populations under the influence of a (polarized) radiation field, and on the other hand, we will present a model on the propagation of the radiation field that is dependent on the state-populations. 

\subsubsection{Evolution of the density operator}
Let us consider a maser transition between two (torsion-)rotational states. The maser medium is permeated by a magnetic field and the two states are coupled by the radiation field. Before considering the interaction of the radiation field and the two (torsion-)rotation states, we will dedicate special attention to hyperfine splitting of the line. When the molecules total nuclear spin, $I$, is nonzero, both (torsion-)rotational levels participating in the maser-transition are split up further by hyperfine interactions in an ensemble of $n_F=2I+1$ hyperfine states,
\[
F_1^{(1)}, F_2^{(1)},\cdots F_{n_F}^{(1)}, \qquad
F_1^{(2)}, F_2^{(2)},\cdots F_{n_F}^{(2)}, 
\]
for the upper and the lower level. As a consequence, the single maser transition splits up in a manifold of hyperfine transitions, where any transition $F_i^{(1)} \to F_j^{(2)}$ is allowed as long as the selection rule $\Delta F = 0,\ \pm 1$ is fulfilled. The hyperfine splitting thus results in a manifold of hyperfine states, of which each upper state is radiatively coupled to multiple other lower states.

However, it turns out that each upper hyperfine level is radiatively coupled strongest to only one lower hyperfine level; dominating other transitions by over an order of magnitude. By virtue of this, we can simplify our problem by decomposing the maser transition into their strongest transitions: $F_i^{(1)} \to F_i^{(2)}$, and neglect all other couplings. In this way, we are left with $n_F$ systems, all independently interacting with the same radiation field.

The Hamiltonian of the $i$'th transition is
\begin{equation}
\hat{H}_i = \begin{pmatrix} \hat{H}_i^{(1)} & \hat{V}_i^{(12)} \\ \hat{V}_i^{(21)} & \hat{H}_i^{(2)} \end{pmatrix}, 
\label{eq:hamrad}
\end{equation}
where the elements of the diagonal matrix elements are defined in the frame where the magnetic field is along the $z$-axis, and are
\begin{align}
\braket{F_i^{(1)} m_F |\hat{H}_i^{(1)}| F_i^{(1)} m_F } &= E_i^{(1)} + g\Omega_i^{(1)} m_F, \nonumber \\
\braket{F_i^{(2)} m_F |\hat{H}_i^{(2)}| F_i^{(2)} m_F } &= E_i^{(2)} + g\Omega_i^{(2)} m_F, 
\end{align} 
where $E_i^{(1,2)}$ are the hyperfine energies of the upper and lower level, and $g\Omega_i^{(1,2)}$ their respective Zeeman splittings. The coupling elements are 
\begin{align}
\hat{V}_i^{(12)} = - \hat{\boldsymbol{d}} \cdot \boldsymbol{E},
\label{eq:perturb}
\end{align}
where $\hat{\boldsymbol{d}}$ is the dipole operator, and $\boldsymbol{E}$ is the electric field. With this decomposition, we can formulate the evolution equation for the states for the $n_F$ independent systems, following the Liouville-von Neumann equation
\begin{align}
\dot{\hat{\rho}}_i = -\frac{i}{\hbar}[\hat{H}_i,\hat{\rho}_i] + \hat{\Lambda}_i - \hat{\Gamma}_i \hat{\rho}_i, 
\end{align}
where we take into account the excitation of both levels, by including a phenomenological term for the pumping of the maser: $\hat{\Lambda}_i$, and the decay of the states by $\hat{\Gamma}_i$. Just as the Hamiltonian of Eq.~(\ref{eq:hamrad}), we can express the density-operator in its parts
\begin{align}
\hat{\rho}_i = \begin{pmatrix} \hat{\rho}_i^{(1)} & \hat{ \rho}_i^{(12)} \\ \hat{\rho}_i^{(21)} & \hat{\rho}_i^{(2)} \end{pmatrix},
\label{eq:density}
\end{align}
so that the evolution of the decomposed density operators is 
\begin{subequations}
\begin{eqnarray}
\dot{\hat{\rho}}_i^{(1)} &= -\frac{i}{\hbar} \left( [\hat{H}_i^{(1)},\hat{\rho}_i^{(1)}] + \hat{V}_i^{(12)} \hat{\rho}_i^{(21)} - \hat{\rho}_i^{(12)} \hat{V}_i^{(21)} \right) \nonumber \\ 
\label{eq:densa}
&- \hat{\Gamma}_i^{(1)} \hat{\rho}_i^{(1)} + \hat{\Lambda}_i^{(1)} \\
\dot{\hat{\rho}}_i^{(2)} &= -\frac{i}{\hbar} \left( [\hat{H}_i^{(2)},\hat{\rho}_i^{(2)}] + \hat{V}_i^{(21)} \hat{\rho}_i^{(12)} - \hat{\rho}_i^{(21)} \hat{V}_i^{(12)} \right) \nonumber \\ 
\label{eq:densb}
&- \hat{\Gamma}_i^{(2)} \hat{\rho}_i^{(2)} + \hat{\Lambda}_i^{(2)} \\
\dot{\hat{\rho}}_i^{(12)} &= -\frac{i}{\hbar} \left( \hat{H}_i^{(1)}\hat{\rho}_i^{(12)} - \hat{\rho}_i^{(12)} \hat{H}_i^{(2)} + \hat{V}_i^{(12)} \hat{\rho}_i^{(2)} - \hat{\rho}_i^{(1)} \hat{V}_i^{(12)} \right) \nonumber \\  
&- \hat{\Gamma}_i^{(1)} \hat{\rho}_i^{(12)}. 
\label{eq:densab}
\end{eqnarray}
\end{subequations}
In D\&W90, it is shown how to integrate out the time-dependence of the off-diagonal elements of Eq.~(\ref{eq:densab}). The solutions of these integrations, are subsequently inserted into the population-equations of Eqs.~(\ref{eq:densa}) and (\ref{eq:densb}). We assume a steady state: $\dot{\hat{\rho}}_i^{(1)} = \dot{\hat{\rho}}_i^{(2)} = 0$. After somewhat involved rearrangements that are analogous to D\&W90, we find the expressions for the upper state-populations
\begin{align}
0  &= -(\Gamma_i + i \omega_{a_ia_i'}) \rho_{a_ia_i'}(v) + \phi (v) \lambda_{a_ia_i'} \nonumber \\
& +\frac{\pi}{c\hbar^2} \left[ \sum_{b_ib_i'} \rho_{b_i b_i'} (v) \left( \braket{\gamma_-^{a_i'b_i'} \zeta^{a_i'b_i,a_i b_i'}}_{\omega} + \braket{\gamma_+^{a_ib_i} \zeta^{a_i'b_i,a_ib_i'}}_{\omega} \right) \right. \nonumber \\
&\left. - \sum_{b_ia_i''} \rho_{a_i''a_i'} (v) \braket{\gamma_-^{a_i'b_i} (\zeta^{a_ib_i,a_i''b_i})^*}_{\omega}  - \sum_{b_ia_i''} \rho_{a_ia_i''} (v) \braket{\gamma_+^{a_ib_i} (\zeta^{a_i'b_i,a_i''b_i})^*}_{\omega} \right],
\label{eq:statepop}
\end{align}
where $a_i$ and $b_i$ are the indices for the magnetic substates of the upper and lower levels, energy difference between magnetic substates are represented by $\hbar \omega_{aa'} = E_a - E_{a'}$. Elements of the pumping operator have been represented as $\Lambda_{aa'} = \phi(v) \lambda_{aa'}$, where $\phi(v)$ stands for the Maxwell-Boltzmann distribution. Furthermore, we have used the simplified notations
\begin{equation}
\zeta^{ij,kl} = I(\omega) \delta_I^{ij,kl} - Q(\omega) \delta_Q^{ij,kl} - iU(\omega) \delta_U^{ij,kl} + V(\omega) \delta_V^{ij,kl}
\label{eq:zeta}
\end{equation}
and $(I,Q,U,V)$ are the Stokes-parameters as defined in D\&W90. The delta-operators are related to the dipole-elements by 
\begin{subequations}
\begin{align}
\delta_I^{ij,kl} &= (d_+^{ij})^* d_+^{kl} + (d_-^{ij})^* d_-^{kl} \\
\delta_Q^{ij,kl} &= (d_+^{ij})^* d_-^{kl} + (d_-^{ij})^* d_+^{kl} \\
\delta_U^{ij,kl} &= (d_+^{ij})^* d_-^{kl} - (d_-^{ij})^* d_+^{kl} \\ 
\delta_V^{ij,kl} &= (d_+^{ij})^* d_+^{kl} - (d_-^{ij})^* d_-^{kl},
\end{align}
\end{subequations}
with explicit elements (D\&W90) 
\begin{equation}
d^{ab}_{\pm} = \pm d_{M=1}^{ab} \frac{1\pm \cos \theta}{2} + i d_{M=0}^{ab} \frac{\sin \theta}{\sqrt{2}} \mp d_{M=-1}^{ab} \frac{1 \mp \cos \theta}{2},
\end{equation} 
where $\theta$ is the angle between the magnetic field and propagation directions. We have also used a simplified notation for the integral
\begin{equation}
\braket{\gamma_{\pm}^{ij} \zeta^{kl,mn}}_{\omega} = \int d\omega \ \gamma_{\pm}^{ij}(\omega, v) \zeta^{kl,mn}(\omega),
\end{equation}
with 
\begin{equation}
\gamma_{\pm}^{a_ib_i} = \frac{1}{\Gamma_i \pm i\left[ \omega_{a_ib_i} - \omega (1 - \frac{v}{c}) \right]}.
\label{eq:gammafun}
\end{equation}
The lower state-populations follow from a similar derivation. The population equations for the upper- and lower-level of the hyperfine-transition $F_i^{(1)} \to F_i^{(2)}$, are mutually dependent, but do not depend directly on other hyperfine-transitions within our approximation, as was motivated in the beginning of this section. We thus have a set of $n_F$-independent population equations for the hyperfine-substates of the rotational transition under investigation.
%
\subsubsection{Evolution of the radiation field}
The evolution of polarized radiation has been derived elsewhere \citep[e.g.][]{goldreich:73,deguchi:90,degl:06}. We will re-iterate the expressions here, while also taking into account the feed of multiple close-lying hyperfine transitions to the radiation field. By using the well-known relation between the propagation of the electric field and the polarization of the medium, one can find the connection between the radiative propagation and the molecular states by expressing the medium polarization in terms of the expectation value of the molecular dipole moment of the ensemble. After expressing the polarization in terms of the molecular states, while maintaining attentive to polarization, one arrives at the propagation relation for the Stokes-parameters \citep{goldreich:73, nedoluha:92} 
\begin{align}
\frac{d}{ds} \begin{pmatrix} I(\omega) \\ Q(\omega) \\ U(\omega) \\ V(\omega) \end{pmatrix} = \begin{pmatrix} A ( \omega ) & B( \omega ) & F( \omega ) & C( \omega ) \\ B( \omega ) & A( \omega ) & E( \omega ) & G( \omega ) \\ F( \omega ) & -E( \omega ) & A( \omega ) & D( \omega ) \\ C( \omega ) & -G( \omega ) & -D( \omega ) & A( \omega ) \end{pmatrix} \begin{pmatrix} I(\omega) \\ Q(\omega) \\ U(\omega) \\ V(\omega) \end{pmatrix},
\label{eq:propstokes}
\end{align}
The expressions for the propagation coefficients are  
\begin{subequations}
\begin{align}
A ( \omega ) &= \frac{-\pi \omega}{c}\sum_i \sum_{a_ib_i} \int dv \ \left[ \sum_{b_i'} \braket{\rho_{b_i'b_i}(\gamma_+^{a_ib_i} + \gamma_-^{a_ib_i'})} \delta_I^{a_ib_i,a_ib_i'} \right. \nonumber \\
&\left. - \sum_{a_i'} \braket{ \rho_{a_ia_i'} (\gamma_+^{a_ib_i} + \gamma_-^{a_ib_i'})} \delta_I^{a_ib_i,a_i'b_i} \right], \\
B ( \omega ) &= \frac{ \pi \omega}{c}\sum_i \sum_{a_ib_i} \int dv \ \left[ \sum_{b_i'} \braket{\rho_{b_i'b_i}(\gamma_+^{a_ib_i} + \gamma_-^{a_ib_i'})} \delta_Q^{a_ib_i,a_ib_i'} \right. \nonumber \\ 
& \left. - \sum_{a_i'} \braket{ \rho_{a_ia_i'} (\gamma_+^{a_ib_i} + \gamma_-^{a_ib_i'})} \delta_Q^{a_ib_i,a_i'b_i} \right], \\
C ( \omega ) &= \frac{-\pi \omega}{c}\sum_i \sum_{a_ib_i} \int dv \ \left[ \sum_{b_i'} \braket{\rho_{b_i'b_i}(\gamma_+^{a_ib_i} + \gamma_-^{a_ib_i'})} \delta_V^{a_ib_i,a_ib_i'} \right. \nonumber \\ 
&\left. - \sum_{a_i'} \braket{ \rho_{a_ia_i'} (\gamma_+^{a_ib_i} + \gamma_-^{a_ib_i'})} \delta_V^{a_ib_i,a_i'b_i} \right], \\ 
D ( \omega ) &= \frac{-i\pi \omega}{c\sum_i }\sum_{a_ib_i} \int dv \ \left[ \sum_{b_i'} \braket{\rho_{b_i'b_i}(\gamma_+^{a_ib_i} - \gamma_-^{a_ib_i'})} \delta_Q^{a_ib_i,a_ib_i'}\right. \nonumber \\ 
&\left. - \sum_{a_i'} \braket{ \rho_{a_ia_i'} (\gamma_+^{a_ib_i} - \gamma_-^{a_ib_i'})} \delta_Q^{a_ib_i,a_i'b_i} \right],\\ 
E ( \omega ) &= \frac{i \pi \omega}{c\sum_i }\sum_{a_ib_i} \int dv \ \left[ \sum_{b_i'} \braket{\rho_{b_i'b_i}(\gamma_+^{a_ib_i} - \gamma_-^{a_ib_i'})} \delta_V^{a_ib_i,a_ib_i'} \right. \nonumber \\
&\left. - \sum_{a_i'} \braket{ \rho_{a_ia_i'} (\gamma_+^{a_ib_i} - \gamma_-^{a_ib_i'})} \delta_V^{a_ib_i,a_i'b_i} \right],\\ 
F ( \omega ) &= \frac{i\pi \omega}{c}\sum_i \sum_{a_ib_i} \int dv \ \left[ \sum_{b_i'} \braket{\rho_{b_i'b_i}(\gamma_+^{a_ib_i} + \gamma_-^{a_ib_i'})} \delta_U^{a_ib_i,a_ib_i'} \right. \nonumber \\
&\left. - \sum_{a_i'} \braket{ \rho_{a_ia_i'} (\gamma_+^{a_ib_i} + \gamma_-^{a_ib_i'})} \delta_U^{a_ib_i,a_i'b_i} \right], \\
G ( \omega ) &= \frac{-\pi \omega}{c}\sum_i \sum_{a_ib_i} \int dv \ \left[ \sum_{b_i'} \braket{\rho_{b_i'b_i}(\gamma_+^{a_ib_i} + \gamma_-^{a_ib_i'})} \delta_U^{a_ib_i,a_ib_i'} \right. \nonumber \\ 
&\left. - \sum_{a_i'} \braket{ \rho_{a_ia_i'} (\gamma_+^{a_ib_i} - \gamma_-^{a_ib_i'})} \delta_U^{a_ib_i,a_i'b_i} \right],  
\end{align}
\label{eq:abc2}
\end{subequations}
where the sum $i$ runs over all hyperfine transitions and $a_i$ and $b_i$ are the magnetic sublevels of the upper, respectively lower level of the $i$'th hyperfine transition.  The tight relation between the molecular states and the feed to the radiation field is reflected also in these equations, as again, the radiative coupling between the two states is represented by the $\delta$-operators. In the method section, we will outline the three approaches to numerically solve Eqs.~(\ref{eq:statepop}) and (\ref{eq:abc2}). 

\subsection{Anisotropic pumping}
We have so far left the expression for the matrix-elements of the pumping-operator general. The pumping-operator is a phenomenological term that, together with the decay-operator, absorbs all the interactions with molecular states that are not participating in the maser-transition. The decay-operator is concerned with the decay of the maser-levels to other states. The pumping-operator encapsulates the collisional and radiative (de-)excitations that will eventually populate our two maser-levels. 

A certain directional alignment may have already krept in the molecular states that will later destine to populate our maser levels. After all, the magnetic field tends to direct all molecular states. Alignment will also manifest itself in a molecular state when a (de-)excitation to it has a preferred direction. An example of a directional excitation would be the directional pumping radiation, like the radiation from central stellar object that leads to the SiO maser \citep{gray:12}. The introduced alignment in the directionally excited molecular state will be transferred (with some depolarization) from state to state in the cascade to our maser levels. The reflection of this partial anisotropy in the pumping operator was already formulated by \citet{nedoluha:90} and \citet{western:83c,western:84}, who defined the elements of a the partial anisotropic pumping operator as 
\begin{align}
\Lambda_{m m'}  = \lambda  \left(1 + \epsilon\left[  \frac{F^2 + F - 1 + m^2}{(2F-1)(2F+3)} - 1 \right] \right)\delta_{m m'},
\label{eq:aniso}
\end{align}
where $\lambda$ is the overall pumping, $F$ is the total angular momentum of the associated state, $m$ is the magnetic quantum number, $\delta_{mm'}$ is the Kronecker-delta and $\epsilon$ is the degree of anisotropy in the pumping. In Eq.~(\ref{eq:aniso}) we have assumed the direction of the anisotropic pumping to be along the magnetic field direction. If the pumping-direction has a different orientation with respect to the magnetic field, the pumping-matrix can be obtained by the simple rotation
\begin{align}
\boldsymbol{\Lambda}' = \boldsymbol{D}^{\dagger} (\alpha' \beta' \gamma') \boldsymbol{\Lambda} \boldsymbol{D} (\alpha' \beta' \gamma'),
\label{eq:anis}
\end{align}
over the Euler-angles $(\alpha' \beta' \gamma')$ that describe the rotation from the pumping-direction to the magnetic field direction.

The partial alignment of the directionally pumped maser will result in the emittance of partially polarized radiation. The polarization will depend not only on the degree of anisotropy in the pumping, $\epsilon$, but will also be dependent on the pumping-efficiency,
\begin{align}
\eta = \frac{\epsilon}{\delta},
\end{align}
where we let $\eta$ be the anisotropy-parameter, and 
\begin{align}
\delta = 2 \frac{\lambda_u - \lambda_l}{\lambda_u + \lambda_l}
\label{eq:delta}
\end{align}
is the pumping-efficiency, with the overall-pumping of the upper and lower level given by $\lambda_{u,l}$. The pumping-efficiency has been investigated for water masers. Estimation of the mean population inversion $\Delta n$ from high-resolution observations of water masers around AGB-stars revealed for most masers $\Delta n \lesssim 0.01$. The most luminous masers had higher degrees of population inversion up to $\delta \sim 0.1$ \citep{richards:11}. It is to be expected that for the more saturated masers that the population inversion will decrease. \citet{richards:11} estimated that most masers in the sample, though, were unsaturated. For unsaturated masers, their mean population inversion reflects the pumping-efficiency $2\Delta n \sim \delta$, thus we estimate $\delta \sim 0.02$. The anisotropy degree $\epsilon$ of anisotropically pumped masers is estimated to be of the same order of magnitude \citep{nedoluha:90}.

\section{Methods}
\subsection{Three numerical approaches}
In this work, we have reproduced and extended the the numerical approximations reported in N\&W90, N\&W92 and N\&W94. The difference between approaches can be traced back to different approximations to the integrals in Eqs.~(\ref{eq:statepop}) and (\ref{eq:abc2}): 
\begin{enumerate}[i)]
  \item In N\&W90, integrals are approximated to peak sharply around the maximum of $\gamma_{\pm}^{ij}$, 
\begin{align}
\int d\omega \ \gamma_{\pm}^{ij}(\omega, v) \zeta^{kl,mn}(\omega) \approx \pi \zeta^{kl,mn} (\omega_{ij}),
\end{align}
where $\omega_{ij}$ is the transition frequency between levels $i$ and $j$. Similarly, the integration over density-matrix elements is 
\begin{align}
\int dv \ \rho_{ij} (v) \gamma_{\pm}^{kl} (\omega,v) = \frac{\pi c}{\omega_0} \rho_{ij} (v_0).
\end{align}  
We only account for one frequency and velocity bin in the Stokes-parameters and density-matrix elements. A consequence of this approximation is that only one hyperfine-transition can be included and that circular polarization is not computed. Within this approximation we are left with the following simplified density equations ($\rho_{ij} = \rho_{ij}(v_0)$ and $\zeta^{ij,kl} = \zeta^{ij,kl}(\omega_0)$)
\begin{align}
0  &= -(\Gamma + i \omega_{aa'}) \rho_{aa'} + \lambda_{aa'} + \frac{2\omega \pi^2}{\hbar^2} \left[ \sum_{bb'} \rho_{b b'} \zeta^{a'b,a b'}  \right. \nonumber \\ 
& \left. - \sum_{ba''} \rho_{a''a'} (\zeta^{ab,a''b})^*  - \sum_{ba''} \rho_{aa''}  (\zeta^{a'b,a''b})^* \right],
\label{eq:statepop-num90}
\end{align}
where we have dropped the $i$-indices because we cannot treat a hyperfine manifold in this method. Similarly, the propagation coefficients are
\begin{subequations}
\begin{align}
A(\omega)  &= \frac{-2\pi^2 \omega}{c} \sum_{ab} \left[ \sum_{b'} \rho_{b'b} \delta_I^{ab,ab'} - \sum_{a'} \rho_{aa'}  \delta_I^{ab,a'b} \right], \\
B(\omega)  &= \frac{2\pi^2 \omega}{c} \sum_{ab} \left[ \sum_{b'} \rho_{b'b} \delta_Q^{ab,ab'} - \sum_{a'} \rho_{aa'}  \delta_Q^{ab,a'b} \right], \\
C(\omega)  &= \frac{-2\pi^2 \omega}{c} \sum_{ab} \left[ \sum_{b'} \rho_{b'b} \delta_V^{ab,ab'} - \sum_{a'} \rho_{aa'}  \delta_V^{ab,a'b} \right], \\
G(\omega)  &= \frac{-2 \pi^2 \omega}{c} \sum_{ab} \left[ \sum_{b'} \rho_{b'b} \delta_U^{ab,ab'} - \sum_{a'} \rho_{aa'}  \delta_U^{ab,a'b} \right],
\end{align}
\end{subequations}
and $D(\omega) = E(\omega) = F(\omega) = 0$.
  \item N\&W92 assume a strong magnetic field. Thus, from Eq.~(\ref{eq:statepop}), under the limiting condition $g\Omega \gg R$, it follows that diagonal elements will dominate the density-populations and that we can neglect off-diagonal elements. Through this simplification, we can assume the Stokes-$U$ component of the radiation absent. Integrals are simplified in the following way 
\begin{subequations}
\begin{align}
\int d\omega\ \ \gamma_{\pm}^{ab}(\omega, v) \zeta^{a'b',a''b''}(\omega) &\approx \pi \zeta^{a'b',a''b''}(\omega_{ab}/(1-v/c)) \\
\int dv \ \gamma_{\pm}^{ab} (\omega,v) \rho_{kk} (v) &\approx \frac{\pi c}{\omega_0} \rho_{kk} (c(\omega - \omega_{ab})/\omega).
\end{align}
\end{subequations}
The populations and $\zeta$-parameters are evaluated for $2N+1$ channels
\[
\boldsymbol{\omega} = \{ \omega_{-N},\ \omega_{-N+1},\cdots,\omega_0, \cdots , \omega_N \}
\]
where $\omega_j = \omega_0 + j\Delta \omega $, and $\Delta \omega$ is the width of the frequency channel. The frequency channels are related to the velocity channels, as $\boldsymbol{v} = \frac{c_0}{\omega_0}\boldsymbol{\omega}$, so that $v_j = j\Delta v = j\frac{c_0}{\omega_0}\Delta \omega$. For each channel, the population and $\zeta$-parameters are
\begin{align}
&\rho_{kk} (c(\omega_j - \omega_{ab})/\omega) \approx \rho_{kk} (v_j) \nonumber \\ 
&- \frac{c}{\omega_0} \left. \frac{\partial \rho_{kk}}{\partial v} \right|_{v_j} (m_a g\Omega_1/2 - m_b g\Omega_2/2) \nonumber \\
&\zeta^{a'b',a''b''}(\omega_{ab}/(1-v_j/c)) \approx \zeta^{a'b',a''b''} (\omega_j) \nonumber \\ 
&+ \left. \frac{\partial \zeta^{a'b',a''b''}}{\partial \omega} \right|_{\omega_j} (m_a g\Omega_1/2 - m_b g\Omega_2/2)  
\end{align}
Leading to simplified density equations as well as simplified propagation coefficients, where $D(\omega) = E(\omega) = F(\omega) = G(\omega) = 0$. Thus, within this method, propagation of the Stokes-$U$ part of the radiation does not occur and can be left out.
  \item If we follow N\&W94, we do not make any of the approximations outlined above. Rather, we will endeavor to solve the Eqs.~(\ref{eq:statepop}) and (\ref{eq:abc2}) by making a numerical approximation to the integral 
\begin{align}
\int d\omega \ \gamma_{\pm}^{ab}(\omega, v_j) \zeta^{a'b',a''b''}(\omega) \nonumber
\end{align}
by dividing $\omega$ and $v$ in $2N+1$-channels as we have done for the N\&W92 method (see above). The first approximation that we make is neglecting all contributions to the integral outside of the boundaries $\omega_{\pm N} \pm \Delta \omega /2$
\begin{align}
&\int d\omega \ \gamma_{\pm}^{ab}(\omega, v_j) \zeta^{a'b',a''b''}(\omega) \approx \nonumber \\ 
&\int_{\omega_{-N}-\Delta \omega /2}^{\omega_N + \Delta \omega /2} d\omega \ \gamma_{\pm}^{ab}(\omega, v_j) \zeta^{a'b',a''b''}(\omega) \nonumber ,
\end{align}
which is a good approximation for $\omega_N \gg \omega_D$ ($\omega_D$ is the Doppler broadening). Then, we divide the integral in their respective channels
\begin{align}
&\int_{\omega_{-N} - \Delta \omega /2 }^{\omega_N + \Delta \omega /2} d\omega \ \gamma_{\pm}^{ab}(\omega, v_j) \zeta^{a'b',a''b''}(\omega) = \nonumber \\ 
&\sum_{i=-N}^{N} \int_{-\Delta \omega /2}^{\Delta \omega / 2} d \omega' \gamma_{\pm}^{ab}(\omega_i + \omega', v_j) \zeta^{a'b',a''b''}(\omega_i + \omega') \nonumber
\end{align}
To solve the individual integrals, we assume that the function $\zeta^{a'b',a''b''}(\omega_i + \omega')$, can be approximated as a Taylor expansion around $\omega_i$, truncated at first-order
\begin{align}
&\zeta^{a'b',a''b''}(\omega_i + \omega') = \sum_{p=0}^{\infty} \frac{\omega'^p}{p!} \left(\frac{d^p \zeta^{a'b',a''b''}}{d \omega'^p} \right)_{\omega_i} \nonumber \\ 
&\approx \zeta^{a'b',a''b''} (\omega_i) + \omega' \left(\frac{d \zeta^{a'b',a''b''}}{d \omega'} \right)_{\omega_i}. \nonumber 
\end{align}
This leads to the approximate expression of the integrals 
\begin{align}
&\sum_{i=-N}^{N} \int_{-\Delta \omega /2}^{\Delta \omega / 2} d \omega' \gamma_{\pm}^{ab}(\omega_i + \omega', v_j) \zeta^{a'b',a''b''}(\omega_i + \omega') \nonumber \\ 
&\approx \sum_{i=-N}^{N} \zeta^{a'b',a''b''}(\omega_i) \int_{-\Delta \omega /2}^{\Delta \omega / 2} d \omega' \gamma_{\pm}^{ab} \nonumber \\ 
&+ \sum_{i=-N}^{N} \left(\frac{d \zeta^{a'b',a''b''}}{d \omega'} \right)_{\omega_i} \int_{-\Delta \omega /2}^{\Delta \omega / 2} d \omega' \omega' \gamma_{\pm}^{ab}. 
\end{align}
The remaining integrals can be solved analytically. From the definition of the $\gamma_{\pm}^{ab}(\omega,v)$ function of Eq.~(\ref{eq:gammafun}), we have the following analytical solutions
\begin{subequations}
\begin{align}
&\int_{-\Delta \omega /2}^{\Delta \omega / 2} d \omega' \ \gamma_{\pm}^{ab} (\omega' + \omega_i,v_j) \nonumber \\
&= - \left( \atan q_{\mathrm{up}}  - \atan q_{\mathrm{down}} \right) \pm \frac{i}{2} \mathrm{log} \left( \frac{\Gamma^2 + q_{\mathrm{up}}^2 }{\Gamma^2 + q_{\mathrm{down}}^2} \right) \nonumber \\
&= g_r (i,j) \pm i g_i (i,j) = g^{\pm} (i,j) \\
&\int_{-\Delta \omega /2}^{\Delta \omega / 2} d \omega' \ \omega' \gamma_{\pm}^{ab} (\omega' + \omega_i,v_j) \nonumber \\  
&= \frac{\Gamma g_i (i,j)}{1-j\Delta v} + \left( \frac{\omega_{ab}}{1-j\Delta v} - (\omega_0 + i\Delta \omega) \right)g_r(i,j) \nonumber \\
& \pm i \left(\left[\frac{\omega_{ab}}{1-j\Delta v} - (\omega_0 + i\Delta \omega) \right]g_i(i,j) + \frac{\Gamma}{1-j\Delta v} g_r (i,j) + \Delta \omega \right) \nonumber \\ 
&= g_{r,\omega} (i,j) \pm i g_{i,\omega} (i,j) = g_{\omega}^{\pm} (i,j)
\end{align}
\label{eq:numin}
\end{subequations}
where
\begin{align}
q_{\mathrm{up}} &= \omega_{ab} - \left[\frac{\Delta \omega}{2} + \omega_0 + i\Delta \omega \right](1- j\Delta v) \nonumber \\
q_{\mathrm{down}} &= \omega_{ab} - \left[-\frac{\Delta \omega}{2} + \omega_0 + i\Delta \omega \right](1- j\Delta v) . \nonumber
\end{align}
We should note that we differ slightly in our approach from N\&W94, because we use the analytical solutions to the integrals of Eq.~(\ref{eq:numin}), instead of the assuming a sharply peaked function. Let us now insert these simplified integrals in formulating the final approximate equation for the integral
\begin{align}
&\int d\omega \ \gamma_{\pm}^{ab}(\omega, v_j) \zeta^{a'b',a''b''}(\omega) \approx \sum_{i = -N}^N \left(\zeta^{a'b',a''b''}(\omega_i) g^{\pm} (i,j) \right. \nonumber \\ 
& \left. + \left(\frac{d \zeta^{a'b',a''b''}}{d \omega'} \right)_{\omega_i} g_{\omega}^{\pm} (i,j) \right),
\label{eq:num94intw}
\end{align}
where the derivatives $\left(\frac{d \zeta^{a'b',a''b''}}{d \omega'} \right)_{\omega_i}$,
can be evaluated via the finite-difference method. The numerical approximation for the integral over $v$, at a particular channel frequency, $\omega_i$
\begin{align}
\braket{\gamma_{\pm}^{ab}(\omega_i) \rho_{kk'}}_v = \int dv \ \gamma_{\pm}^{ab}(\omega_i, v) \rho_{k k'} (v) \nonumber
\end{align}
is obtained in a similar way, and yields
\begin{align}
&\int dv \ \gamma_{\pm}^{ab}(\omega_i, v) \rho_{kk'}(v) \approx \sum_{j = -N}^N \left( \frac{c_0}{\omega_0}\rho_{kk'} (v_j) g^{\pm} (i,j) \right. \nonumber \\ 
& \left. - \left( \frac{c_0}{\omega_0} \right)^2 \left(\frac{d \rho_{kk'}}{d v'} \right)_{v_j} g_{\omega}^{\pm} (i,j) \right),
\label{eq:num94intv}
\end{align}
where the derivatives $\left(\frac{d \rho_{kk'}}{d v'} \right)_{v_j}$,
again, can be evaluated via the finite-difference method. We have evaluated the accuracy of the truncated Taylor expansion, and found that adding higher-order terms had minimal effect. Using the numerical expressions of Eqs.~(\ref{eq:num94intw}) and (\ref{eq:num94intv}) for the integrals, the density-equations and propagation-matrix can be set up. The latter will contain all 7 propagation-coefficients. Solving the density-equations will be the subject of the next subsection. 
\end{enumerate}
Note that above, we have assumed that different frequency-components of the radiation field are uncorrelated, which is a standard assumption in maser theory \citep{gray:12}. The same goes for the different velocity-components of the molecular states. Numerical simulation of maser polarization propagation can be made using these formalisms, by (i) computing the state-populations for a given radiation-field (see next paragraph) with the use of Eq.~(\ref{eq:statepop}) and (ii) computing the propagation coefficients using Eq.~(\ref{eq:abc2}) and the newly found state-populations. Subsequently, the radiation field is propagated using Eq.~(\ref{eq:propstokes}), where, for small enough $\Delta s$, the propagated vector of Stokes-parameters can be approximated by $\boldsymbol{I} (s + \Delta s, \omega) = e^{\Delta s \boldsymbol{K}(s,\omega)} \boldsymbol{I} (s,\omega)$, where $\boldsymbol{K}(s,\omega)$ stands for the matrix of propagation-coefficients (see Eq.~\ref{eq:propstokes}). The initial radiation field may be black-body radiation, and the initial guess for the state-populations $\sim \Lambda / \Gamma$. In the following paragraph, we will put extra emphasis on the computation of the state-populations. 
\subsection{Solving the density-equations}
Because convergence issues have been known to arise for the density-equations of N\&W90 and N\&W94 at high maser saturation, we will explicitly comment on our used method of solving the density-equations. In the following, we will consider the density-equations for N\&W90, but similar methodology was used for the other approaches. From Eq.~(\ref{eq:statepop-num90}), we have $n_{F_1}^2 + n_{F_2}^2$ coupled equations for the density-matrix (for N\&W92, the dimensionality is reduced to $n_{F_1} + n_{F_2}$). To ensure hermicity of the solutions, it is convenient to separate the density-matrix elements in their real and imaginary parts,
\begin{align}
\rho_{aa'} = \mathrm{Re} (\rho_{aa'}) + i\mathrm{Im} (\rho_{aa'}),
\end{align} 
and we require $\mathrm{Re} (\rho_{aa'}) = \mathrm{Re} (\rho_{a'a})$ as well as $\mathrm{Im} (\rho_{aa'}) = -\mathrm{Im} (\rho_{a'a})$. We will bundle the unique elements in the vector $\boldsymbol{\rho} = [ \boldsymbol{\rho}_a,\boldsymbol{\rho}_b]^T$, where
\begin{align}
\boldsymbol{\rho}_a = [\rho_{11}^{(a)},\rho_{22}^{(a)},\cdots \rho_{n_{F_1} n_{F_1}}^{(a)},\mathrm{Re} (\rho_{12}^{(a)}),\mathrm{Im} (\rho_{12}^{(a)}),\cdots, \mathrm{Im}(\rho_{n_{F_1}-1,n_{F_1}}) ]
\end{align}
and $\boldsymbol{\rho}_b$ is the analogous population-vector for the lower-state. We take the real and imaginary parts from Eq.~(\ref{eq:statepop-num90}) and find
\begin{align}
\mathrm{Re}(\lambda_{aa'}) &= -\Gamma \mathrm{Re}(\rho_{aa'})  + \mathrm{Im}( \rho_{aa'}) \omega_{aa'} + \frac{2\omega \pi^2}{\hbar^2} \nonumber \\ 
& \times \left[ \sum_{bb'} ( \mathrm{Re} (\rho_{b b'}) \mathrm{Re}( \zeta^{a'b,a b'}) - \mathrm{Im}( \rho_{b b'}) \mathrm{Im}(\zeta^{a'b,a b'}))  \right. \nonumber \\
 &- \sum_{ba''} \left. ( \mathrm{Re}(\rho_{a''a'}) \mathrm{Re}(\zeta^{ab,a''b})^* - \mathrm{Im}(\rho_{a''a'}) \mathrm{Im}(\zeta^{ab,a''b})^* ) \right. \nonumber \\
&- \left. \sum_{ba''} ( \mathrm{Re} ( \rho_{aa''})   \mathrm{Re}((\zeta^{a'b,a''b})^*) - \mathrm{Im}(\rho_{aa''} ) \mathrm{Im}((\zeta^{a'b,a''b})^*) ) \right], \nonumber \\
 &= \boldsymbol{a}^{aa'} \boldsymbol{\rho}, \\
\mathrm{Im}(\lambda_{aa'}) &= -\Gamma \mathrm{Im}(\rho_{aa'})  - \mathrm{Re}( \rho_{aa'}) \omega_{aa'} + \frac{2\omega \pi^2}{\hbar^2} \nonumber \\ 
&\times \left[ \sum_{bb'} ( \mathrm{Re} (\rho_{b b'}) \mathrm{Im}( \zeta^{a'b,a b'}) + \mathrm{Im}( \rho_{b b'}) \mathrm{Re}(\zeta^{a'b,a b'}))  \right. \nonumber \\
 &- \sum_{ba''} \left. ( \mathrm{Re}(\rho_{a''a'}) \mathrm{Im}(\zeta^{ab,a''b})^* + \mathrm{Im}(\rho_{a''a'}) \mathrm{Re}(\zeta^{ab,a''b})^* ) \right. \nonumber \\
&- \left. \sum_{ba''} ( \mathrm{Re} ( \rho_{aa''})   \mathrm{Im}((\zeta^{a'b,a''b})^*) + \mathrm{Im}(\rho_{aa''} ) \mathrm{Re}((\zeta^{a'b,a''b})^*) ) \right], \nonumber \\
 &= \boldsymbol{b}^{aa'} \boldsymbol{\rho}.
\end{align}
The collection of density-equations can thus be formulated in the following matrix-equation
\begin{align}
\boldsymbol{\lambda} = \boldsymbol{M} \boldsymbol{\rho},
\end{align}  
where all matrices and vectors are real, and we have the elements of the matrix 
\begin{align}
\boldsymbol{M} =  [\boldsymbol{a}_{11},\boldsymbol{a}_{22},\cdots \boldsymbol{a}_{n_{F_1} n_{F_1}}, \boldsymbol{a}_{12},\boldsymbol{b}_{12},\cdots, \boldsymbol{b}_{n_{F_1}-1,n_{F_1}},\cdots]^T, 
\end{align}
where the last part is omitted, but is comprised of the analogous density-equations of the lower level. We can solve for all densities by 
\begin{align}
\boldsymbol{\rho} = \mathrm{inv}(\boldsymbol{M}) \boldsymbol{\lambda}.
\end{align}
The matrix-inversion is performed using an LQ-decomposition, taken from the standard LAPACK-libraries \citep{LAPACK}. This method is very robust, as exemplified by the fact that these density-equations are solvable for arbitrary angular momentum transitions (matrix-dimensionality) and maser saturation. This is in contrast to N\&W90 and N\&W94, where convergence problems were reported for transitions of $J>3$ \citep{nedoluha:90}.  
\subsection{Experiments}
We present the developed methods by using them to analyze masers with a non-paramagnetic Zeeman effect that have shown polarization in their emission. In the following, we only present results from the most rigorous N\&W94 method. We consider all Stokes parameters, high rates of stimulated emission and non-Zeeman polarizing mechanisms.

We report our calculations mainly through contour maps of the linear polarization degree, $p_L = \sqrt{Q_0^2 + U_0^2}/I_0$, polarization angle, $p_a = \mathrm{atan}(U_0/Q_0)/2$ and circular polarization degree, $p_V = (V_{\mathrm{max}} - V_{\mathrm{min}})/I_0$. The circular polarization degree is taken to be negative if $V_{\mathrm{max}}$ occurs at a frequency $\omega < \omega_0$. The Stokes-parameters $I_0$, $Q_0$ and $U_0$ are taken at the peak of $I(\omega)$. The polarization angle is relative to the rejection of the magnetic field direction from the propagation direction, i.e.~the magnetic field direction projected onto the plane of the sky. 
\subsubsection{SiO masers}
We analyze the polarization of SiO masers by a magnetic field. We run simulations at various magnetic field strengths, angular momentum transitions, and propagation angles $\theta$. The molecular parameters that are used in the simulation are given in Table \ref{tab:sio}. We perform calculations for the SiO masers in the vibrational state $v=1$. SiO masers also occur in higher vibrational states. The results we present can roughly be taken to be similar to higher vibrational states. Only the different isotropic decay rates, that scale roughly as $\Gamma \approx 5 v \ \mathrm{s}^{-1}$ \citep{elitzur:92}, will lead to a different ratio $g\Omega/\Gamma$ which will have a small impact on the presented results. 

Maser polarization properties converge for $\omega_D \gg g\Omega$. To ensure $\omega_D \gg g\Omega$, we use a thermal maser width of $\Delta \omega_{\mathrm{th}} = 1000 \times g\Omega \times J$, where $J$ is the angular momentum of the upper level. This thermal maser width corresponds to $v_{\mathrm{th}} \approx \frac{0.83 \times J^2}{B(\mathrm{G})} \ \mathrm{km/s}$. We perform studies with 
\begin{itemize}
  \item isotropic pumping, where the pumping matrix is $\boldsymbol{\Lambda} = \lambda \boldsymbol{1}$
  \item polarized incident seed radiation, with isotropic pumping, but with seed radiation of $U/I = 0.1$ and $U/I = 0.5$.
  \item anisotropic pumping, where the pumping matrix characterized by Eq.~(\ref{eq:anis}). We run simulations for with moderate, $\eta =0.1$ and high $\eta = 0.5$ degrees of anisotropy. We run simulations for three anisotropy-directions, namely (i) parallel to the magnetic field, (ii) perpendicular to the magnetic field and propagation direction, (iii) at $45^o$ from the magnetic field in the plane perpendicular to the propagation direction. 
\end{itemize}
\begin{table*}[t]
\centering
\caption{Molecular parameters for $v=1$ SiO masers}
\begin{tabular}{l c c c c c }
\hline \hline
 & $\nu_0$ (GHz) & $g \Omega_{\mathrm{up}}/B$ (s$^{-1}/$mG) & $g \Omega_{\mathrm{down}}/B$ (s$^{-1}/$mG) & $A_{ij}$ (s$^{-1}$) & $\Gamma$ (s$^{-1}$)  \\ \hline 
$J=1-0$ & $ 43.122$ & $0.75$  & $0.75$ & $3.024\times 10^{-6}$ & $5$ \\
$J=2-1$ & $ 86.243$ & $0.75$  & $0.75$ & $2.903\times 10^{-5}$ & $5$ \\
$J=3-2$ & $129.363$ & $0.75$  & $0.75$ & $1.050\times 10^{-4}$ & $5$ \\ 
$J=4-3$ & $172.481$ & $0.75$  & $0.75$ & $2.580\times 10^{-4}$ & $5$ \\ 
$J=5-4$ & $215.595$ & $0.75$  & $0.75$ & $5.134\times 10^{-4}$ & $5$ \\ 
\hline \hline 
\end{tabular}
\label{tab:sio}
\end{table*}

\subsubsection{Water masers}
We present the polarization of water masers in the parameter space relevant to observations. From maser observations, we know that the strongest water masers do not exceed $T_b\Delta \Omega = 10^{13}$ Ksr \citep{garay:89, sobolev:18}, and that magnetic field estimates range from $B=1\ \mathrm{mG} - 1\ \mathrm{G}$. As was shown in N\&W92, the thermal width of the maser-molecules affects the maser polarization, so we will analyze the water masers excited at different temperatures. Preferred hyperfine pumping is a possibility for this maser specie, so we will analyze a range of relevant cases. Also, we will explore the effect of alternative polarization mechanisms on the polarization of water masers.

\begin{table*}[t]
\centering 
\caption{Molecular parameters for the $22.235$ GHz water maser}
\begin{tabular}{c l l l l l }
\hline \hline
 & $\Delta \nu_{\mathrm{hyp}}$ (kHz) & $g \Omega_{\mathrm{up}}/B$ (s$^{-1}/$mG) & $g \Omega_{\mathrm{down}}/B$ (s$^{-1}/$mG) & $A_{ij}$ (s$^{-1}$) & $\Gamma$ (s$^{-1}$)  \\ \hline 
$F=5-4$ & $-33.38$ & $ -0.79$ & $-1.34$ & $1.789 \times 10^{-9}$ & $1$ \\
$F=6-5$ & $0     $ & $ 3.71 $ & $4.12 $ & $1.806 \times 10^{-9}$ & $1$ \\
$F=7-6$ & $43.018$ & $ 6.51 $ & $7.24 $ & $1.860 \times 10^{-9}$ & $1$ \\ 
\hline \hline 
\end{tabular}
\label{tab:water}
\end{table*}


\begin{figure*}
    \centering
    \begin{subfigure}[b]{0.45\textwidth}
       \includegraphics[width=\textwidth]{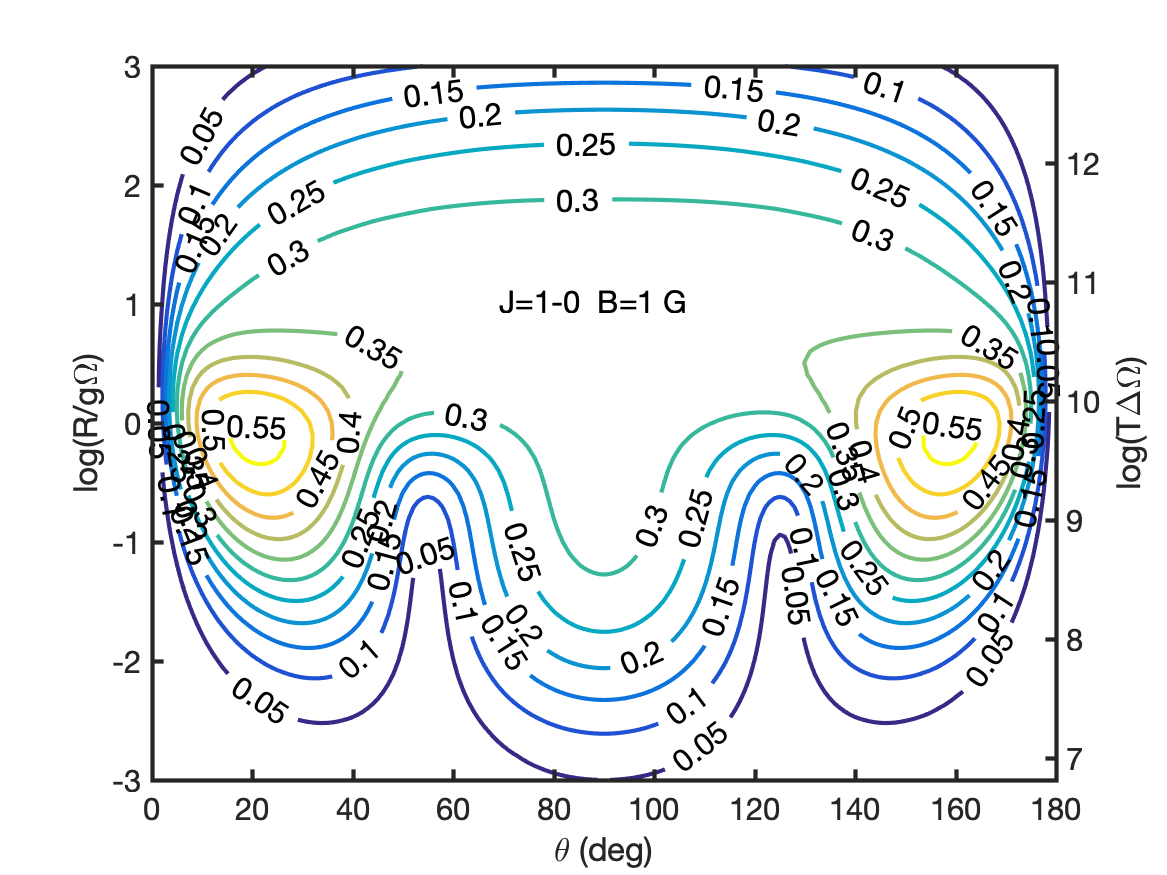}
       \caption{}
    \end{subfigure}
    ~
    \begin{subfigure}[b]{0.45\textwidth}
       \includegraphics[width=\textwidth]{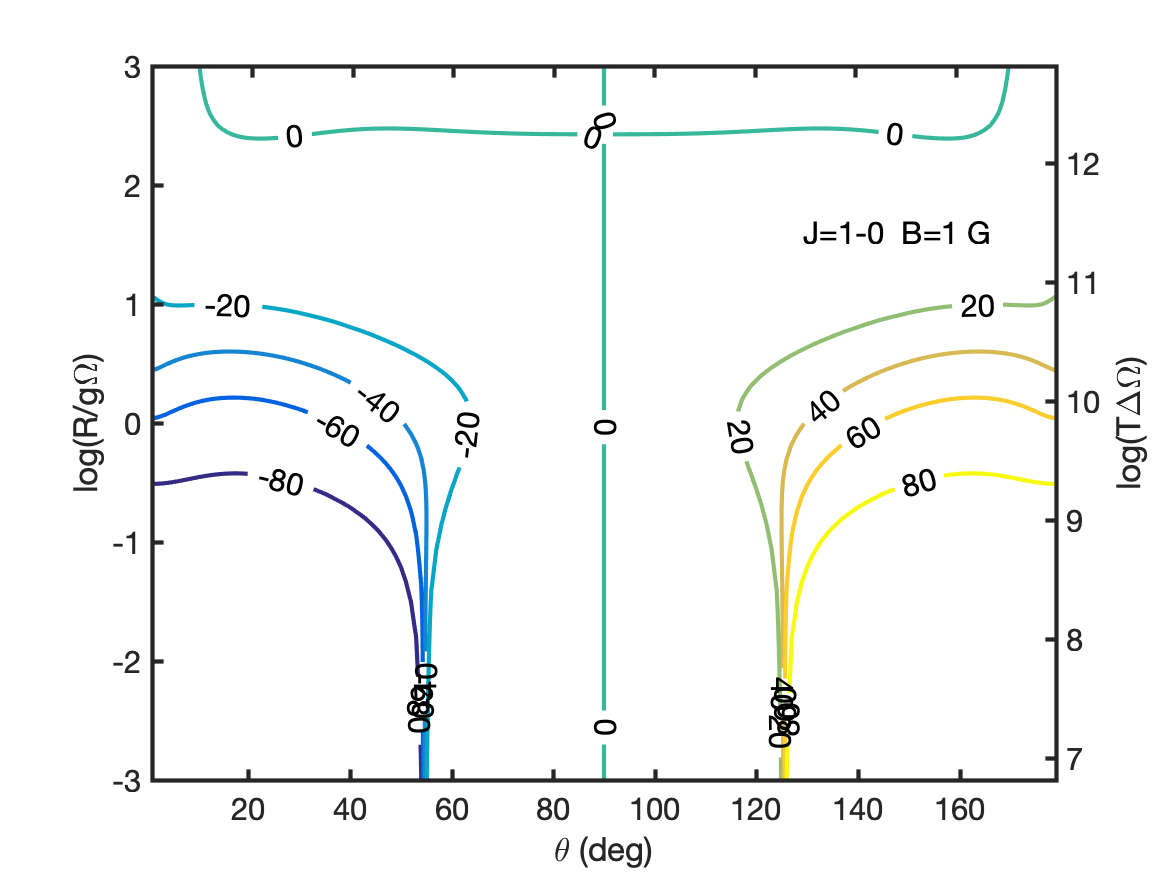}
       \caption{}
    \end{subfigure}
     ~
    \begin{subfigure}[b]{0.45\textwidth}
      \includegraphics[width=\textwidth]{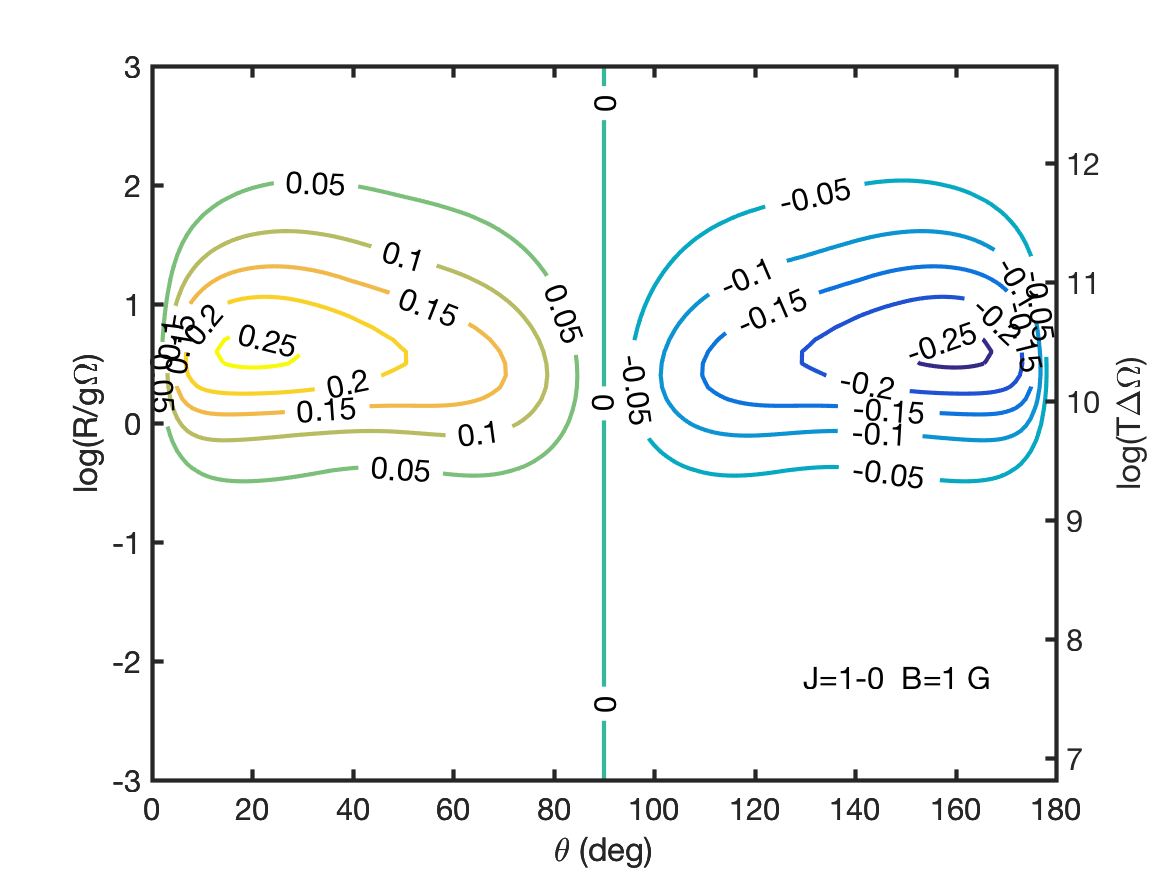}
      \caption{}
    \end{subfigure}
  \caption{Contour polarization plots of $v=1$ SiO masers. The linear polarization fraction (a), angle (b) and circular polarization fraction (c) are plotted as a function of the propagation angle $\theta$ and the rate of stimulated emission. Magnetic field strength and transition angular momentum are denoted inside the figure. For simulations with $J_{\mathrm{up}}>1$ and other magnetic field strengths, see Figs.~A.1-3}
  \label{fig:contour_iso}
\end{figure*}

\section{Results} 
We report here the results of representative numerical simulations to several $v=1$ SiO masers and the $22$ GHz water maser. Results are only reported for the most rigorous N\&W94 approach. We divide up this section into experiments on SiO and water masers, and will further compartmentalize experiments of: isotropically pumped masers, masers with polarized seed radiation, and anisotropically pumped masers. The results are graphically summarized as polarization landscapes, dependent on maser luminosity and propagation-magnetic field angle. A large part of the results are placed in the Appendix. In this section, we lay out observable patterns in the reported polarization landscapes. Then, in the following section, we discuss the physical processes that give rise to these patterns. 
\subsection{SiO masers}
\subsubsection{Isotropic pumping}
Simulations of a $J=1-0$ SiO maser in a $1$ G magnetic field with varying luminosity 
and magnetic field angle are given in Fig.~\ref{fig:contour_iso}. 
Simulations of higher angular momentum and at different magnetic fields are given in Figs.~A.1-3. 
The only polarizing
entity in these simulations is the magnetic field and its interaction with the directional maser radiation. 
We observe, regardless of the magnetic field strength or angular 
momentum of the transition, a peak in the linear polarization fraction
around $\mathrm{log}(R/g\Omega)=0$ that is, in the region where the rate of
magnetic precession ($g\Omega$) and stimulated emission rate ($R$) become comparable in size. 
The peak of the linear polarization fraction is 
in the order of the GKK73-estimate of linear polarization fraction, but can 
exceed it by $10\%$. This excess of polarization is associated with significant polarization in the
Stokes-$U$ spectrum, and is most pronounced for strong magnetic fields and around
$\theta = 20^o$. The linear polarization fraction increases
with the magnetic field strength, and decreases with the angular momentum $J$,
of the transition. A large region, around
$\mathrm{log}(R/g\Omega)=$ $0-0.5$, $0-1.5$ and $0-2.5$, for $B=100$ mG, $1$ G and
$10$ G has a stable polarization fraction of about $p_L=1/3$ (for the $J=1-0$ transition) for a large range
of angles. The stability of the polarization fraction over 
$R / g\Omega$ correlates with the
propagation angle and magnetic field strength. For $\theta$ close to
$90^o$, and strong magnetic fields, the polarization fraction is stable for
a large range of $R / g\Omega$. Significant
polarization occurs for a much greater region of $R$ and $\theta$ when the magnetic
field strength is increased. We note that the
polarization fraction function fulfills the symmetry-relation:
$p_L(\theta) = p_L(180^o-\theta)$. The polarization angle and circular polarization flip according to
$p_a(\theta) = -p_a(180^o-\theta)$ and $p_V(\theta) = -p_V(180^o-\theta)$. 
An interesting feature is found near the magic angle, where for $\mathrm{log}(R / g\Omega) \lesssim 0$, a sharp drop in the polarization fraction is observed that
becomes more pronounced with decreasing $\mathrm{log}(R / g\Omega)$. Polarization
around the magic angle for $\mathrm{log}(R / g\Omega)\lesssim -2$ is mostly absent.

\begin{figure}
    \centering
    \begin{subfigure}[b]{0.45\textwidth}
       \includegraphics[width=\textwidth]{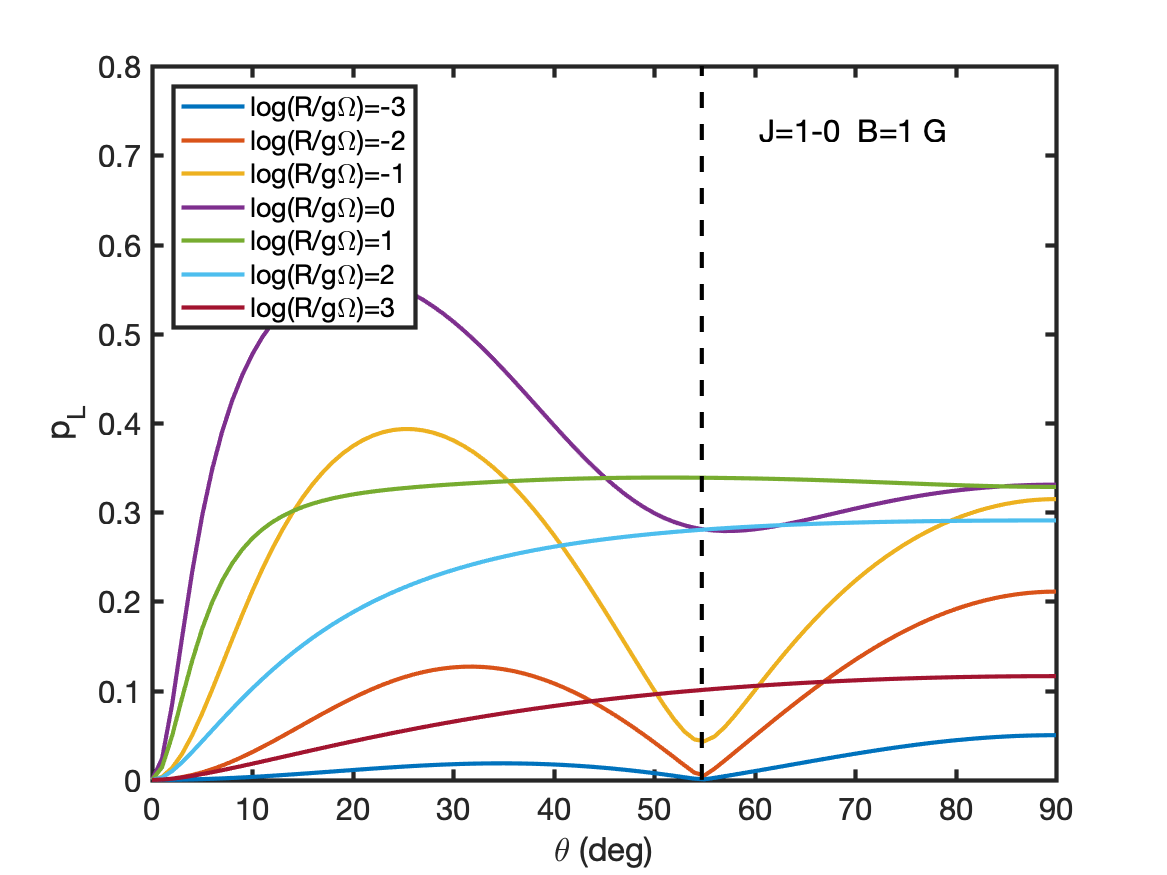}
       \caption{}
    \end{subfigure}
    ~
    \begin{subfigure}[b]{0.45\textwidth}
       \includegraphics[width=\textwidth]{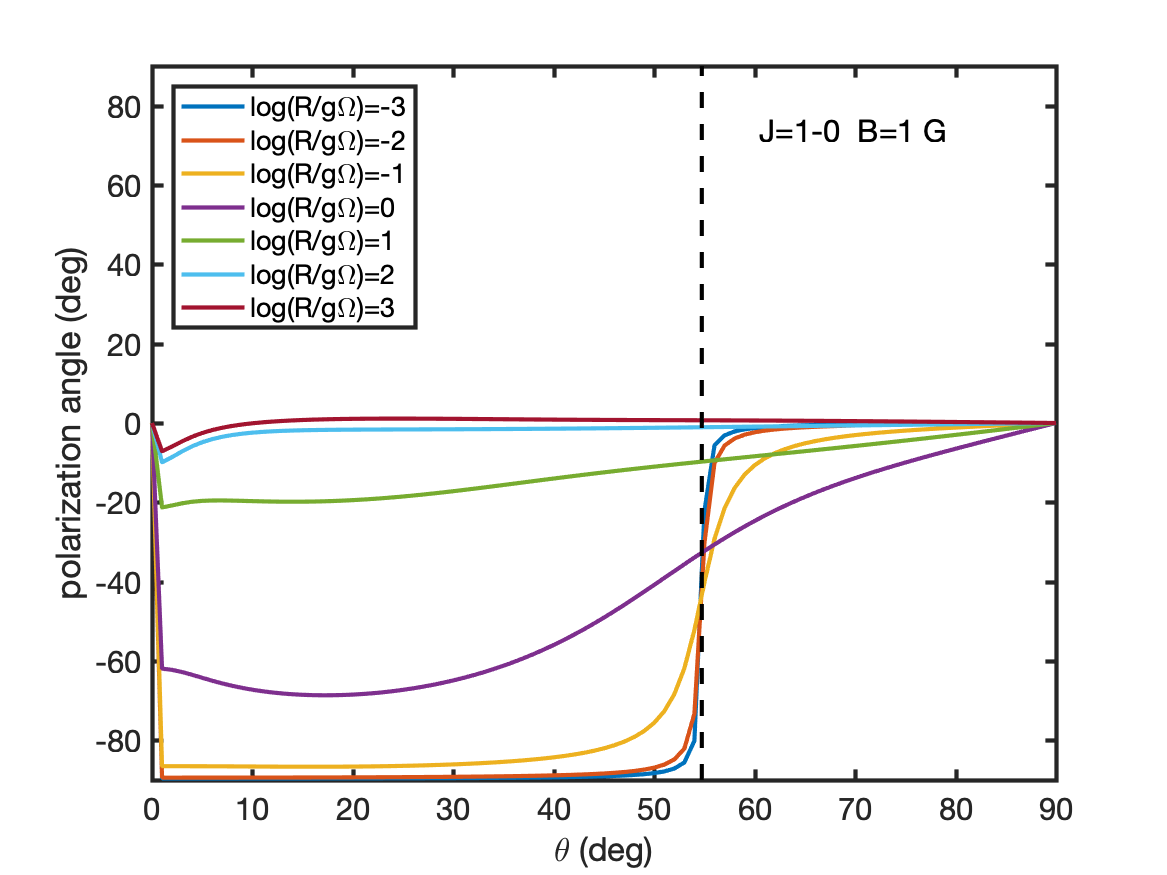}
       \caption{}
    \end{subfigure}
  \caption{Plots of the linear polarization fraction (a) and angle (b) of isotropically pumped $v=1$ SiO masers as a function of the propagation angle $\theta$ for different saturation rates. Note the sharp 90$^o$-flip of the polarization at the magic angle (denoted with the black-dotted line) that bluntens and disappears with increasing levels of saturation. Magnetic field strength and transition angular momentum are denoted inside the figure.}
  \label{fig:magiccross}
\end{figure}

Directing our attention to the polarization angles, we observe that the $90^o$-flip of the polarization angle can
be produced by crossing the magic angle, $\theta_m$, as well as the transition
from $\mathrm{log}(R/g\Omega)\ll 0$ to $\mathrm{log}(R/g\Omega)\gg 0$.
The $\theta_m$-crossing polarization angle flip becomes sharper
with $B$, and manifests itself only for
$\mathrm{log}(R/g\Omega)<-1$. For higher $\mathrm{log}(R/g\Omega)$ 
the flip will get less sharp. These features are particularly clear in Fig.~\ref{fig:magiccross}. 
In the intermediate region around $\mathrm{log}(R/g\Omega)$, the region of 
highest linear polarization, arbitrary 
polarization angles can be produced.  
Overall, apart from the
sharper $90^o$-flip at $\theta_m$, the polarization angle as a function of
$\mathrm{log}(R/g\Omega)$ and $\theta$ is very consistent for the different magnetic field strengths and different transitions. 
At $\mathrm{log}(R/g\Omega) \gg 1$, the polarization
vectors will be aligned ($\theta=0$) with the (projected) magnetic field direction at any propagation angle $\theta$. 

We continue by analyzing the landscape of circular polarization. We observe that the highest circular polarization fractions occur around $\theta = 20^o$, and is associated with the region of maximal linear polarization fractions. However, for circular polarization maximal polarization occurs at slightly higher $R$. Circular polarization is most significant between log$(R/g\Omega)>-1$ and log$(R/g\Omega)<2.5$, and quickly drops to zero for $\theta \to 90^o$. Circular polarization contours for other magnetic field strengths (Appendix) show similar circular polarization landscapes. The maximum circular polarization fraction does not change much for stronger magnetic field strengths, although the region of significant polarization becomes larger. We saw an analogue effect for the linear polarization. Reversely, lower magnetic field strength does decrease the maximum circular polarization fraction, and also decreases region of significant circular polarization. For these simulations, we have chosen a thermal width, $\Delta \omega$, so that $\Delta \omega = 1000 g\Omega$ ($v_{\mathrm{th}}^{\mathrm{SiO}\ J=1-0} = 0.033 \frac{B}{\mathrm{mG}}\ \mathrm{km/s}$), and found that variations in the thermal width did not yield significantly different circular polarization as long as the requirement $\Delta \omega \gg g\Omega$ was fulfilled\footnote{We should note that these remarks are concerned with the polarizing mechanism around $-2<\mathrm{log}(R/g\Omega)<2$. As will be discussed later, circular polarization can be introduced via pure spectral decoupling of the $\Delta m=\pm 1$ transitions. Circular polarization via such a mechanism is dependent on the line-width and thus maser thermal width.}.

Simulations of the $J=2-1$ SiO maser-transition reveal a sharp drop in both linear and circular polarization fractions with respect to the
$J=1-0$ transitions. The maxima of the polarization fractions are $p_{Q_{\mathrm{max}}} = 0.20$ and $p_{V_{\mathrm{max}}} = 0.10$ for $B=1$ G,
constituting a 60\% loss in polarization with respect to the $J=1-0$ transition. The general shapes of the contour maps are retained, although 
the weaker polarization does
entail that the area of polarization is smaller. The 90$^o$-flip, caused by increase in $R$, characteristic for the $\theta < \theta_m$ masers,
is observed to be a less sharp, and occurs at higher $\mathrm{log}(R/g\Omega)$. Going to higher angular momentum transitions, the changes become
less pronounced with respect to the $J=2-1$ transition, although we do observe a minor but steady loss in polarizing strength of the maser with increasing $J$.

\begin{figure}
    \centering
    \begin{subfigure}[b]{0.45\textwidth}
        \includegraphics[width=\textwidth]{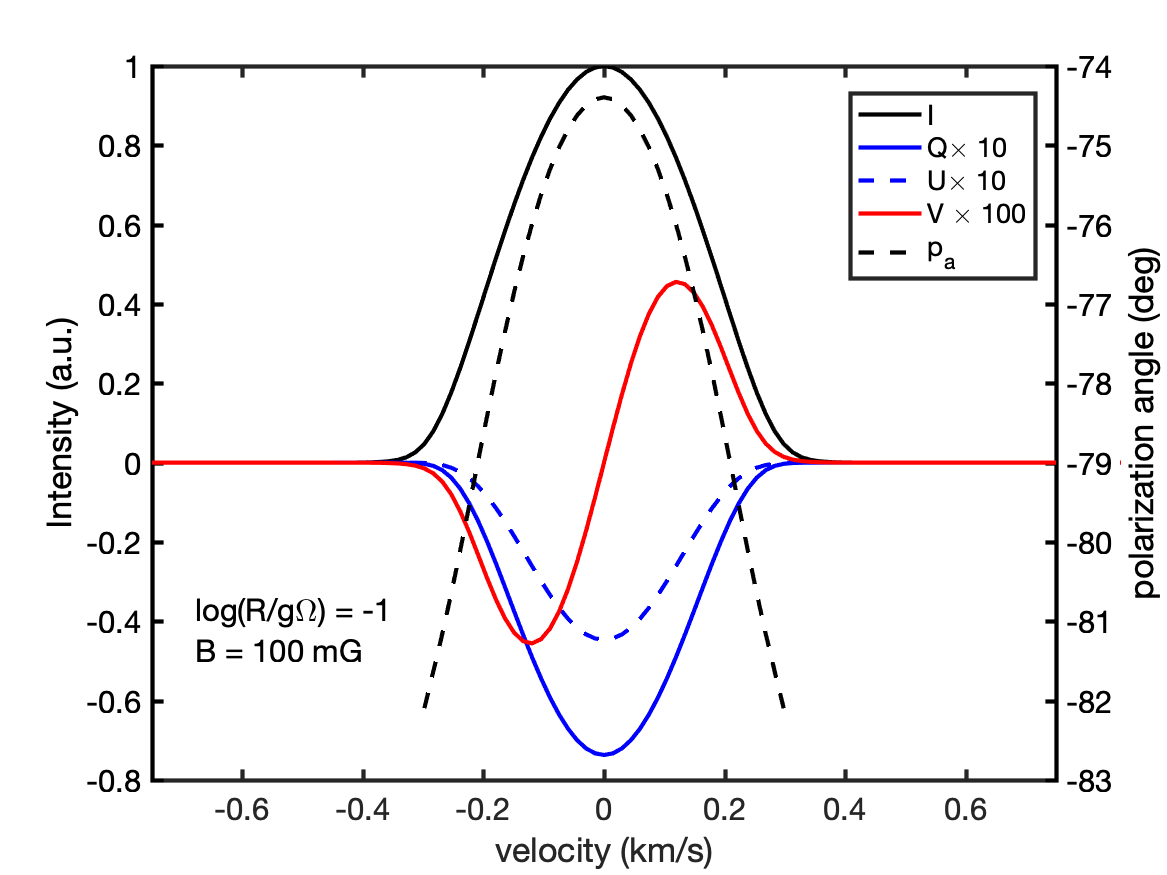}
        \caption{}
    \end{subfigure}
    ~
    \begin{subfigure}[b]{0.45\textwidth}
        \includegraphics[width=\textwidth]{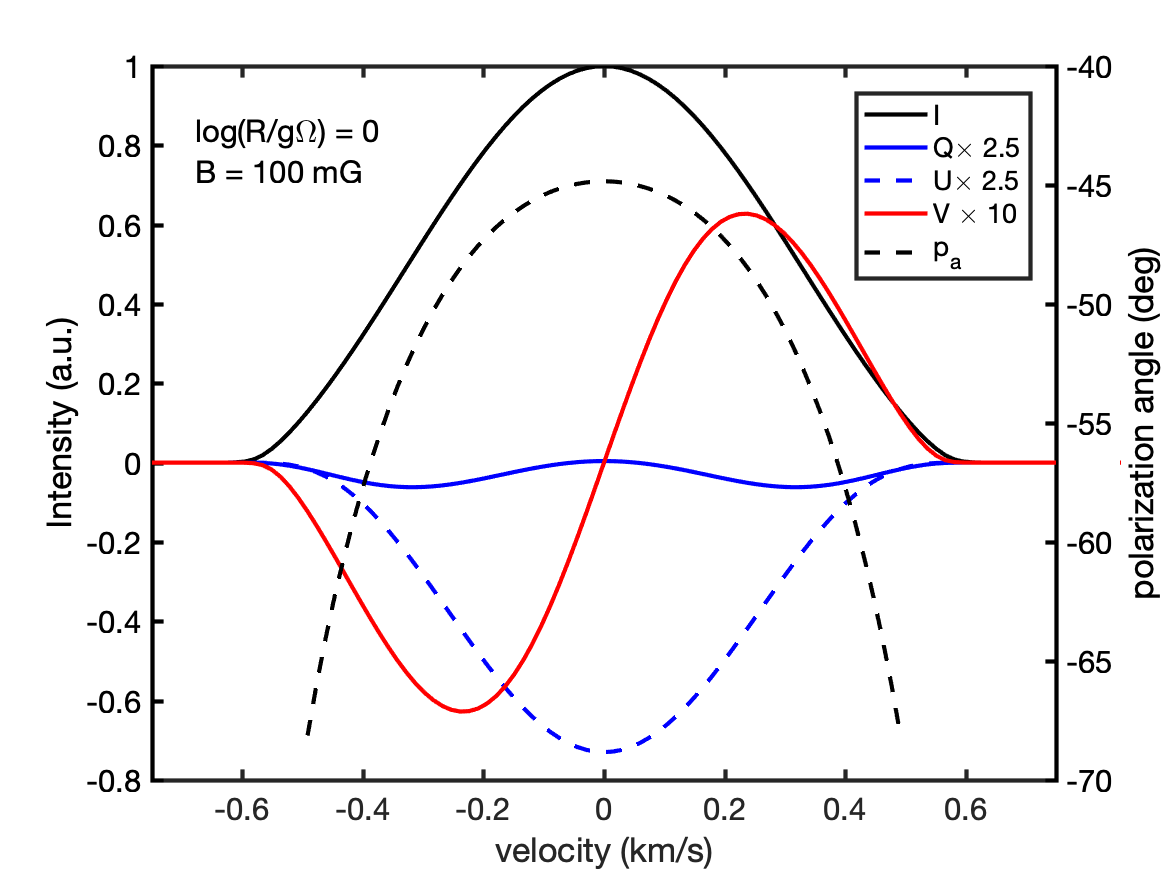}
        \caption{}
    \end{subfigure}
    ~
    \begin{subfigure}[b]{0.45\textwidth}
        \includegraphics[width=\textwidth]{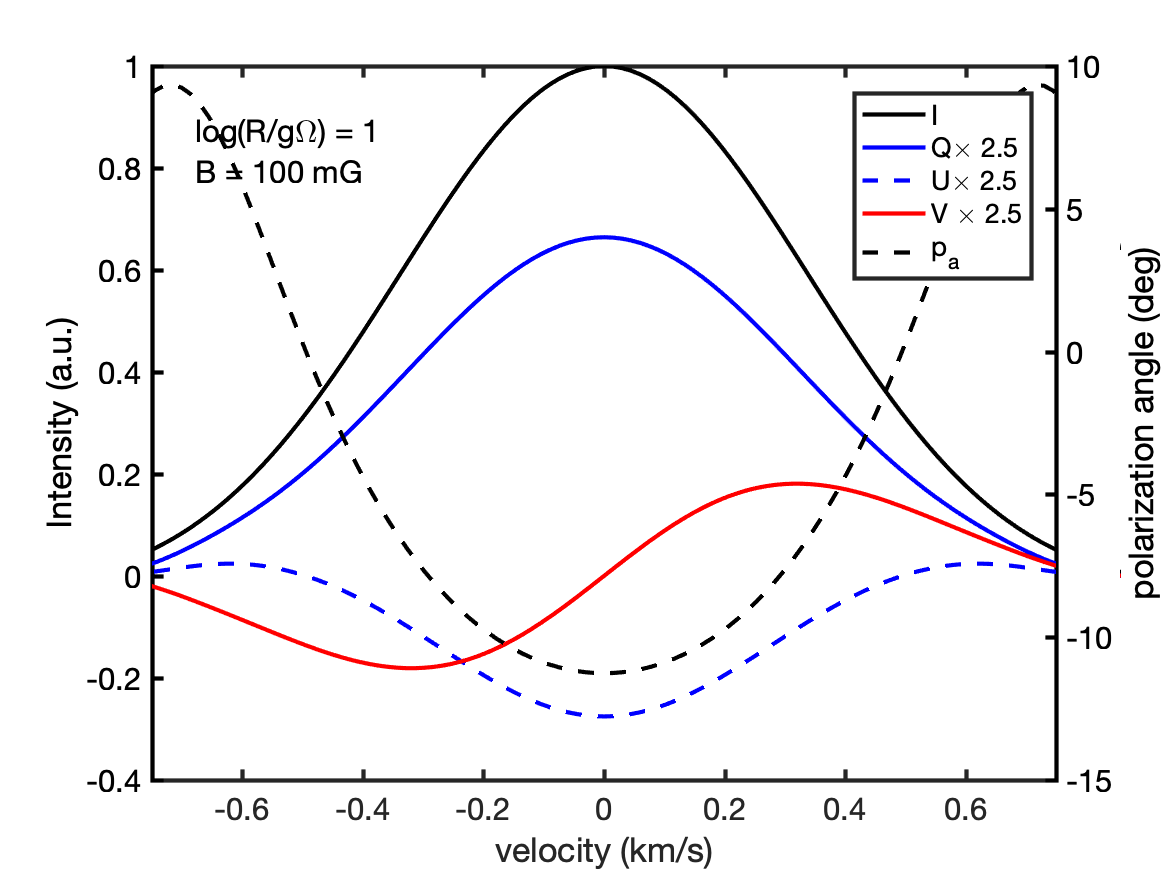}
        \caption{}
    \end{subfigure}
    \caption{SiO $J=1-0$ maser spectra for different levels of saturation. We plot all Stokes parameters (left $y$-axis) as well as the polarization angle (right $y$-axis). The polarization angle is defined with respect to the magnetic field direction projected on the plane of the sky. Simulations were carried out at $B=100$ mG, with a magnetic field propagation angle of $\theta = 45^o$.}
    \label{fig:sio_spectra}
\end{figure}

We also investigate the spectral properties of the SiO maser polarization. In Fig.~\ref{fig:sio_spectra}, we report three spectra of $J=1-0$, $B=100$ mG, isotropically pumped SiO masers at log$(R/g\Omega)=-1,0$ and $1$. In the figure, all Stokes parameters are plotted, as well as the polarization angle across the spectrum. We note that the spectrum is broadening with $R$. Because we have already passed the saturation level at log$(R/g\Omega)=-1$. With the broadening, though, the Stokes-$V$ fraction does not decrease as would be expected from an LTE analysis. The linear polarization follows roughly the same spectral form as the Stokes-$I$ spectrum and the polarization angle can change with up to $\sim 30^o$ across the spectrum. We note also the perfect anti-symmetrical nature of the Stokes-$V$ spectrum, as is expected from an LTE analysis, which is retained for all $R$.

%

\subsubsection{Polarized incident radiation}
\begin{figure*}
    \centering
    \begin{subfigure}[b]{0.45\textwidth}
        \includegraphics[width=\textwidth]{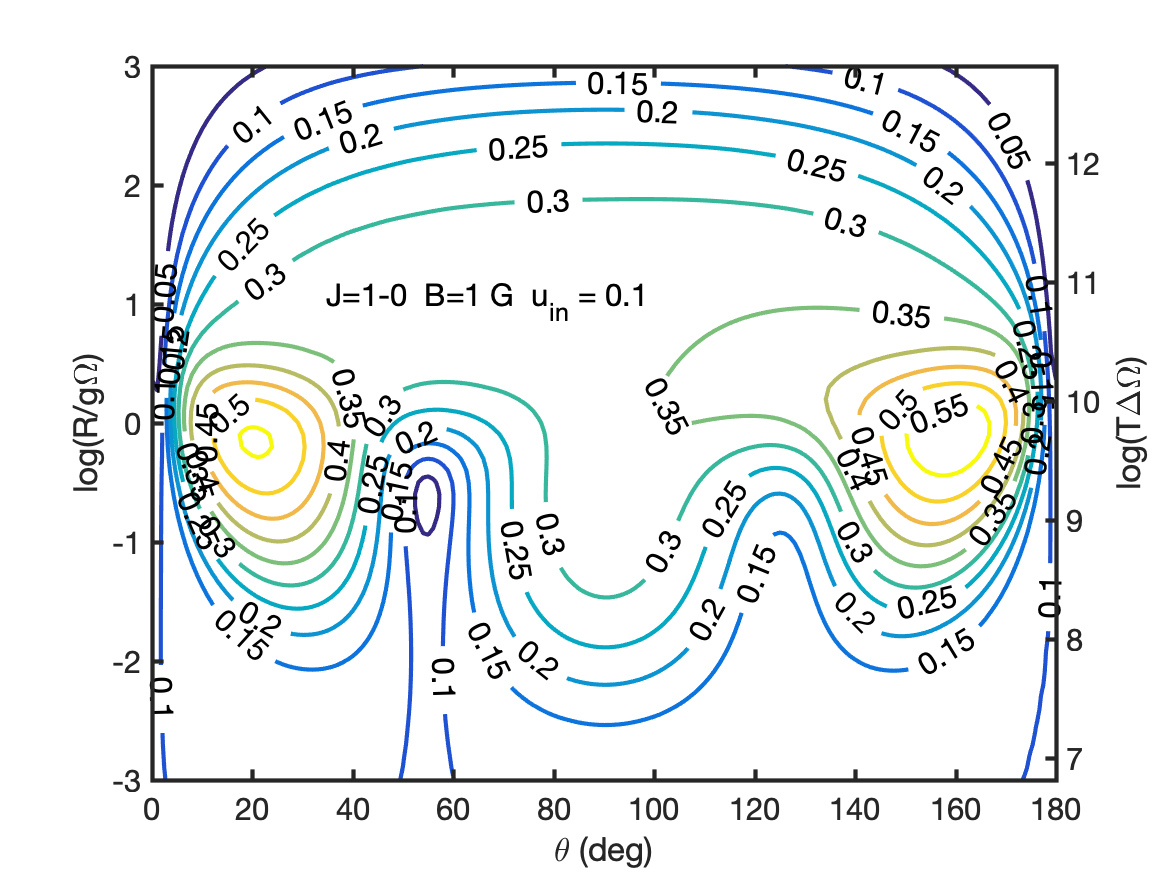}
        \caption{}
    \end{subfigure}
    ~
    \begin{subfigure}[b]{0.45\textwidth}
        \includegraphics[width=\textwidth]{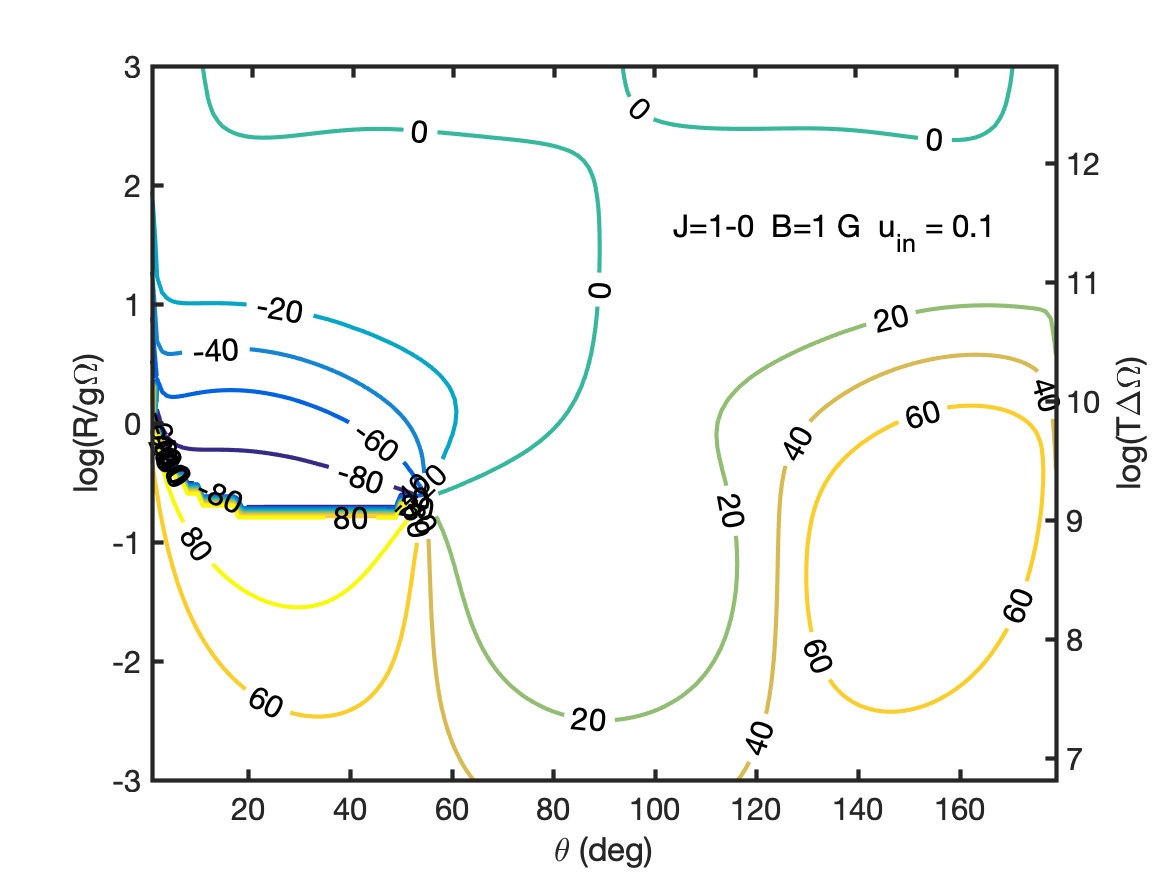}
        \caption{}
    \end{subfigure}
    ~
    \begin{subfigure}[b]{0.45\textwidth}
        \includegraphics[width=\textwidth]{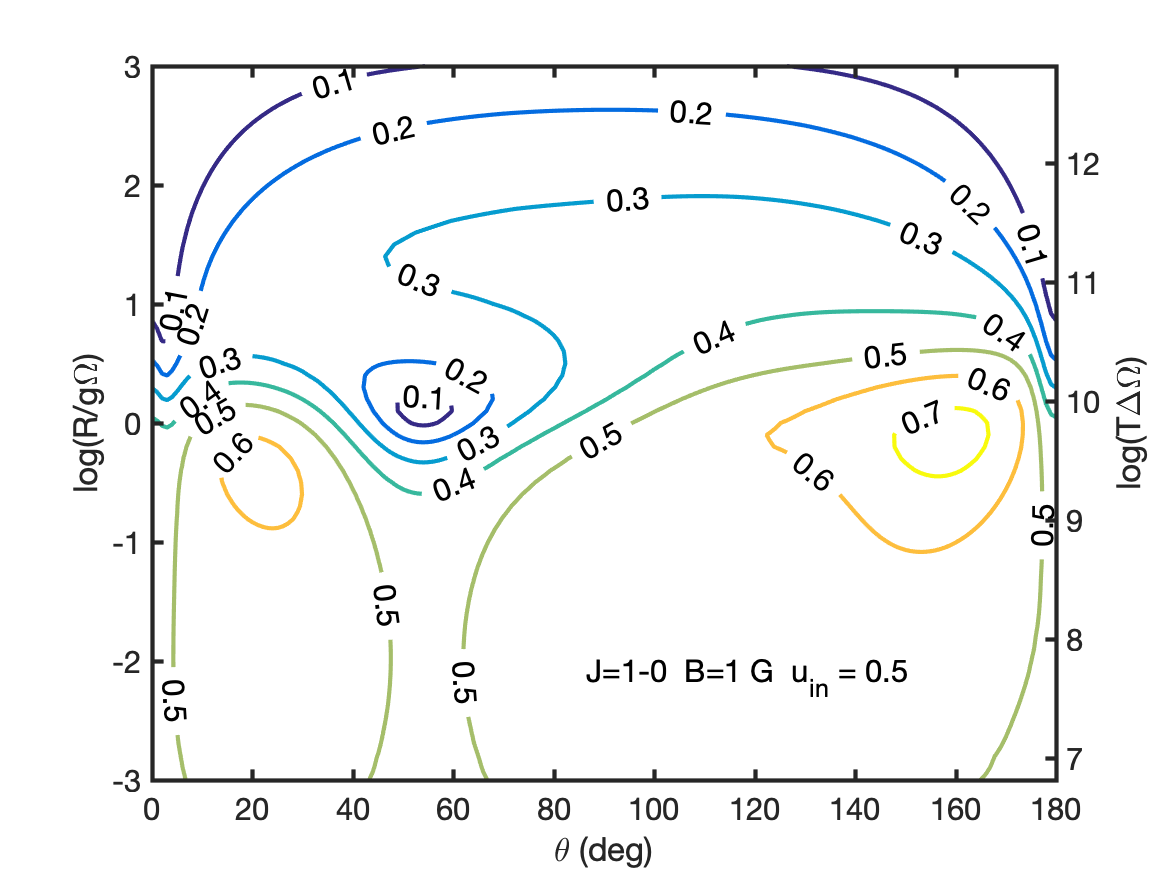}
        \caption{}
    \end{subfigure}
    ~
    \begin{subfigure}[b]{0.45\textwidth}
        \includegraphics[width=\textwidth]{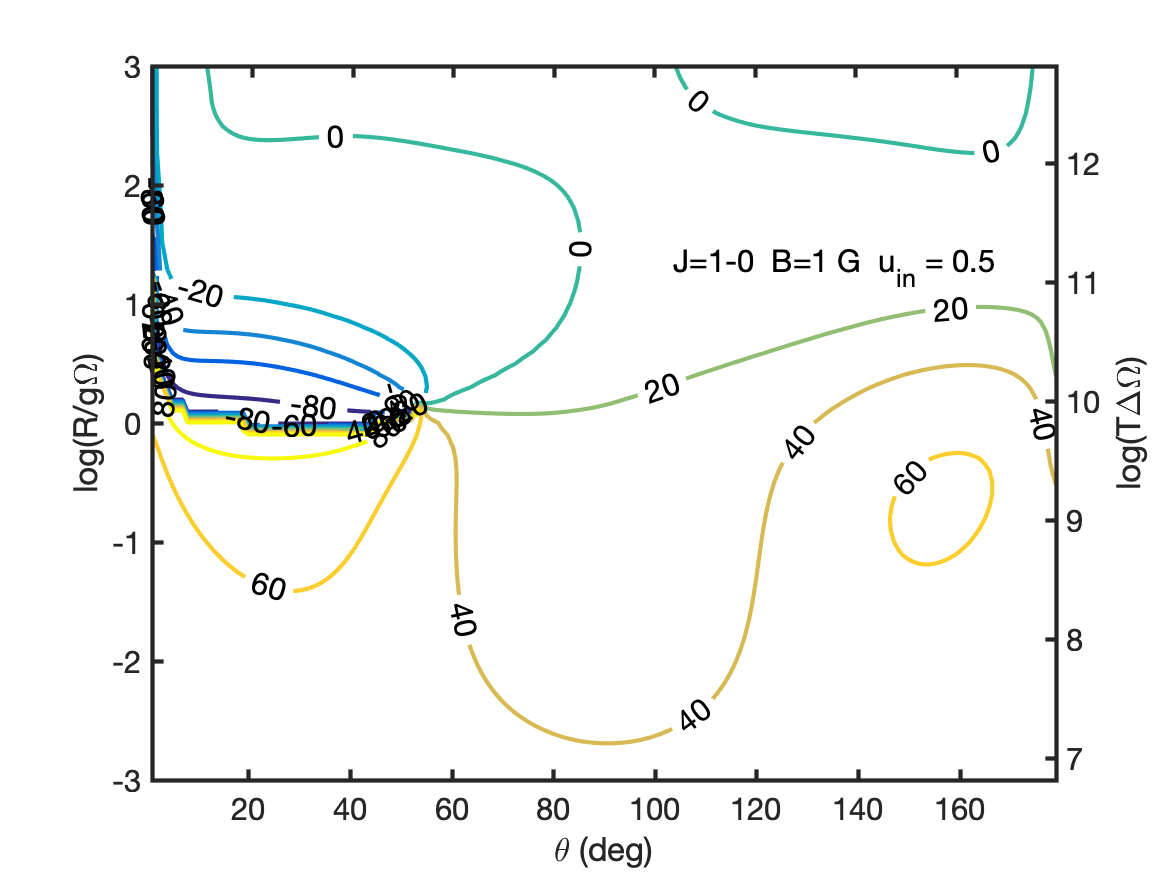}
        \caption{}
    \end{subfigure}
    \caption{Contour plots of the linear polarization fraction and angle of an SiO maser as a function of the propagation angle $\theta$ and the rate of stimulated emission. Maser simulations performed with incident polarized radiation of (a,b) $U/I=0.1$ and (c,d) $U/I=0.5$. Magnetic field strength and transition angular momentum are denoted inside the figure. For simulations with $J_{\mathrm{up}}>1$ and other magnetic field strengths, see Figs.~A.4-9 in the Appendix.}
    \label{fig:contour_lin_in}
\end{figure*}

\begin{figure}
    \centering
    \begin{subfigure}[b]{0.45\textwidth}
        \includegraphics[width=\textwidth]{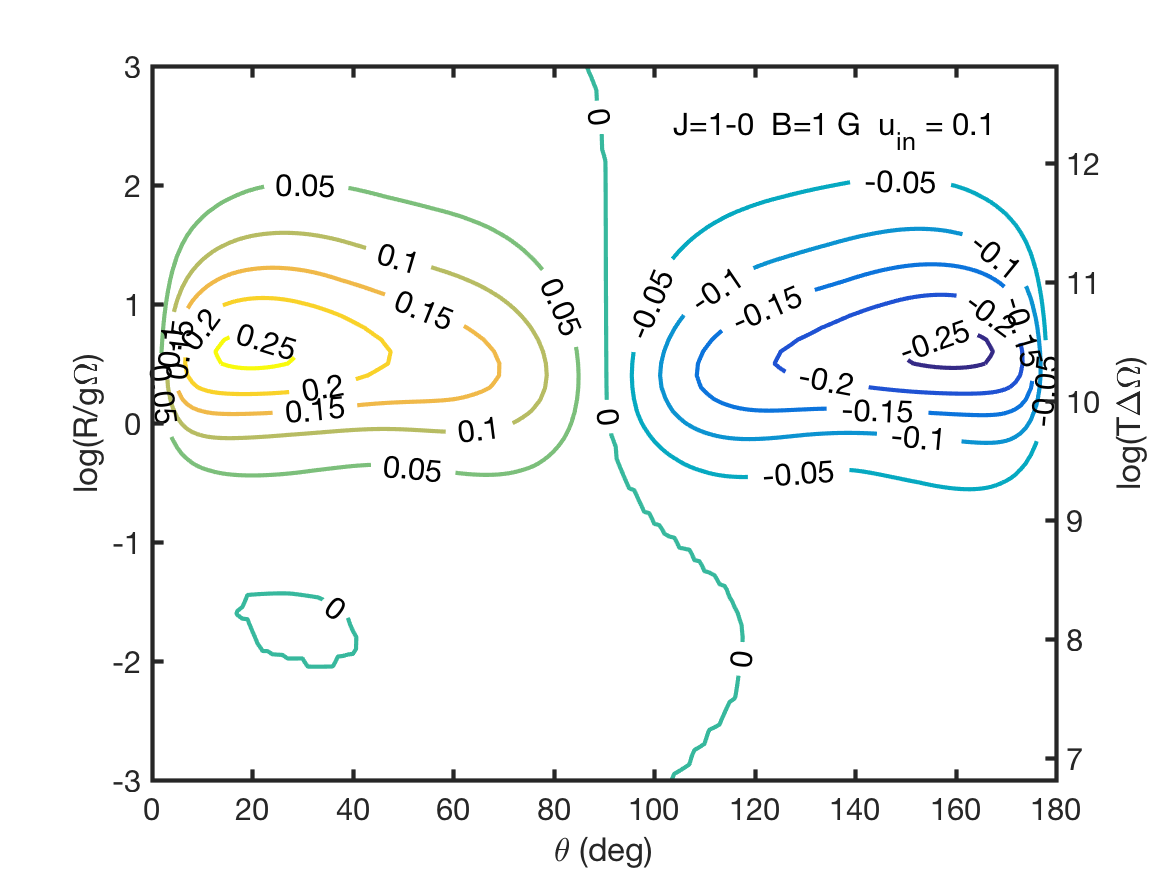}
        \caption{}
    \end{subfigure}
    ~
    \begin{subfigure}[b]{0.45\textwidth}
        \includegraphics[width=\textwidth]{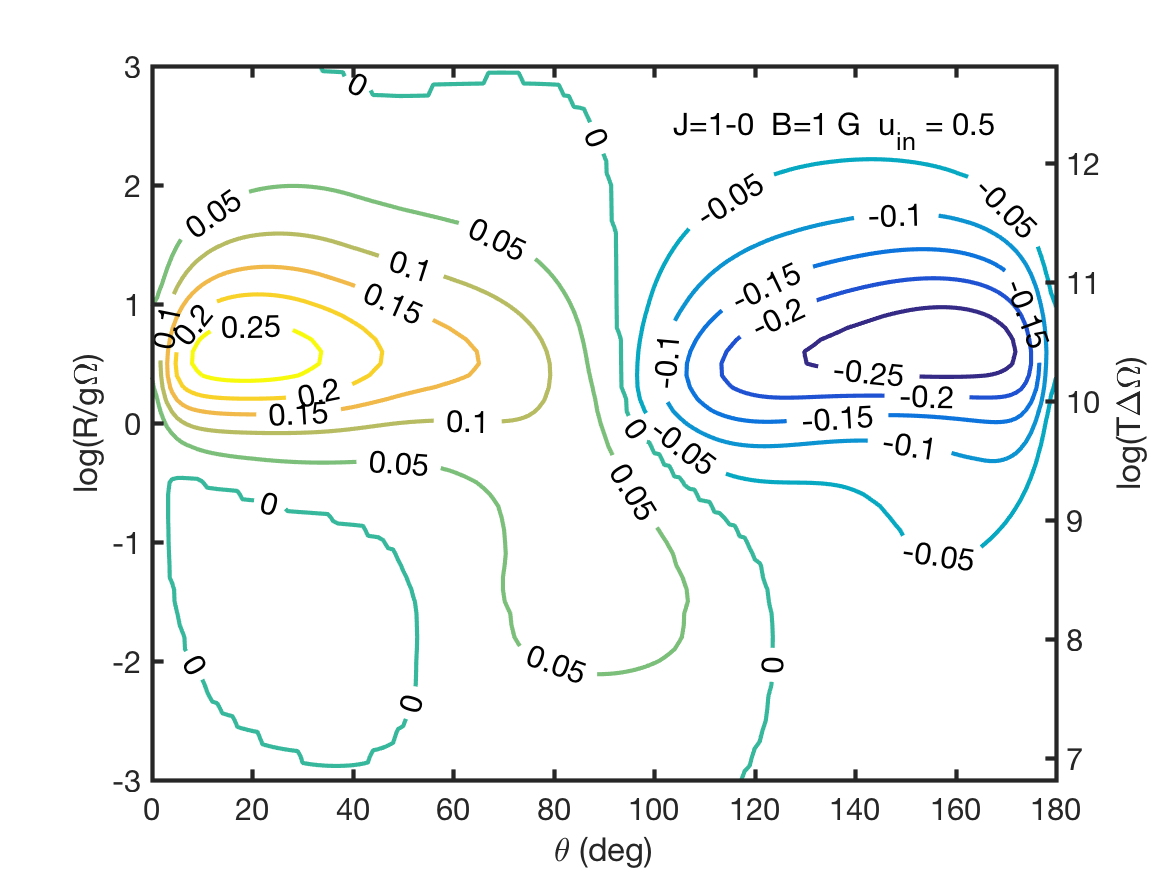}
        \caption{}
    \end{subfigure}
    \caption{Contour plots of the circular polarization fraction of the SiO maser as a function of the propagation angle $\theta$ and the rate of stimulated emission. Maser simulations performed with incident polarized radiation of (a) $U/I=0.1$ and (b) $U/I=0.5$. Magnetic field strength and transition angular momentum are denoted inside the figure.}
    \label{fig:contour_circ_in}
\end{figure}

Simulations of the polarization of a $J=1-0$ SiO maser at $B=1$ G with partially polarized seed radiation are reported in 
Figs.~\ref{fig:contour_lin_in} and \ref{fig:contour_circ_in}. Simulations of higher angular momentum and at different magnetic fields are given in Figs.~A.4-9. In analyzing these types of masers, we should make the 
distinction between the regime of weak maser emission, where the rate of stimulated emission 
is significantly weaker than the magnetic field ($\mathrm{log}(R/g\Omega)<-2$), and the regime of 
strong maser emission, where the two 
quantities are comparable in size. In the weak maser-regime, the incident polarized radiation 
is simply amplified and the fractional polarization from the incident radiation is retained, 
along with the polarization angle of the incident radiation. In the strong maser-regime, we notice
distinct differences in the polarization landscapes, between the strongly ($U/I=0.5$) and the weakly 
($U/I=0.1$) polarized incident radiation. The linear and circular polarization landscapes of 
the weakly polarized incident radiation, above $\mathrm{log}(R/g\Omega)>0$ look very similar to the 
landscapes generated from isotropic seed radiation. In contrast, the linear polarization landscape of
the strongly polarized incident seed radiation looks completely different, and only converges to the
landscape of isotropic seed radiation for $\mathrm{log}(R/g\Omega)>2$. Interestingly, the effects 
on the circular polarization landscapes are rather small, even for the strongly polarized incident 
seed radiation. Although the effects are small, we observe an increase in circular polarization 
fraction with the polarized incident seed radiation.   

Around the magic angle, $\theta_m$, the incident polarization fraction is retained for the highest
$R$. The strongest linear polarization fraction is found around $\theta = 20^o$, and where 
$R\sim g\Omega$, just as we have seen for isotropic seed radiation. Although we should note that 
the maximum linear polarization fraction occurs for somewhat lower $R$, which is an effect most 
pronounced at the strongly polarized seed radiation. We should also note that the symmetry around 
$\theta = 90^o$ that characterizes the simulations with isotropic seed radiation, is not retained 
by these simulations. The preferred direction of the incident radiation breaks the symmetry. 
This is perhaps most strongly reflected in the polarization angle maps. 
Here, a feature is seen in the maps for both strong and weakly polarized incident radiation, at 
the magic angle, $\theta=\theta_m$, and around $g\Omega \sim R$ 
where a range of different angles come together. Additionally, for $\theta < \theta_m$, a large 
and sharp polarization angle change is seen around $\mathrm{log}(R/g\Omega) \sim -1$. Further 
inspection of these fluctuations in the polarization angle reveal that in this region, the initially
positive Stokes-$U$ element of the radiation drops and changes sign. For $\theta < \theta_m$, the 
Stokes-$Q$ coefficient initially builds up as negative, but will turn positive after $\mathrm{log}(R/g\Omega) \sim 0$.
For $\theta > \theta_m$, the Stokes-$Q$ coefficient will not become negative. For angles $\theta > 90^o$, 
the Stokes-$U$ element of the radiation will retain its positive sign throughout the propagation.  

At different magnetic field strengths, similar general features are observed that were also pointed 
out in the isotropic seed-radiation simulations. For instance, we observe that the magnetic field 
strength is 
correlated to the area ($\theta$ vs.~$R$) of significant polarization. An interesting feature, is that the lower
magnetic field-strength simulations seem to be more affected by the incoming radiation than the stronger
magnetic field-strength simulations that retain more of the general structure also observed for the 
isotropic seed radiation. Just as for the isotropic seed-radiation masers, the higher angular momentum 
transitions are significantly less polarized. However, for the higher angular momentum
contours, the general structure of polarization contours is strongly influenced by the incoming polarized
radiation. The simulations with strongly polarized incoming radiation, have nearly no general dependence 
on $\theta$, as the incoming (linear) polarization fraction smoothly deteriorates from $\mathrm{log}(R/g\Omega)>0$, 
to nullify around $\mathrm{log}(R/g\Omega)\sim 3$. These effects are also reflected in the landscape of 
circular polarization, which is affected for the highly polarized incoming radiation. Although the effects are not as pronounced as in the linear polarization contours, and do not cause high fractions of circular 
polarization.   

\subsubsection{Anisotropic pumping} 
\begin{figure}
    \centering
    \begin{subfigure}[b]{0.45\textwidth}
        \includegraphics[width=\textwidth]{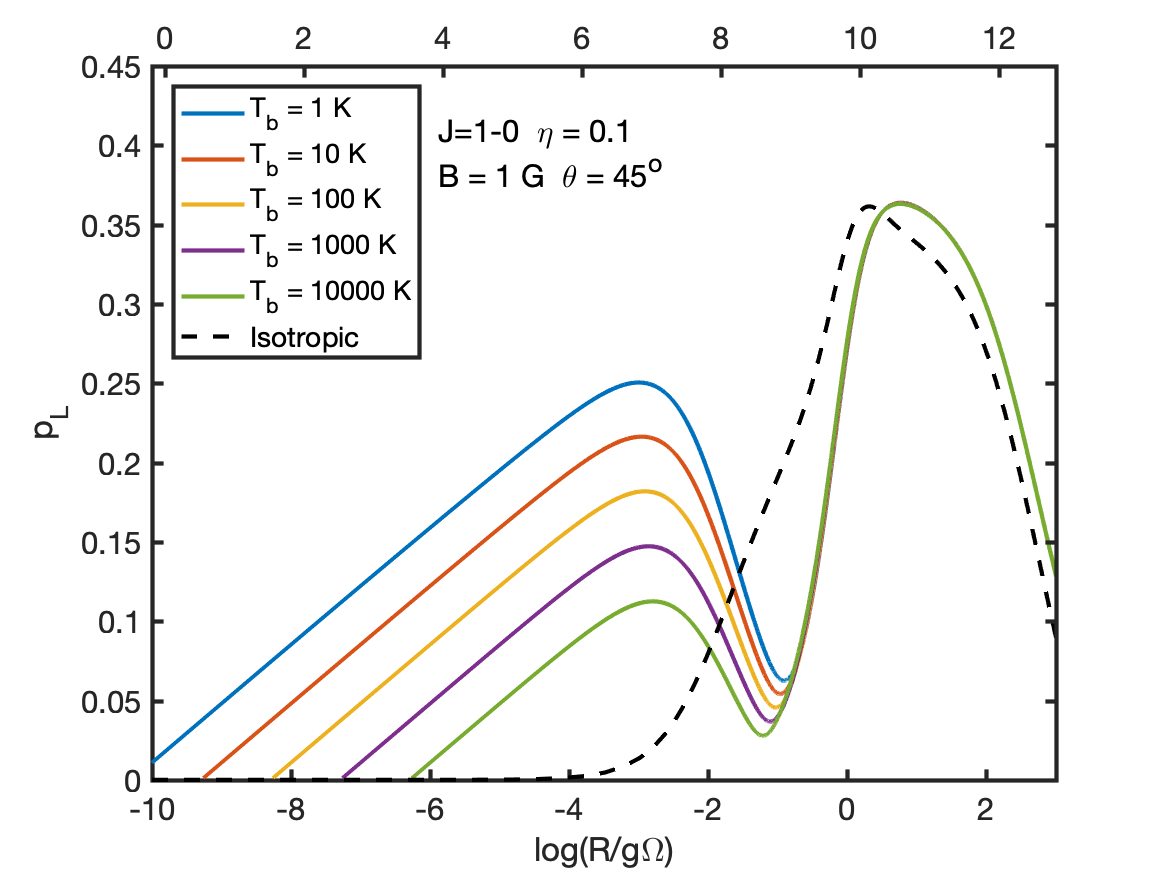}
        \caption{}
    \end{subfigure}
    ~
    \begin{subfigure}[b]{0.45\textwidth}
        \includegraphics[width=\textwidth]{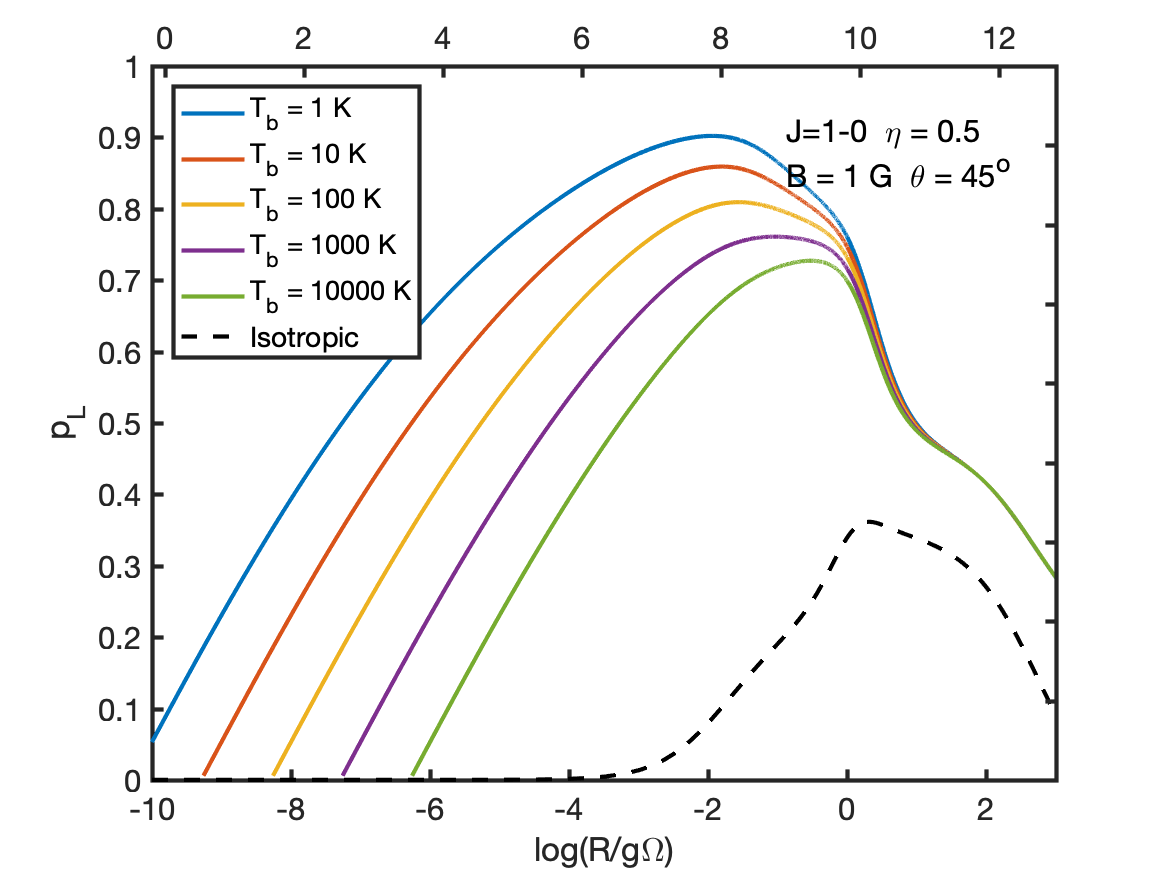}
        \caption{}
    \end{subfigure}
    \caption{Plots of the linear polarization fraction of anisotropically pumped SiO masers as a function of the rate of stimulated emission. Maser simulations performed with anisotropy-parameters of (a) $\eta=0.1$ and (b) $\eta=0.5$. Simulations are performed for a range of incident radiation-strength, as indicated inside the figure. Magnetic field strength, propagation angle and transition angular momentum are denoted inside the figure.}
    \label{fig:plot_tvar}
\end{figure}

When anisotropic pumping is in play, one should make the distinction between strong masers, where the radiation significantly influences the direction of the molecule ($R\gtrsim g\Omega$), and weak masers, where this is not the case. Weak masers propagating through an anisotropically pumped medium, will accrue polarization monotonically. The polarization will rise until the point where the radiative interaction becomes stronger than the degree of anisotropic pumping. After this point, the polarization degree will drop, and the standard magnetic field-polarization mechanism will take over as the main source of polarization. Fig.~\ref{fig:plot_tvar} shows the polarization of anisotropically pumped SiO masers with varying intensity of seed radiation as a function of the rate of stimulated emission. 

The polarization of weak masers is independent of the magnetic field strength, but will be highly dependent on the intensity of the seed radiation, as well as the anisotropy of the pumping, $\eta$. Strong masers have as their main polarization mechanism the magnetic-field interaction, but are still minorly influenced by the anisotropic pumping, especially in the transitory period between the weak and strong maser. The polarization of the strongest masers is independent of the intensity of the incoming radiation. 

\begin{figure*}
    \centering
    \begin{subfigure}[b]{0.45\textwidth}
        \includegraphics[width=\textwidth]{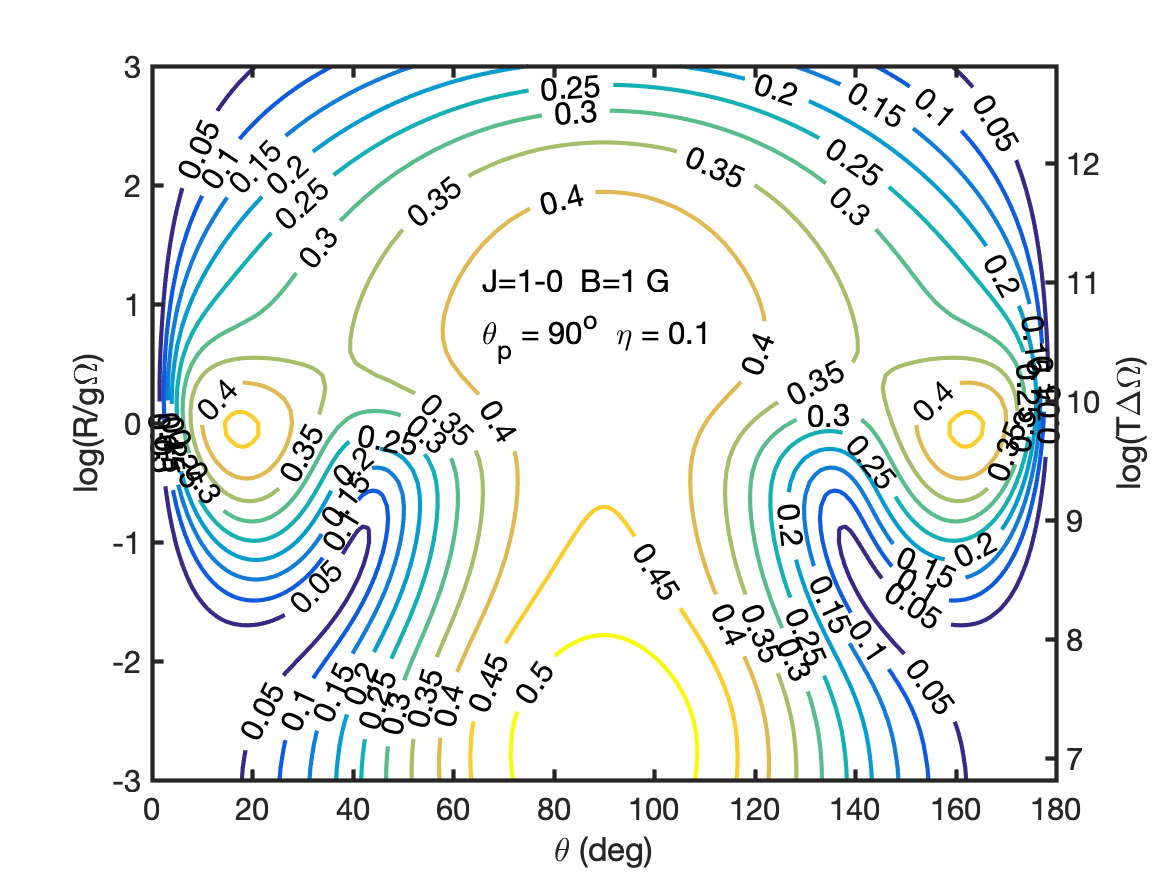}
        \caption{}
    \end{subfigure}
    ~
    \begin{subfigure}[b]{0.45\textwidth}
        \includegraphics[width=\textwidth]{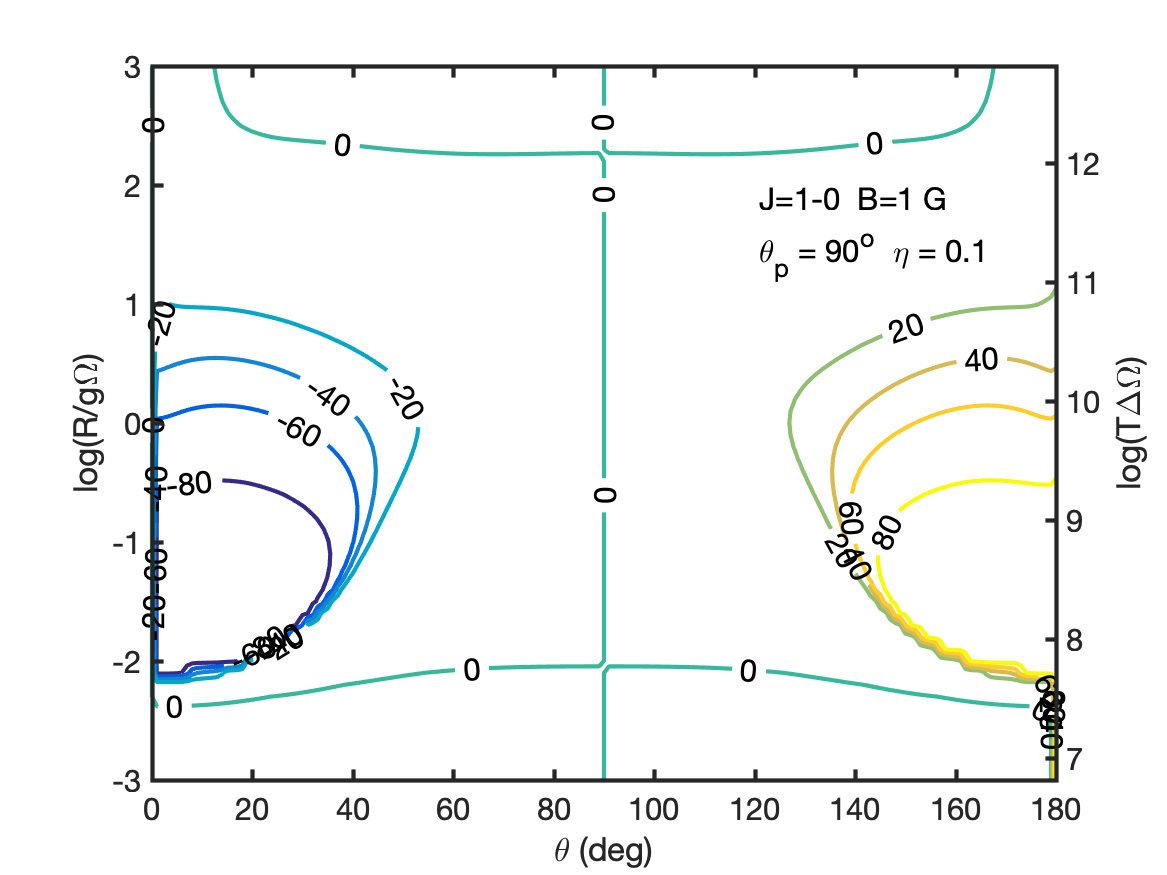}
        \caption{}
    \end{subfigure}
    ~
    \begin{subfigure}[b]{0.45\textwidth}
        \includegraphics[width=\textwidth]{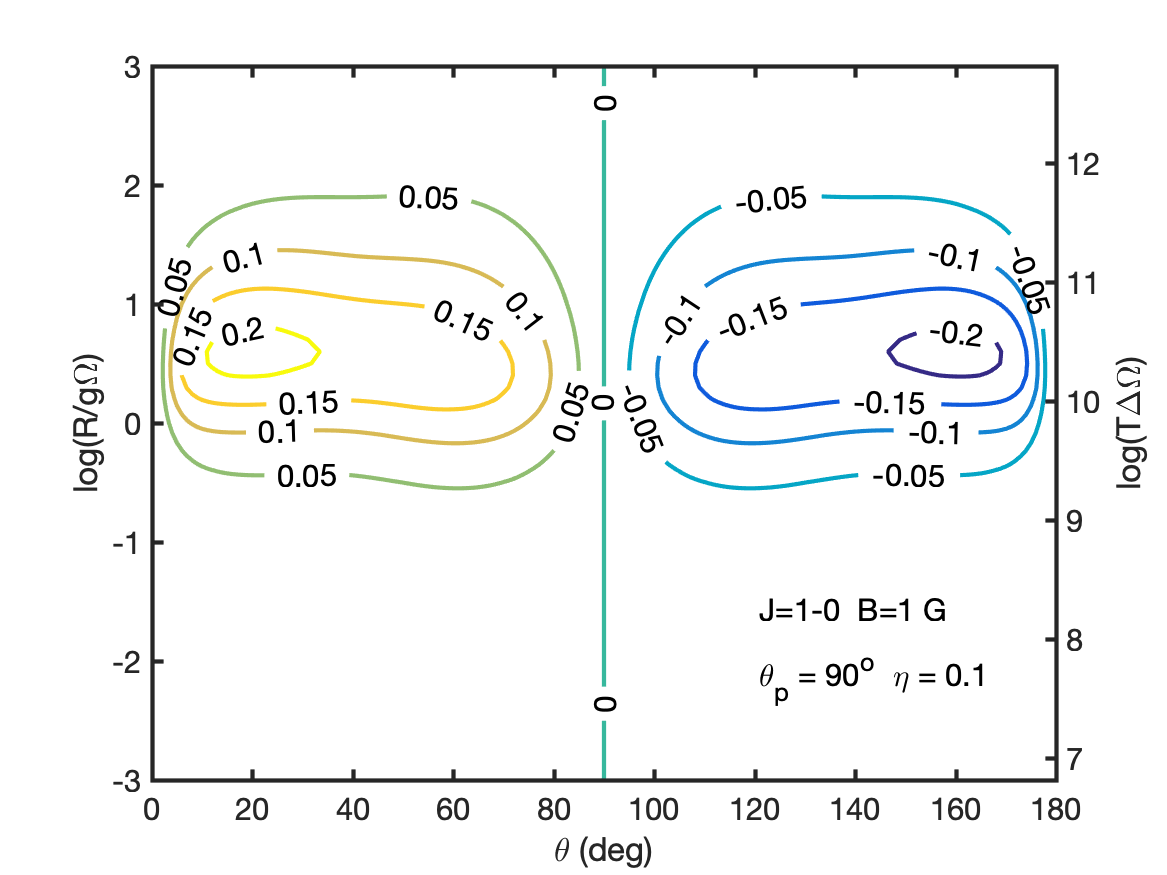}
        \caption{}
    \end{subfigure}
  \caption{Contour plot of the (a) linear polarization fraction, (b) polarization angle and (c) circular polarization fraction of anisotropically pumped SiO masers as a function of the rate of stimulated emission and the angle $\theta$. Maser simulations were performed with anisotropy-parameters of $\eta=0.1$ and at an anisotropy angle perpendicular to the magnetic field and propagation direction. Simulations are performed for an incident radiation-strength of $T_b = 0.1$ K, but we highlight here only a small part of the region that is sensitive to this parameter. Magnetic field strength and transition angular momentum are denoted inside the figure. For simulations with $J_{\mathrm{up}}>1$ and other magnetic field strengths, see Figs.~A.10-18 in the Appendix.}
    \label{fig:contour_anis}
\end{figure*}

The polarization landscape of an anisotropically pumped SiO maser at $B = 1$ G is plotted in Fig.~\ref{fig:contour_anis}. Simulations of higher angular momentum and at different magnetic fields are given in Figs.~A.10-18. If we examine the weak-maser region, we notice directly a strong decline of polarization for $\theta \to 0$. For higher rates of stimulated emission, at $\mathrm{log}(R/g\Omega)>1$, we notice that the polarization is largely similar to polarization generated by an isotropically pumped maser (Fig.~\ref{fig:contour_iso}), although we observe additional polarization in the regions around $\theta = 90^o$ and $R \sim g\Omega$. Also, we actually observe a decrease in polarization in the region around $\theta=20^o$ and $R \sim g\Omega$ with respect to the isotropically pumped maser. However, if the anisotropy-parameter is increased the resemblance to the isotropically pumped maser will vanish rapidly and arbitrarily high polarization can be achieved. 

We observe that for increasing the angular momentum of the transition, the same anisotropy parameter, $\eta$, will yield a weaker polarization build-up in the weak maser-regime. Still, though, large fractional linear polarization can be achieved for the higher angular momentum transitions as a result of the anisotropic pumping. A sufficiently large anisotropy-parameter can yield polarization as high as $100\%$.

The orientation of the anisotropy in Fig.~\ref{fig:contour_anis}, is perpendicular to both the magnetic field direction, $\vec{b}$, and propagation direction, $\vec{s}$. In this orientation, the polarization maps are symmetric so that $p_L (\theta) = p_L(180^o - \theta )$, $p_a (\theta) = -p_a(180^o - \theta )$ and $p_V (\theta) = -p_V(180^o - \theta )$. This symmetry will however be broken when the direction of the anisotropy orients itself in the plane that $\vec{b}$ spans with $\vec{s}$ (see Appendix).

We direct our attention to the circular polarization of the anisotropically pumped SiO maser. We can discern some influence of the anisotropic pumping on the circular polarization, but the structure is mostly similar to the one obtained from isotropic pumping, and the enhancement of polarization is not as strong as it was for the linear polarization analogues. Comparing two orientations of the anisotropy directions, $\vec{a}_1 \perp \vec{b},\vec{s}$ and $\vec{a}_2 \parallel \vec{b}$, we find that the anisotropic pumping in the $\vec{a}_1$ direction actually lowers the circular polarization, while pumping in the $\vec{a}_2$ direction enhances it. In the weak-maser regime, there is no large circular polarization fraction, nor does the fraction depend on the brightness of the seed radiation. 
 
\subsection{H$_2$O masers}
\subsubsection{Isotropic pumping}
\begin{figure*}
    \centering
    \begin{subfigure}[b]{0.45\textwidth}
	\includegraphics[width=\textwidth]{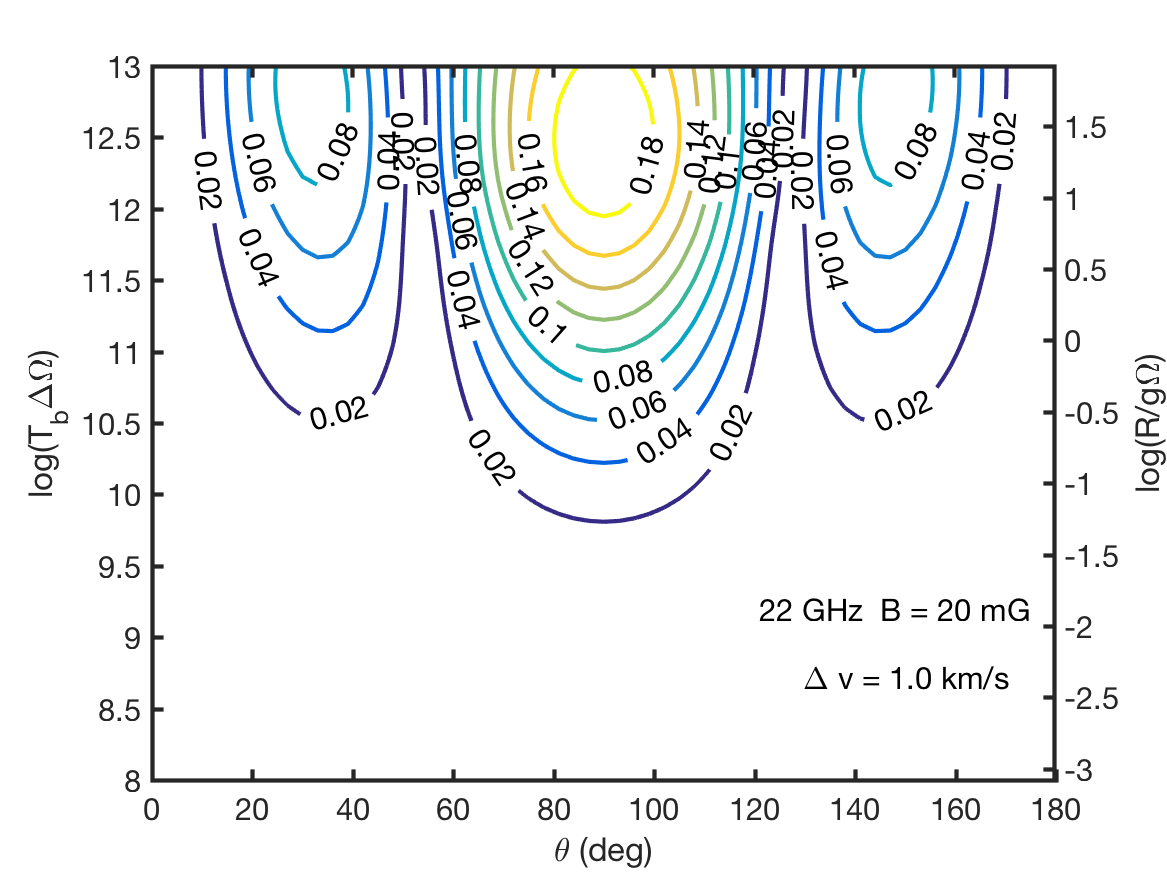} 
	\caption{} 
    \end{subfigure}
    ~ 
    \begin{subfigure}[b]{0.45\textwidth}
	\includegraphics[width=\textwidth]{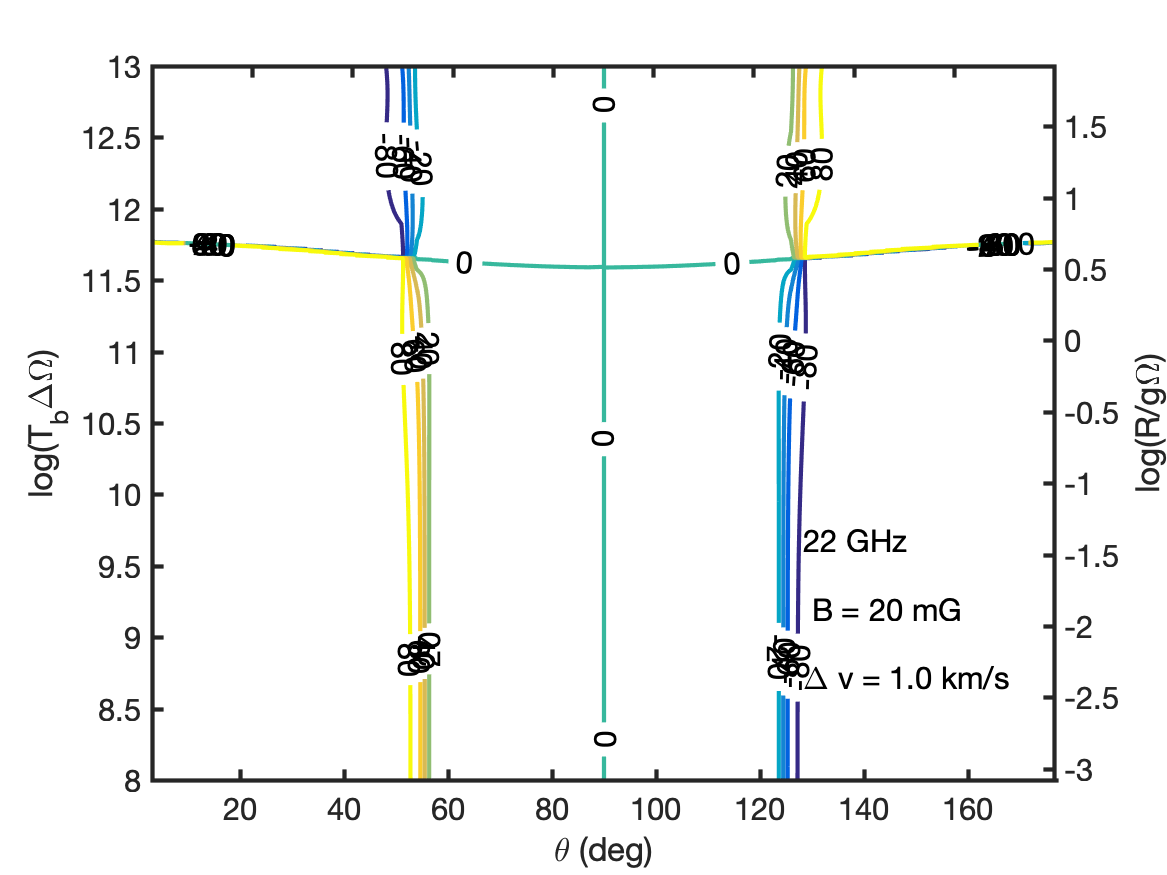} 
	\caption{}
    \end{subfigure}
    ~ 
    \begin{subfigure}[b]{0.45\textwidth}
	\includegraphics[width=\textwidth]{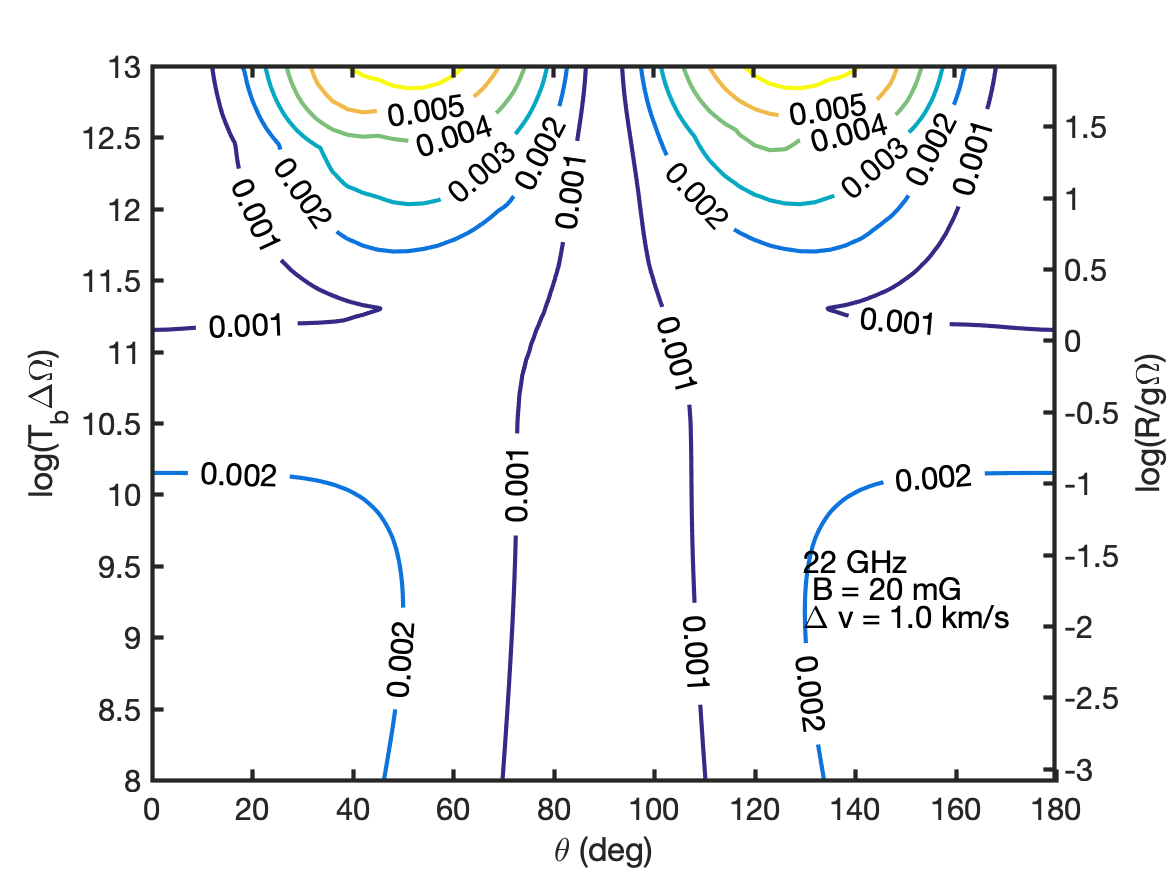} 
	\caption{}
    \end{subfigure}
    \caption{Contour plots of the linear polarization fraction (a), angle (b) and circular polarization fraction (c) of water masers as a function of the propagation angle $\theta$ and the rate of stimulated emission. Magnetic field strength and thermal width denoted inside the figure. For simulations of other magnetic field strengths and thermal widths, see Figs.~A.19-20 in the Appendix.}
    \label{fig:contour_water_iso}
\end{figure*}

We examined the regime of magnetic fields from $B=20$ mG to $B=100$ mG, at $v_{th}=0.6$ km/s 
($T=260$ K) to $v_{th} = 3.0$ km/s ($T=6500$ K). We summarize the results of these simulations 
in Fig.~\ref{fig:contour_water_iso}, and 
further results can be found in Figs.~A.19-20. 
The linear polarization fraction for these 
water masers is only appreciable from about $T_b\Delta\Omega=10^{10}$ Ksr, or 
$\mathrm{log}(R/g\Omega)>-1.5$, where the strongest masers display the strongest polarization. 
The magnetic field interaction term is not strong enough to facilitate the large overshoot in 
polarization around $\theta =20^o$ that we have seen earlier. Rather, the maximum linear polarization 
is found around $\theta \to 90^o$. In the range of $B=20$ mG to $B=100$ mG, the linear 
polarization of the water masers does not change significantly, although there is a slight 
general increase in linear polarization fraction. For simulations at higher thermal widths, 
$v_{th} > 1$ km/s, there is no significant effect on the linear polarization fraction. 
For $v_{th} < 1$ km/s, we observe minor effects, as lines are not completely blended 
anymore. For these simulations, polarization will start at higher maser intensity, but will soon 
converge to the landscape of the other 
$v_{th}$ solutions, as broadening of the maser blends the individual lines. 
Analysis of the polarization angle maps reveal no significant difference between different 
magnetic field strengths, as well as different thermal widths. The most striking feature of
the polarization angle maps are the sharp $90^o$-flips, associated with crossing the 
magic angle that are general for any $T_b \Delta \Omega$. We observe another sharp 
angle-flip, around $\mathrm{log}(R/g\Omega)\sim 0.75$ for $\theta < \theta_m$, but this 
concerns a $180^o$-flip.

The circular polarization maps present a rather complicated landscape of circular polarization, 
never quite reaching high degrees of circular polarization. Weaker masers with 
$T_b\Delta\Omega \ll 10^{11}$ Ksr, follow roughly the LTE estimate of the circular polarization  
$p_V \propto 2A_{FF'} B_{\mathrm{Gauss}}\cos \theta / \Delta v_F (\mathrm{km/s})$ \citep{fiebig:89}. 
Indeed, for these masers we observe the 
strongest circular polarization for $\theta \to 0^o$ and low $v_{th}$, which gradually 
diminishes for higher $v_{th}$ and angles $\theta \to 90^o$. When $T_b\Delta\Omega > 10^{9}$ Ksr,
the simulation results for circular polarization depart from the LTE estimates. For the strongest 
masers, around $T_b\Delta\Omega \sim 10^{13}$ Ksr, we find (for $B=20$ mG) highest circular polarization, 
that can get up to $0.55 \%$ around $\theta \sim 60^o$. Circular polarization in this region has only a 
minor dependence on the magnetic field strength and maser thermal width. 

We have already touched upon the complicating multi-transitional nature of the water-maser. It 
is very well possible that asymmetries occur in the pumping of the different hyperfine 
transitions \citep[see][]{walker:84,lankhaar:18}. To further investigate this, we plot for a number of preferred 
hyperfine pumping ratios $\lambda = \lambda_{F=7-6}/\lambda_{\mathrm{other}}$, the fractional 
circular and linear polarizations of a water maser at $\theta = 45^o$ as a function of the maser luminosity. 
The $F=7-6$ transition is the strongest hyperfine transition and, incidentally, 
also the transition with the highest Zeeman-coefficient. 
From Fig.~\ref{fig:plot_water_varpump_iso}, quite surprisingly, 
we observe a negative correlation between the generated linear polarization and the favoring 
of the $F=7-6$ transition. However, the circular polarization does increase as a result of 
the preferred pumping of the $F=7-6$ transition. Another interesting 
feature that was not apparent from the contour maps, are the discontinuities in both the linear 
and circular polarization fractions. Discontinuities in these functions arise because of the 
complex nature of the multi-transitional lines---and indeed do not occur for the most preferably 
pumped masers. 

\begin{figure}
    \centering
    \begin{subfigure}[b]{0.45\textwidth}
        \includegraphics[width=\textwidth]{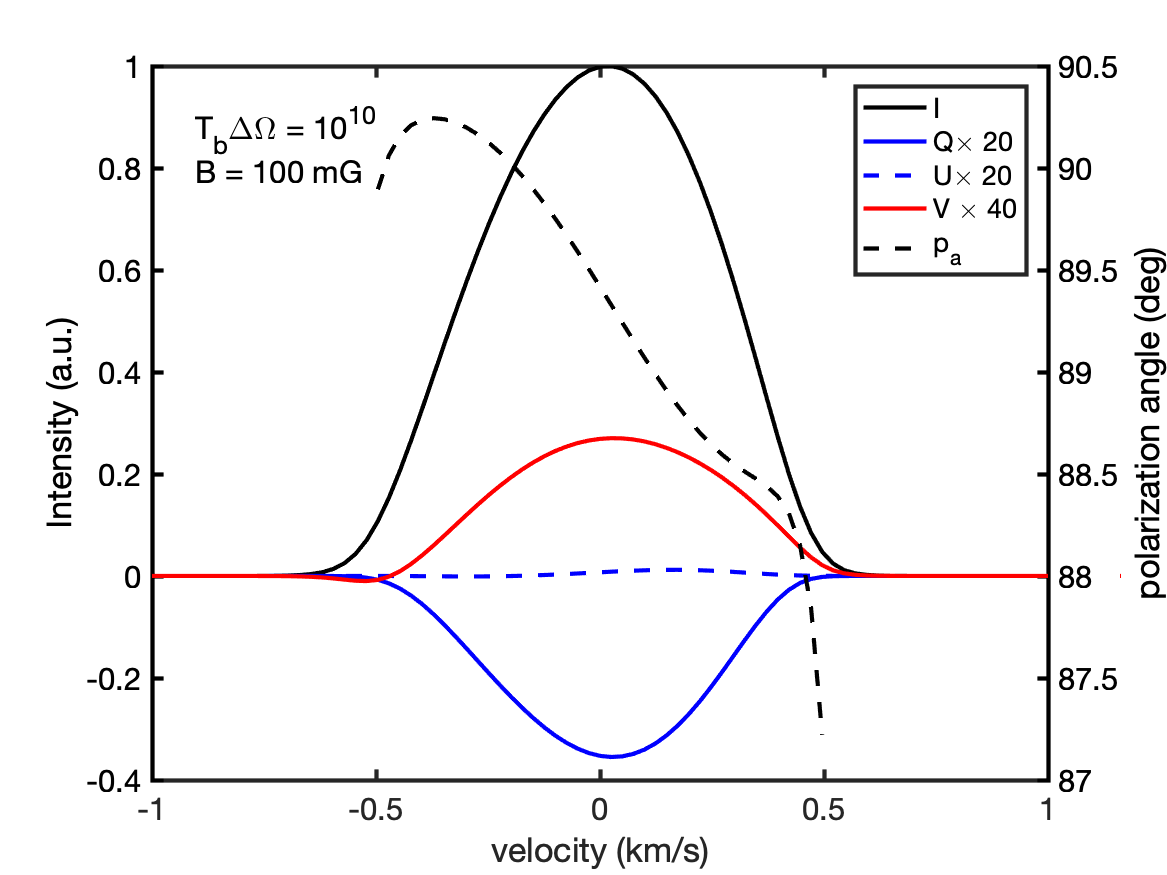}
        \caption{}
    \end{subfigure}
    ~
    \begin{subfigure}[b]{0.45\textwidth}
        \includegraphics[width=\textwidth]{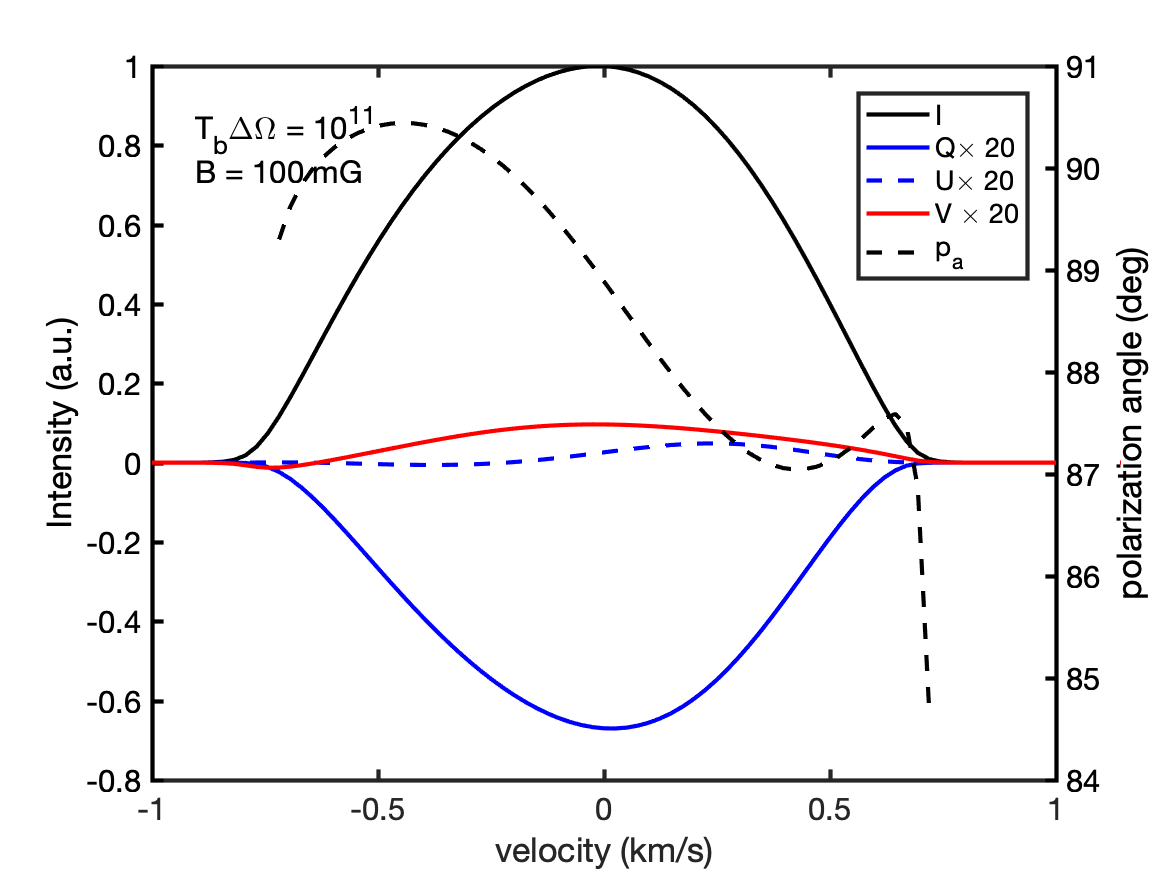}
        \caption{}
    \end{subfigure}
    ~
    \begin{subfigure}[b]{0.45\textwidth}
        \includegraphics[width=\textwidth]{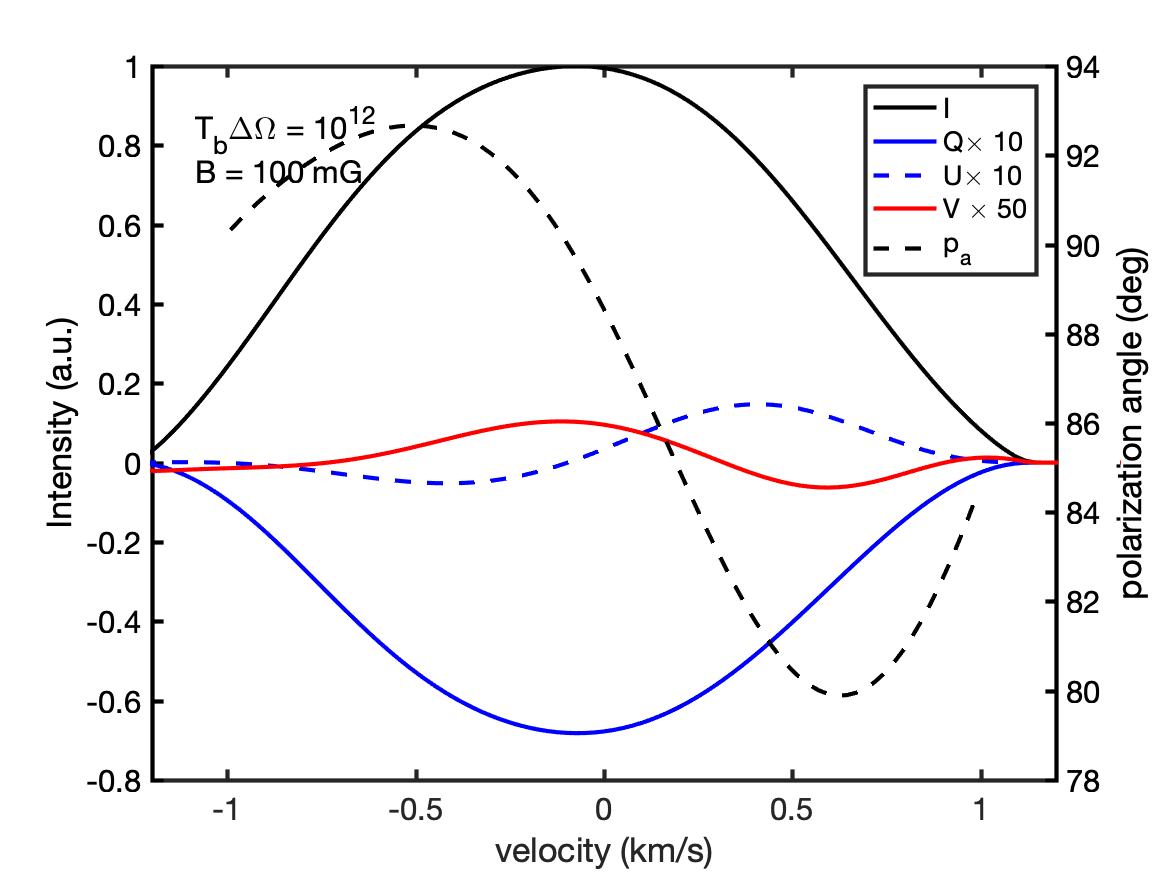}
        \caption{}
    \end{subfigure}
    \caption{Water $22$ GHz maser spectra for different levels of saturation. We plot all Stokes parameters (left $y$-axis) as well as the polarization angle (right $y$-axis). The polarization angle is defined with respect to the magnetic field direction projected on the plane of the sky. Simulations were carried out at $B=100$ mG, $v_{th}=1$ km/s and with a magnetic field propagation angle of $\theta = 45^o$.}
    \label{fig:water_spectra}
\end{figure}

\begin{figure}
    \centering
    \begin{subfigure}[b]{0.45\textwidth}
        \includegraphics[width=\textwidth]{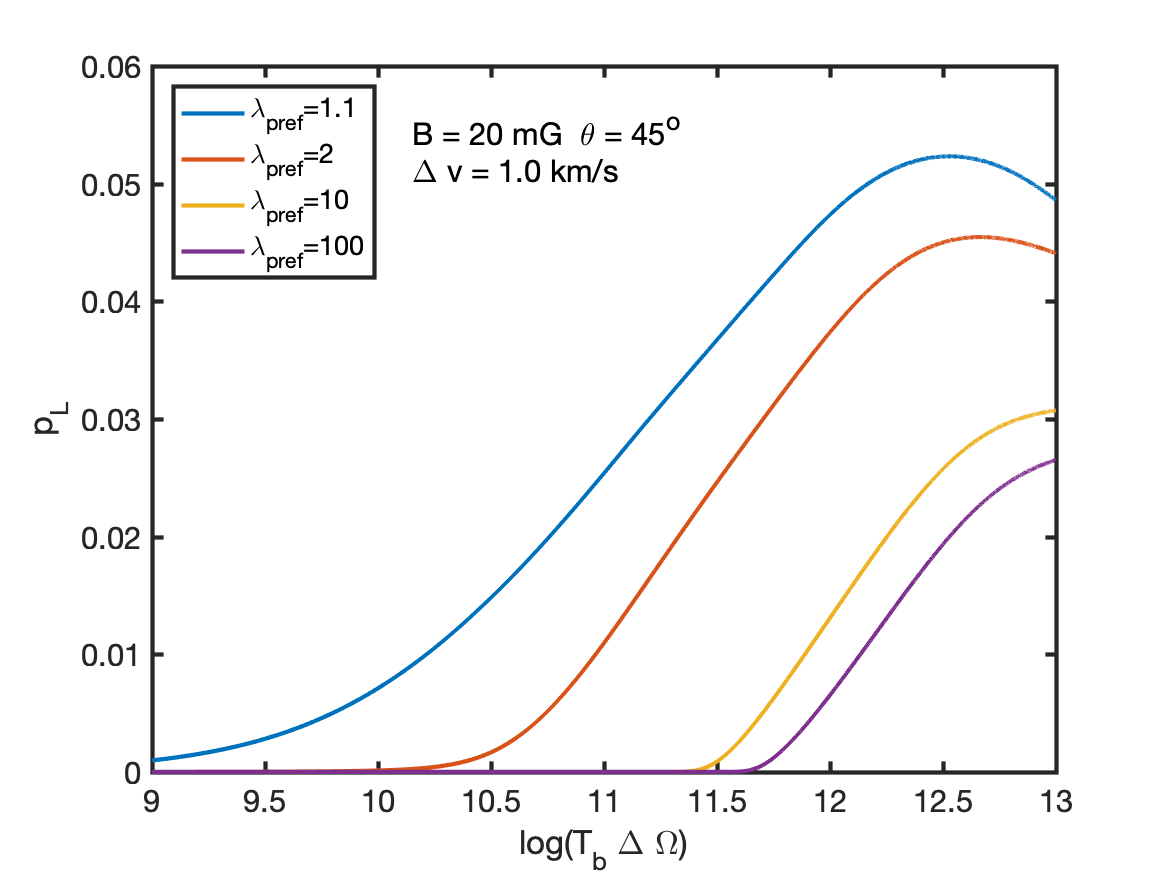}
        \caption{}
    \end{subfigure}
    ~
    \begin{subfigure}[b]{0.45\textwidth}
        \includegraphics[width=\textwidth]{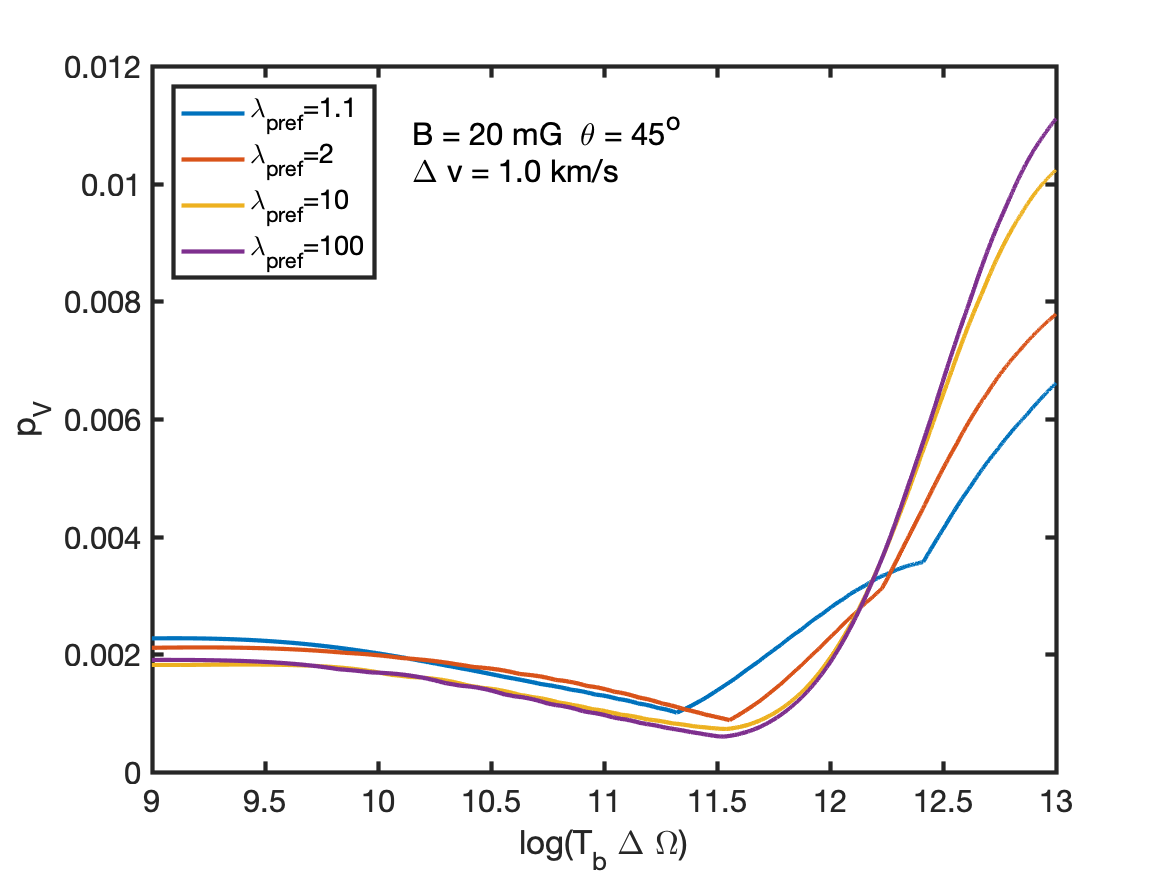}
        \caption{}
    \end{subfigure}
    \caption{Plots of the (a) linear and (b) circular polarization fraction of water masers as a function of the maser luminosity. Different degrees of preferred pumping are plotted. Magnetic field strength, angle $\theta$ and thermal width denoted inside the figure.}
    \label{fig:plot_water_varpump_iso}
\end{figure}

In Fig.~\ref{fig:water_spectra}, we present the $22$ GHz water maser spectra for different levels of saturation. It is immediately obvious that for all levels of saturation, the Stokes-$I$ spectra are slightly asymmetric because of the multiple hyperfine components of this maser. This asymmetry is also seen in the linear polarization, that follows roughly the total intensity spectrum. We should note that circular polarization profiles are not the anti-symmetric S-shaped signals we observed for the single-transition SiO-masers. Through the contributions from multiple hyperfine components an asymmetric circular polarization spectrum arises \citep{nedoluha:92, vlemmings:01}. A preferably pumped water maser will however show the characteristic S-shaped circular polarization signal.

\subsubsection{Polarized incident radiation}
We already observed in the SiO masers that for the higher-angular momentum contours, the general structure of polarization contours is strongly influenced by the incoming polarized radiation. This is thus also the case for water masers, who generally also show weaker magnetic field interactions. The simulations with strongly polarized incoming radiation, have nearly no general dependence on $\theta$, as the incoming (linear) polarization fraction smoothly deteriorates from $T_b \Delta \Omega > 10^{12}$ Ksr. The weakly polarized incident radiation has a less pronounced effect on the polarization landscape, although it strongly dominates the landscape for $T_b \Delta \Omega < 10^{10}$ Ksr. 

These effects are also reflected in the landscape of circular polarization, which is strongly affected for the highly polarized incoming radiation, in contrast to the weak effects incident polarized seed radiation had on the SiO maser. Incident polarized seed radiation can cause relatively high fractions of circular polarization, especially in the region around $\theta = 90^o$---interestingly the region where isotropic incoming radiation leads to no circular polarization---, where for the highly polarized incoming radiation, the circular polarization can get up to 5\% (1\% for weakly polarized incoming radiation).
\begin{figure*}
    \centering
    \begin{subfigure}[b]{0.45\textwidth}
        \includegraphics[width=\textwidth]{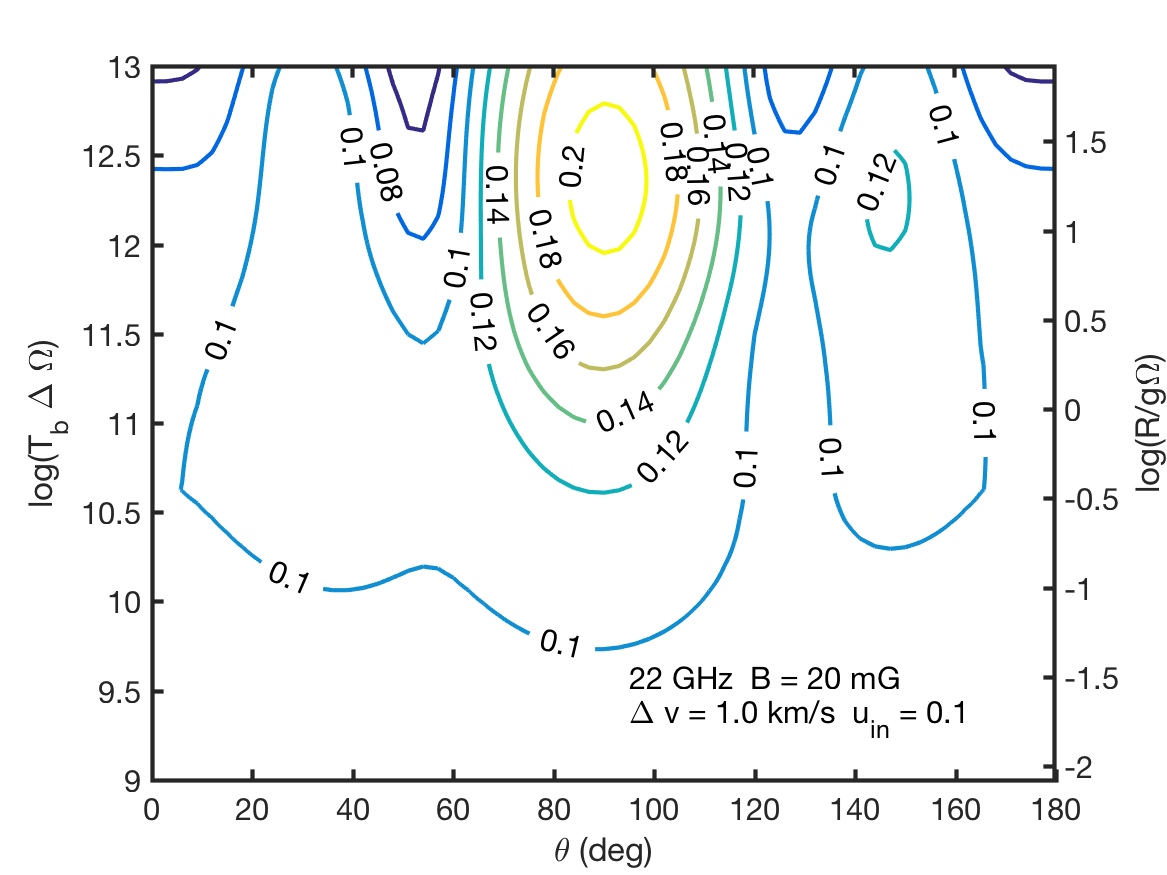}
        \caption{}
    \end{subfigure}
    ~
    \begin{subfigure}[b]{0.45\textwidth}
        \includegraphics[width=\textwidth]{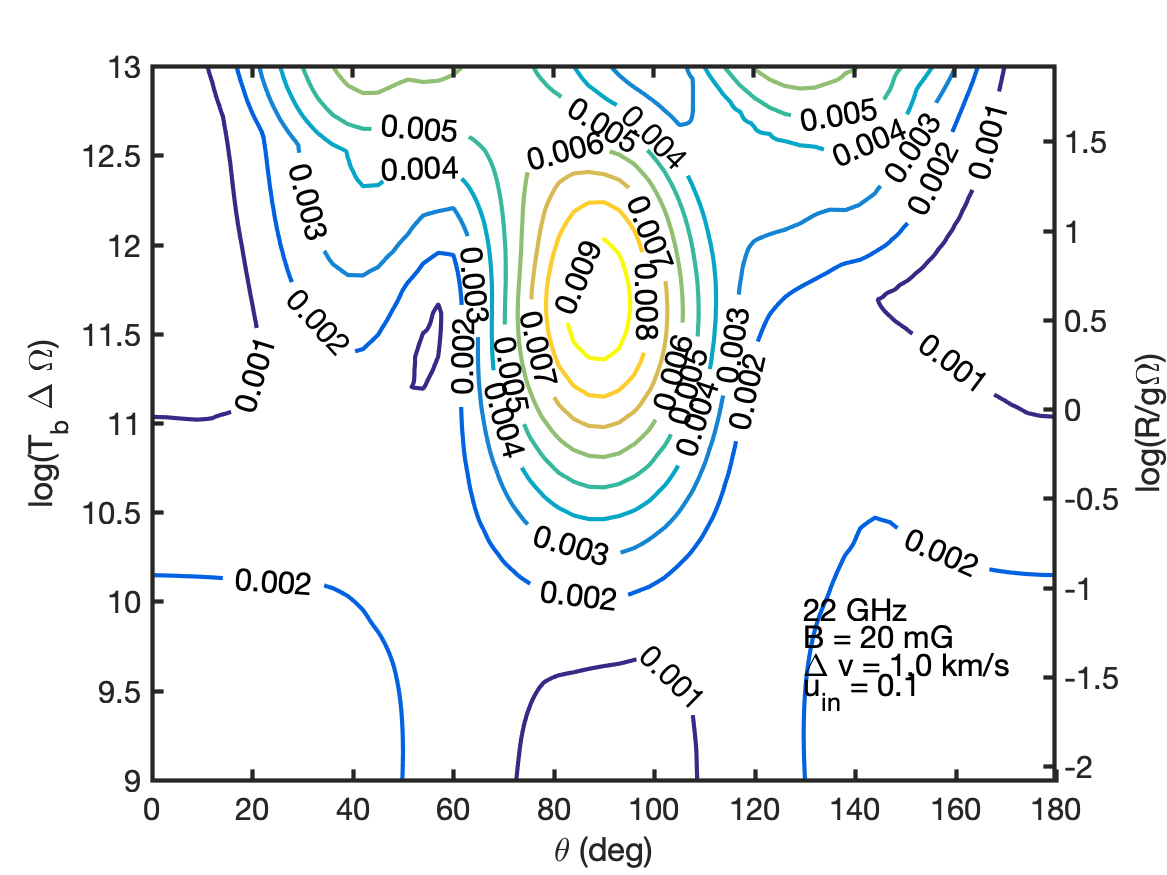}
        \caption{}
    \end{subfigure}
    ~
    \begin{subfigure}[b]{0.45\textwidth}
        \includegraphics[width=\textwidth]{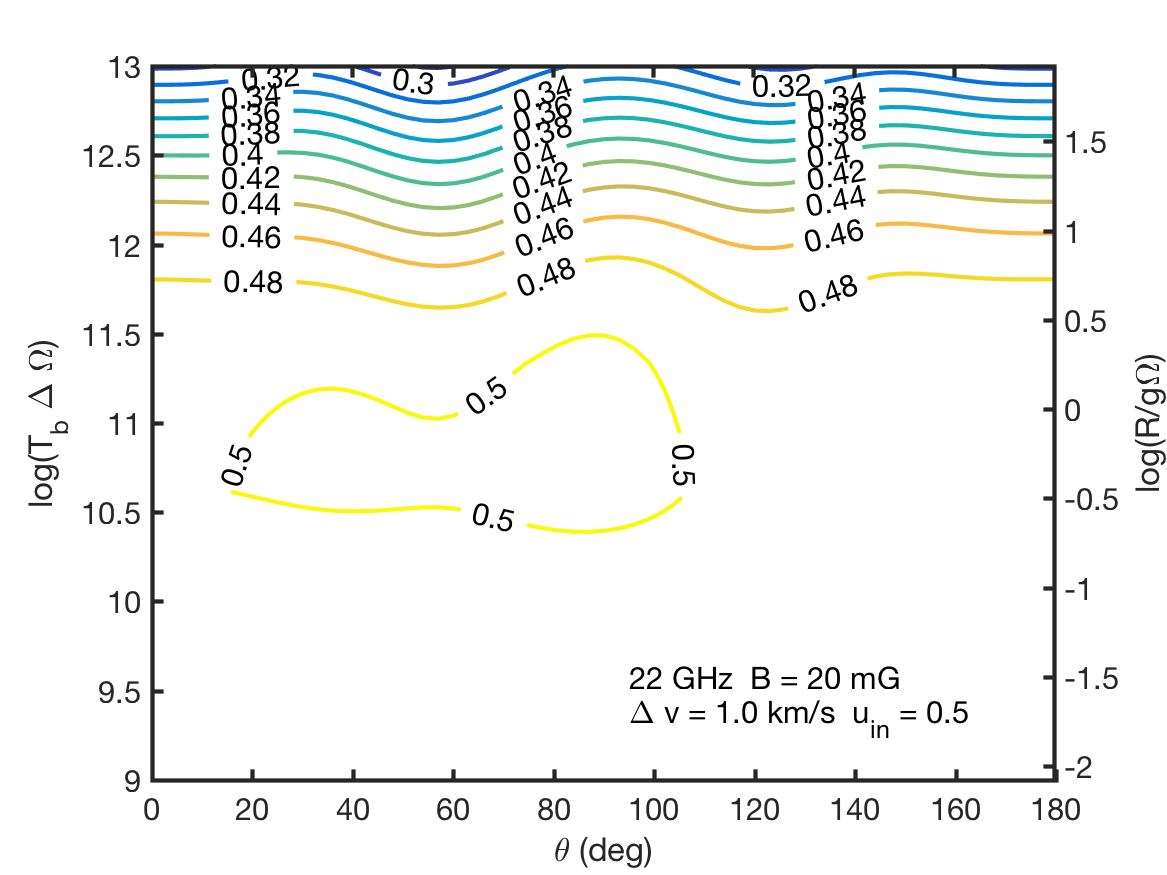}
        \caption{}
    \end{subfigure}
    ~
    \begin{subfigure}[b]{0.45\textwidth}
        \includegraphics[width=\textwidth]{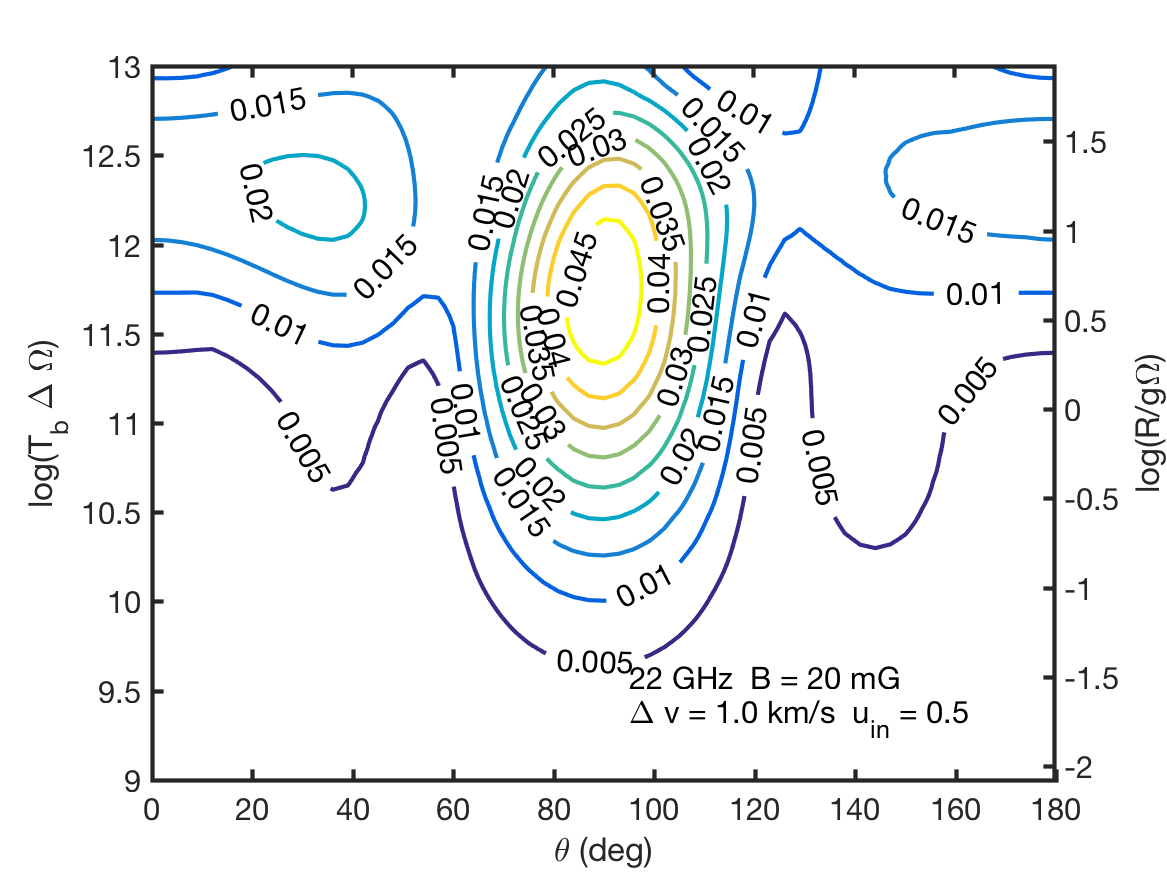}
        \caption{}
    \end{subfigure}
    \caption{Contour plots of the (a,c) linear and (b,d) circular polarization fraction of a water maser as a function of the propagation angle $\theta$ and the rate of maser luminosity. Maser simulations performed with incident polarized radiation of (a,b) $U/I=0.1$ and (c,d) $U/I=0.5$. Magnetic field strength and thermal width are denoted inside the figure. For simulations with other $v_{th}$ and other magnetic field strengths, see Figs.~A.21-22 in the Appendix.}
    \label{fig:contour_water_in}
\end{figure*}

\subsubsection{Anisotropic pumping}
\begin{figure*}
    \centering
    \begin{subfigure}[b]{0.45\textwidth}
        \includegraphics[width=\textwidth]{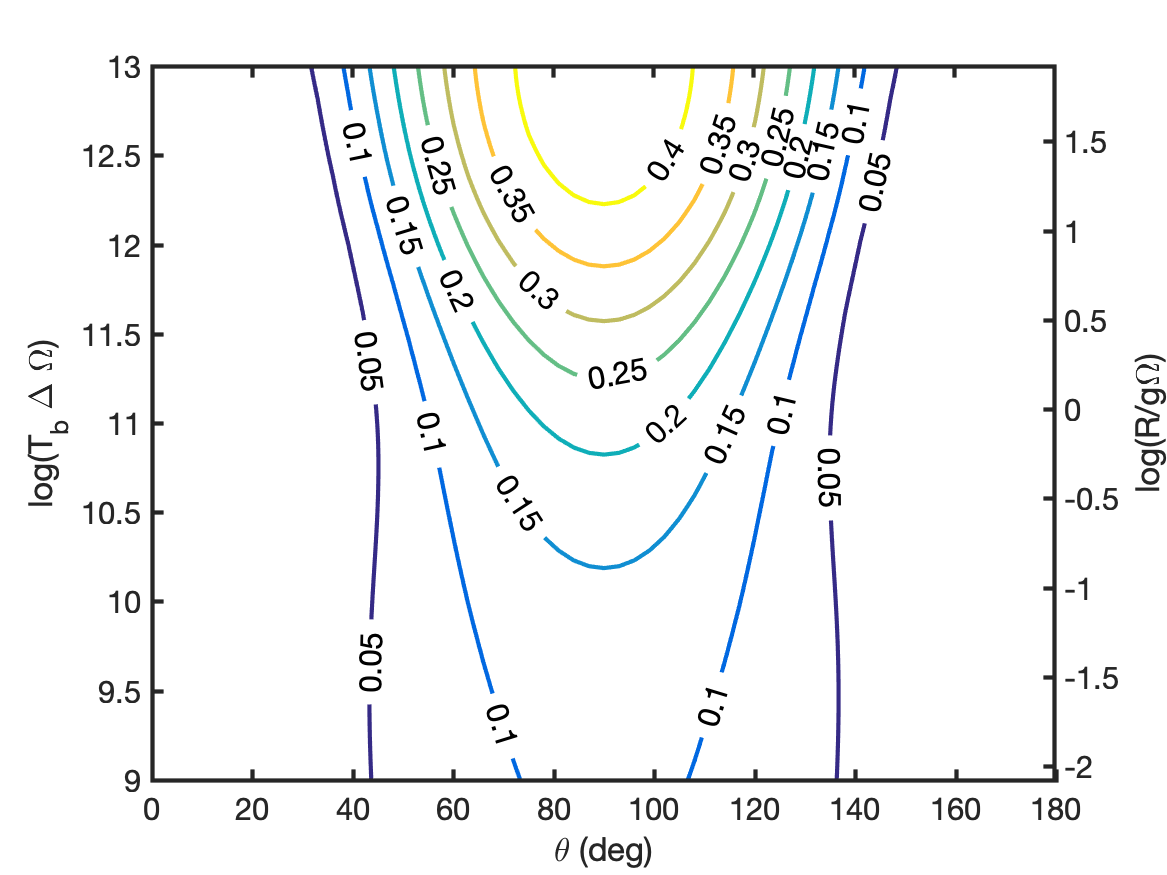}
        \caption{}
    \end{subfigure}
    ~
    \begin{subfigure}[b]{0.45\textwidth}
        \includegraphics[width=\textwidth]{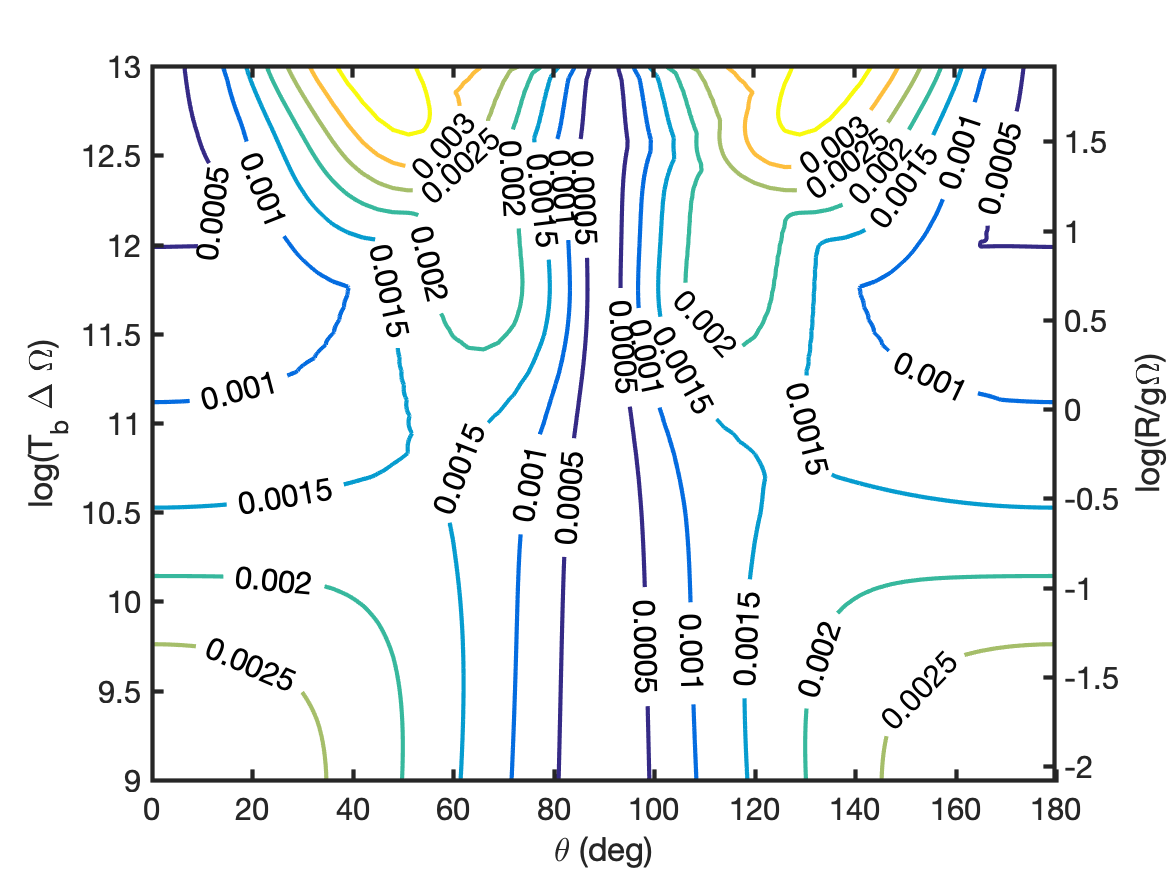}
        \caption{}
    \end{subfigure}
    ~
    \begin{subfigure}[b]{0.45\textwidth}
        \includegraphics[width=\textwidth]{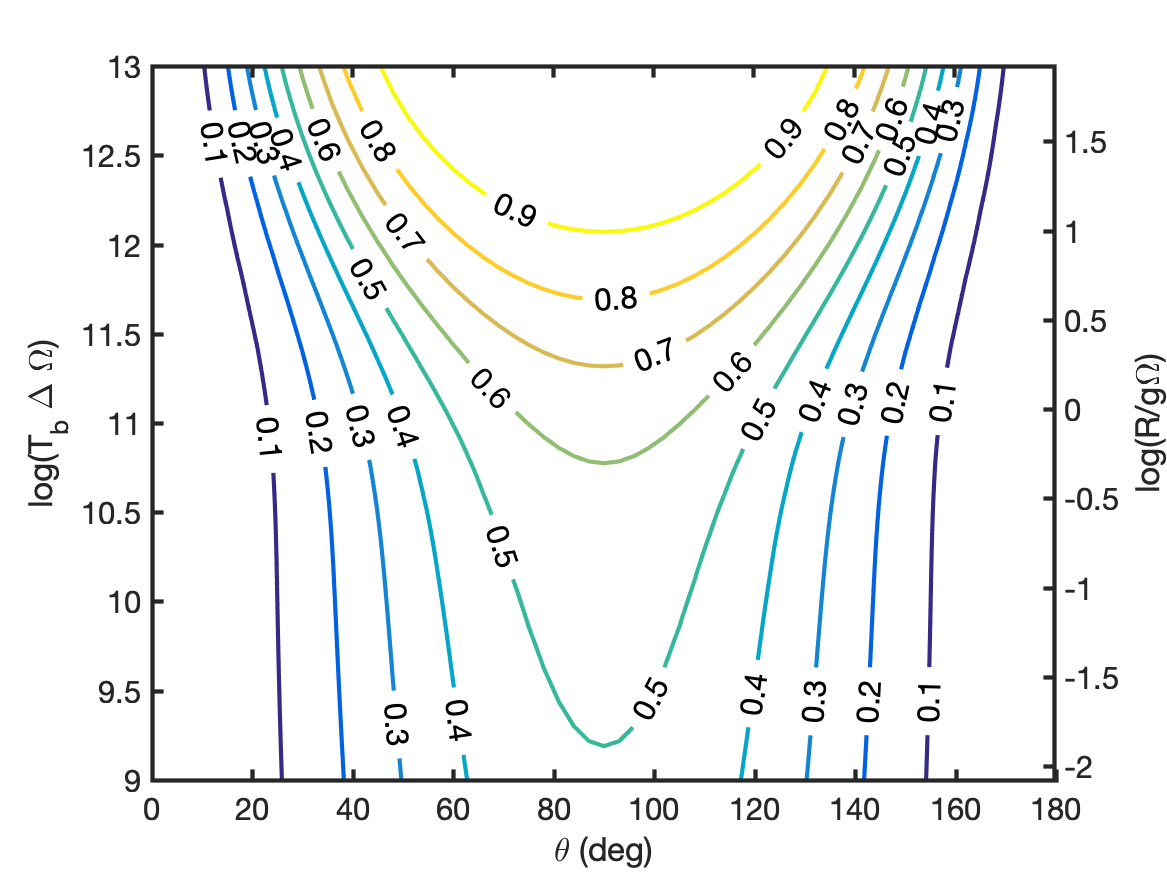}
        \caption{}
    \end{subfigure}
    ~
    \begin{subfigure}[b]{0.45\textwidth}
        \includegraphics[width=\textwidth]{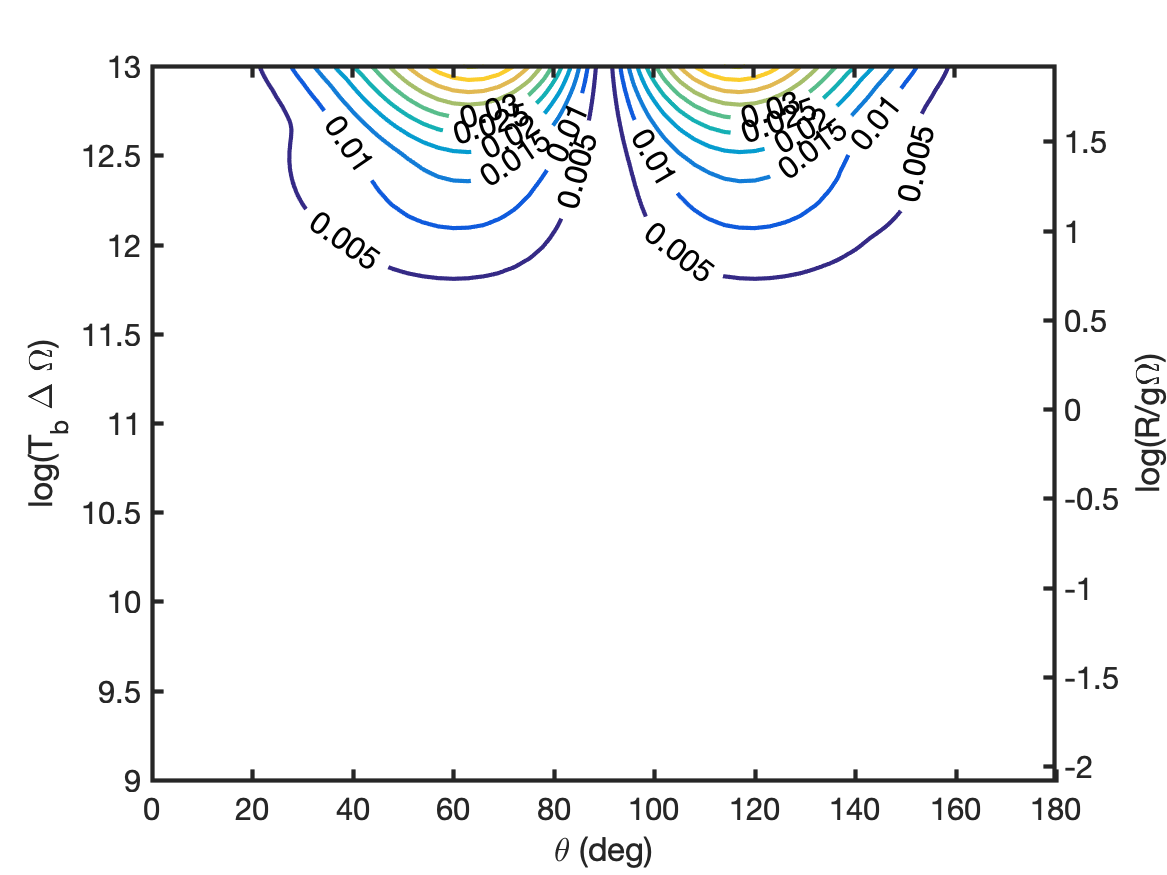}
        \caption{}
    \end{subfigure}
    \caption{Contour plots of the (a,c) linear and (b,d) circular polarization fraction of a water maser as a function of the propagation angle $\theta$ and the rate of maser luminosity. Maser simulations performed with anisotropic pumping perpendicular to both propagation and magnetic field direction, with anisotropy parameters of (a,b) $\eta = 0.1$ and (c,d) $\eta = 0.5$, and seed radiation of $T_b = 0.1 K$. We used a magnetic field strength for these simulations of $B=20$ mG and thermal width $v_{th} = 1$ km/s.}
    \label{fig:contour_water_anis}
\end{figure*}

As a consequence of the shocked material that water masers occur in, photons that are associated with the radiative relaxation from the collisionally excited water molecules, may have a preferred escape direction. This can lead to a small anisotropy in the maser pumping. Analyzing our simulations of the anisotropically pumped water maser, we notice that the linear accrual of polarization with the maser brightness is also characteristic of these masers. We notice for the perpendicularly pumped water masers from Fig.~\ref{fig:contour_water_anis} that masers of $\theta \to 90^o$ gather the most linear polarization from the propagation. In fact, for the water masers it seems that the standard magnetic-field polarization mechanism has barely any effect on the polarization maps of both weak and strong anisotropy, as signified by the symmetry of the linear polarization landscapes. The polarization of these masers are almost independent of the magnetic field strength, but will be highly dependent on the intensity of the seed radiation, as well as the anisotropy of the pumping, $\eta$. Anisotropic pumping can generate arbitrary linear polarization fractions for the water masers. 

High circular polarization fractions are only weakly associated with the drastically higher linear polarization from anisotropic pumping. Only the brightest of the strongly anisotropically pumped masers show significantly higher circular polarization, but not exceeding $5\%$. 

\section{Discussion}
We will divide the discussion up in two parts. First, we will discuss the 
results we have presented in the previous section, and lay out the physical
mechanisms behind the phenomena we have observed from the simulations. 
In the second part of the discussion, we will discuss these results in the 
context of previous SiO and water maser polarization observations.
\subsection{SiO masers}
\subsubsection{Simulations}
\paragraph{90$^o$-flip of the polarization angle.} 
We have observed two processes that can give rise to a 90$^o$-flip in the 
polarization angle: an increase in rate of stimulated emission over two 
orders of magnitude, or the crossing of the magic angle, $\theta_m$. 
When $g\Omega \gtrsim 100 R$, the 
magnetic field determines the symmetry axis of the molecule. When this 
condition is fulfilled, and for the propagation radiation at an angle
with the magnetic field smaller than $\theta_m$, the polarization will
be oriented perpendicular to the magnetic field. For angles greater than
$\theta_m$, polarization will be oriented parallel to the magnetic field.
Thus, when we cross the magic angle and the condition 
$g\Omega \gtrsim 100 R$ is fulfilled, we see a sharp $90^o$-flip in the 
polarization angle across $\theta_m$. For stronger masers, where 
$100 \gtrsim g\Omega / R \gtrsim 0$, 
we also observed a flip in the polarization angle, but this flip is 
rather gradual (over $\sim 10^o$), and does not predict zero polarization 
at the magic angle. 
The 90$^o$-flip feature of $J=1-0$ SiO masers has recently be investigated by 
\citet{tobin:19}. \citet{tobin:19} analyze the changing polarization fraction 
and angle of SiO maser spots across a clump. They assume a gradually changing 
propagation angle with the projected angular distance. From an analysis based on 
GKK73, they fit the observed polarization fraction and angle. 
Indeed, a $90^o$-flip is observed around the magic angle, but 
the $90^o$-flip is rather gradual. According to their analysis, this is due to the 
free $K$-parameter that arises in the GKK73 models. Usually, this parameter is 
assumed zero on the grounds of symmetry.  
According to our analysis, one need not invoke such a free parameter. Because as we 
have seen in our simulations (Fig.~\ref{fig:magiccross}), 
a blunt $90^o$-flip around the magic angle is characteristic of masers where the rate 
of stimulated emission is in the same order as the magnetic precession rate. Indeed, 
\citet{tobin:19} estimate $\mathrm{log}(R/g\Omega)\sim -1$, and our simulations of a 
magic angle flip at these conditions (Fig.~\ref{fig:magiccross}, $\mathrm{log}(R/g\Omega)=-1$)
show a similar blunted magic angle flip in the polarization angle. We should note that our 
analysis underestimates the polarization fraction with respect to the observations, 
and one needs to invoke non-Zeeman polarizing mechanisms to reach the observed polarization 
fractions. 


Sometimes, it is stated in the literature that
in the limit $R \gg g\Omega$, maser polarization will be randomly oriented \citep{plambeck:03}. This is not 
the case. Even though the radiation field determines the alignment of 
the molecules, its interaction with the magnetic field through the maser medium is still the 
polarizing mechanism. It is therefore that the magnetic field determines the polarization direction. A 
$90^o$-flip across $\theta_m$, however, will not occur in the case of $R\gg g\Omega$  
as the orientation of 
the polarization is invariably parallel to the magnetic field. 
This is also associated with the alternative mechanism that leads to a 
$90^o$-polarization angle flip. When $R \ll g\Omega$ and
the propagation angle is smaller than $\theta_m$, maser polarization
will be oriented perpendicular to the magnetic field direction. However,
if the rate of stimulated emission would increase, or the magnetic field
strength would decrease, and the condition $g\Omega \gg R$ would not be 
fulfilled anymore, the polarization would gradually align itself 
parallel to the magnetic field. A change in two orders of magnitude of
$R$ or $g\Omega$ can cause a $90^o$-flip in the polarization angle.

\paragraph{A peak in polarization at $g\Omega \sim R$.} 
Invariably, the highest linear and circular polarization fractions are observed for the case that the magnetic 
field strength is of the same order of magnitude as the rate of stimulated 
emission. This effect seems to be most pronounced for angles smaller than the magic angle,
specifically around the propagation angle $\theta = 20^o$. The extra polarization is coming
from a strongly enhanced Stokes-$U$ component in the radiation, and significant off-diagonal
state density elements. The effect is absent for $90^o$-propagation, because off-diagonal
elements need not be invoked in these masers.

\paragraph{Absence of polarization below $R=1$ s$^{-1}$ ($T_b \Delta \Omega = 9 \times 10^6$ Ksr).}
Considering an isotropically pumped maser, and when $R$ is so small that $R \ll g\Omega$, we recognize from Eq.~(\ref{eq:statepop}) that the radiation field has only a small influence on the populations of the magnetic substates of SiO, and will be minimally polarized because of this. Also, because the (isotropic) decay of the states, described by the term $\Gamma$, is larger than $R$, the polarization of the states will be drastically lowered through the depolarizing decay. 

\paragraph{The circular polarization of SiO masers}
Just as for linear polarization, the highest circular polarization fraction was found in the region $R\sim g\Omega$. The polarization fraction in this region is not dependent on the maser thermal width. The high degree of circular polarization found here, is due to an effect that was earlier described as ``intensity-dependent circular polarization'' \citep{nedoluha:94}. Circular polarization is associated with the changing of the molecular symmetry axis that in the transition from $R<g\Omega$ to $R>g\Omega$, changes from parallel to the magnetic field, to parallel to the propagation direction. 

A version of the above described effect is also responsible for the circular polarization that will be generated by a randomly oriented magnetic field that is  strong enough to align the molecule. \citet{wiebe:98} investigated the propagation of polarized radiation through a medium with a randomly oriented magnetic field along ($128\times128$) maser propagation paths. Along the path, linear polarization builds up. However, this linear polarization would not be aligned with the orientation of the molecules along the changing magnetic field. Locally, the linearly polarized radiation is rotated towards the (local) molecular alignment axis, with the associated production of circular polarization. In this way, relatively high degrees ($<3\%$) of circular polarization could be generated already from magnetic fields of ($\sim 30$ mG) \citep{wiebe:98}. Because circular polarization is generated from the linear polarization, the circular polarization should not exceed a certain linear polarization-dependent limit. Through analyzing this relation, \citet{cotton:11} found that the polarizing effects described by \citet{wiebe:98} could not explain the high degrees of circular polarization found in their observations of SiO $J=1-0$ masers. The circular polarization effects we have included in our models alone can also not fully explain the observations of \citet{cotton:11} (see also our discussion of the maser line-profiles that follows).

\paragraph{Slow convergence to the GKK73 solutions.} With a magnetic precession rate of $g\Omega = 1500 \times B (\mathrm{G}) \ \mathrm{s}^{-1}$ and an isotropic decay rate of $\Gamma = 5 \ \mathrm{s}^{-1}$, the SiO maser generally fulfills the condition $g\Omega \gg \Gamma$. For the GKK73 solutions to maser linear polarization to apply, we furthermore have a constraint on the rate of stimulated emission so that $g\Omega \gg R \gg \Gamma$. For an $R$ in the range from $\Gamma$ to $g\Omega$, this requirement cannot be fulfilled for magnetic field strenghts expected around SiO masers. This is confirmed by our calculations, where we do not find the GKK73 solutions in the relevant parameter space. Convergence to the GKK73 solutions only occurs for unphysically strong magnetic fields and unphysically luminous masers. 

\paragraph{Dependence of polarization on the angular momentum, $J$, of the transition.}
The difference in polarization fraction between the $J=1-0$ and $J=2-1$ transitions is very
large. For higher $J$-transitions, the polarization decrease with $J$ is less drastic. 
This phenomenon has already
been observed by D\&W90 and N\&W90, and can be explained by the inability for the
$J=0$-state to get polarized. The radiation field couples directly to the (in irreducible 
tensor terms) rank-0, 1 and 2 elements. Coupling to higher rank elements is mediated by higher
order effects and is therefore orders of magnitude weaker. The maximum rank of the 
elements of a certain state is $2J+1$. Therefore, all the polarization modes of the radiation 
field can couple directly to 
states of $J\geq 1$. Direct coupling of the polarization thus exists for all transitions but 
$J=1-0$, leading to this transition to be highly polarized. The further consistent 
polarization decrease with $J$ can be explained by the introduction of more higher 
rank irreducible population terms, to where some of the polarization leaks away to, and which 
do not couple directly to the radiation field.

\paragraph{Incident polarized seed radiation as a polarization mechanism.} 
One principal result of the simulations with polarized seed radiation was contained in the 
distinction between a weak-maser regime and a strong-maser regime. We observed that the in the 
weak-maser regime, the incident polarization was retained, and in the strong-maser regime 
the polarization would converge to the polarization obtained with isotropic seed radiation. 
In the weak-maser regime, the magnetic field is defining the symmetry axis. Because the  
radiation field is so weak, there is no appreciable influence of it on the molecular states. 
In fact, we can consider the states to be unpolarized. That means that amplification is 
characterized by a dominant $A_{\omega}$-term (see Eqs.~\ref{eq:abc2}). Thus, radiation is 
amplified and not altered in terms of polarization until it becomes a significant entity 
that can align and polarize the molecular states. After the weak-maser regime, at about 
$\mathrm{log}(R/g\Omega) = -1$, a transition regime can be recognized, where both the 
initial polarization, as well as the overall radiation have an appreciable influence on 
the molecular states. The feedback of the polarized molecular states in the propagation of 
the polarized radiation causes the radiation to converge to a polarization that is general 
for the system (in terms of $R$, $g\Omega$ and $\theta$), invariable of its initial 
conditions, which is what we call the strong-maser regime. Convergence is attained later 
for strongly polarized seed radiation, and lower magnetic fields. High degrees of 
polarization can be obtained in the transition regime. Later in this discussion, we will 
comment on the effect these high degrees of linear polarization have on the circular 
polarization.
 
\paragraph{Anisotropic pumping as a polarization mechanism.} 
For the anisotropically pumped maser, we have a weak-maser regime and a strong-maser 
regime as well. We should however note that these regimes carry a different meaning with 
respect to the regimes of the masers with polarized seed radiation of the same name. 
The weak-maser regime of the anisotropically pumped maser is characterized by a 
linear growth of the polarization with maser luminosity. This growth can continue to 
arbitrary degrees of (linear) 
polarization until the radiation becomes strong enough to align the molecular states. 
In the weak-maser regime, because the pumping is anisotropic---where the anisotropic 
part can be represented by a second-rank irreducible tensor---polarization is 
pumped into the molecular states; causing a feed to the radiation field via the 
propagation coefficients $B_{\omega}$ and $F_{\omega}$ (see Eqs.~\ref{eq:abc2}). 
The build-up of polarization is thus dependent on the relative anisotropy in the 
pumping, $\epsilon$, but also on the relative size of $A_{\omega}$, given by 
$\delta$ (Eq.~\ref{eq:delta}), leading to the anisotropy-parameter $\eta = \epsilon / \delta$. 
The build-up of polarization in the weak-maser regime is independent of the magnetic 
field and is not associated with circular polarization. But it is dependent on the 
brightness of the seed radiation.

When radiative interactions become strong enough to influence the alignment of the 
molecule, a transition regime begins and, generally, the strongly polarized radiation 
begins to lose most of its polarization. The alignment of the molecular states is 
countering the large overshoot in polarization left from the weak-maser regime and 
converges in the strong-maser regime to a polarization that is a function of the 
anisotropy of the pumping (including direction), $R$ and $\theta$ that is independent 
of the incoming radiation.


\paragraph{Maser line-profiles.}
Maser line-profiles are often much narrower than their LTE counterparts because of the stimulated emission mechanism. This is most manifest when the rate of stimulated emission is near the isotropic decay rate $R\sim \Gamma$. After that point, broadening of the line starts and increases with $R$. From analyzing the polarized spectra, we observe that linear polarization spectra roughly follow the Stokes-$I$ spectrum, which is expected because the molecular states get polarized by the directional intensity field. The difference in polarizing intensity also leads to a variable polarization angle across the spectrum. This is particularly present for rates of stimulated emission $R\sim g\Omega$. The degree of change of the polarization angle across the maser-line can therefore be taken as a proxy for the saturation level. 

We observe that the polarizing mechanism under investigation in our simulations produce perfect anti-symmetrical S-shaped spectra for the Stokes-$V$ component of the radiation field. Such anti-symmetric spectra are often seen in astrophysical maser spectra \citep{amiri:12}. However, asymmetric Stokes-$V$ spectra are observed regularly as well. \citet{cotton:11} report the observation of many strongly asymmetrically circularly polarized SiO masers. Our models do not produce such asymmetrical spectra in the absence of hyperfine multiplicity, but would need to include alternative effects. A velocity gradient across the maser column or the presence of strong anisotropic resonant scattering in either a foreground cloud or as a part of the maser action itself, are known to be able to produce asymmetric Stokes-$V$ spectra \citep{houde:14}. Kinematic effects coming from other polarized background maser sources could also explain the asymmetric signals.
 
Interesting evidence for kinematic effects can be found by analyzing some individual maser line-spectra \citep{cotton:11}. The polarization spectra of the maser spot of figure~5, row $1$, from \citet{cotton:11}, show for the polarization angle similar variation across the spectrum as our Fig.~\ref{fig:sio_spectra} of the spectral polarization of SiO masers. Indeed, this maser shows an S-shaped antisymmetric Stokes-$V$ spectrum. Analyzing then rows $3$ and $4$ of the same figure in \citet{cotton:11}, we see a variation in the polarization angle across the maser line that is more reminiscent of the $22$ GHz water maser spectra of Fig.~\ref{fig:water_spectra}. Indeed, also the circular polarization of these signals is similar to our spectral models of the water masers (Fig.~\ref{fig:water_spectra}). The different hyperfine components in water masers can reasonably be considered to emulate kinematic effects as they would occur for an SiO maser. A deeper analysis of such effects is beyond the scope of this paper, but we can suggest that asymmetric circular polarization signals can be the product of kinematic effects.  
\paragraph{Alternative polarizing mechanisms and circular polarization.}
An interesting result of our investigations to the effects of anisotropic pumping and 
polarized incident radiation is the rather marginal effects these polarizing 
mechanisms have on the circular polarization fraction of the maser. This can be 
best explained in a tensorial picture of the matter-radiation interactions. In a 
tensorial picture of the polarized radiation, Stokes-$Q$ and -$U$ (and -$I$) are expressed 
as second-rank components of the irreducible radiation tensor, while Stokes-$V$ is a
first rank component of this tensor \citep{degl:06}. Direct polarization of the molecular states by 
linearly polarized radiation will thus only affect the second-rank populations. It 
is also the second-rank populations that are pumped by the anisotropic pumping. Thus, 
for incident polarized radiation and anisotropic pumping, there exist no direct 
coupling to the first rank populations, and thus no direct coupling to the Stokes-$V$ 
radiation. Indeed, the Stokes-$V$ will only be slightly enhanced by higher-order effects, 
such as anisotropic resonant scattering \citep{houde:13} that will be more pronounced with high 
linear polarization of the radiation.

\begin{figure*}
    \centering
    \begin{subfigure}[b]{0.45\textwidth}
       \includegraphics[width=\textwidth]{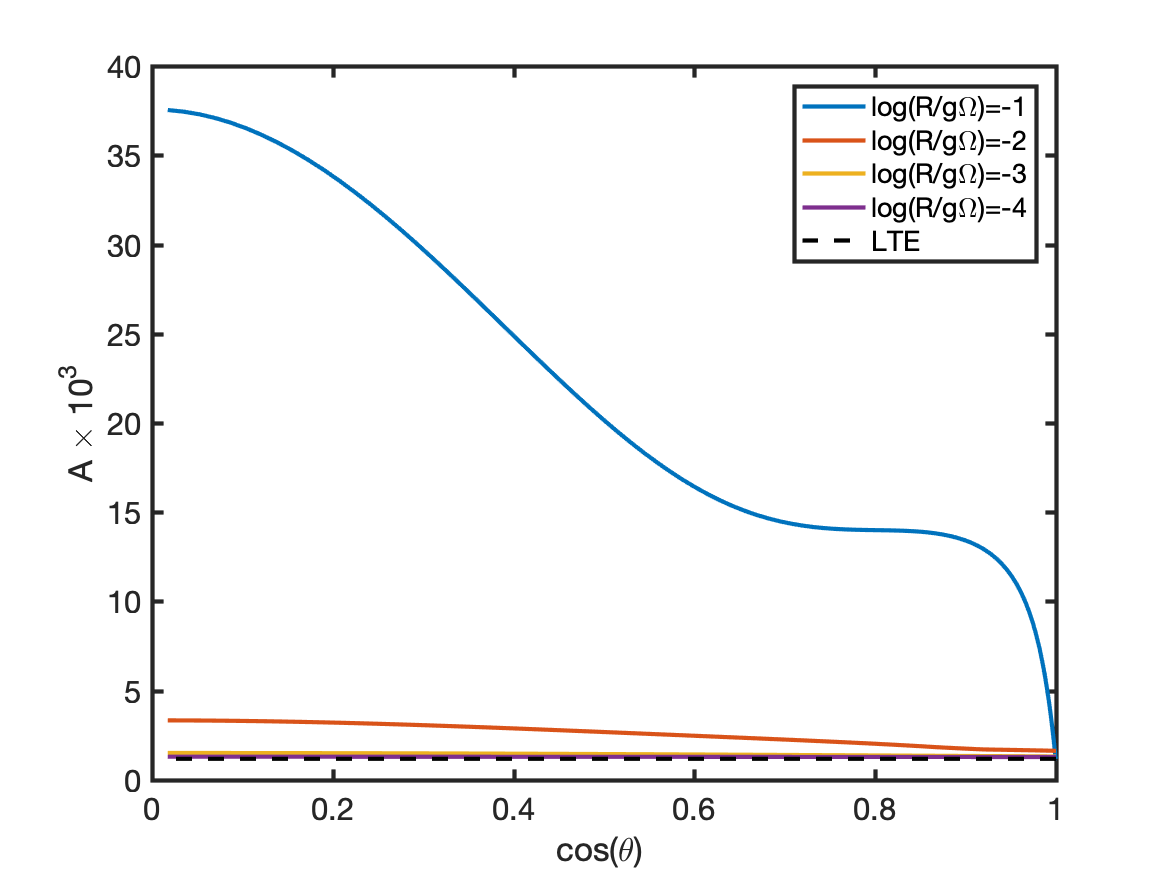} 
       \caption{}
    \end{subfigure}
    ~ 
    \begin{subfigure}[b]{0.45\textwidth}
       \includegraphics[width=\textwidth]{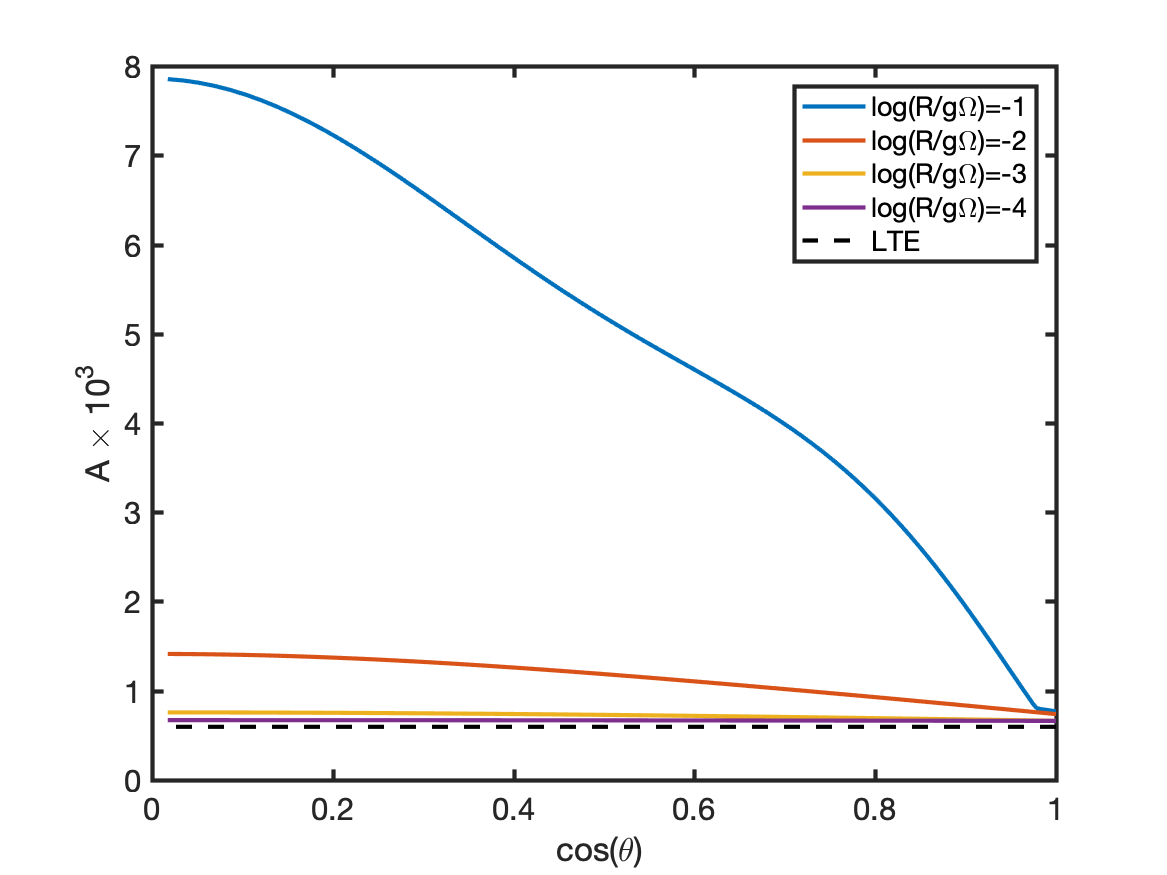} 
       \caption{}
    \end{subfigure}
     ~ 
    \begin{subfigure}[b]{0.45\textwidth}
      \includegraphics[width=\textwidth]{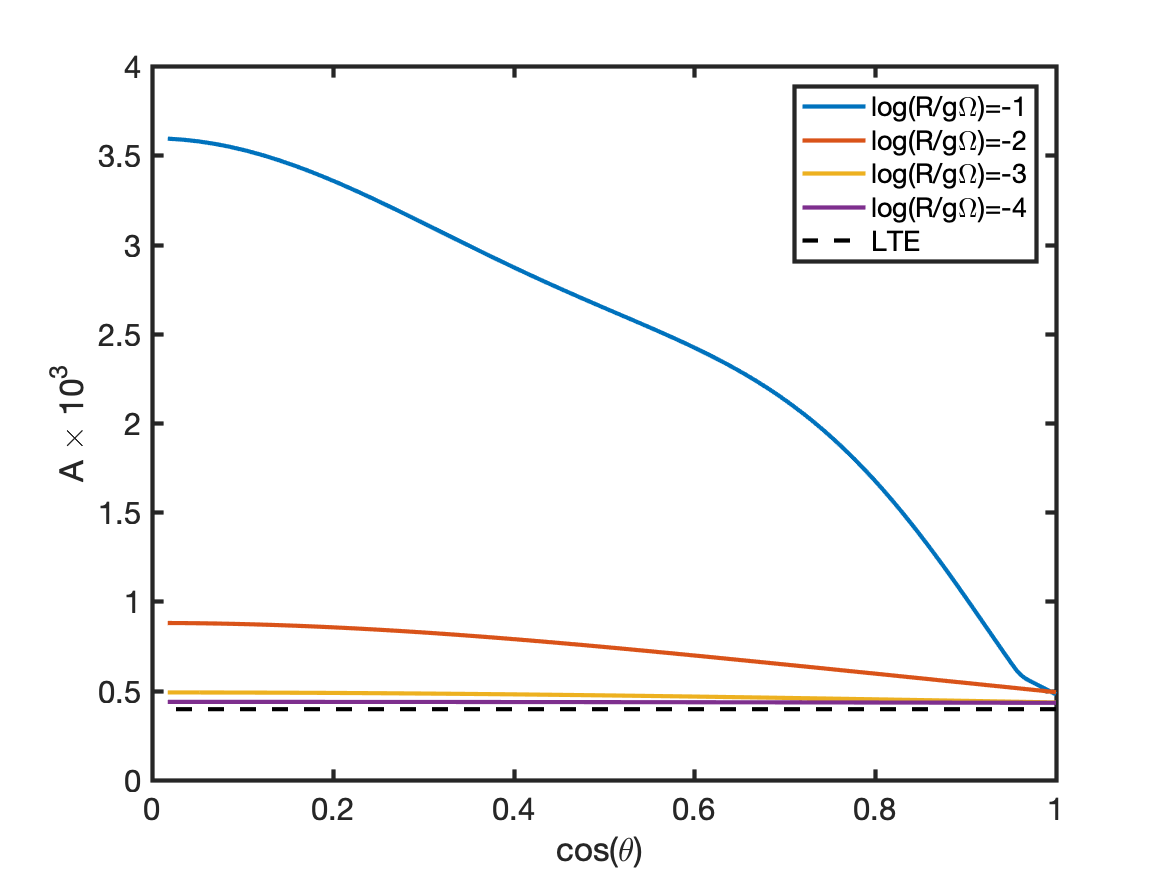}
      \caption{}
    \end{subfigure}
  \caption{The $A_{JJ'}$ coefficients of an isotropically pumped SiO maser at $B=1$ G as a function of the magnetic field-propagation direction angle $\cos \theta$. The different subfigures give the (a) $J=1-0$, (b) $J=2-1$ and (c) $J=3-2$ transitions. Plots are given for different $\mathrm{log}(R/g\Omega)$. The LTE solutions (constant over $\cos \theta$) are denoted with a dotted line.}
  \label{fig:aff-const}
\end{figure*}

\paragraph{Observational heuristics.}
Generally, we can recognize different regimes that are connected to the maser luminosity that show particular behavior regarding maser polarization. We therefore define characteristic maser luminosities that will simplify the analysis. The maser luminosity at which the rate of stimulated emission is equal to the rate of magnetic precession is defined as
\begin{align}
\left(T_b \Delta \Omega \right)_{\mathrm{mag.\ sat.}} = \frac{4\pi \omega_0 (g\Omega) }{A_{ij} k_B}
\end{align}
where $\omega_0$ is the masers natural frequency, $k_B$ is the Boltzmann constant and 
$A_{ij}$ is the Einstein coefficient. Furthermore, we define the luminosity after which the maser will start broadening because of saturation
\begin{align}
\left(T_b \Delta \Omega \right)_{\mathrm{sat.}} = \frac{4\pi \omega_0 \Gamma }{A_{ij} k_B}.
\end{align}
Table \ref{tab:sio_temps} gives these luminosities for the different SiO masers. 
Already at weak magnetic fields of $B>10$ mG, $\left(T_b \Delta \Omega \right)_{\mathrm{mag.\ sat.}} > \left(T_b \Delta \Omega \right)_{\mathrm{sat.}}$.
For the weakest masers, where $T_b \Delta \Omega < (T_b \Delta \Omega)_{\mathrm{sat}}$, linear polarization is 
mostly absent in the emission, because of the depolarizing effect of the isotropic decay. 
Circular polarization is generated through the Zeeman effect. Because of the Zeeman effect, 
the $\sigma^{\pm}$ ($\Delta m = \pm 1$) transitions will have a slight spectral disposition, 
that, if subtracted from each other, will yield the S-shaped Stokes-$V$ spectrum. It can be 
shown via an LTE analysis that the circular polarization will follow \citep{fiebig:89,watson:01}
\begin{align} 
p_V = \frac{2 A_{JJ'} B_{\mathrm{Gauss}} \cos \theta }{ \Delta v_L (\mathrm{km/s})},
\label{eq:lte_circ}
\end{align}
where $A_{JJ'}$ is a transition-dependent constant and $\Delta v_L$ is the FWHM of the maser profile. 
The LTE estimates for the constant $A_{JJ'}$ of SiO transitions
are
\[
A_{JJ'} = \frac{1.1807 \times 10^{-3}}{J} ,
\]
where $J$ is the rotational quantum number of the upper-state. It is usual to employ an LTE analysis of the circular polarization of weak masers, since the maser circular polarization mechanism for these masers is similar to the LTE mechanism. To check the validity of this analysis, we plot the results of our simulations for the $A_{JJ'}$-constants for three transitions at $B = 1$ G in Fig.~\ref{fig:aff-const}. For $T_b \Delta \Omega \lesssim (T_b \Delta \Omega)_{\mathrm{mag.\ sat}}/1000$, the $A_{JJ'}$ coefficient obtained from our simulations is similar to the LTE estimate. However, already for  $T_b \Delta \Omega \sim (T_b \Delta \Omega)_{\mathrm{mag.\ sat}}/100$, we find that the $A_{JJ'}$-constants from our simulations are twice that of the LTE estimate, meaning that an LTE analysis of the magnetic field strength would lead to an overestimation by a factor of $2$. 

For masers $T_b \Delta \Omega \ll \left(T_b \Delta \Omega \right)_{\mathrm{mag.\ sat.}}$, the highest circular polarization will be found for the masers that haven't start broadening yet ($T_b \Delta \Omega \sim (T_b \Delta \Omega)_{\mathrm{sat}}$). After $T_b \Delta \Omega > (T_b \Delta \Omega)_{\mathrm{sat}}$, the maser starts saturating with the associated
broadening. So long as the magnetic precession rate remains far greater than the rate of stimulated emission, 
$T_b \Omega \ll (T_b \Delta \Omega)_{\mathrm{mag.\ sat.}}$, the circular polarization will decrease because
of this broadening. Linear polarization starts to build up, either oriented parallel
($\theta > \theta_m$) or perpendicular ($\theta < \theta_m$) to the projected magnetic field
direction. Linear polarization will rise steadily with the maser luminosity, until it reaches
the GKK73 solution for the specific propagation angle. However, (long) before the GKK73 solution is reached, when the maser luminosity approaches
$(T_b \Delta \Omega)_{\mathrm{mag.\ sat.}}$, alternative polarization effects will take over.


\begin{table}[t]
\centering
\caption{Characteristic maser luminosities temperatures for $v=1$ SiO masers}
\begin{tabular}{l c c }
\hline \hline
Transition & $(T_b \Delta \Omega)_{\mathrm{sat}}$ (Ksr) & $(T_b \Delta \Omega)_{\mathrm{mag.\ sat.}}/B$ (Ksr/mG) \\ \hline
$J=1-0$ & $4.35 \times 10^{7}$  & $6.52 \times 10^6$ \\
$J=2-1$ & $9.00 \times 10^{6}$  & $1.35 \times 10^6$ \\
$J=3-2$ & $3.73 \times 10^{6}$  & $5.60 \times 10^5$ \\
$J=4-3$ & $2.03 \times 10^{6}$  & $3.04 \times 10^5$ \\
$J=5-4$ & $1.27 \times 10^{6}$  & $1.90 \times 10^5$ \\
\hline \hline
\end{tabular}
\label{tab:sio_temps}
\end{table}
%

In the regime of $T_b \Delta \Omega \sim (T_b \Delta \Omega)_{\mathrm{mag.\ sat.}}$, polarization associated with 
the change in molecular alignment will manifest itself in the emission spectrum. Linear polarization 
in this regime can therefore exceed the GKK73 solutions by $\sim 10\%$. For $\theta<\theta_m$, 
the polarization vector will change from perpendicular to parallel between 
$T_b \Delta \Omega \sim (T_b  \Delta \Omega)_{\mathrm{mag.\ sat.}} / 10$ and 
$T_b \Delta \Omega \sim 10(T_b \Delta  \Omega)_{\mathrm{mag.\ sat.}}$, 
and will have intermediate polarization angles within this range. With the gradual changing of the polarization angle  
a lot of circular polarization is associated. This is reflected in the high $A_{JJ'}$ constants for 
the circular polarization (see Fig.~\ref{fig:aff-const}). Constancy of $A_{JJ'}$ over $\theta$ is also lost.
For the lower angular momentum transitions, there is a large overshoot of the Zeeman circular 
polarization. Already for weak magnetic fields, high degrees of circular polarization can be 
generated and the Zeeman analysis cannot be applied directly. Extraction of the magnetic field strength 
from masers in the regime $T_b \Delta \Omega \sim (T_b \Delta \Omega)_{\mathrm{mag.\ sat.}}$ can be achieved by a 
simultaneous analysis of both the linear and circular polarization of the radiation, which will be 
demonstrated later on.  


 
Alternative polarizing mechanisms such as anisotropic pumping can enhance the polarization of masers 
to arbitrarily high degrees. 
The presence of anisotropic pumping could be discerned by analyzing the weaker masers 
($T_b \Delta \Omega \ll (T_b \Delta \Omega)_{\mathrm{mag.\ sat.}}$) for their polarization. The (linear) 
polarization degree of these masers should be proportional to their luminosity. When the anisotropically pumped maser 
approaches the luminosity $(T_b \Delta \Omega)_{\mathrm{mag.\ sat.}}$, linear polarization will drop as the standard 
polarizing mechanisms take over. Indeed, \cite{richter:16} find in their VLBA observations of VY CMa the strongest 
polarization for the weaker masers, and observe a drop in polarization after a certain maser luminosity threshold. 
Turning to polarized seed radiation, in the regime ($T_b \Delta \Omega \ll (T_b \Delta \Omega)_{\mathrm{mag.\ sat.}}$), the polarization 
is simply that of the seed radiation, and has no dependence on the maser luminosity. Circular 
polarization is only slightly enhanced for alternatively polarized masers.

Finally, we should make a note that the polarization properties are a function of the maser luminosity $T_b \Delta \Omega (\propto R)$, which cannot be measured directly. To estimate the maser luminosity from observations, one requires knowledge about the maser beaming solid angle $\Delta \Omega$. Direct observations of $\Delta \Omega$ have proven difficult to date, 
but have been performed with VLBA measurements to SiO around AGB stars \citep{assaf:13}. In these observations, \citet{assaf:13} measure, with a sizable error margin due to (relatively) low resolution, $\Delta \Omega \sim 5\times 10^{-2}$ sr. This maser beaming solid angle is independent of its brightness when the amplification is matter bounded (most easily approximated by the cylindrical maser) \citep{elitzur:92b}. When the maser is amplification bounded (most easily approximated by the spherical maser) the beaming solid angle drops with increasing maser brightness. To the best of our knowledge, no investigations have been done to the geometrical nature of the maser amplification of SiO masers.

\subsubsection{SiO maser polarization observations}
Many SiO maser polarization observations have been performed. VLBI observations have shown that SiO masers orient themselves in a ring-like structure around the central stellar object. The polarization of these SiO masers, irrespective of their angular momentum transition, show well-ordered polarization vectors with respect to this structure \citep{kemball:97,cotton:04,plambeck:03,vlemmings:11b,vlemmings:17}. This is taken to be an indicator of an ordered magnetic field. The linear polarization fraction of individual masers can be arbitrarily high, but median values are much lower. The $J=1-0$-transition has median linear polarization fractions of $\sim 25\%$ \citep{kemball:97}. Analyzing the angular momentum dependence of the linear polarization fraction, we note the general trend of lower degrees of polarization for the higher-angular momentum transitions. This is not to say that high ($>50\%$) fractions of linear polarization do not occur for high $J$ SiO maser-transitions. It is almost certain that the most strongly polarized masers are the product of anisotropically pumped maser action, as incident polarized radiation at these fractions is unlikely; and should lead to the same effect for the high-$J$ masers. The hypothesis of anisotropic pumping could be further supported by correlating maser-brightness for the weaker masers ($R<g\Omega$) to linear polarization. 

The relationship between maser-brightness and polarization fraction is unfortunately not well-documented. However, \citet{barvainis:87} meticulously tabulated their observations, from which we could construct a scatter plot that indicated lowest fractions of polarization for the strongest masers. This is in line with the simulations we have delineated above, where we have seen that above $R \sim g\Omega$, polarization fractions start to drop. 

\begin{figure}
    \centering
    \begin{subfigure}[b]{0.45\textwidth}
      \includegraphics[width=\textwidth]{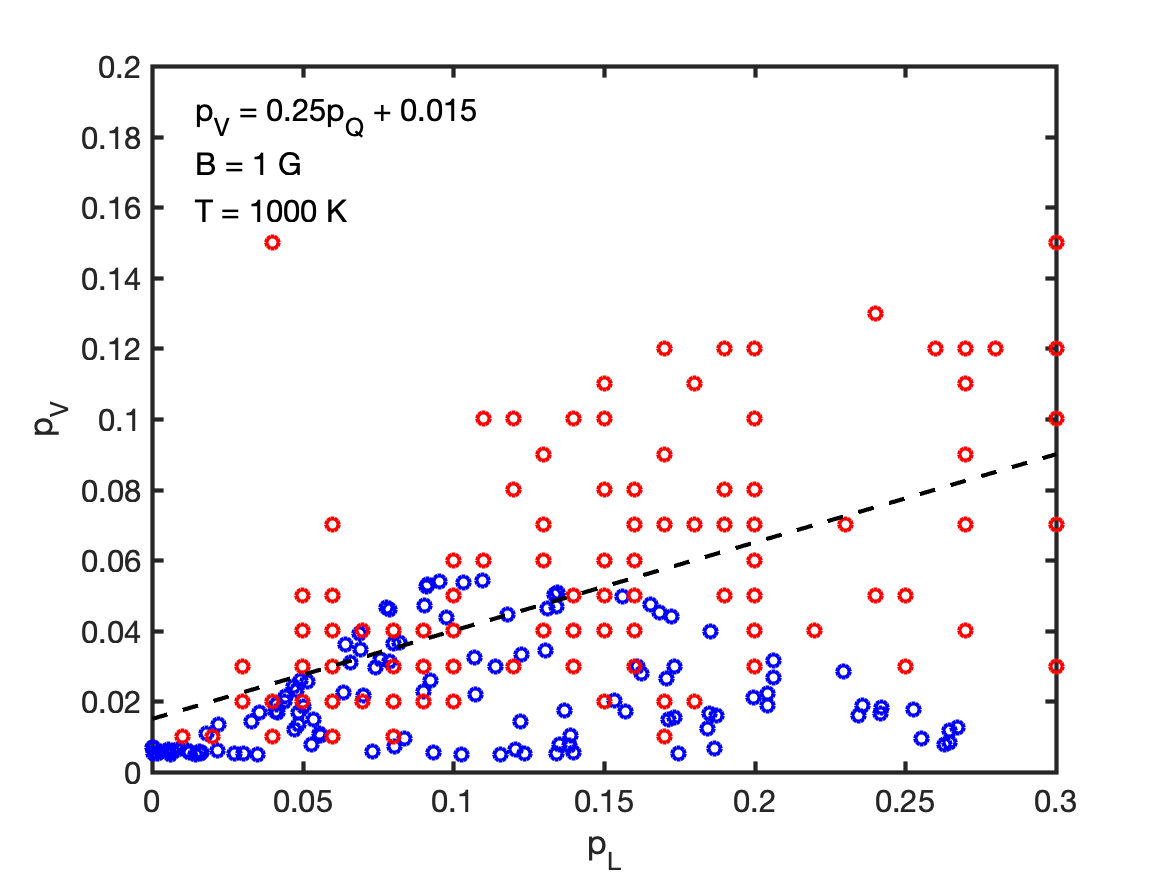}
      \caption{}
    \end{subfigure}
    \begin{subfigure}[b]{0.45\textwidth}
      \includegraphics[width=\textwidth]{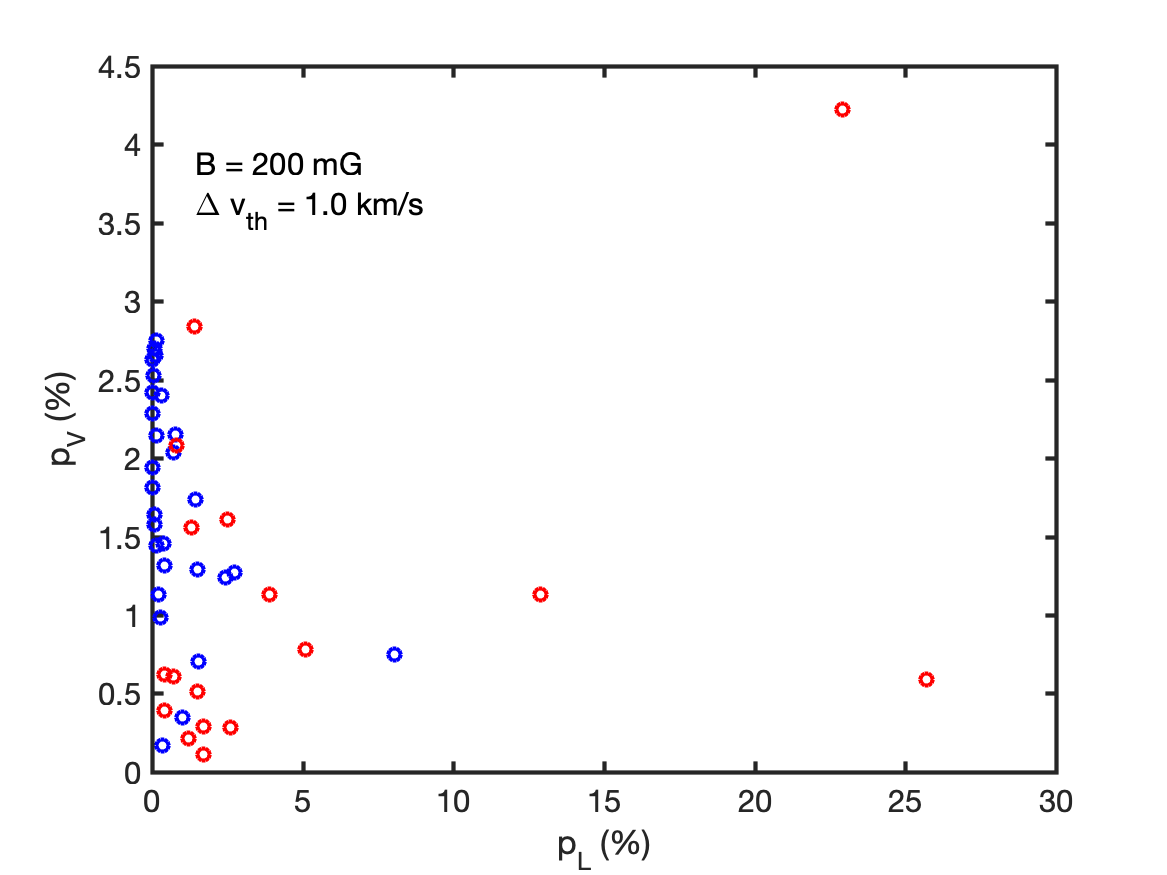}
      \caption{}
    \end{subfigure}
    \caption{Scatter plot for the linear to circular polarization fraction relation $p_L - p_V$. In red, the observations of (a) \citet{herpin:06}  and (b) \citet{surcis:11} are reported. In subfigure (a), the blue points come from our simulations of the $J=2-1$ transition at $B=1$ G, with the kinetic temperature of the maser $T=1000$ K. Subfigure (b), the blue points come from our simulations of the isotropically pumped water maser at $v_{\mathrm{th}} = 1.0$ km/s and $B=200$ mG. To generate these scatters, we have computed the polarization fractions from (a) $200$ (b: 30) isotropically pumped masers with a randomly selected a luminosity between (a) $T_b \Delta \Omega = 10^6 - 10^{11}$ (b: $T_b \Delta \Omega = 10^{8.5} - 10^{11}$) and a randomly selected propagation angle $\theta$. In the scatter plot, we do not include masers that show polarization $<0.5 \%$ (b: $<0.1 \%$. Inside the figure (a), we also report the linear regression analysis result from \citet{herpin:06}.}
    \label{fig:lincirc}
\end{figure}

\citet{herpin:06} have been able to derive an interesting relation between the circular polarization and linear polarization of SiO $J=2-1$ masers. In a large survey of a number of evolved stars, they analyzed, among other things, the correlation between linear and circular polarization fractions of the SiO masers. Even though the correlation was highly scattered, a clear linear relation was observed between linear and circular polarization (figure 4, \citet{herpin:06}). Also, invariably, high circular polarization was associated with high linear polarization. To simulate their observations, we have used CHAMP to compute the linear and circular polarization fractions of $200$ isotropically pumped SiO masers, at randomly selected luminosities between $T_b \Delta \Omega = 10^6 - 10^{11}$ Ksr and randomly selected propagation angles $\theta$. We plot the results for SiO $J=2-1$ masers pumped at $T=1000$ K, and magnetic field of $B=1$ G in Fig.~\ref{fig:lincirc}. \citet{herpin:06} found a rough linear relation between the linear and circular polarization, $p_V = 0.25p_L + 0.015$, which we plot in the figure. 

Only for lower degrees of linear polarization we find a reasonable agreement between our simulations and the observations of \citet{herpin:06}. Our simulations seems to underestimate the circular polarization with respect to the observations of \citet{herpin:06}. This is especially true for the strongly linearly polarized masers. One factor that could play a role here is the enhancement of circular polarization by the presence of a velocity gradient along the propagation path of the SiO-maser. N\&W94 have shown that this can enhance the circular polarization. Another explanation of the high circular polarization might be the anisotropic resonant scattering of maser radiation by a foreground cloud of non-masing SiO \citep{houde:13, houde:14}. Via anisotropic resonant scattering, linearly polarized radiation can be converted to circularly polarized radiation. Anisotropic resonant scattering will not necessarily produce the anti-symmetric S-shaped Stokes-V spectrum profile, characteristic for circular polarization generated by the Zeeman effect, but it can arise from scattering of a cloud outside the velocity-range of the maser. Indeed, non anti-symmetric Stokes-V spectra were observed by \citet{herpin:06}, but these can also be explained by a velocity gradient along the propagation path of the maser, or the lack of spatial resolution from the single-dish observations.
 
\subsection{H$_2$O masers}
\subsubsection{Simulations}
The relevant characteristic maser luminosities are tabulated in Table \ref{tab:water_temps}. We tabulate the relevant luminosities for individual hyperfine transitions as well as the blended line. Compared to the SiO maser, radiative interactions remain relatively weak with respect to magnetic interactions up to high maser luminosities. This is due to the much smaller line strength of this maser-transition. Because of this, the Zeeman effect will be the dominating polarizing mechanism up to high maser luminosities, and will thus follow Eq.~(\ref{eq:lte_circ}) up to high maser brightness. Linear polarization will also remain rather low because the isotropic decay will be a dominant de-polarizing entity up to $(T_b \Delta \Omega)_{\mathrm{sat}}$ (Ksr) at about $\sim 10^{10}$ Ksr. Strong linear polarization is thus only seen for the strongest masers. 

\begin{table}[t]
\centering
\caption{Characteristic maser luminosities for the $22$ GHz water maser}
\begin{tabular}{l c c }
\hline \hline
Transition & $(T_b \Delta \Omega)_{\mathrm{sat}}$ (Ksr) & $(T_b \Delta \Omega)_{\mathrm{mag.\ sat.}}/B$ (Ksr/mG) \\ \hline
$F=7-6$ & $7.2 \times 10^{9}$  & $3.1 \times 10^{10}$ \\
$F=6-5$ & $7.4 \times 10^{9}$  & $2.0 \times 10^{10}$ \\
$F=5-4$ & $7.5 \times 10^{9}$  & $2.3 \times 10^9$ \\
blend & $7.4 \times 10^{9}$  & $6.0 \times 10^9$ \\
\hline \hline
\end{tabular}
\label{tab:water_temps}
\end{table}

\begin{figure*}
    \centering
    \begin{subfigure}[b]{0.45\textwidth}
       \includegraphics[width=\textwidth]{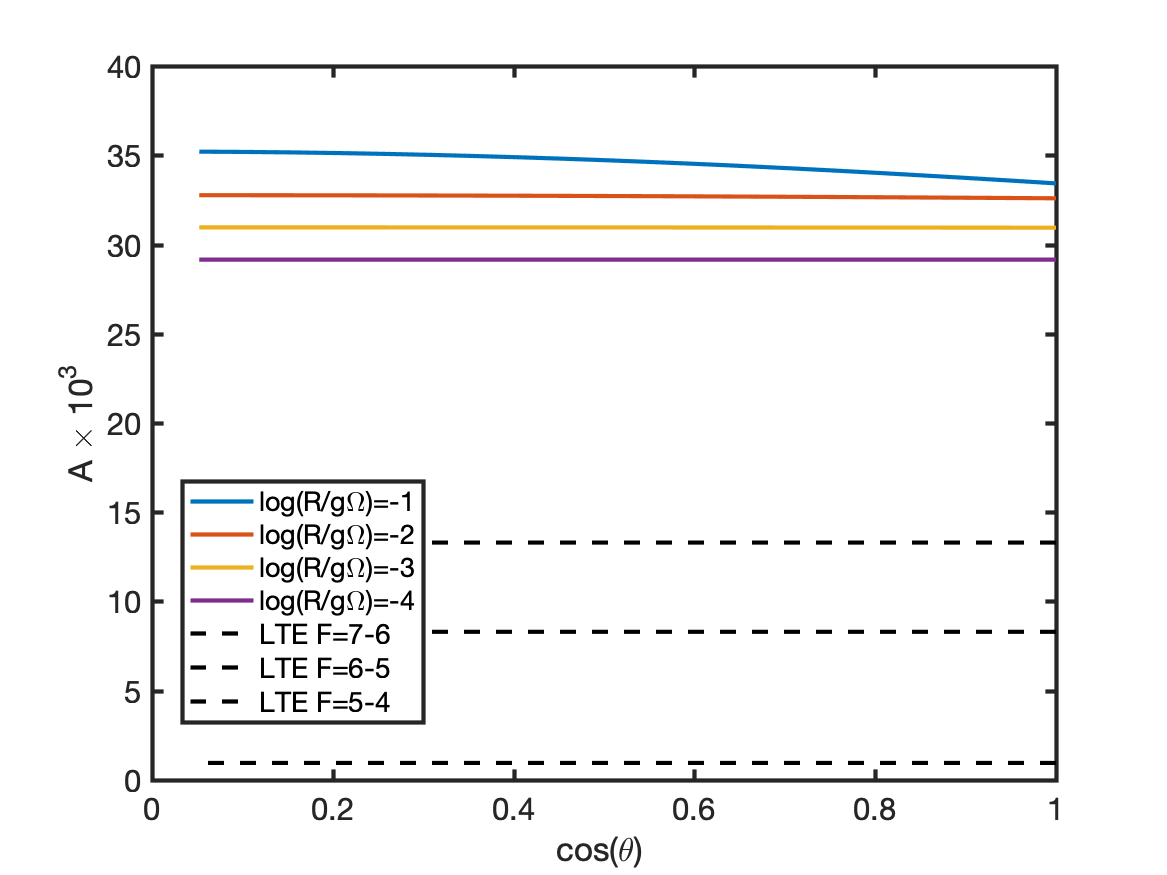} 
       \caption{}
    \end{subfigure}
    ~ 
    \begin{subfigure}[b]{0.45\textwidth}
       \includegraphics[width=\textwidth]{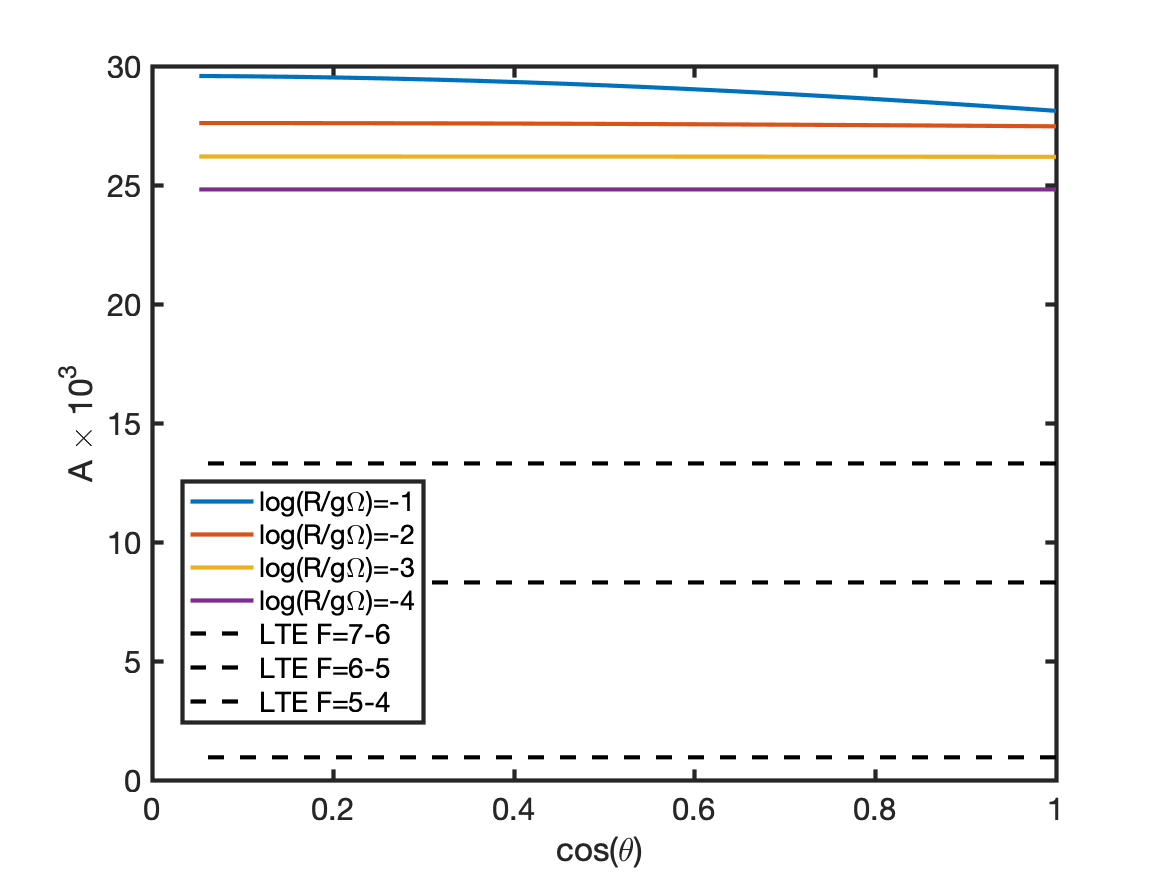} 
       \caption{}
    \end{subfigure}
     ~ 
    \begin{subfigure}[b]{0.45\textwidth}
       \includegraphics[width=\textwidth]{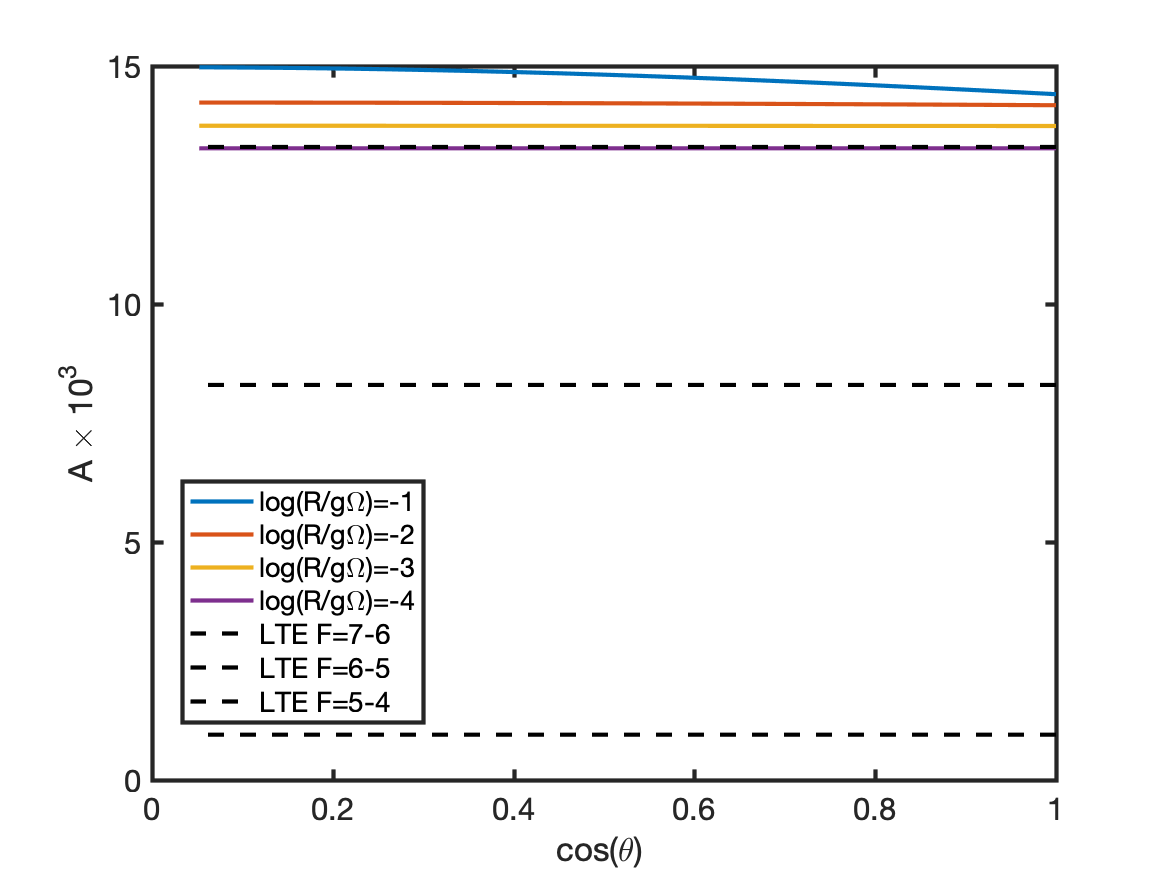}
       \caption{}
    \end{subfigure}
  \caption{The $A_{FF'}$ coefficients of an isotropically pumped water maser at $B=20$ G as a function of the magnetic field-propagation direction angle $\cos \theta$. Subfigures denote different thermal widths $v_{\mathrm{th}}=$ (a) $0.6$ km/s, (b) $1.0$ km/s and (c) $2.0$ km/s. Plots are given for different $\mathrm{log}(R/g\Omega)$. The LTE solutions (constant over $\cos \theta$) of the different hyperfine sub-transitions are denoted with a dotted line.}
  \label{fig:aff-water-const}
\end{figure*}

For the regime $T_b \Delta \Omega \ll (T_b \Delta \Omega)_{\mathrm{mag.\ sat}}$, the maser circular polarization can be described by Eq.~(\ref{eq:lte_circ}). An LTE analysis of the constant $A_{FF'}$, will give for the individual hyperfine transitions, $A_{76} = 13.3$, $A_{65} = 8.3$ and $A_{54} = 1.0$. An LTE analysis of a completely blended water-maser line gives $A_{\mathrm{blend}} = 8.2$. Fig.~\ref{fig:aff-water-const} shows the results of our full radiative transfer analysis of the circular polarization constants. Apart from the standard maser line-profile, water masers are further broadened by their hyperfine structure. This will lead to an overestimation of $\Delta v_L$. The dominant Zeeman effect though, will come from a single hyperfine transition. This produces higher Zeeman $A_{FF'}$ coefficients with respect to an LTE analysis. We observe that this effect is most pronounced for masers pumped at $v_{th} = 0.6$ km/s, where the hyperfine transitions are minimally mixed. At $v_{th} = 2.0$ km/s, the hyperfine broadening is negligible and the LTE value for the $A_{FF'}$ coefficient of the $F=7-6$ hyperfine transition is returned for the weakest masers. 

Paradoxically, the preferred pumping of the hyperfine component with the strongest Zeeman effect 

precession rate effectively increases, and the $(T_b \Delta \Omega)_{\mathrm{mag.\ sat.}}$ is out of reach. 
Thus, the change in molecular symmetry-axis that is associated with the production of linear 
polarization, occurs only for the strongest masers. It is therefore, that 
transitions with weaker Zeeman interactions
are associated with higher degrees of linear-polarization in the relevant brightness regime for
the water maser. We should note that this effect is not as 
pronounced for the high-temperature masers, where the broadening of the lines causes the other 
transitions to blend in more. The maser circular polarization is proportional to the strength 
of the Zeeman effect as expected from Eq.~(\ref{eq:lte_circ}). 



\subsubsection{Water maser polarization observations}
We start this subsection with a note on the water maser beaming solid angle. \citet{richards:11} performed water maser observations around AGB stars with e-MERLIN. For the brightest masers, a beaming solid angle in the order of $\Delta \Omega \sim 1.5 \times 10^{-3}$ sr was found. For some AGB stars, the geometrical masing mechanism seemed to be amplification bounded, but also hints of matter-bounded amplification were found for some sources. Line-profile analysis by \citet{vlemmings:05} revealed a $\Delta \Omega \sim 10^{-2}-10^{-3}$ sr for water masers around AGB stars. A line-profile analysis of the extremely strong water masers around Orion-KL, yielded beaming solid angles as low as $\Delta \Omega \sim 10^{-5}$ sr \citep{nedoluha:91}.

Water masers have been observed for their polarization on many occasions, both around evolved stars \citep{vlemmings:06a} and around star forming regions \citep{garay:89}. The most striking observations were the early observations of the flaring, very strong ``super" water maser \citep{garay:89, fiebig:89}. \citet{garay:89} report the 7-year monitoring of the polarization characteristics of the most powerful water maser feature of Orion-KL. Brightness temperatures over $T_b = 10^{15}$ K were observed with associated maser fluxes of $T_b \Delta \Omega \leq 10^{10}$ \citep{nedoluha:91}. High degrees of linear polarization up to $75\%$ were observed. Analysis of the relation between the polarization fraction and the maser brightness for the highly polarized strongest feature, shows a decline of polarization with the maser-intensity. This is in line with an anisotropically pumped maser at high brightness beyond $(T_b \Delta \Omega)_{\mathrm{mag.\ sat}}$. 

Circular polarization up to $\sim 2 \%$ for these strong maser flares was also detected \citep{fiebig:89}. \citet{fiebig:89} also included the masers of a number of other star-forming regions in their sample. Stokes-$V$ spectra show the characteristic S-shaped spectra, which is an anomaly for water masers that we found only to occur for preferably pumped water masers, where one hyperfine transition dominates the others. The spread in circular polarization fractions can be explained by the variable projection of the magnetic field ($10-100$ mG) onto the propagation axis, and variable magnetic fields in the sources. \citet{vlemmings:01,vlemmings:02} investigated the circular polarization of masers occurring in the circumstellar envelopes of late type stars. Magnetic fields around these masers are expected to be strong ($\sim$ G), and circular polarization should thus be detectable in the stronger maser features. Circular polarization up to $13\%$ is found, but this concerns a single outlier. The weaker masers show circular polarization up to $6\%$, which can be generated by a magnetic field of $\sim 400$ mG. Generally, circular polarization seems to decline with increasing maser-brightness, but this might an effect of the of the detection limit.  

A large sample of polarization observations of water masers comes from VLBI measurements around the high-mass star forming region W75N \citep{surcis:11}. Here, for 17 maser features, significant linear and circular polarization is found. Linear polarization tends to be small $<10\%$, but relatively high circular polarization ($<3\%$) is found. In part, the large fraction of highly circularly polarized masers is due to observational bias against weakly polarized masers. A similar scatter analysis as performed for the SiO maser sample of \citet{herpin:06}, assuming that the water maser is pumped isotropically with no hyperfine-preference, at a thermal width of $\Delta v_{\mathrm{th}} =1$ km/s, and the magnetic field is randomly oriented per maser, shows that a magnetic field of $\sim 200$ mG best reproduces the obtained linear-to-circular polarization distribution (see Fig.~\ref{fig:lincirc}). 

\section{Conclusions}
In this paper, we present CHAMP, a program that performs one-dimensional numerical maser polarization simulations of non-paramagnetic molecules. Simulations are possible for masers with arbitrary high angular momentum transitions. Also, multiple close-lying hyperfine transitions that contribute to the same maser can be included in our modeling. Simulation of the polarization of complex and highly excited masers will become more relevant in the era of ALMA and its full polarization capabilities.

Illustrative calculations of the SiO and water masers reveal the following general observations about the polarization of masers 
\begin{itemize}
  \item Linear polarization is mostly absent when the rate of stimulated emission is smaller than the isotropic decay ($T_b \Delta \Omega < (T_b \Delta \Omega)_{\mathrm{sat}}$). If polarization occurs for such weak masers, alternative polarizing mechanisms are in play. Circular polarization however, is present for such weak masers and comes from the Zeeman effect. An LTE analysis of the Zeeman effect will give a reasonable estimate of the polarizing effects, but this approximation worsens with the maser brightness.
  \item The 90$^o$ polarization angle flip at the magic angle, $\theta_m$, predicted by GKK73, is sharp only in the limit, $g\Omega \gg R$, when the magnetic precession rate is far greater than the rate of stimulated emission. However, for $g\Omega / R < 100$, the 90$^o$-flip bluntens and significant polarization is found also at propagation at the magic angle, $\theta_m$.  
  \item Anisotropic pumping of a maser can lead to arbitrarily high linear polarization fractions, but will only be weakly associated with circular polarization. Characteristic for an anisotropically pumped weak maser, is a linear growth in linear polarization fraction as a function of the maser-brightness. 
  \item Incident polarized seed radiation will maintain its polarization degree up until the rate of stimulated emission becomes comparable to the magnetic precession rate. From here, it will slowly converge to the standard isotropic polarization solution. 
  \item Circular polarization fractions are highest in the region where the rate of stimulated emission is in the same order as the magnetic precession rate. Circular polarization in this regime is associated with high linear polarization. Weak masers are weakly polarized, with a polarizing effect akin to thermal polarization.
  \item Overall polarization will drop strongly between the $J=1-0$ and $J=2-1$ transitions. The polarization of transitions with increasing angular momentum will gradually deteriorate.
\end{itemize} 
A cursory overview of existing maser polarization observations leads to a reinforcement of the idea that highly-polarized SiO masers are the product of anisotropic pumping. A similar mechanism probably underlies the highly polarized water ``super'' maser at Orion that also showed a drop in polarization with maser-brightness, as predicted by our theories. We show that comparing randomly ($\theta$ and $T_b \Delta$) generated (at fixed $B$) $p_L-p_V$ scatter-plots to the observationally obtained $p_L-p_V$ scatter, can be a promising method to ascertain the overall magnetic field strength of a region with a large number of masers. Finally, we have found the variation of the polarization angle across a maser spectrum can be used as a proxy for the rate of stimulated emission. This would be an important additional measure to determine the maser saturation level and beaming angle, which are difficult to observe directly. 
\begin{acknowledgements} Support for this work was provided by the the Swedish Research Council (VR), and by the European Research Council under the European Union's Seventh Framework Programme (FP7/2007-2013), through the ERC consolidator grant agreement nr. 614264. Simulations were performed on resources at Chalmers Centre for Computational Science and Engineering (C3SE) provided by the Swedish National Infrastructure for Computing (SNIC).
 \end{acknowledgements}
\bibliography{../texlibs/lib}

\begin{thebibliography}{57}
\expandafter\ifx\csname natexlab\endcsname\relax\def\natexlab#1{#1}\fi

\bibitem[{Amiri {et~al.}(2012)Amiri, Vlemmings, Kemball, \&
  Van~Langevelde}]{amiri:12}
Amiri, N., Vlemmings, W., Kemball, A., \& Van~Langevelde, H. 2012, Astron.
  Astrophys., 538, A136

\bibitem[{Anderson {et~al.}(1999)Anderson, Bai, Bischof, Blackford, Demmel,
  Dongarra, Du~Croz, Greenbaum, Hammarling, McKenney, \& Sorensen}]{LAPACK}
Anderson, E., Bai, Z., Bischof, C., {et~al.} 1999, {LAPACK} Users' Guide, 3rd
  edn. (Philadelphia, PA: Society for Industrial and Applied Mathematics)

\bibitem[{Assaf {et~al.}(2013)Assaf, Diamond, Richards, \& Gray}]{assaf:13}
Assaf, K., Diamond, P., Richards, A., \& Gray, M. 2013, Mon. Not. R. Astron.
  Soc., 431, 1077

\bibitem[{Barvainis {et~al.}(1987)Barvainis, McIntosh, \&
  Predmore}]{barvainis:87}
Barvainis, R., McIntosh, G., \& Predmore, C.~R. 1987, Nature, 329, 613

\bibitem[{Baudry \& Diamond(1998)}]{baudry:98}
Baudry, A. \& Diamond, P. 1998, Astron. Astrophys., 331, 697

\bibitem[{Cotton {et~al.}(2004)Cotton, Mennesson, Diamond, Perrin, du~Foresto,
  Chagnon, van Langevelde, Ridgway, Waters, Vlemmings, {et~al.}}]{cotton:04}
Cotton, W., Mennesson, B., Diamond, P., {et~al.} 2004, Astron. Astrophys., 414,
  275

\bibitem[{Cotton {et~al.}(2011)Cotton, Ragland, \& Danchi}]{cotton:11}
Cotton, W., Ragland, S., \& Danchi, W. 2011, Astrophys. J., 736, 96

\bibitem[{Degl'Innocenti \& Landolfi(2006)}]{degl:06}
Degl'Innocenti, M.~L. \& Landolfi, M. 2006, Polarization in spectral lines,
  Vol. 307 (Springer Science \& Business Media)

\bibitem[{Deguchi \& Watson(1990)}]{deguchi:90}
Deguchi, S. \& Watson, W.~D. 1990, Astrophys. J., 354, 649

\bibitem[{Elitzur(1991)}]{elitzur:91}
Elitzur, M. 1991, Astrophys. J., 370, 407

\bibitem[{Elitzur(1992)}]{elitzur:92}
Elitzur, M. 1992, Annu. Rev. Astron. Astrophys., 30, 75

\bibitem[{Elitzur(1993)}]{elitzur:93}
Elitzur, M. 1993, Astrophys. J., 416, 256

\bibitem[{Elitzur(1995)}]{elitzur:95}
Elitzur, M. 1995, arXiv preprint astro-ph/9508007

\bibitem[{Elitzur(1998)}]{elitzur:98}
Elitzur, M. 1998, Astrophys. J., 504, 390

\bibitem[{Elitzur {et~al.}(1992)Elitzur, Hollenbach, \& McKee}]{elitzur:92b}
Elitzur, M., Hollenbach, D.~J., \& McKee, C.~F. 1992, Astrophys. J., 394, 221

\bibitem[{Fiebig \& G{\"u}sten(1989)}]{fiebig:89}
Fiebig, D. \& G{\"u}sten, R. 1989, Astron. Astrophys., 214, 333

\bibitem[{Fish \& Reid(2006)}]{fish:06}
Fish, V.~L. \& Reid, M.~J. 2006, Astrophys. J. Supp. S., 164, 99

\bibitem[{Garay {et~al.}(1989)Garay, Moran, \& Haschick}]{garay:89}
Garay, G., Moran, J., \& Haschick, A. 1989, Astrophys. J., 338, 244

\bibitem[{Goldreich {et~al.}(1973)Goldreich, Keeley, \& Kwan}]{goldreich:73}
Goldreich, P., Keeley, D.~A., \& Kwan, J.~Y. 1973, Astrophys. J., 179, 111

\bibitem[{Gray(2003)}]{gray:03}
Gray, M. 2003, Mon. Not. R. Astronom. Soc., 343, L33

\bibitem[{Gray(2012)}]{gray:12}
Gray, M. 2012, Maser Sources in Astrophysics, Vol.~50 (Cambridge University
  Press)

\bibitem[{Gray \& Field(1995)}]{gray:95}
Gray, M. \& Field, D. 1995, Astron. Astrophys., 298, 243

\bibitem[{Herpin {et~al.}(2006)Herpin, Baudry, Thum, Morris, \&
  Wiesemeyer}]{herpin:06}
Herpin, F., Baudry, A., Thum, C., Morris, D., \& Wiesemeyer, H. 2006, Astron.
  Astrophys., 450, 667

\bibitem[{Houde(2014)}]{houde:14}
Houde, M. 2014, Astrophys. J., 795, 27

\bibitem[{Houde {et~al.}(2013)Houde, Hezareh, Jones, \& Rajabi}]{houde:13}
Houde, M., Hezareh, T., Jones, S., \& Rajabi, F. 2013, Astrophys. J., 764, 24

\bibitem[{Kemball \& Diamond(1997)}]{kemball:97}
Kemball, A. \& Diamond, P. 1997, Astrophys. J. Lett., 481, L111

\bibitem[{Kemball {et~al.}(2009)Kemball, Diamond, Gonidakis, Mitra, Yim, Pan,
  \& Chiang}]{kemball:09}
Kemball, A.~J., Diamond, P.~J., Gonidakis, I., {et~al.} 2009, Astrophys. J.,
  698, 1721

\bibitem[{Lankhaar {et~al.}(2016)Lankhaar, Groenenboom, \& van~der
  Avoird}]{lankhaar:16}
Lankhaar, B., Groenenboom, G.~C., \& van~der Avoird, A. 2016, J. Chem. Phys.,
  145, 244301

\bibitem[{Lankhaar {et~al.}(2018)Lankhaar, Vlemmings, Surcis, van Langevelde,
  Groenenboom, \& van~der Avoird}]{lankhaar:18}
Lankhaar, B., Vlemmings, W., Surcis, G., {et~al.} 2018, Nat. Astron., 2, 145

\bibitem[{Nedoluha \& Watson(1990)}]{nedoluha:90}
Nedoluha, G.~E. \& Watson, W.~D. 1990, Astrophys. J., 354, 660

\bibitem[{Nedoluha \& Watson(1991)}]{nedoluha:91}
Nedoluha, G.~E. \& Watson, W.~D. 1991, Astrophys. J., 367, L63

\bibitem[{Nedoluha \& Watson(1992)}]{nedoluha:92}
Nedoluha, G.~E. \& Watson, W.~D. 1992, Astrophys. J., 384, 185

\bibitem[{Nedoluha \& Watson(1994)}]{nedoluha:94}
Nedoluha, G.~E. \& Watson, W.~D. 1994, Astrophys. J., 423, 394

\bibitem[{P{\'e}rez-S{\'a}nchez \& Vlemmings(2013)}]{perez:13}
P{\'e}rez-S{\'a}nchez, A. \& Vlemmings, W. 2013, Astron. Astrophys., 551, A15

\bibitem[{Plambeck {et~al.}(2003)Plambeck, Wright, \& Rao}]{plambeck:03}
Plambeck, R., Wright, M., \& Rao, R. 2003, Astrophys. J., 594, 911

\bibitem[{Richards {et~al.}(2011)Richards, Elitzur, \& Yates}]{richards:11}
Richards, A., Elitzur, M., \& Yates, J. 2011, Astron. Astrophys., 525, A56

\bibitem[{Richter {et~al.}(2016)Richter, Kemball, \& Jonas}]{richter:16}
Richter, L., Kemball, A., \& Jonas, J. 2016, Mon. Not. R. Astron. Soc., 461,
  2309

\bibitem[{Robishaw {et~al.}(2008)Robishaw, Quataert, \& Heiles}]{robishaw:08}
Robishaw, T., Quataert, E., \& Heiles, C. 2008, Astrophys. J., 680, 981

\bibitem[{Sobolev {et~al.}(2018)Sobolev, Moran, Gray, Alakoz, Imai, Baan,
  Tolmachev, Samodurov, \& Ladeyshchikov}]{sobolev:18}
Sobolev, A., Moran, J., Gray, M., {et~al.} 2018, Astrophys. J., 856, 60

\bibitem[{Surcis {et~al.}(2011)Surcis, Vlemmings, Curiel, Kramer, Torrelles, \&
  Sarma}]{surcis:11}
Surcis, G., Vlemmings, W., Curiel, S., {et~al.} 2011, Astron. Astrophys., 527,
  A48

\bibitem[{Tobin {et~al.}(2019)Tobin, Kemball, \& Gray}]{tobin:19}
Tobin, T., Kemball, A., \& Gray, M. 2019, Astrophys. J., 871, 189

\bibitem[{Trung(2009)}]{trung:09}
Trung, D.-V. 2009, Mon. Not. R. Astronom. Soc., 399, 1495

\bibitem[{Vlemmings(2008)}]{vlemmings:08}
Vlemmings, W. 2008, Astron. Astrophys., 484, 773

\bibitem[{Vlemmings {et~al.}(2001)Vlemmings, Diamond, \&
  Van~Langevelde}]{vlemmings:01}
Vlemmings, W., Diamond, P., \& Van~Langevelde, H. 2001, Astron. Astrophys.,
  375, L1

\bibitem[{Vlemmings {et~al.}(2002)Vlemmings, Diamond, \&
  Van~Langevelde}]{vlemmings:02}
Vlemmings, W., Diamond, P., \& Van~Langevelde, H. 2002, Astron. Astrophys.,
  394, 589

\bibitem[{Vlemmings {et~al.}(2011{\natexlab{a}})Vlemmings, Humphreys, \&
  Franco-Hern{\'a}ndez}]{vlemmings:11b}
Vlemmings, W., Humphreys, E., \& Franco-Hern{\'a}ndez, R. 2011{\natexlab{a}},
  Astrophys. J., 728, 149

\bibitem[{Vlemmings {et~al.}(2017)Vlemmings, Khouri, Mart{\'\i}-Vidal, Tafoya,
  Baudry, Etoka, Humphreys, Jones, Kemball, O’Gorman,
  {et~al.}}]{vlemmings:17}
Vlemmings, W., Khouri, T., Mart{\'\i}-Vidal, I., {et~al.} 2017, Astron.
  Astrophys., 603, A92

\bibitem[{Vlemmings {et~al.}(2010)Vlemmings, Surcis, Torstensson, \&
  Van~Langevelde}]{vlemmings:10}
Vlemmings, W., Surcis, G., Torstensson, K., \& Van~Langevelde, H. 2010, Mon.
  Not. R. Astron. Soc., 404, 134

\bibitem[{Vlemmings {et~al.}(2011{\natexlab{b}})Vlemmings, Torres, \&
  Dodson}]{vlemmings:11a}
Vlemmings, W., Torres, R., \& Dodson, R. 2011{\natexlab{b}}, Astron.
  Astrophys., 529, A95

\bibitem[{Vlemmings {et~al.}(2006{\natexlab{a}})Vlemmings, Diamond, \&
  Imai}]{vlemmings:06b}
Vlemmings, W.~H., Diamond, P.~J., \& Imai, H. 2006{\natexlab{a}}, Nature, 440,
  58

\bibitem[{Vlemmings \& van Langevelde(2005)}]{vlemmings:05}
Vlemmings, W.~H. \& van Langevelde, H.~J. 2005, Astron. Astrophys., 434, 1021

\bibitem[{Vlemmings {et~al.}(2006{\natexlab{b}})Vlemmings, Diamond, van
  Langevelde, \& Torrelles}]{vlemmings:06a}
Vlemmings, W. H.~T., Diamond, P.~J., van Langevelde, H.~J., \& Torrelles, J.~M.
  2006{\natexlab{b}}, Astron. Astrophys., 448, 597

\bibitem[{Walker(1984)}]{walker:84}
Walker, R. 1984, Astrophys. J., 280, 618

\bibitem[{Watson \& Wyld(2001)}]{watson:01}
Watson, W. \& Wyld, H. 2001, Astrophys. J. Lett., 558, L55

\bibitem[{Western \& Watson(1983)}]{western:83c}
Western, L. \& Watson, W. 1983, Astrophys. J., 275, 195

\bibitem[{Western \& Watson(1984)}]{western:84}
Western, L. \& Watson, W. 1984, Astrophys. J., 285, 158

\bibitem[{Wiebe \& Watson(1998)}]{wiebe:98}
Wiebe, D. \& Watson, W. 1998, Astrophys. J. Lett., 503, L71

\end{thebibliography}
\begin{appendix}
\section{Appended figures}
\begin{figure*}[h!]
    \centering
    \begin{subfigure}[b]{0.45\textwidth}
       \includegraphics[width=\textwidth]{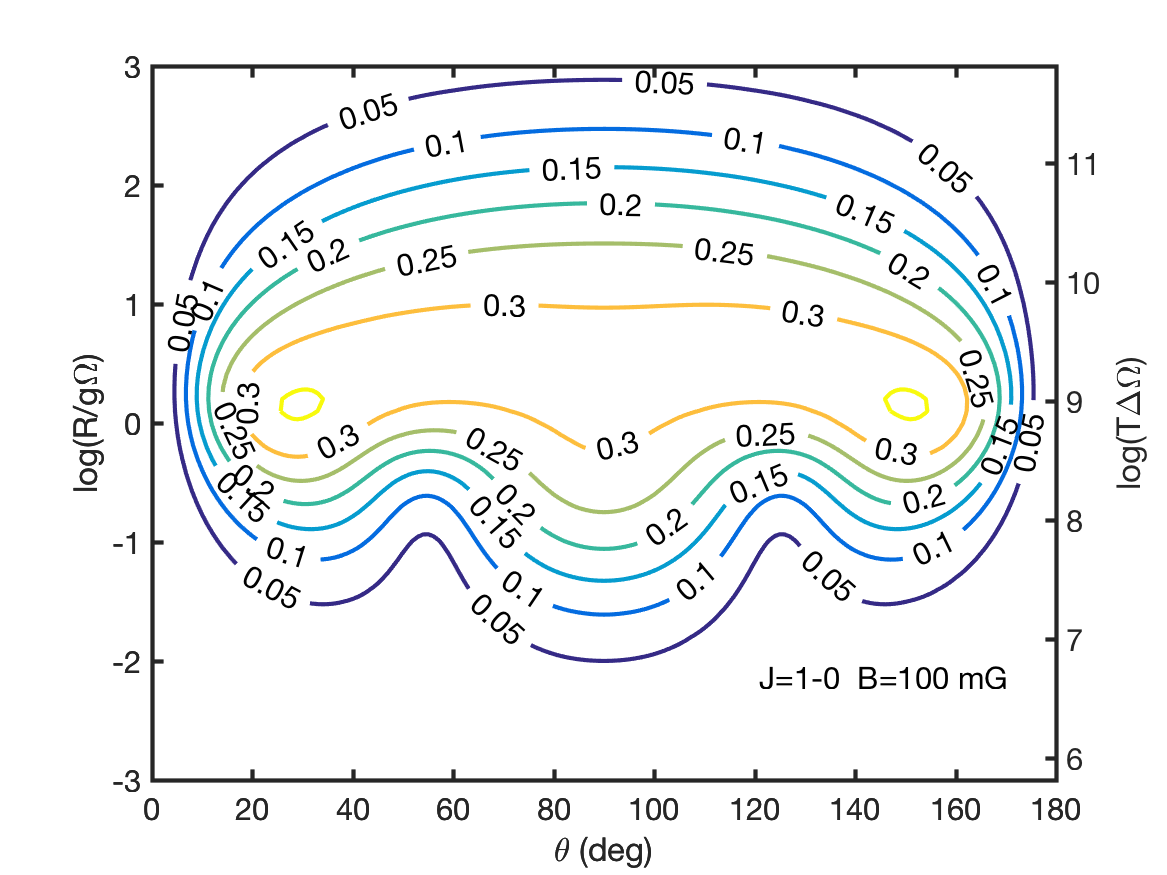} 
       \caption{}
    \end{subfigure}
    ~ 
    \begin{subfigure}[b]{0.45\textwidth}
       \includegraphics[width=\textwidth]{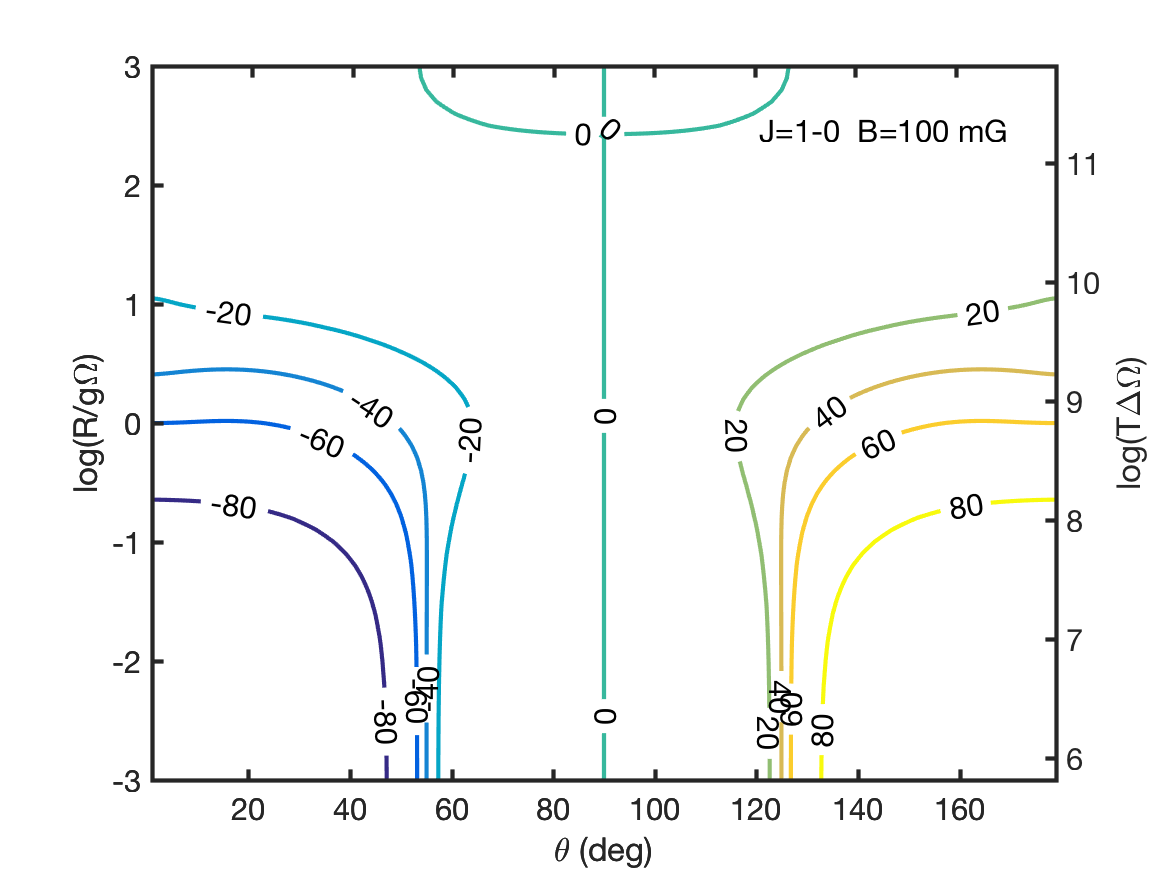} 
       \caption{}
    \end{subfigure}
     ~ 
    \begin{subfigure}[b]{0.45\textwidth}
      \includegraphics[width=\textwidth]{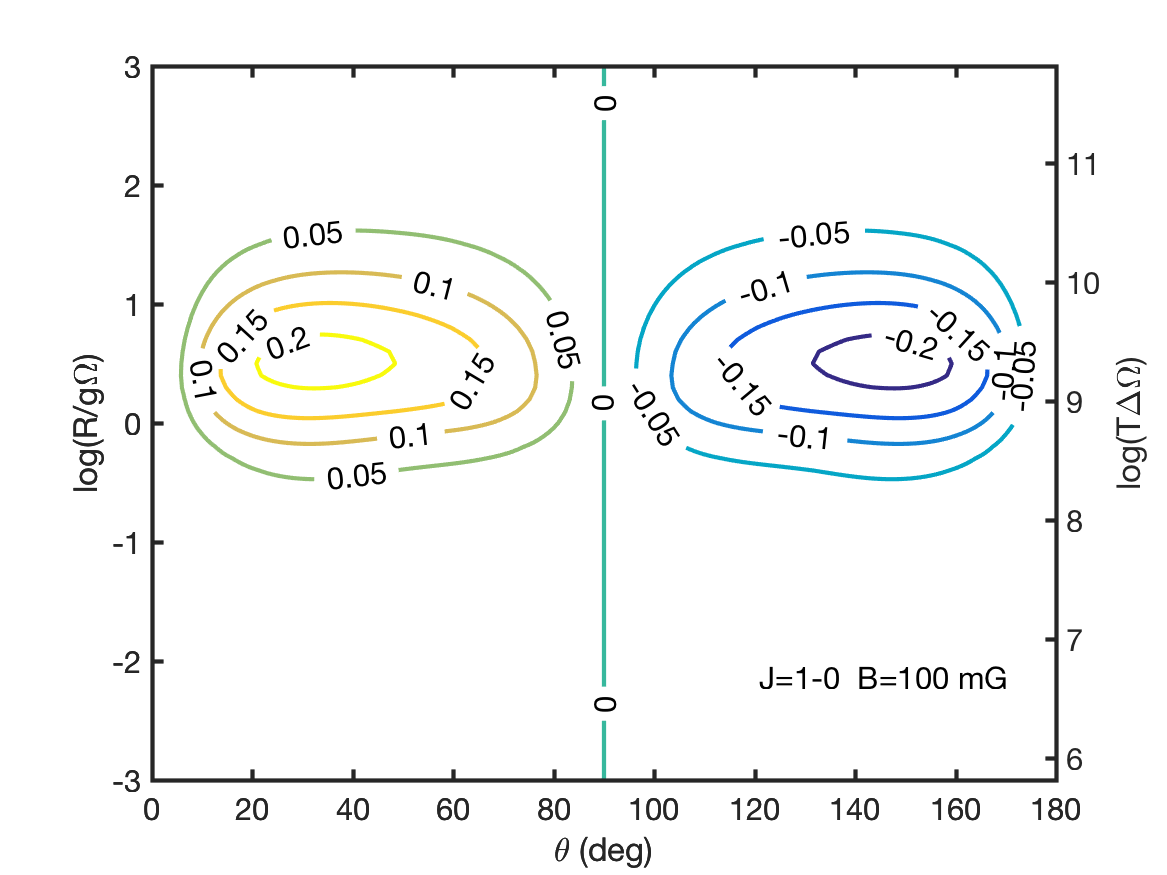}
      \caption{}
    \end{subfigure}
    ~ 
    \begin{subfigure}[b]{0.45\textwidth}
       \includegraphics[width=\textwidth]{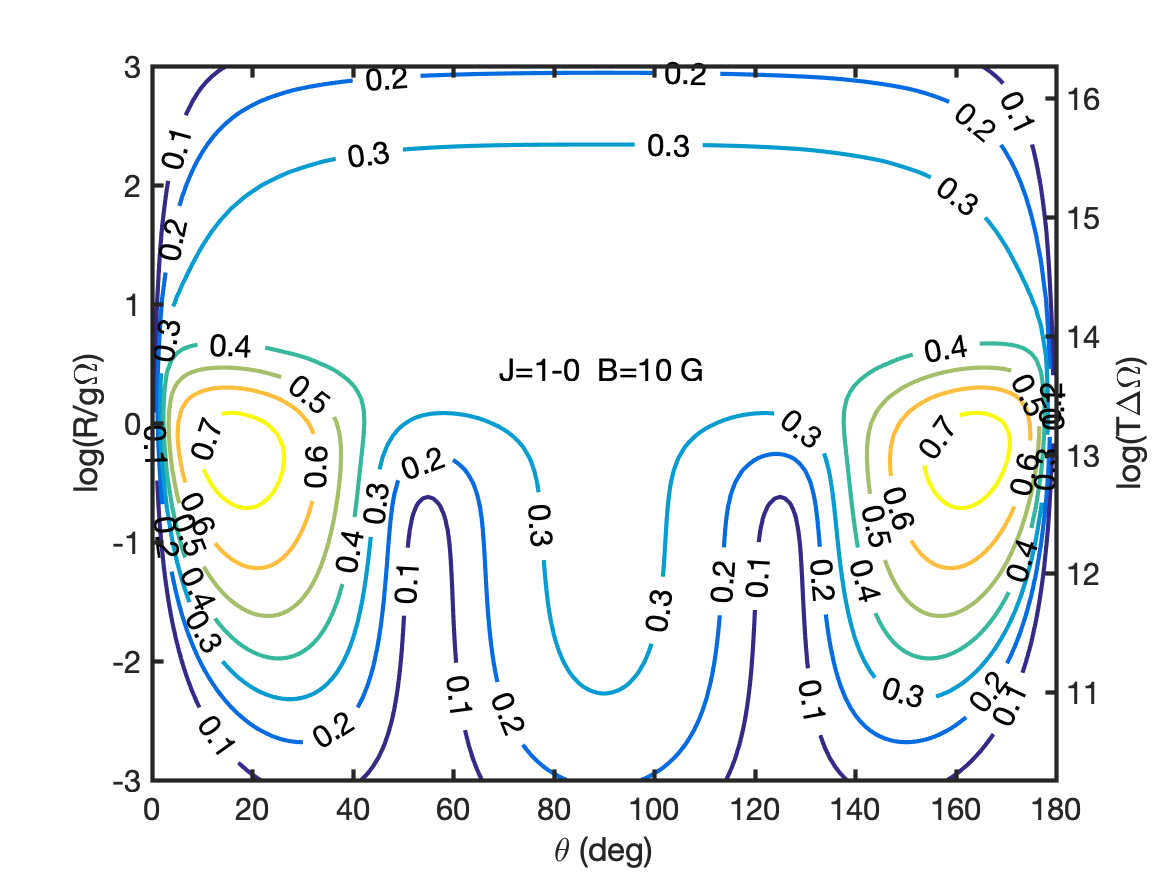} 
       \caption{}
    \end{subfigure}
    ~ 
    \begin{subfigure}[b]{0.45\textwidth}
       \includegraphics[width=\textwidth]{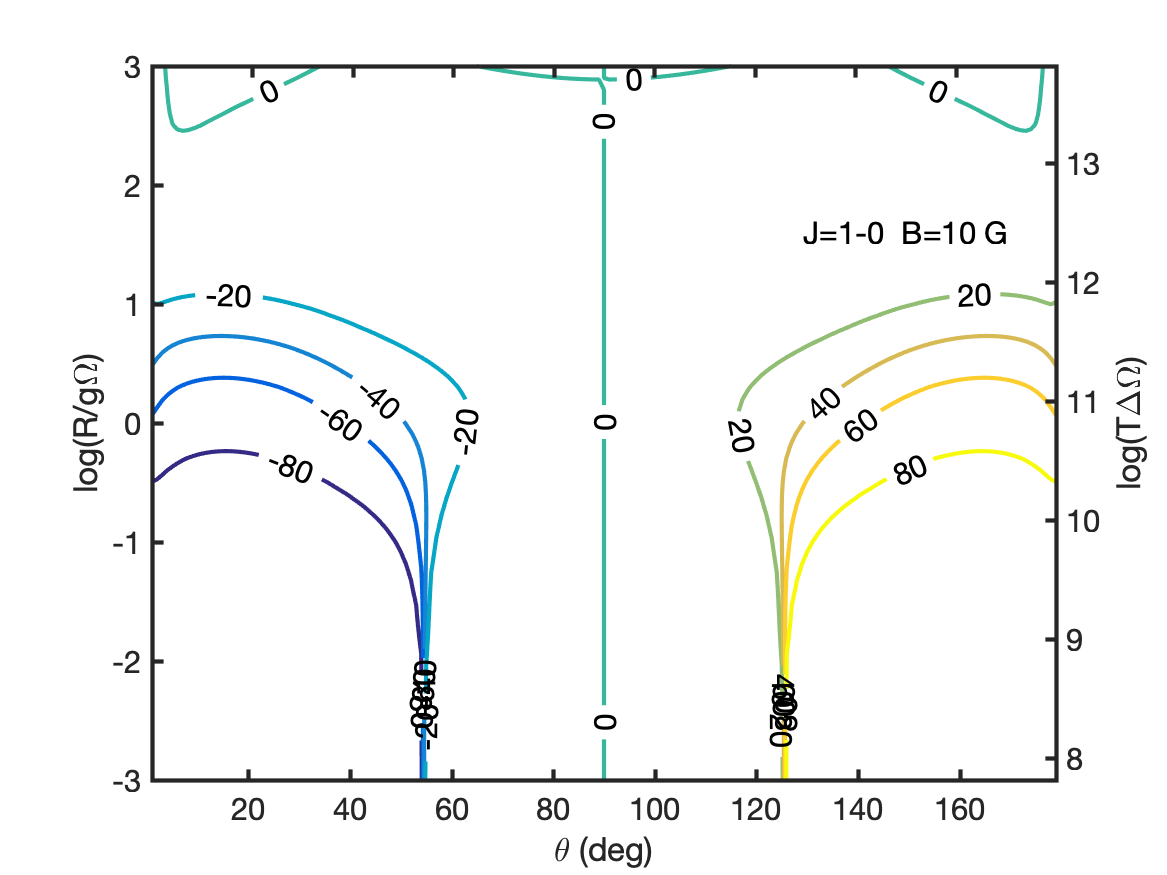} 
       \caption{}
    \end{subfigure}
     ~ 
    \begin{subfigure}[b]{0.45\textwidth}
      \includegraphics[width=\textwidth]{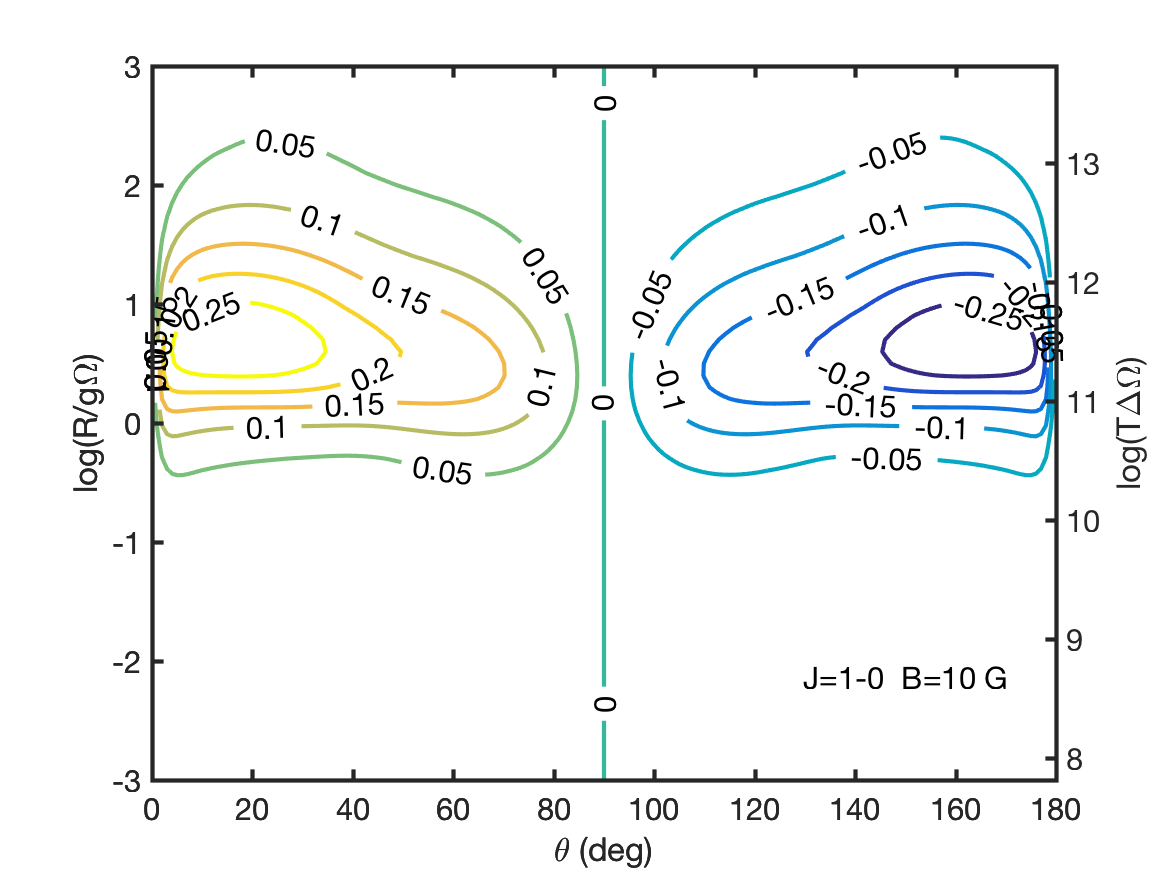}
      \caption{}
    \end{subfigure}
  \caption{Simulations of an isotropically pumped SiO maser. Linear polarization fraction (a,d) and angle (b,e) and circular polarization fraction (c,f). Magnetic field strength and transition angular momentum are denoted inside the figure.}
\end{figure*}

\begin{figure*}[h!]
  \centering
    \begin{subfigure}[b]{0.32\textwidth}
       \includegraphics[width=\textwidth]{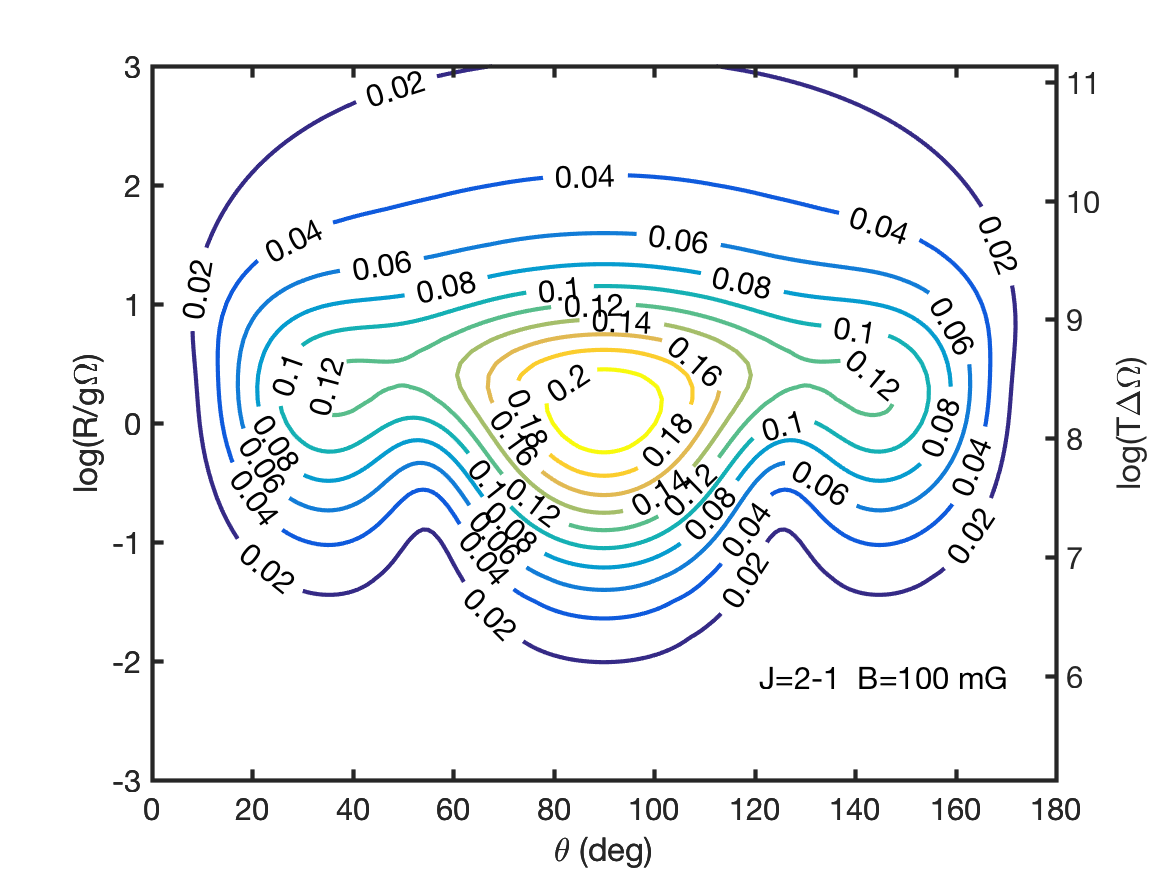}
       \caption{}
    \end{subfigure}
    ~
    \begin{subfigure}[b]{0.32\textwidth}
       \includegraphics[width=\textwidth]{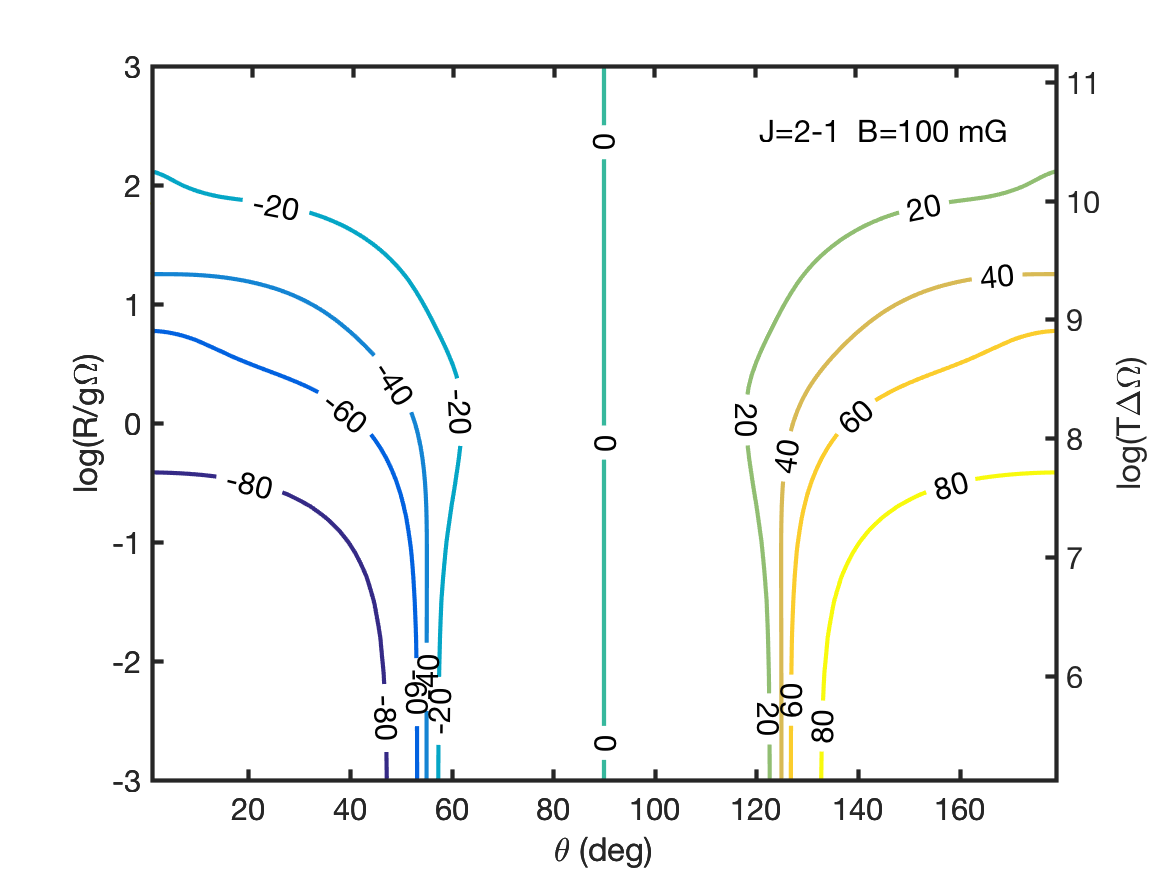}
       \caption{}
    \end{subfigure}
     ~
    \begin{subfigure}[b]{0.32\textwidth}
      \includegraphics[width=\textwidth]{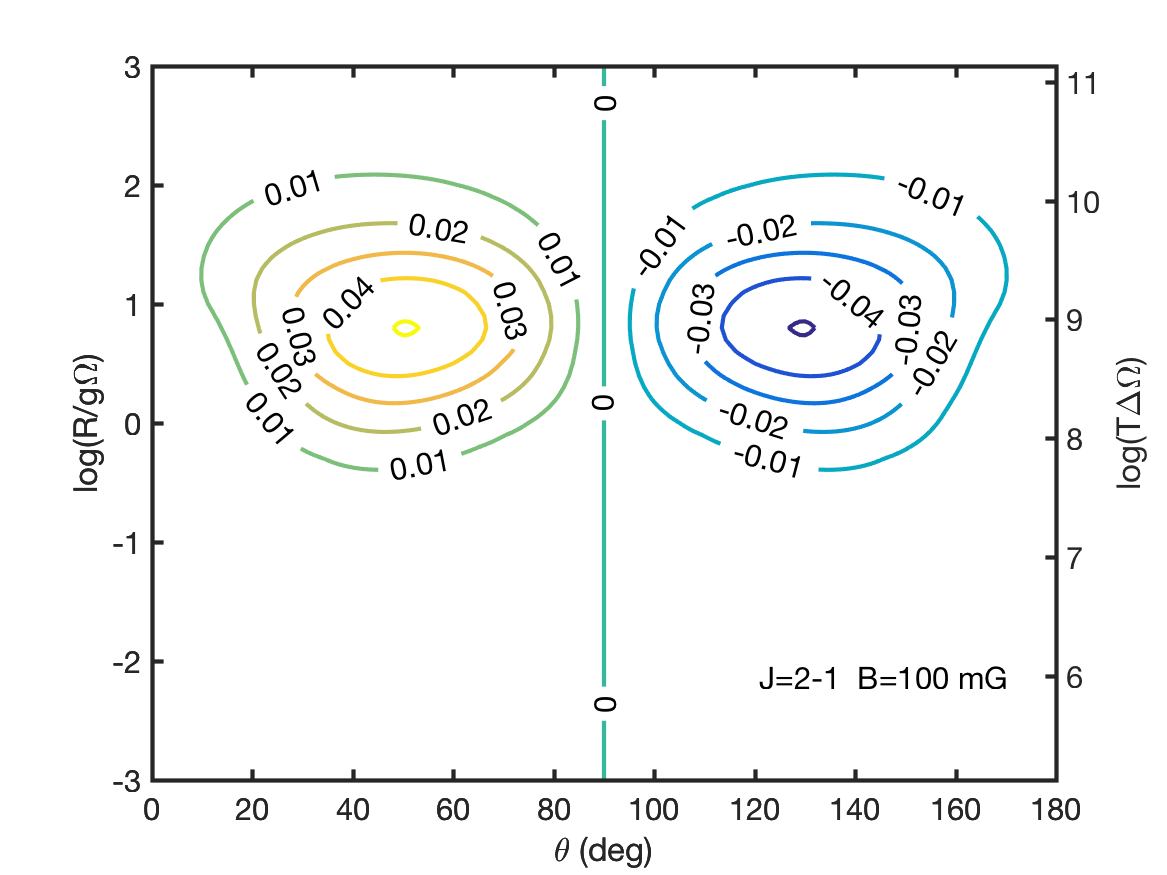}
      \caption{}
    \end{subfigure}
     ~
    \begin{subfigure}[b]{0.32\textwidth}
       \includegraphics[width=\textwidth]{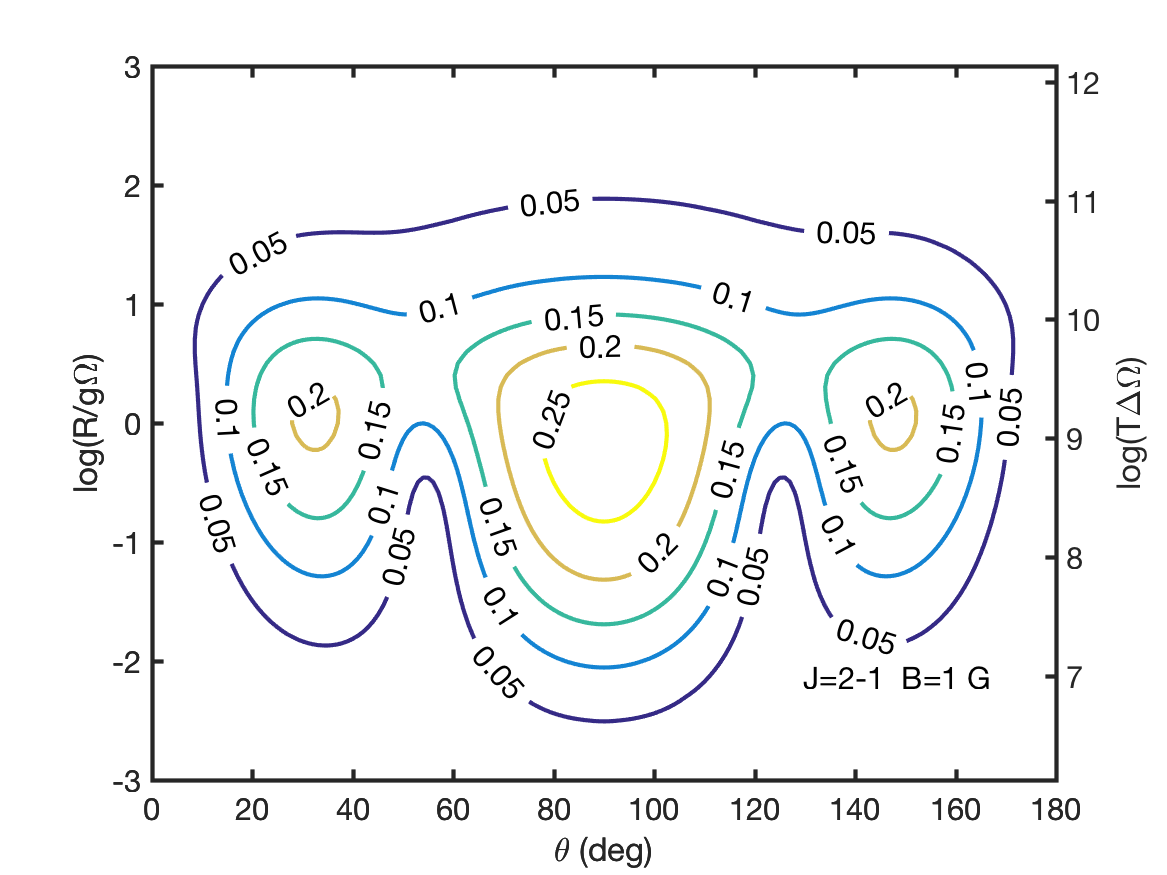}
       \caption{}
    \end{subfigure}
    ~
    \begin{subfigure}[b]{0.32\textwidth}
       \includegraphics[width=\textwidth]{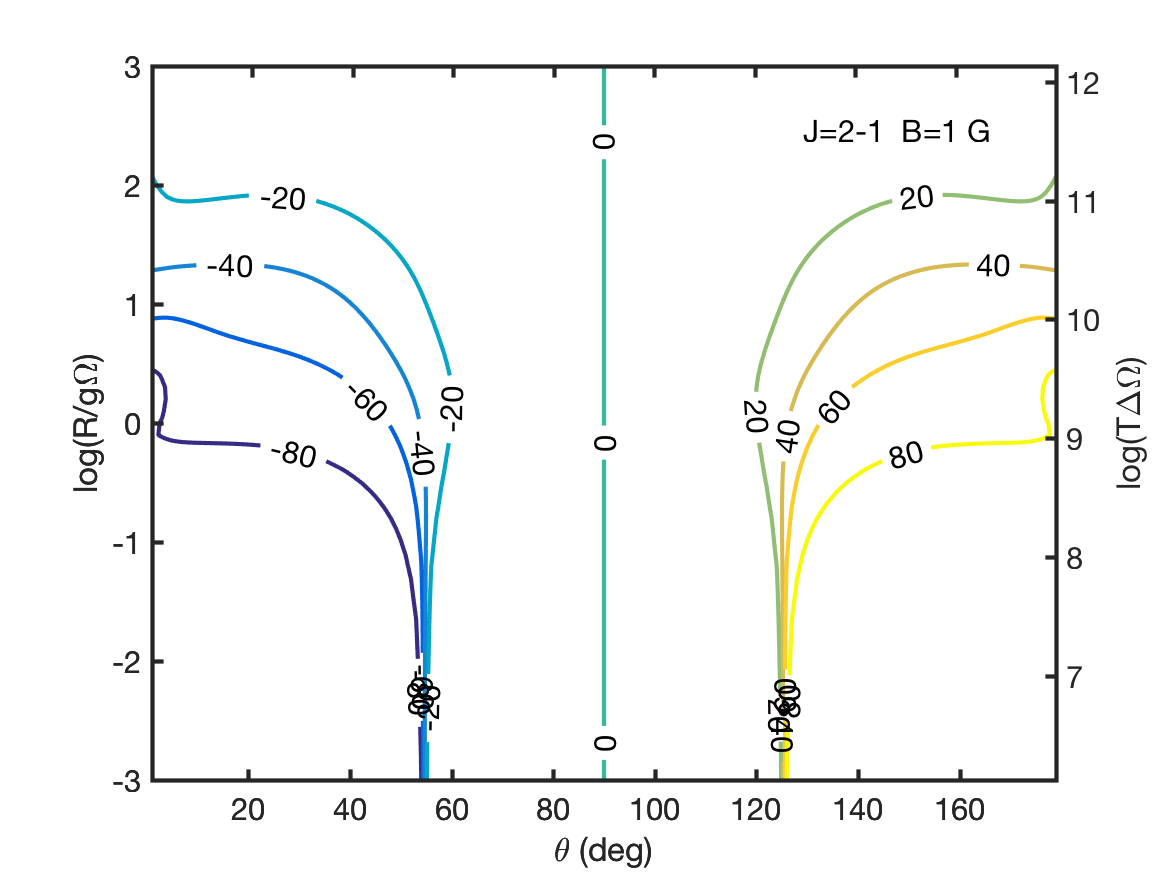}
       \caption{}
    \end{subfigure}
     ~
    \begin{subfigure}[b]{0.32\textwidth}
      \includegraphics[width=\textwidth]{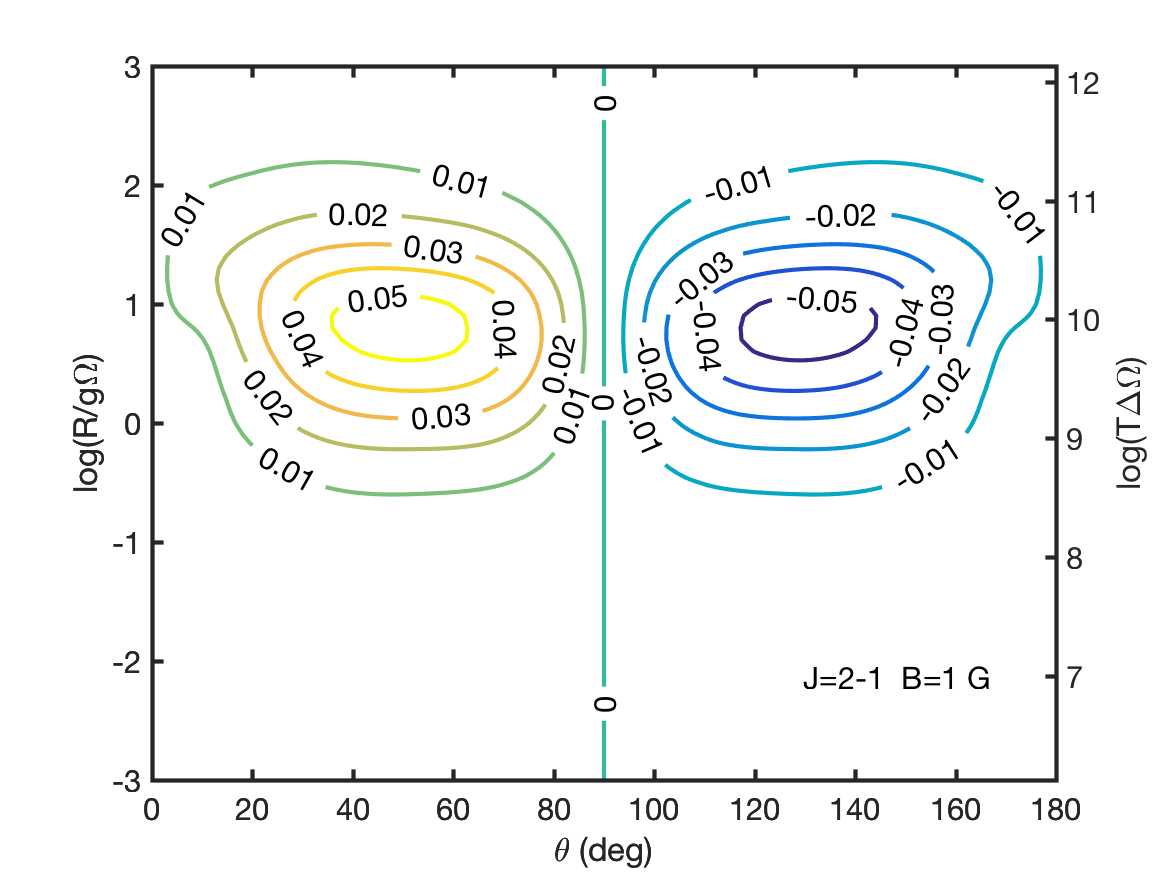}
      \caption{}
    \end{subfigure}
     ~
    \begin{subfigure}[b]{0.32\textwidth}
       \includegraphics[width=\textwidth]{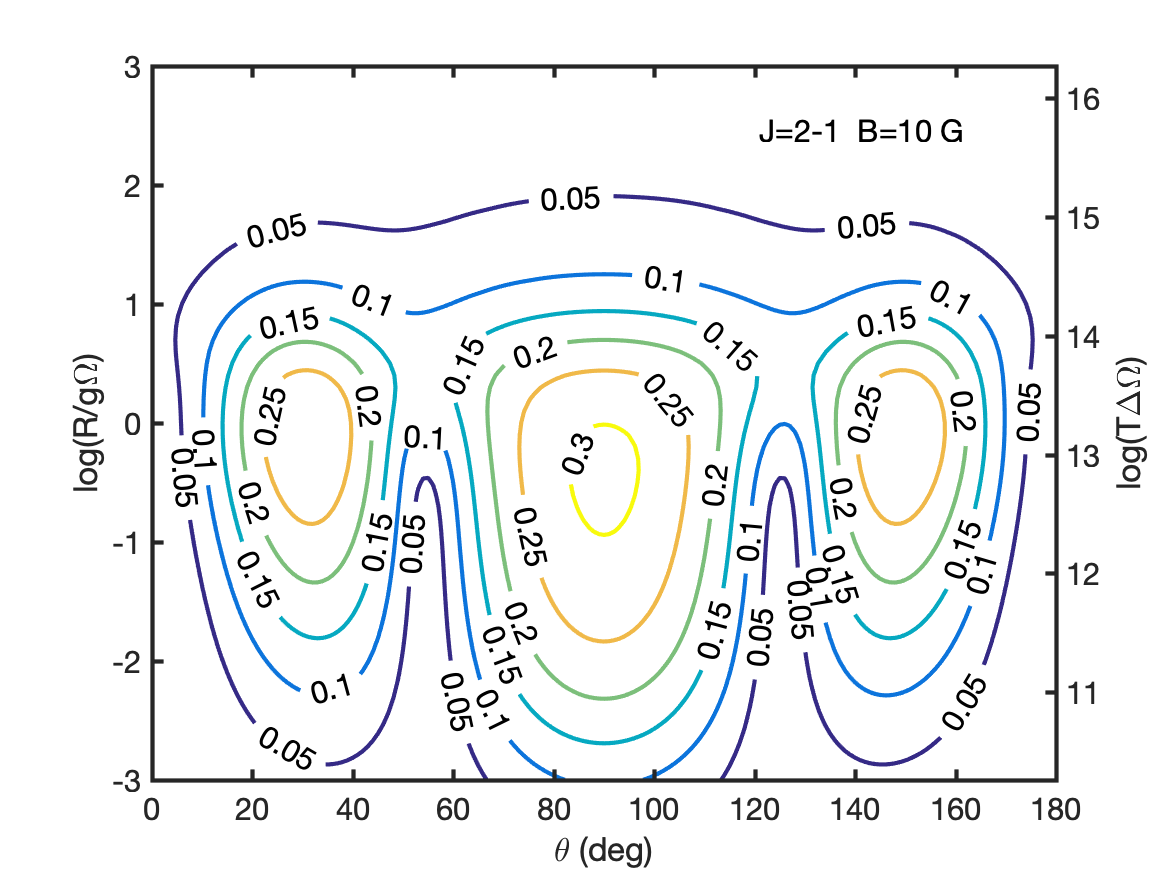}
       \caption{}
    \end{subfigure}
    ~
    \begin{subfigure}[b]{0.32\textwidth}
       \includegraphics[width=\textwidth]{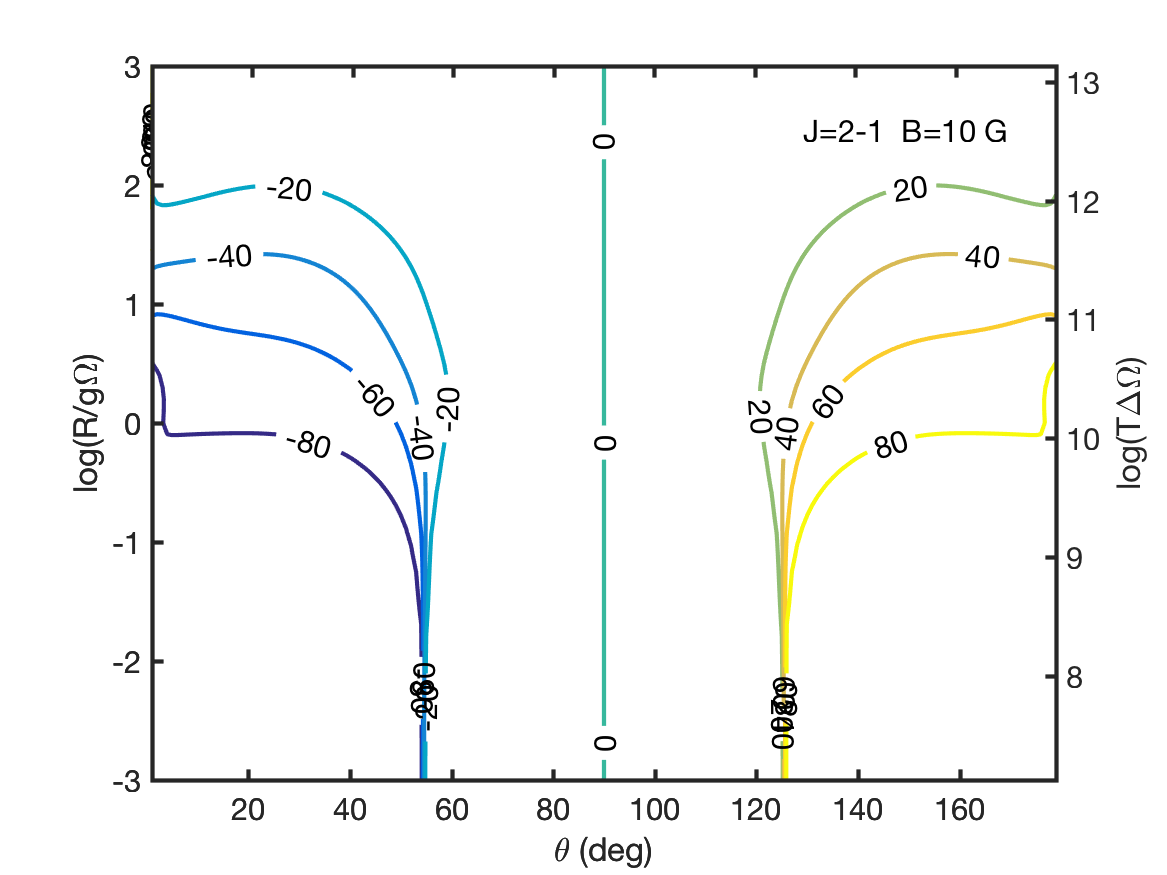}
       \caption{}
    \end{subfigure}
     ~
    \begin{subfigure}[b]{0.32\textwidth}
      \includegraphics[width=\textwidth]{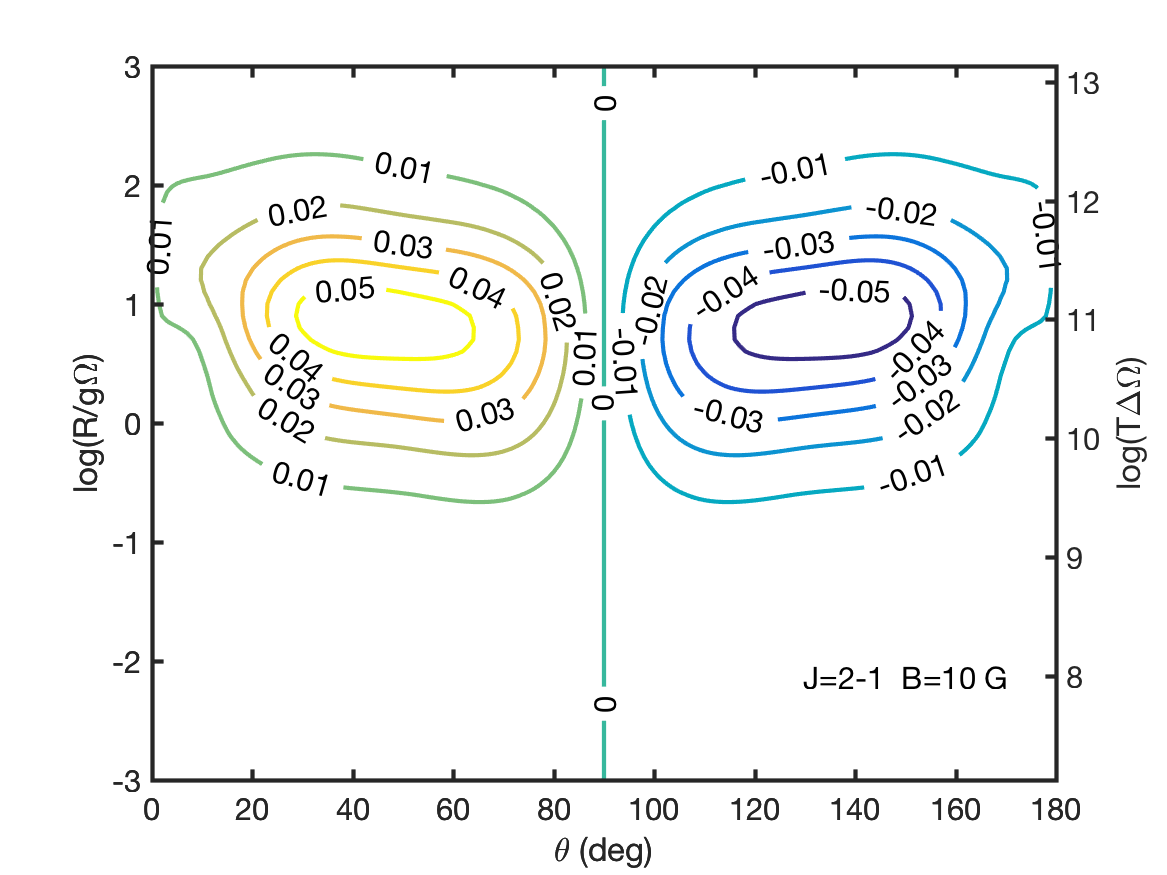}
      \caption{}
    \end{subfigure}
  \caption{Simulations of an isotropically pumped SiO maser. Linear polarization fraction (a,d,g) and angle (b,e,h) and circular polarization fraction (c,f,i). Magnetic field strength and transition angular momentum are denoted inside the figure.}
\end{figure*}

\begin{figure*}[h!]
    \centering
    \begin{subfigure}[b]{0.32\textwidth}
       \includegraphics[width=\textwidth]{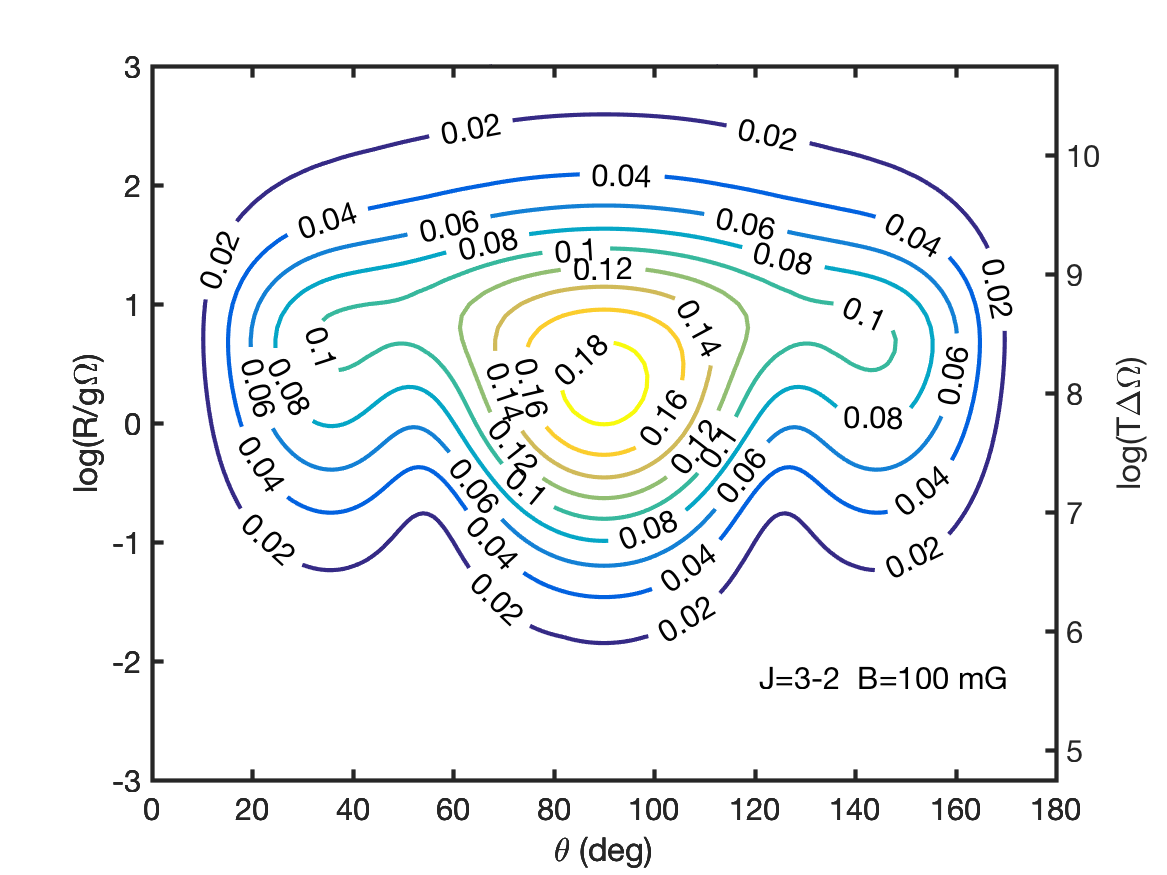}
       \caption{}
    \end{subfigure}
    ~
    \begin{subfigure}[b]{0.32\textwidth}
       \includegraphics[width=\textwidth]{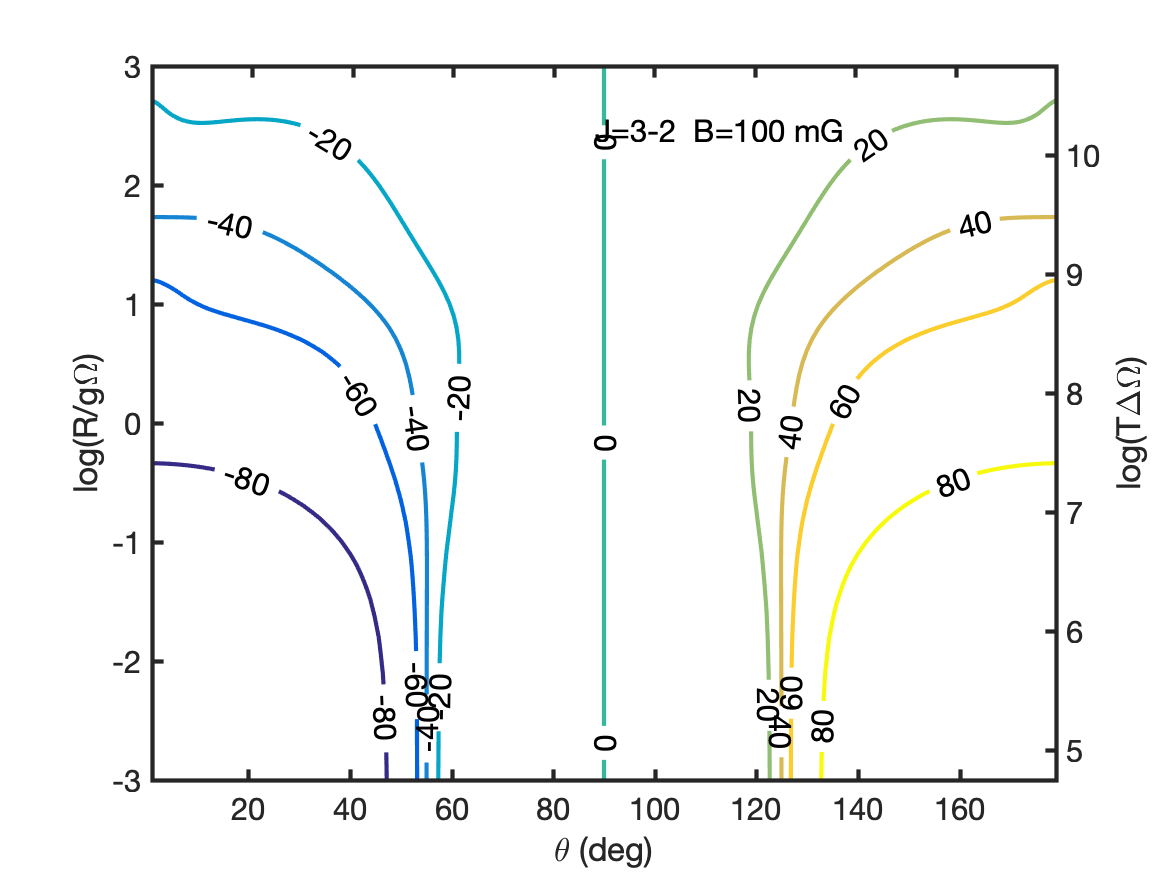}
       \caption{}
    \end{subfigure}
     ~
    \begin{subfigure}[b]{0.32\textwidth}
      \includegraphics[width=\textwidth]{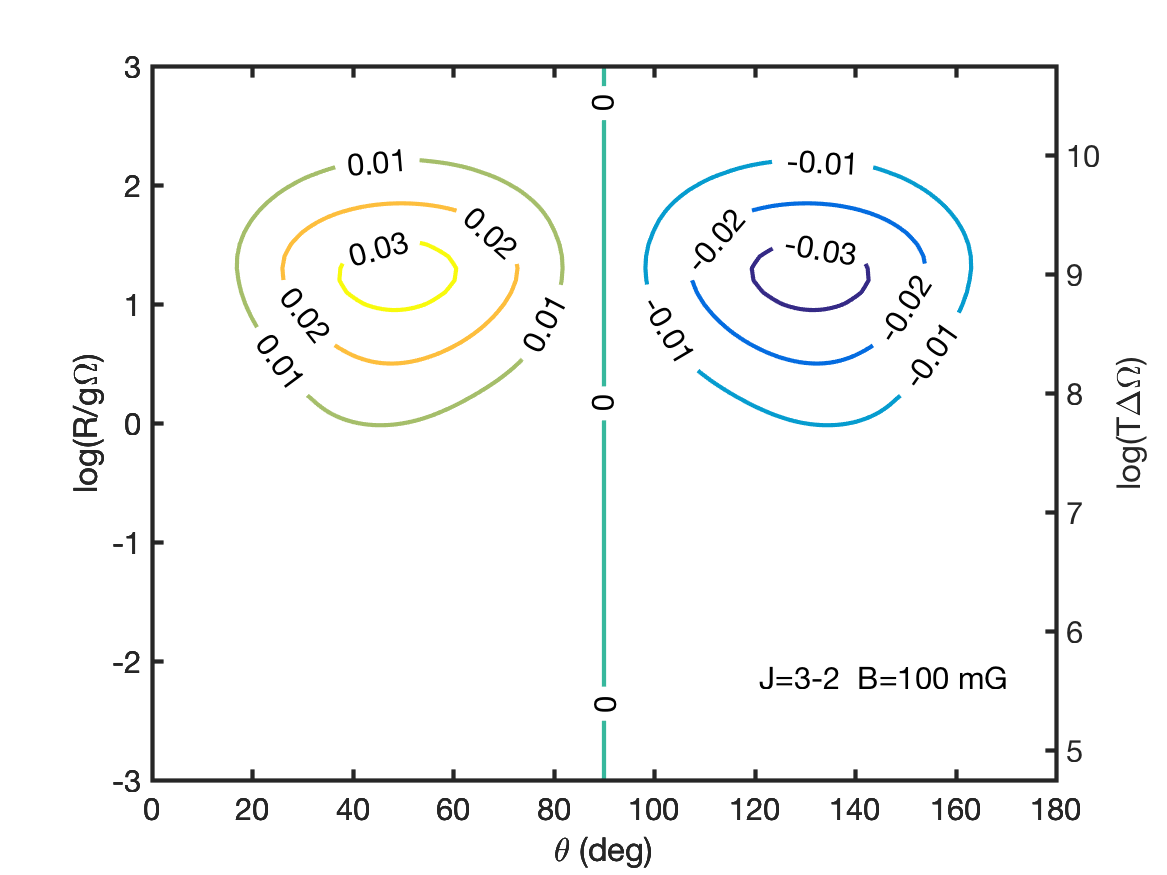}
      \caption{}
    \end{subfigure}
    ~
    \begin{subfigure}[b]{0.32\textwidth}
       \includegraphics[width=\textwidth]{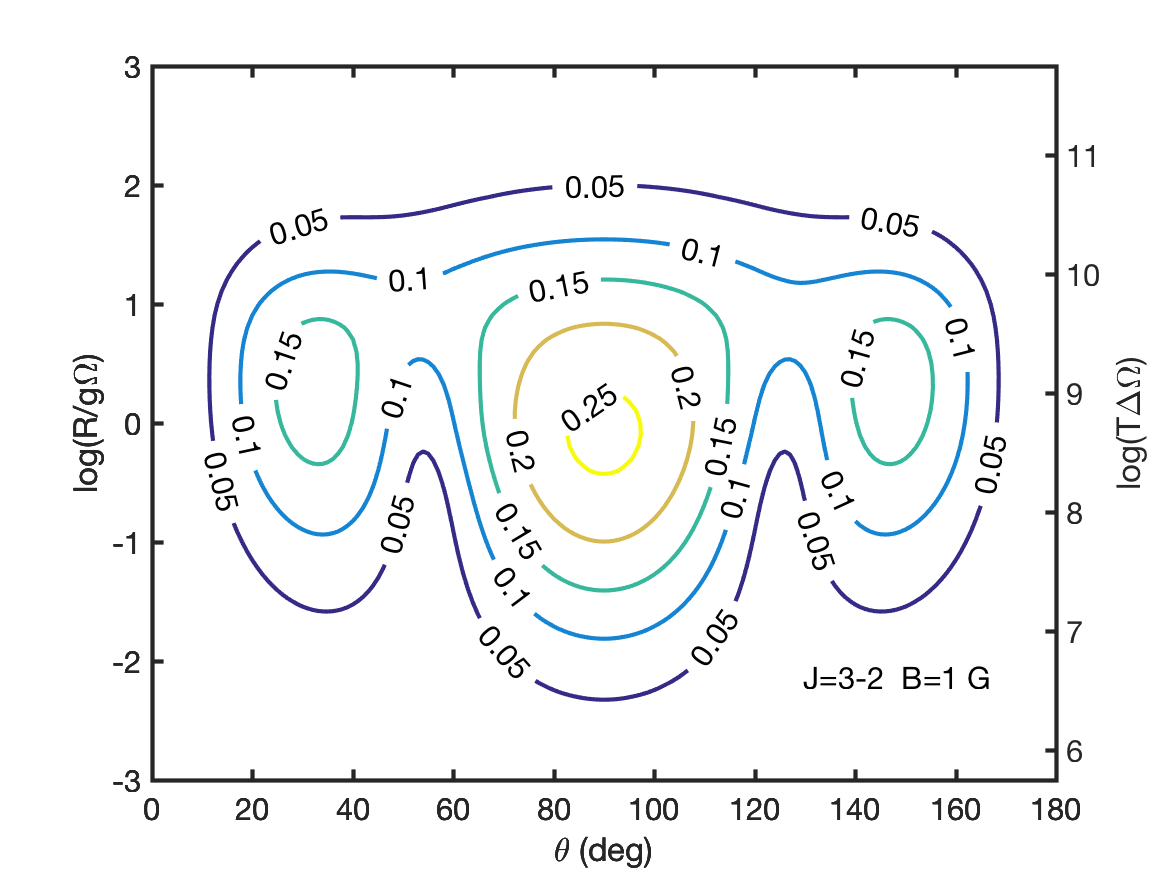}
       \caption{}
    \end{subfigure}
    ~
    \begin{subfigure}[b]{0.32\textwidth}
       \includegraphics[width=\textwidth]{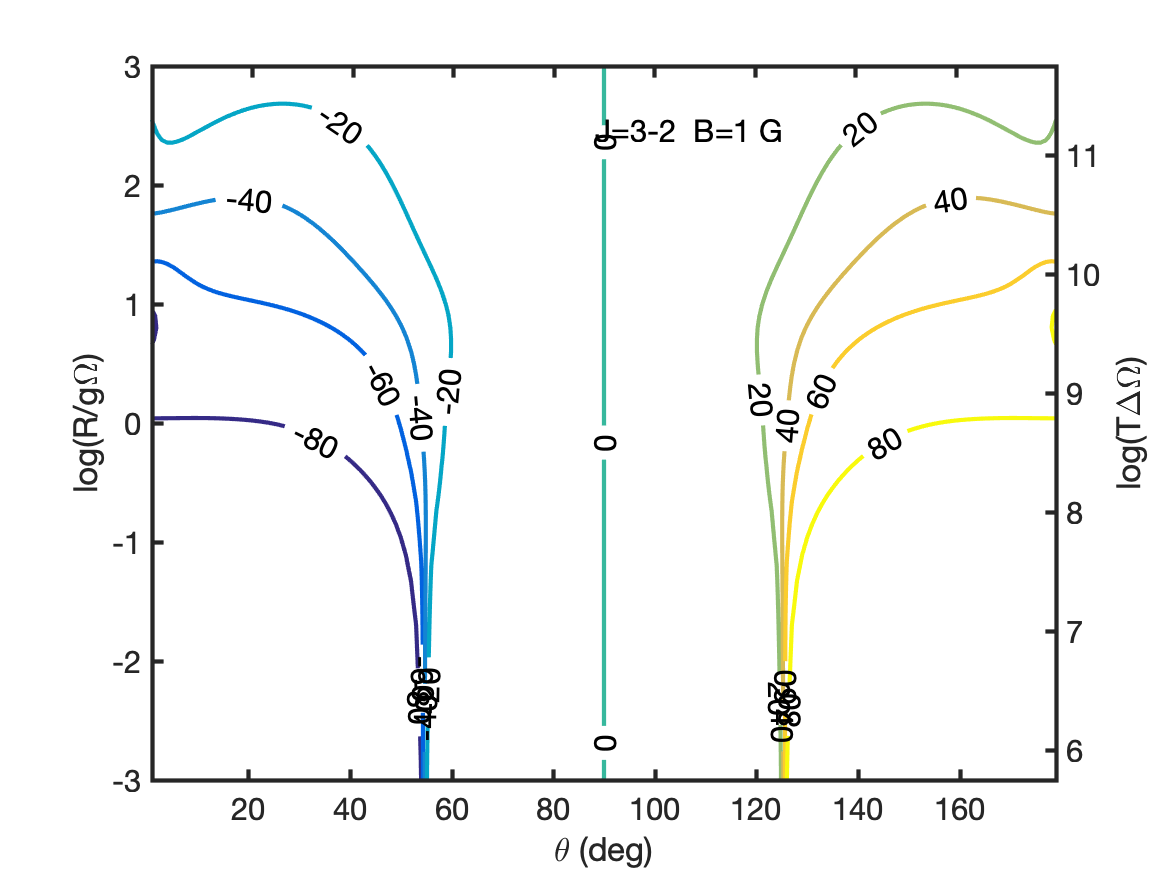}
       \caption{}
    \end{subfigure}
     ~
    \begin{subfigure}[b]{0.32\textwidth}
      \includegraphics[width=\textwidth]{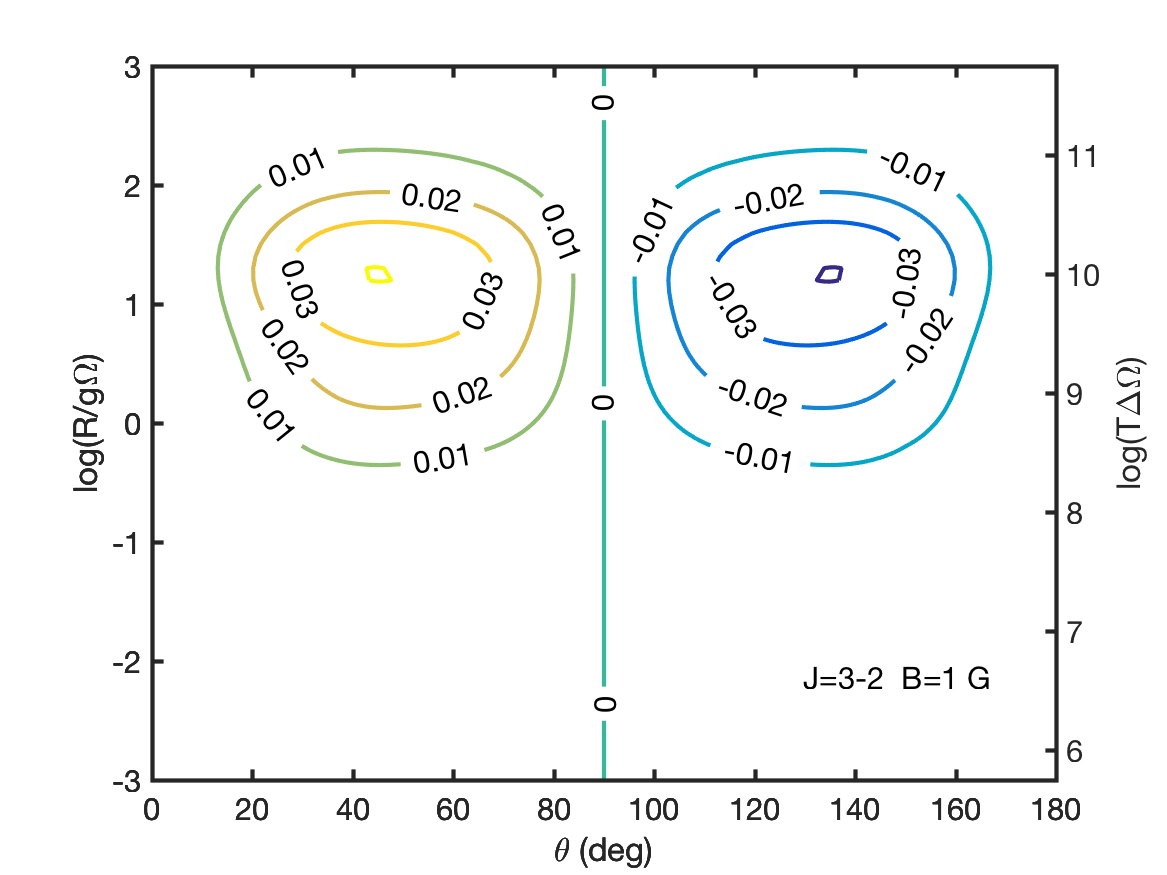}
      \caption{}
    \end{subfigure}
    ~
    \begin{subfigure}[b]{0.32\textwidth}
       \includegraphics[width=\textwidth]{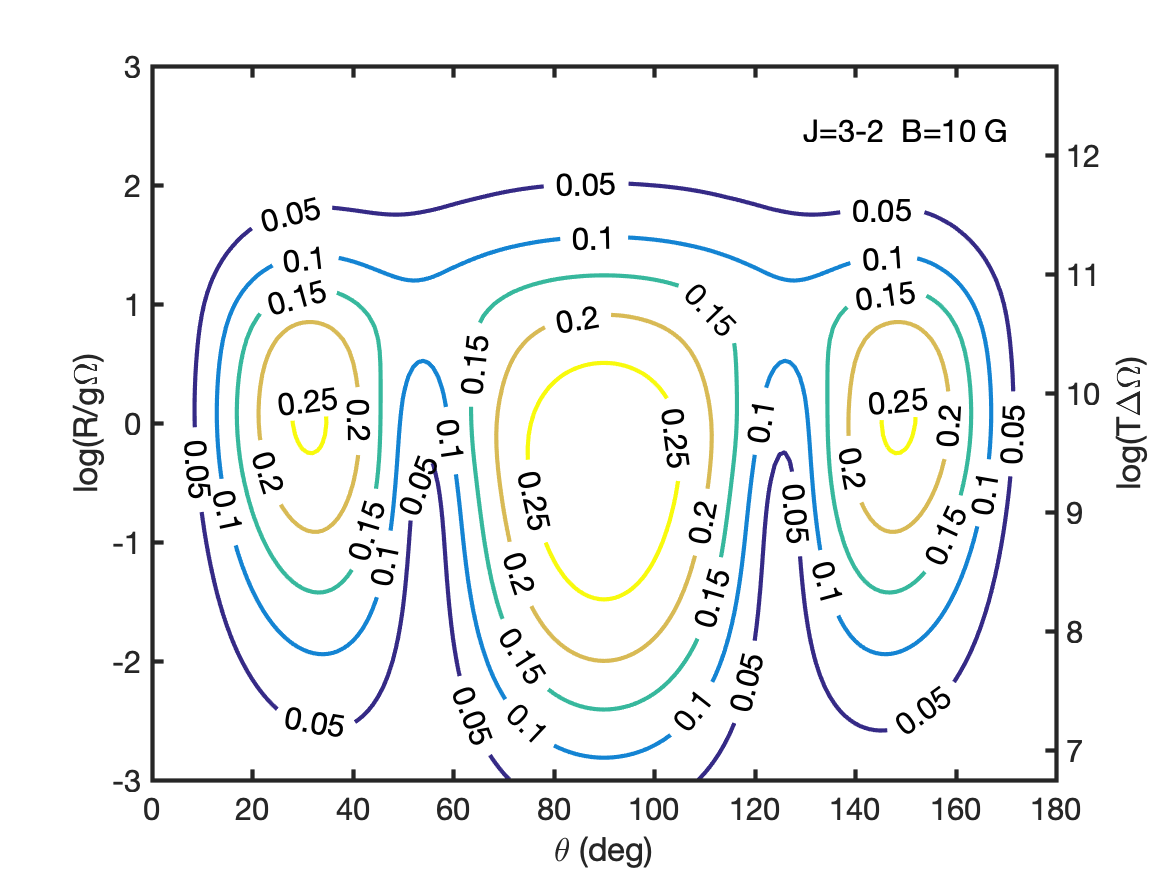}
       \caption{}
    \end{subfigure}
    ~
    \begin{subfigure}[b]{0.32\textwidth}
       \includegraphics[width=\textwidth]{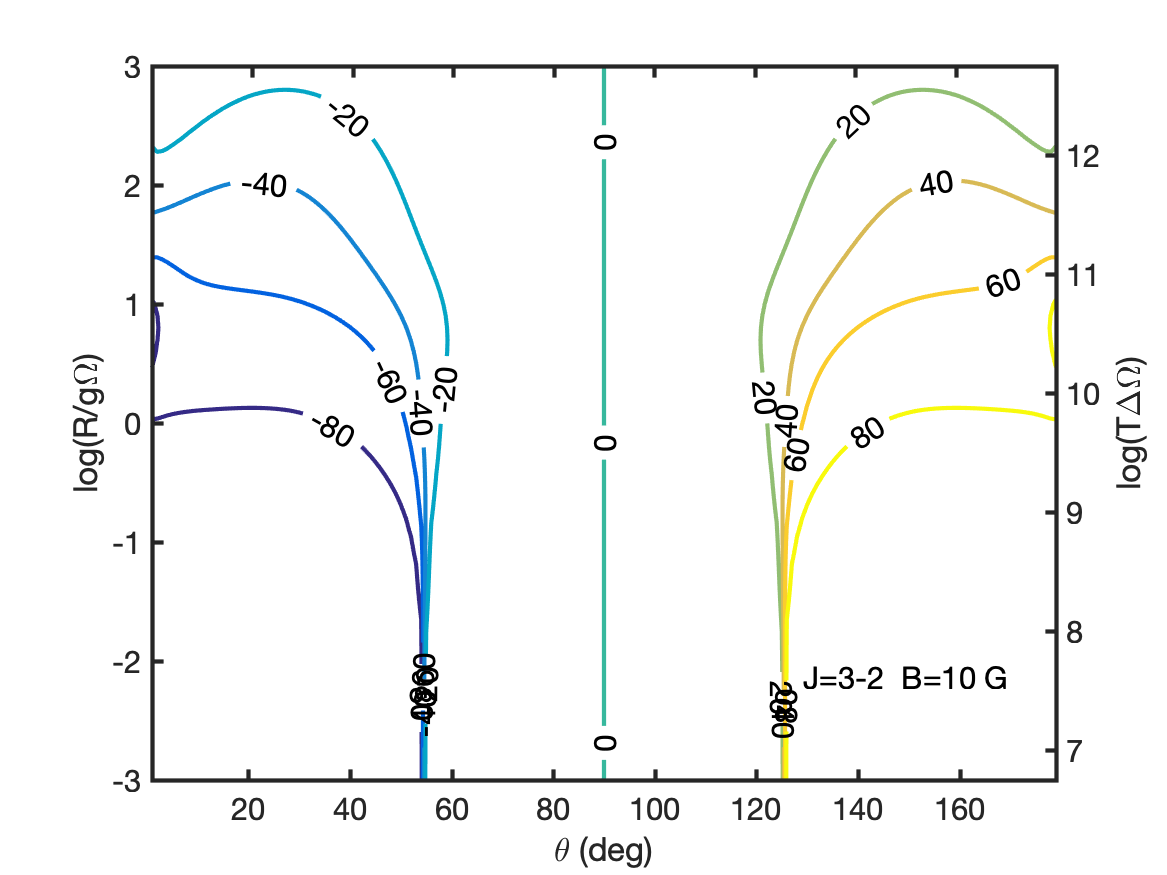}
       \caption{}
    \end{subfigure}
     ~
    \begin{subfigure}[b]{0.32\textwidth}
      \includegraphics[width=\textwidth]{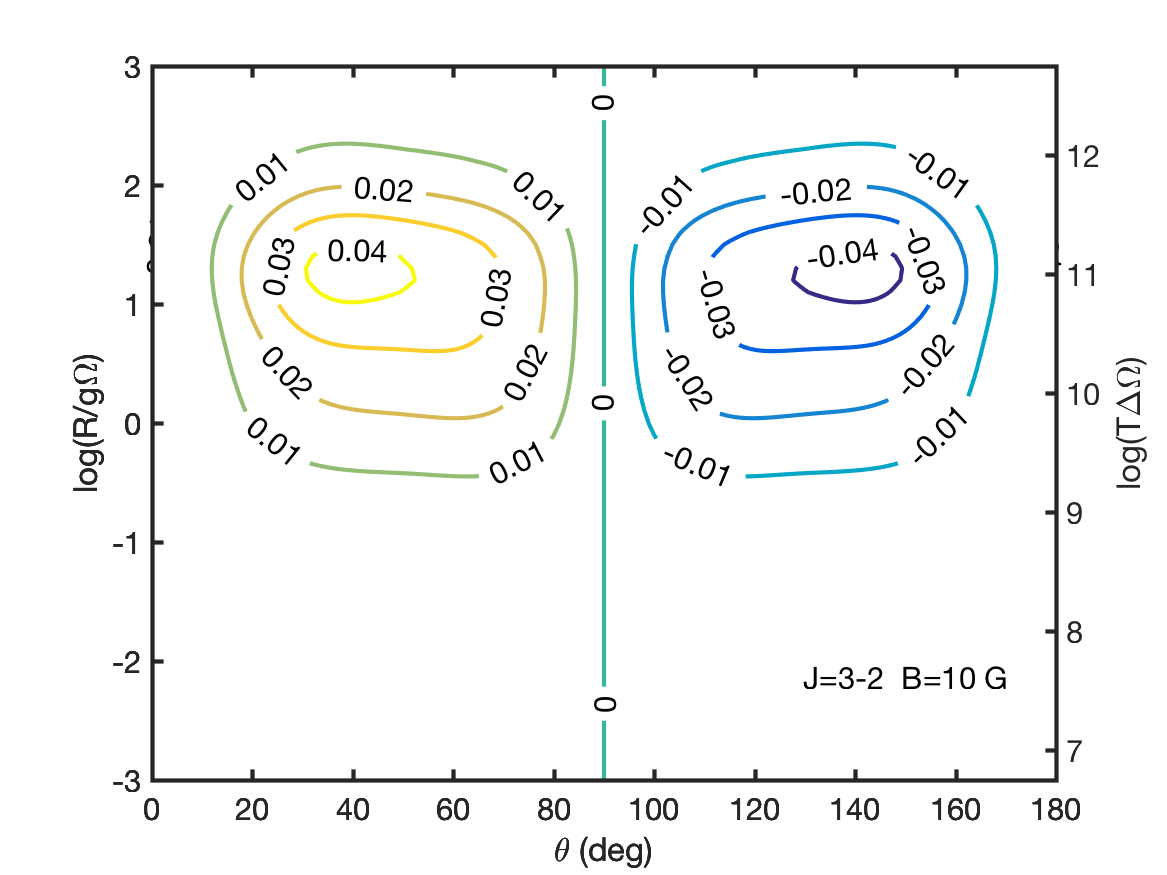}
      \caption{}
    \end{subfigure}
  \caption{Simulations of an isotropically pumped SiO maser. Linear polarization fraction (a,d,g) and angle (b,e,h) and circular polarization fraction (c,f,i). Magnetic field strength and transition angular momentum are denoted inside the figure.}
\end{figure*}

\begin{figure*}[h!]
    \centering
    \begin{subfigure}[b]{0.45\textwidth}
       \includegraphics[width=\textwidth]{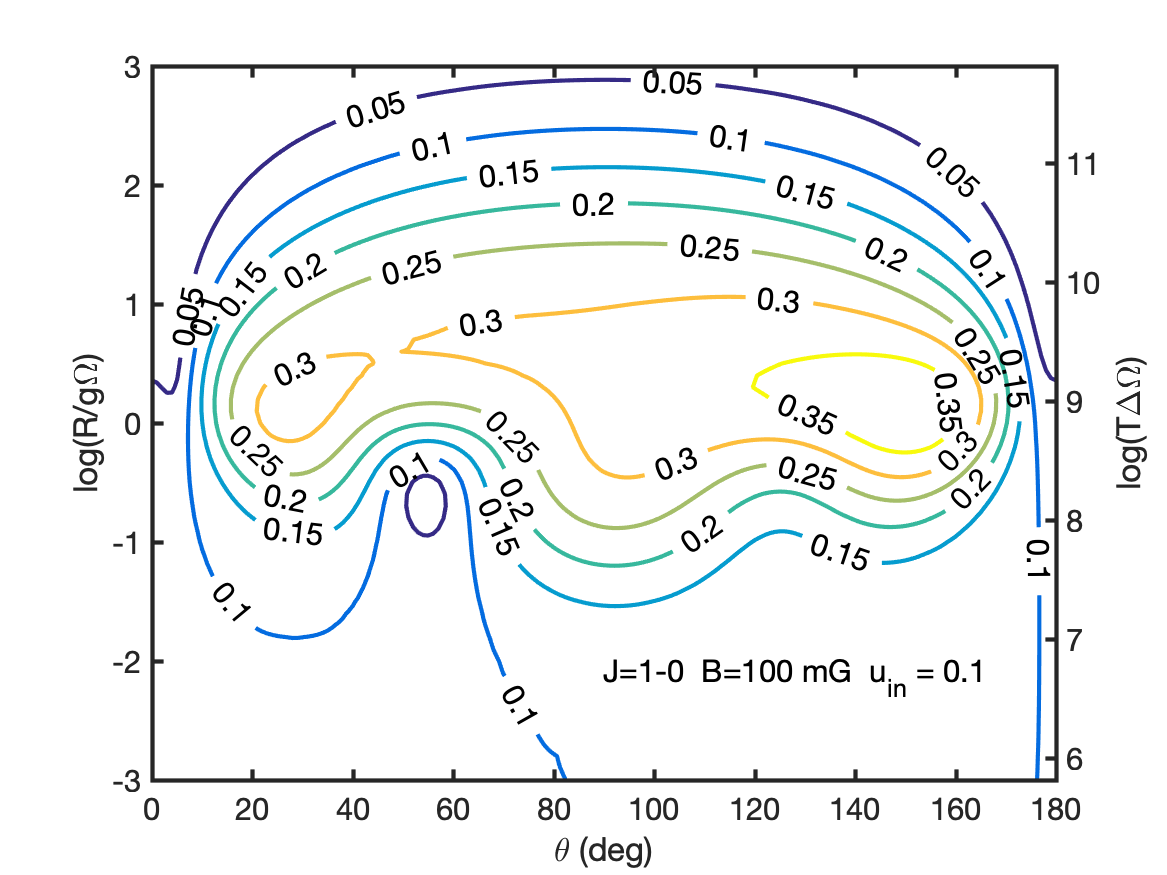} 
       \caption{}
    \end{subfigure}
    ~ 
    \begin{subfigure}[b]{0.45\textwidth}
       \includegraphics[width=\textwidth]{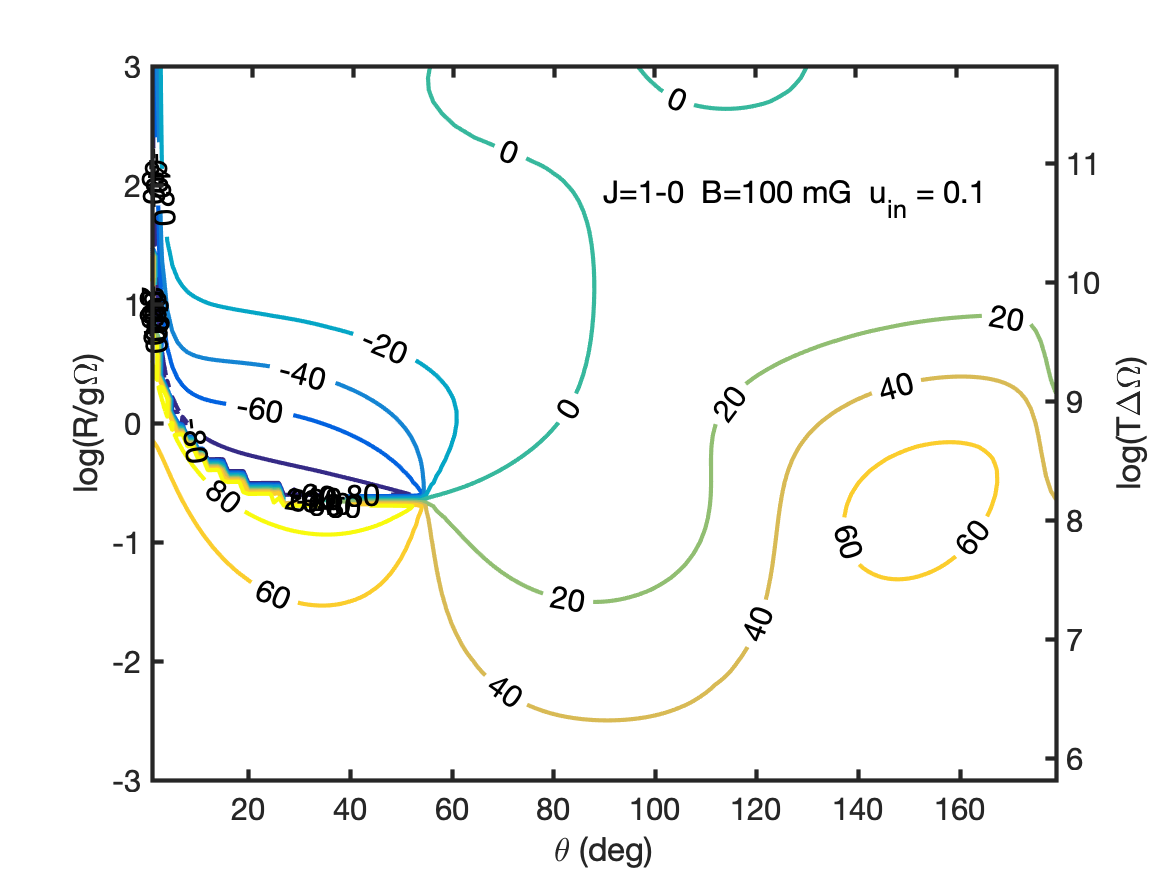} 
       \caption{}
    \end{subfigure}
     ~ 
    \begin{subfigure}[b]{0.45\textwidth}
      \includegraphics[width=\textwidth]{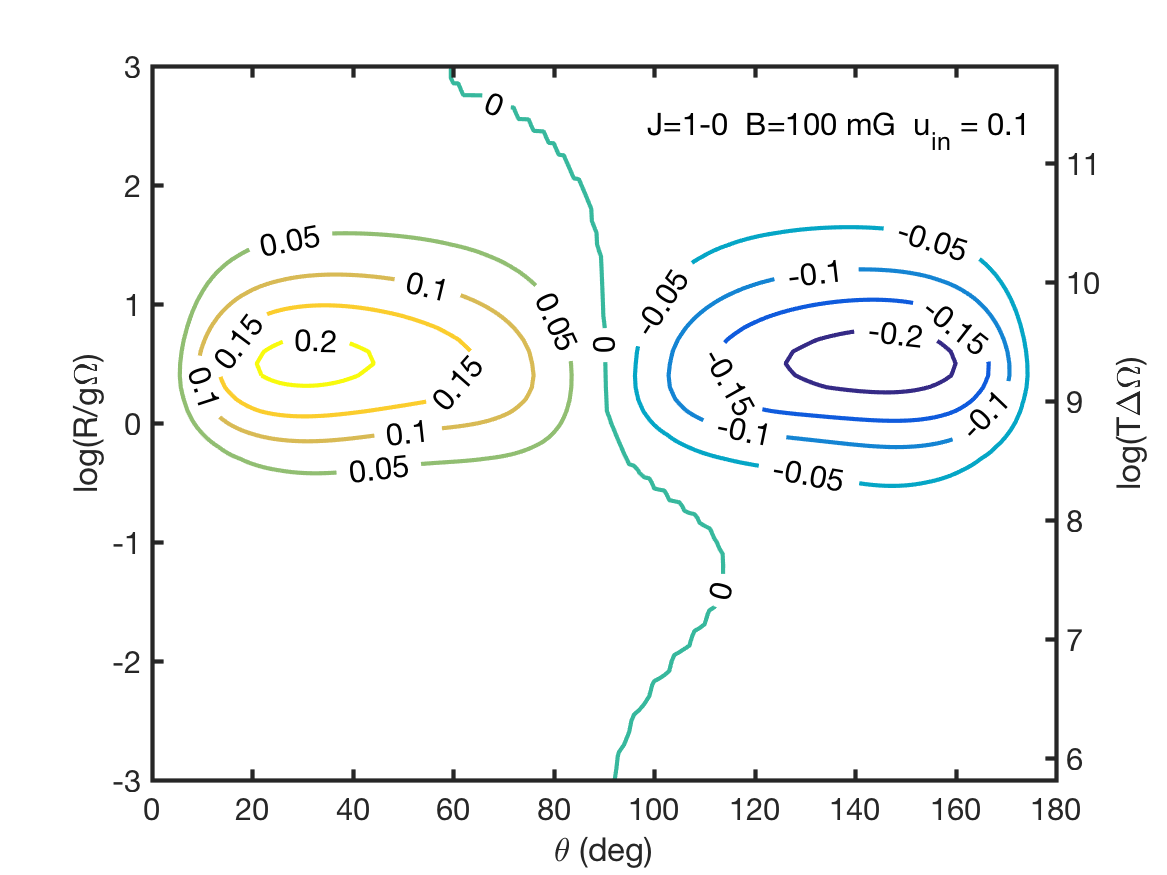}
      \caption{}
    \end{subfigure}
    ~
    \begin{subfigure}[b]{0.45\textwidth}
       \includegraphics[width=\textwidth]{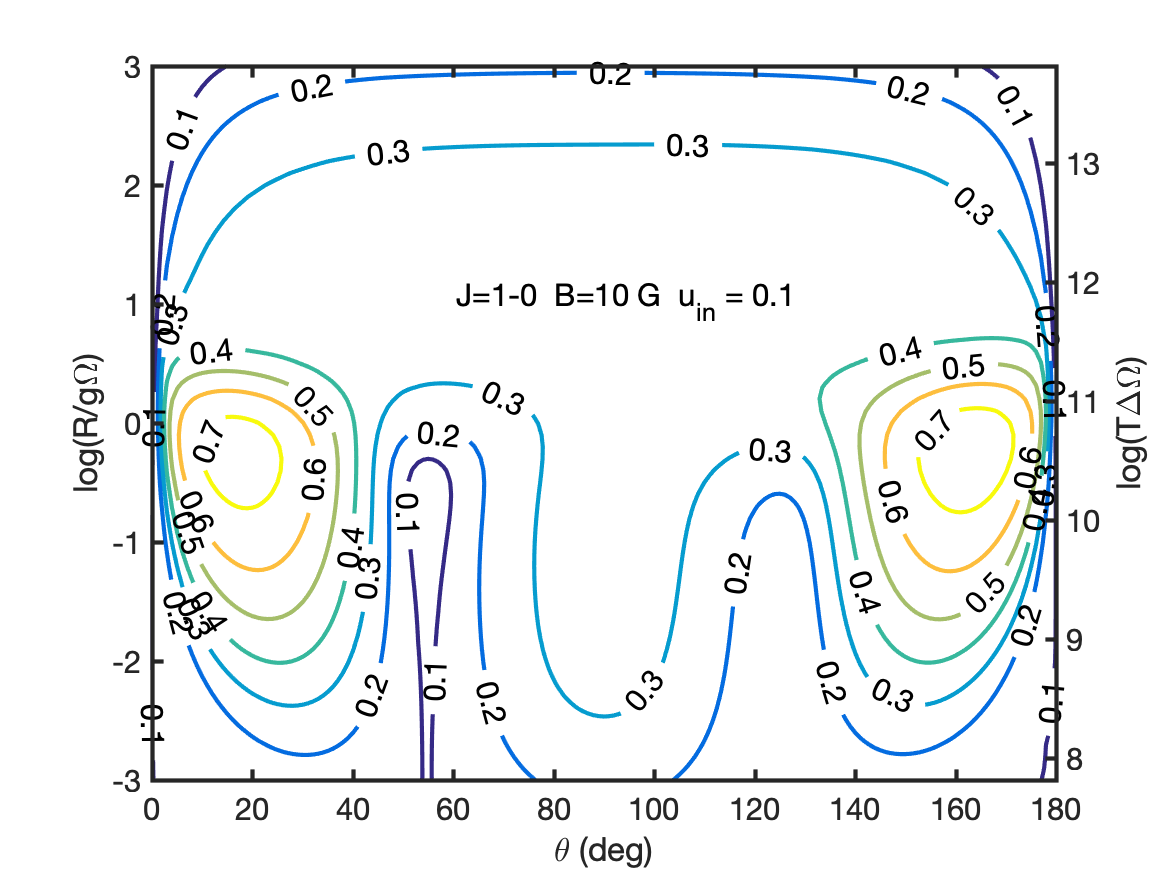} 
       \caption{}
    \end{subfigure}
    ~ 
    \begin{subfigure}[b]{0.45\textwidth}
       \includegraphics[width=\textwidth]{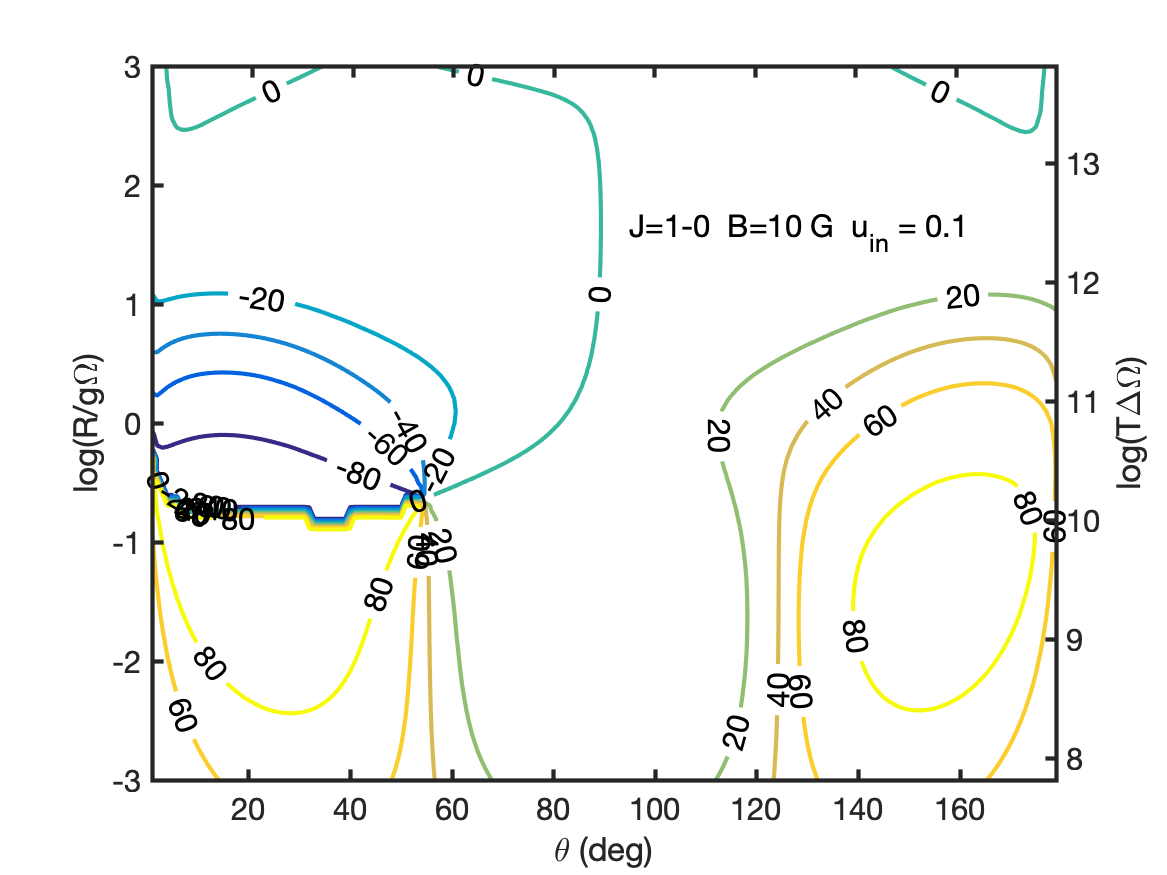} 
       \caption{}
    \end{subfigure}
     ~ 
    \begin{subfigure}[b]{0.45\textwidth}
      \includegraphics[width=\textwidth]{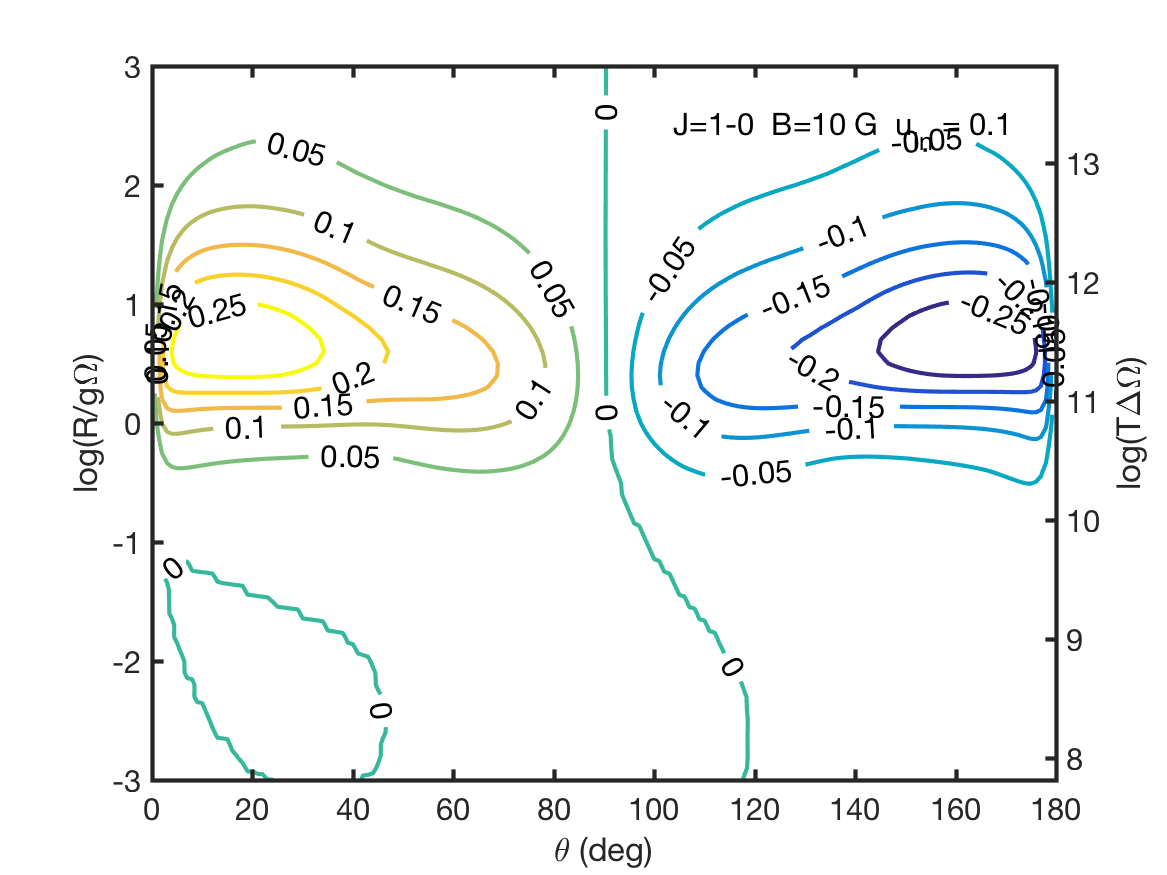}
      \caption{}
    \end{subfigure}
  \caption{Simulations of a SiO maser with $10\%$ polarized seed radiation. Linear polarization fraction (a,d) and angle (b,e) and circular polarization fraction (c,f). Magnetic field strength and transition angular momentum are denoted inside the figure.}
\end{figure*}
\begin{figure*}
    \centering
    \begin{subfigure}[b]{0.32\textwidth}
       \includegraphics[width=\textwidth]{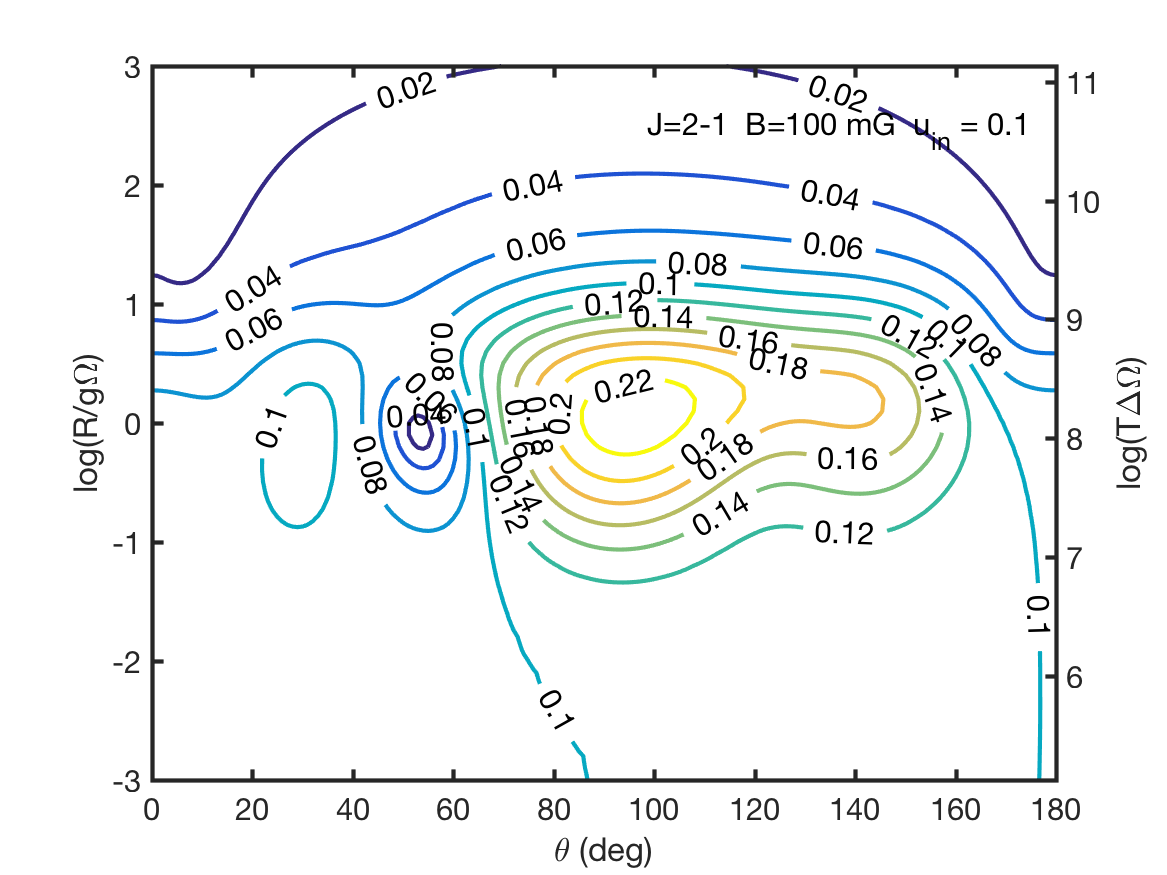}
       \caption{}
    \end{subfigure}
    ~
    \begin{subfigure}[b]{0.32\textwidth}
       \includegraphics[width=\textwidth]{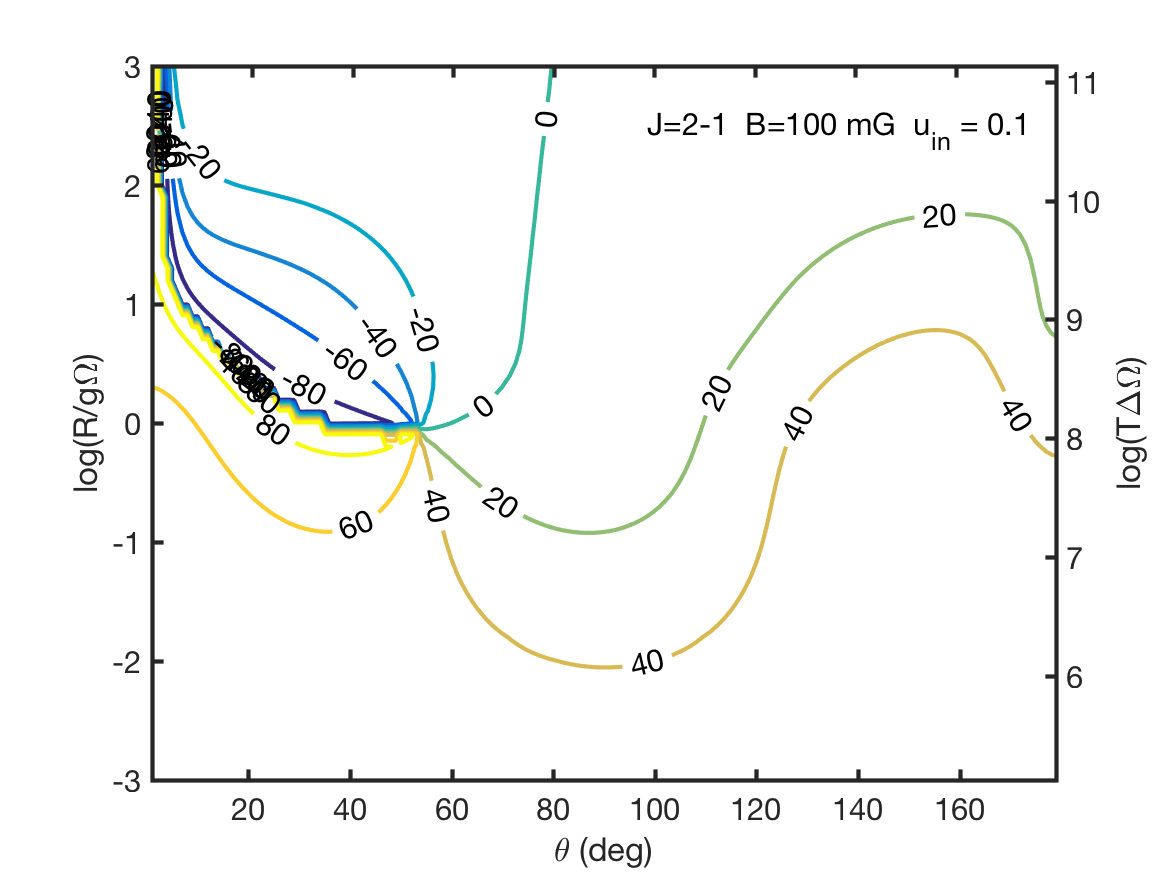}
       \caption{}
    \end{subfigure}
    ~
    \begin{subfigure}[b]{0.32\textwidth}
      \includegraphics[width=\textwidth]{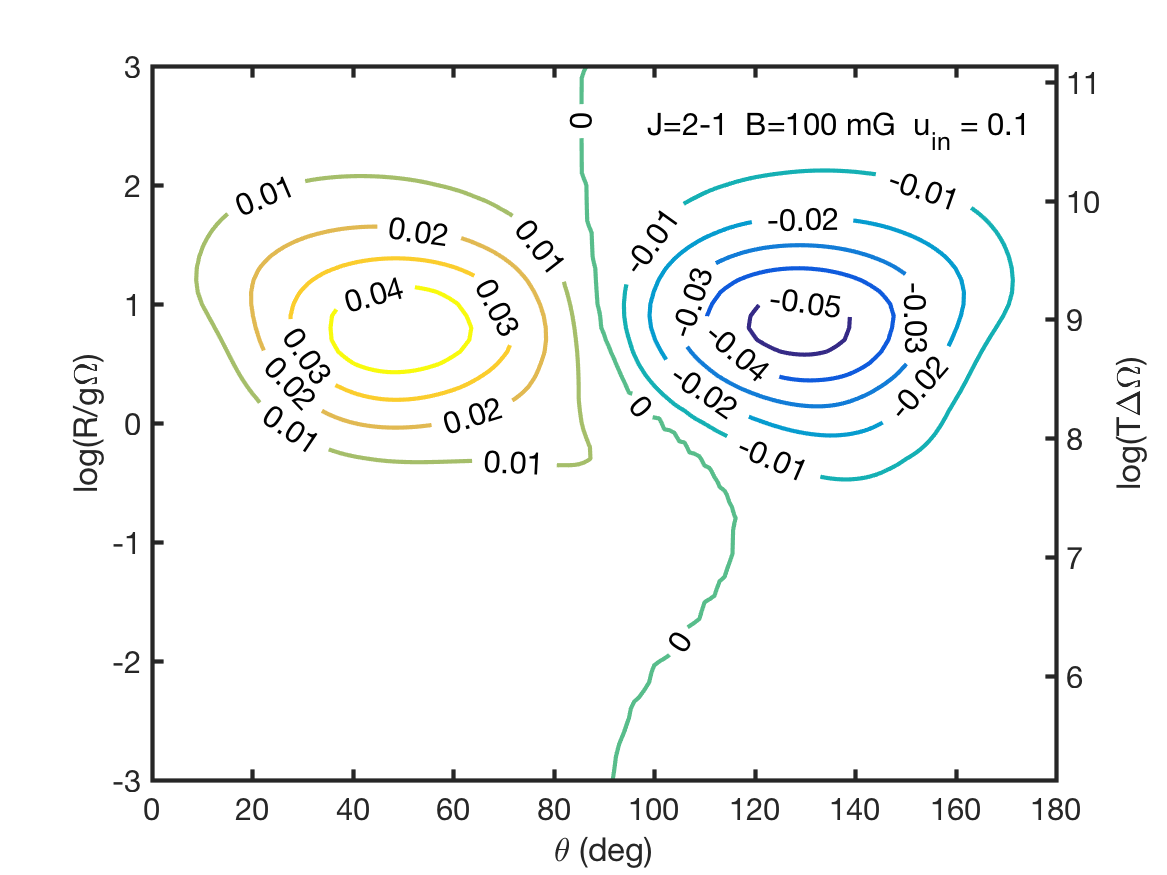}
      \caption{}
    \end{subfigure}
    ~
    \begin{subfigure}[b]{0.32\textwidth}
       \includegraphics[width=\textwidth]{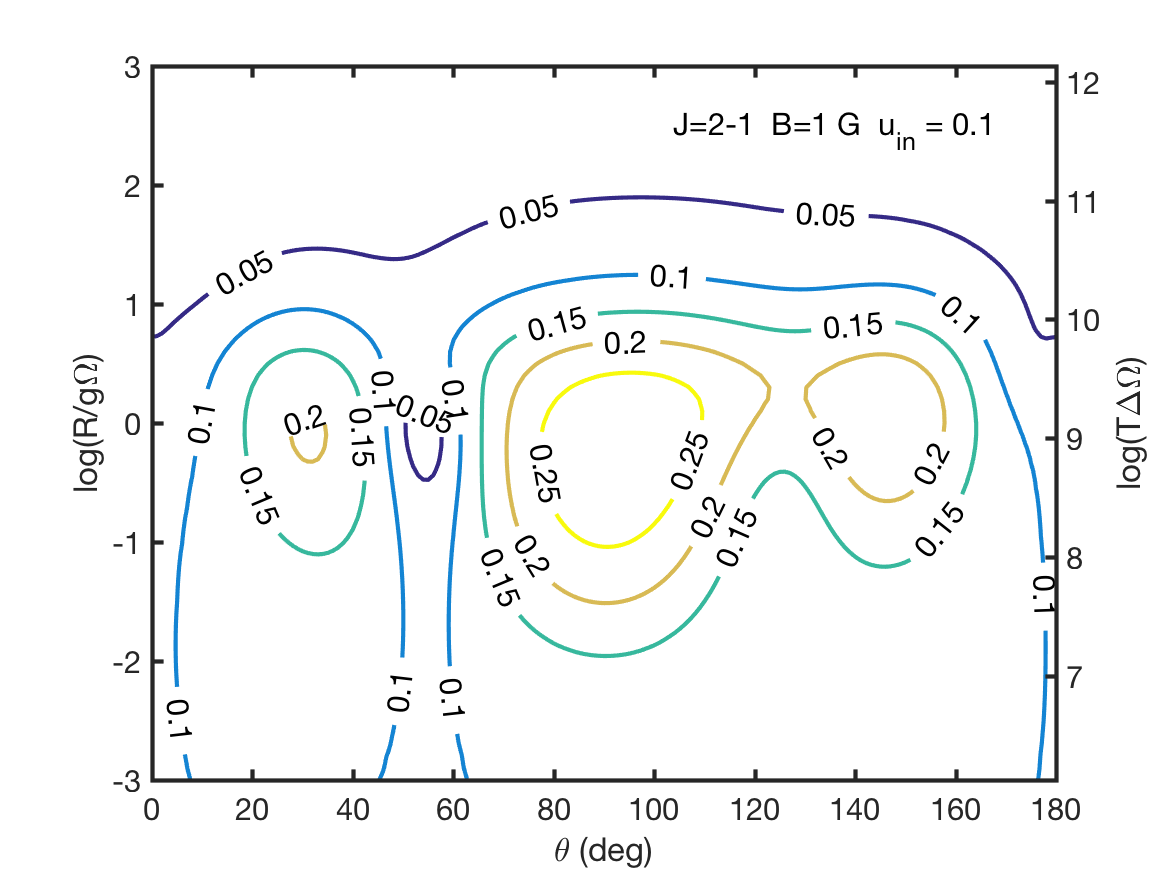}
       \caption{}
    \end{subfigure}
    ~
    \begin{subfigure}[b]{0.32\textwidth}
       \includegraphics[width=\textwidth]{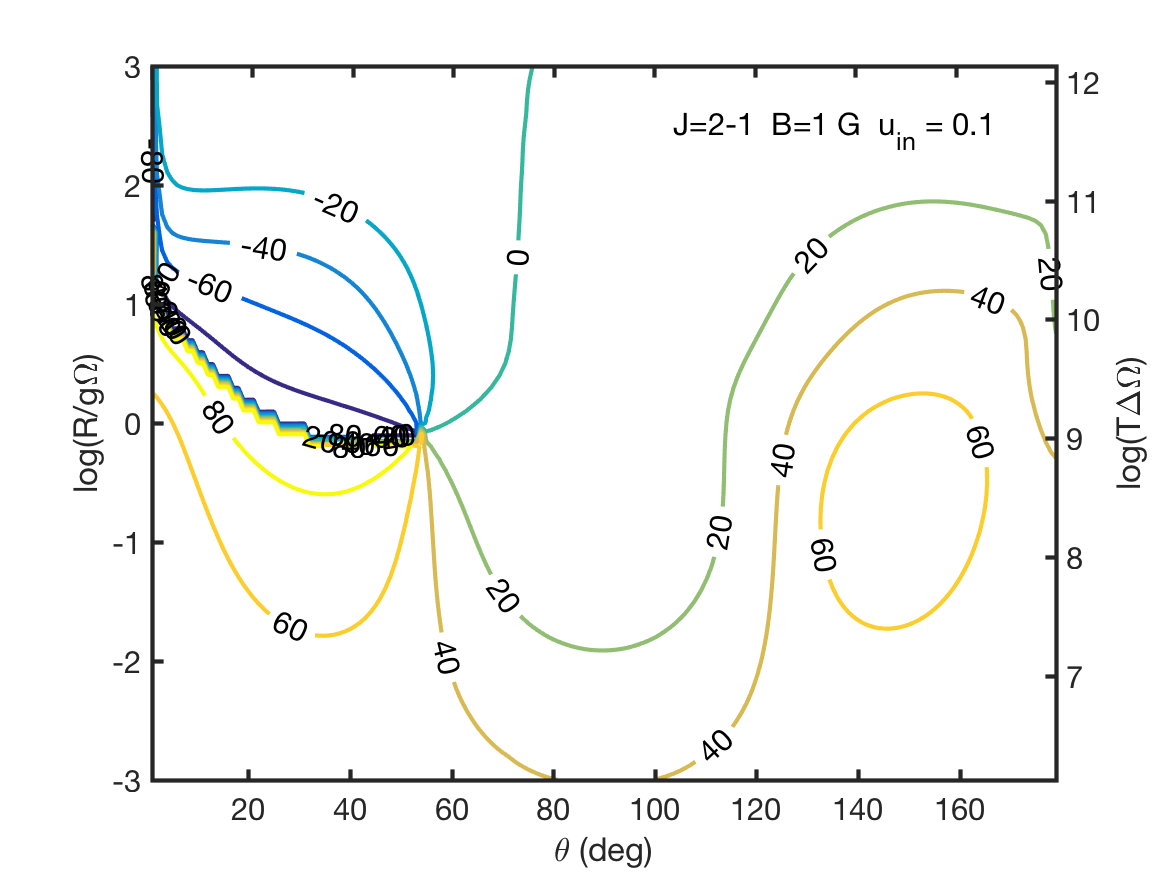}
       \caption{}
    \end{subfigure}
     ~
    \begin{subfigure}[b]{0.32\textwidth}
      \includegraphics[width=\textwidth]{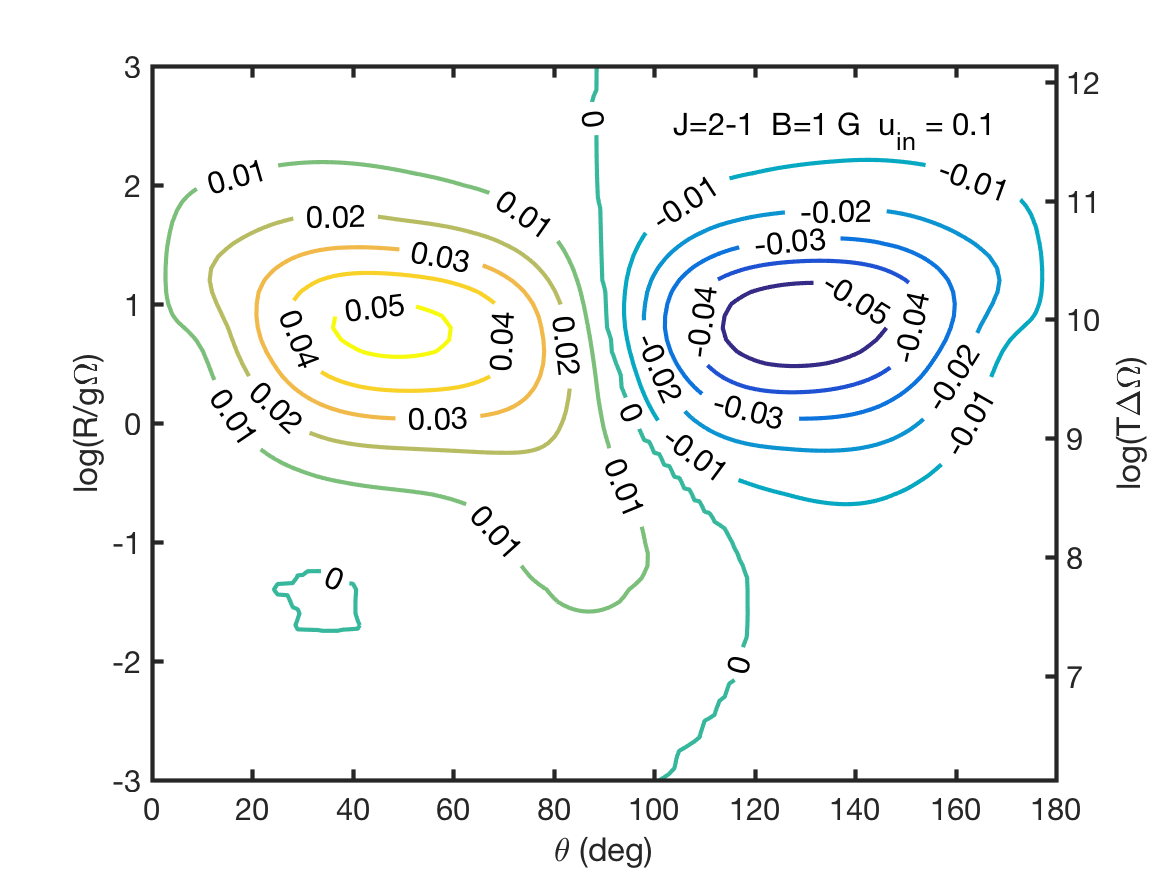}
      \caption{}
    \end{subfigure}
     ~
    \begin{subfigure}[b]{0.32\textwidth}
       \includegraphics[width=\textwidth]{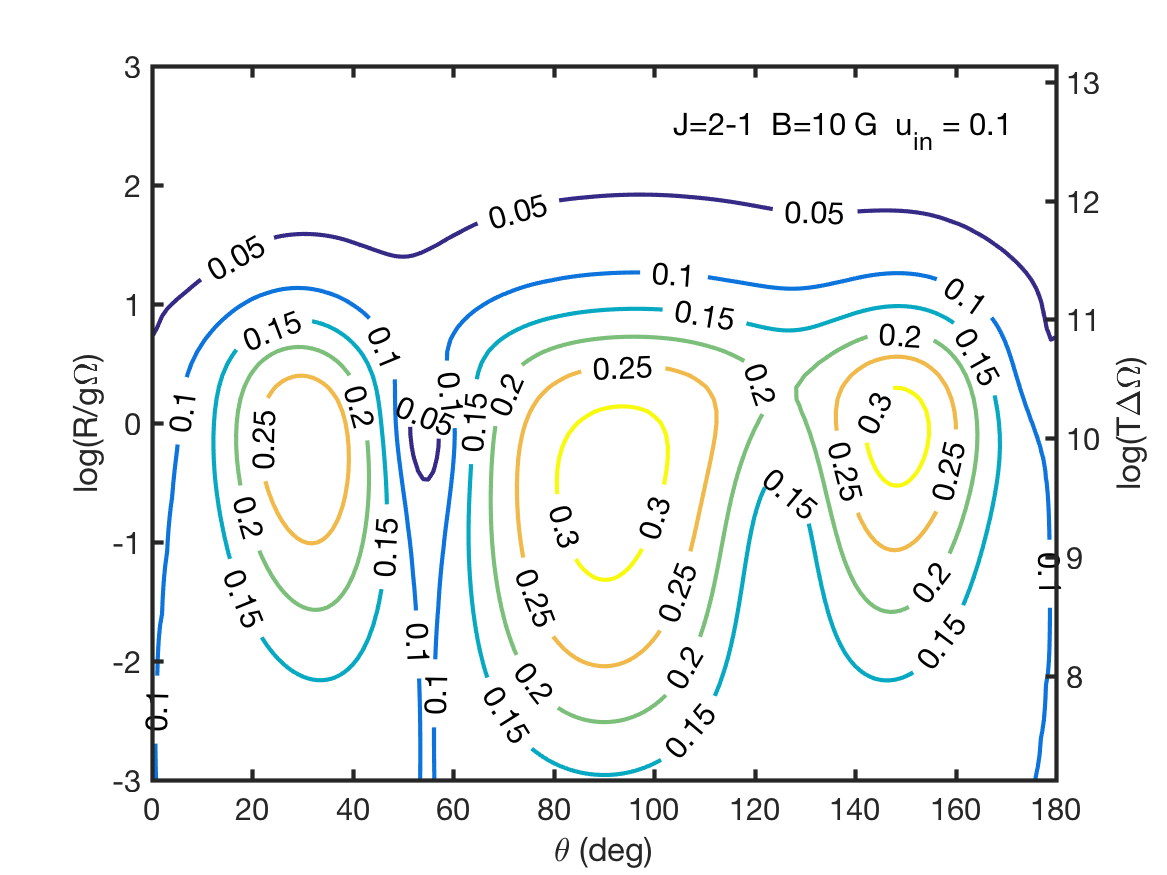}
       \caption{}
    \end{subfigure}
    ~
    \begin{subfigure}[b]{0.32\textwidth}
       \includegraphics[width=\textwidth]{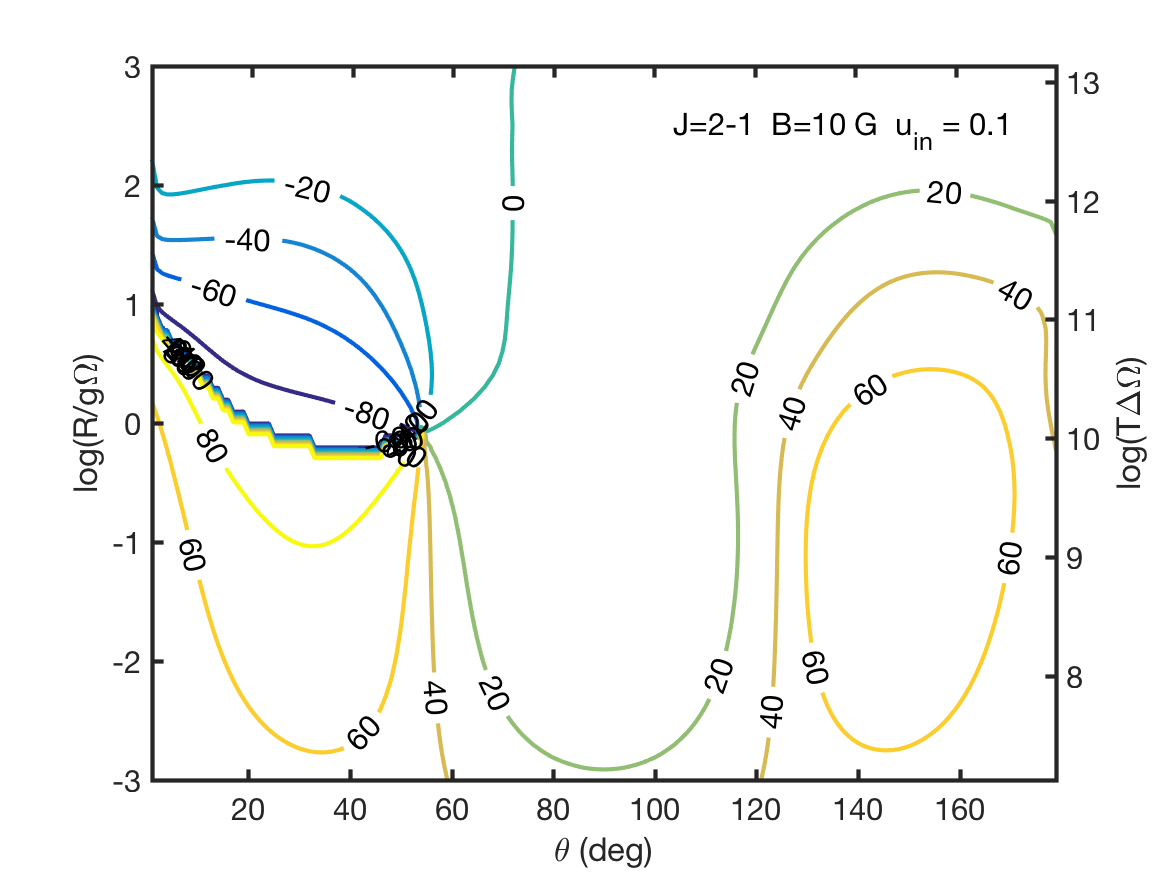}
       \caption{}
    \end{subfigure}
     ~
    \begin{subfigure}[b]{0.32\textwidth}
      \includegraphics[width=\textwidth]{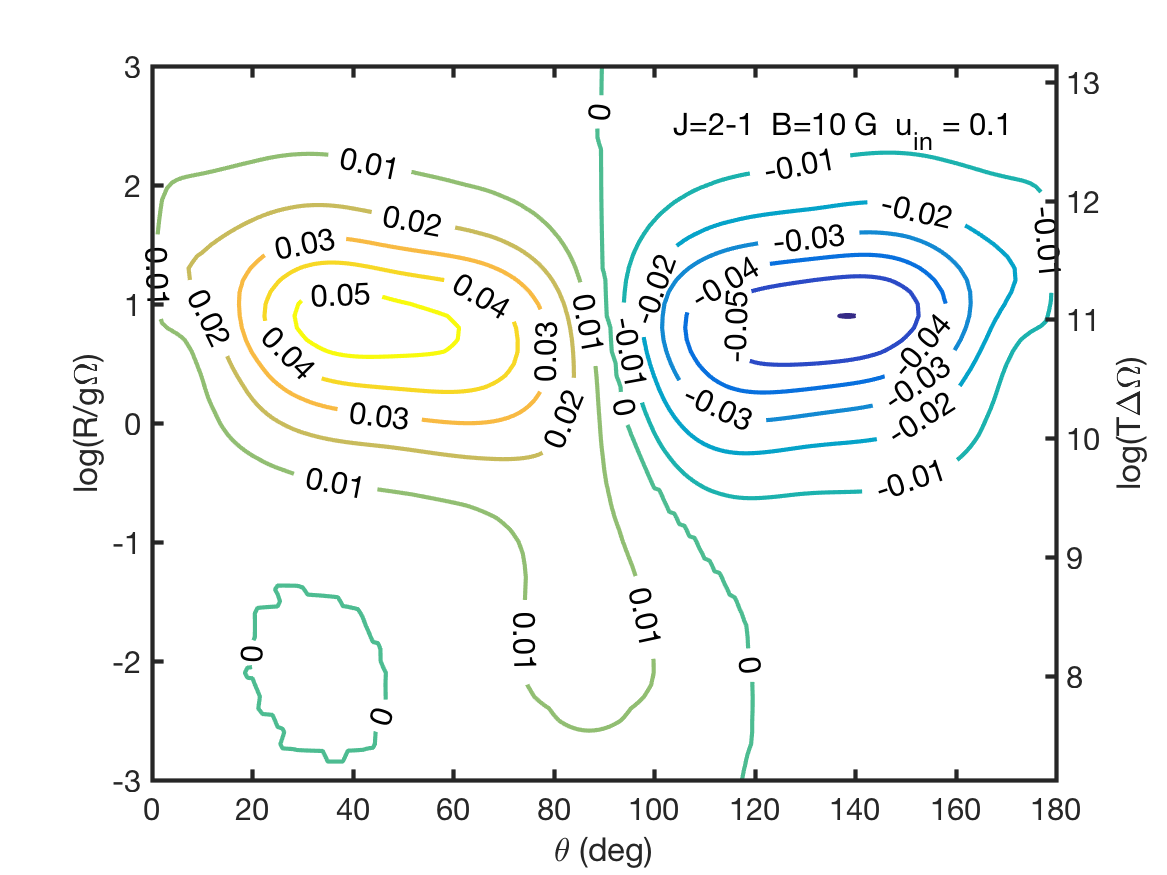}
      \caption{}
    \end{subfigure}
  \caption{Simulations of a SiO maser with $10\%$ polarized seed radiation. Linear polarization fraction (a,d,g) and angle (b,e,h) and circular polarization fraction (c,f,i). Magnetic field strength and transition angular momentum are denoted inside the figure.}
\end{figure*}

\begin{figure*}
    \centering
    \begin{subfigure}[b]{0.32\textwidth}
       \includegraphics[width=\textwidth]{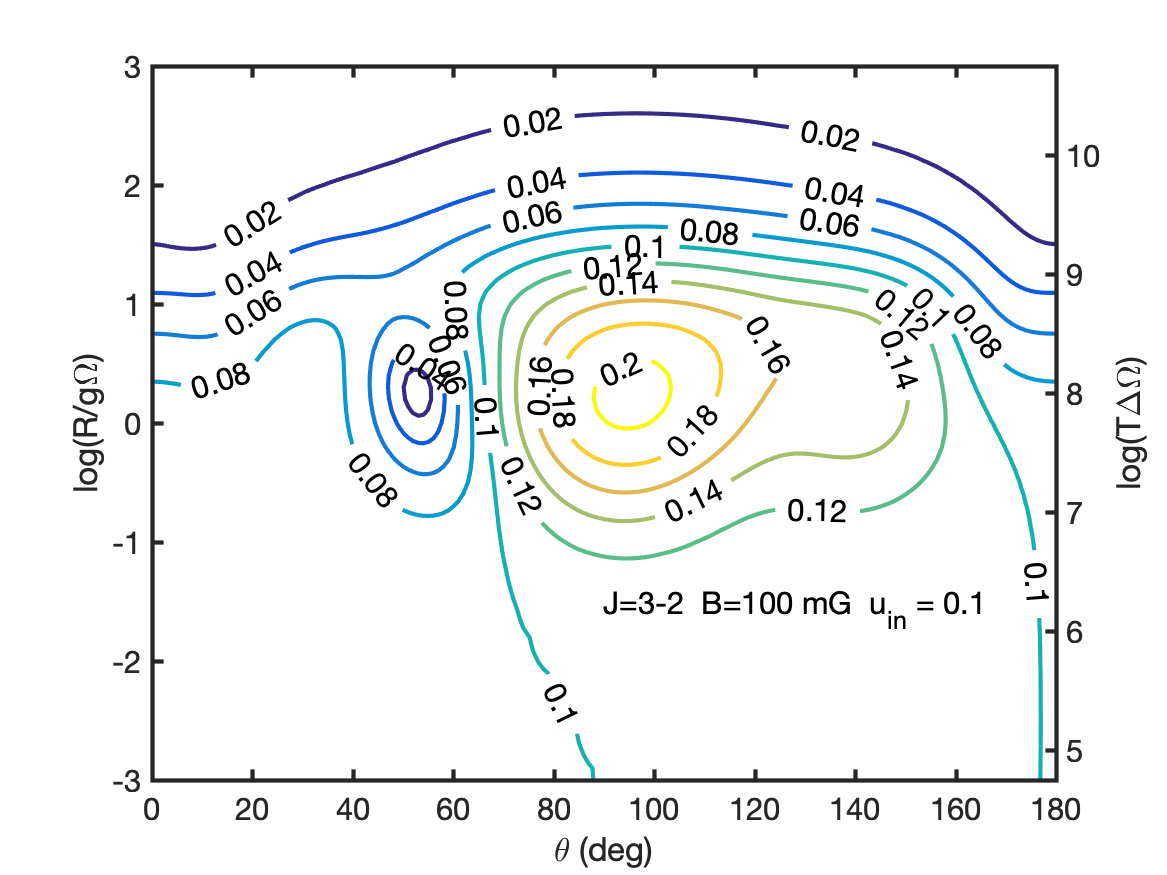}
       \caption{}
    \end{subfigure}
    ~
    \begin{subfigure}[b]{0.32\textwidth}
       \includegraphics[width=\textwidth]{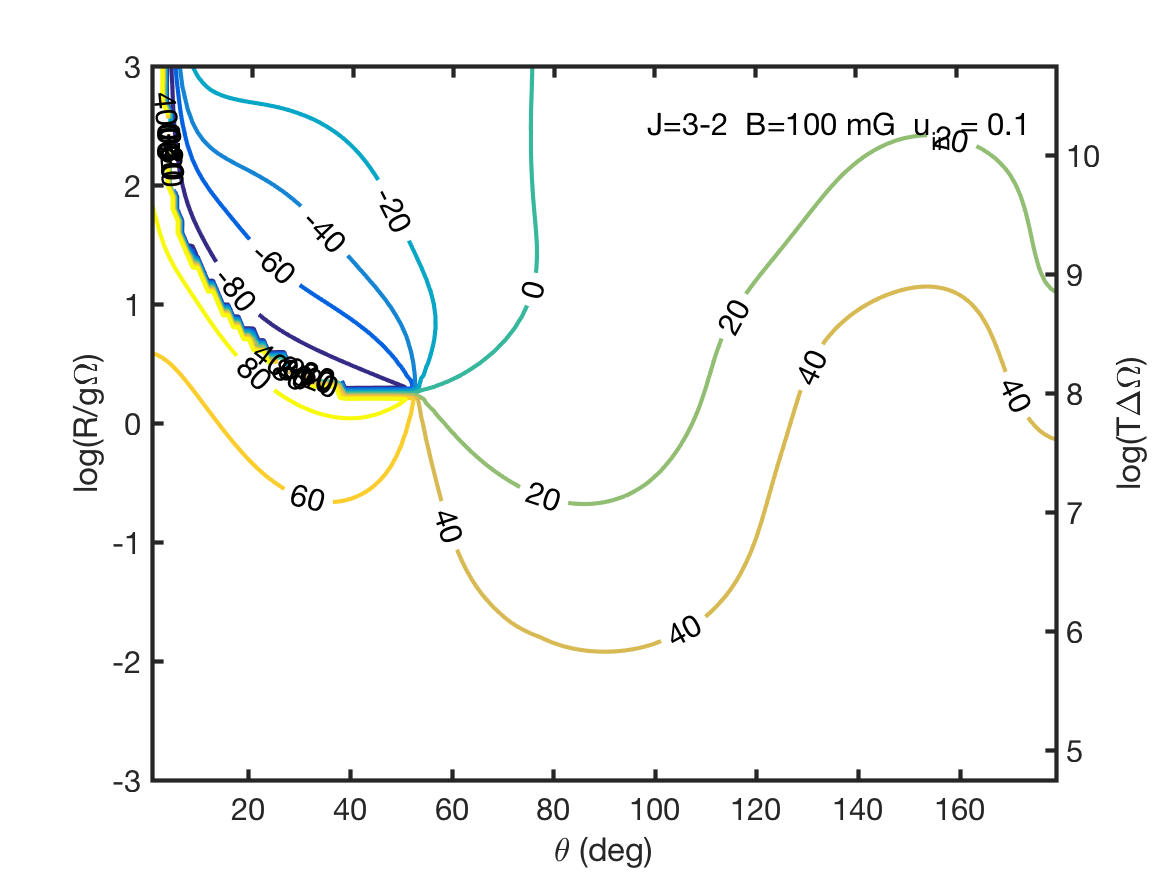}
       \caption{}
    \end{subfigure}
     ~
    \begin{subfigure}[b]{0.32\textwidth}
      \includegraphics[width=\textwidth]{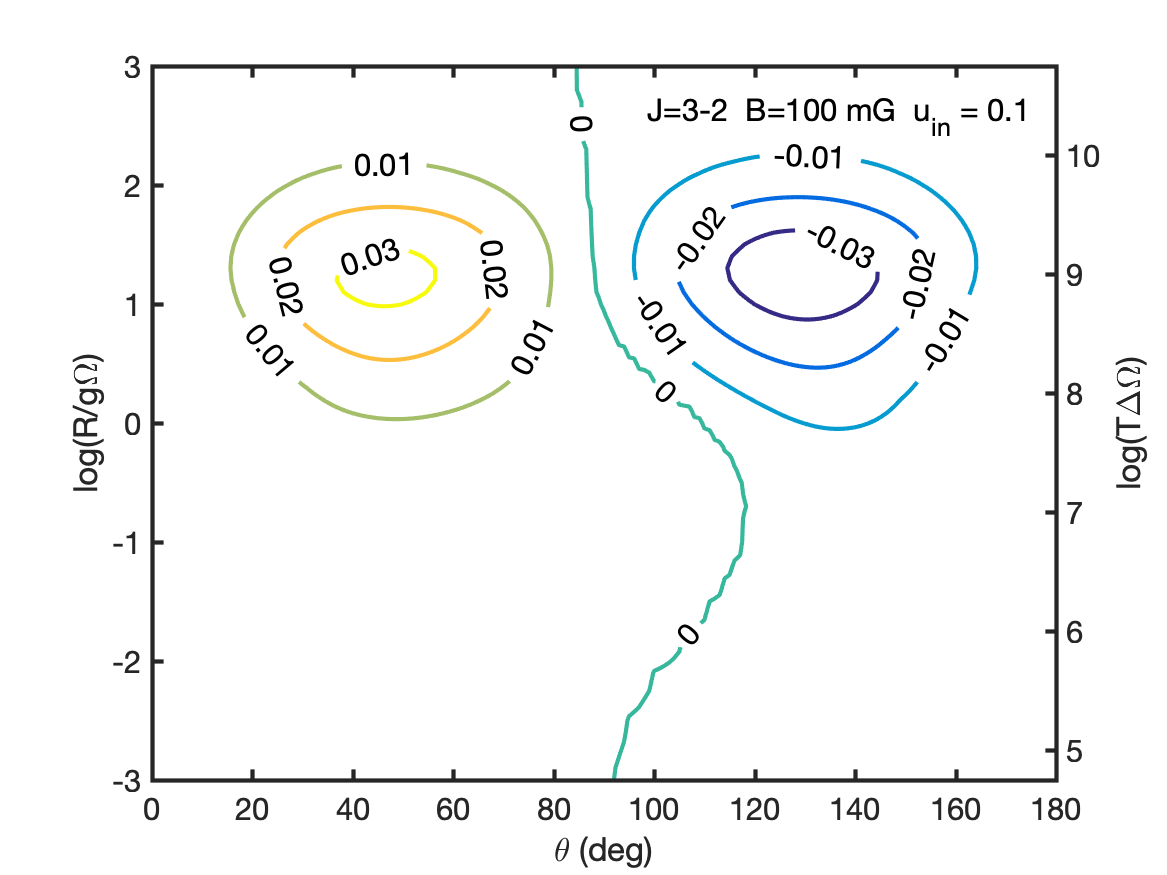}
      \caption{}
    \end{subfigure}
    ~
    \begin{subfigure}[b]{0.32\textwidth}
       \includegraphics[width=\textwidth]{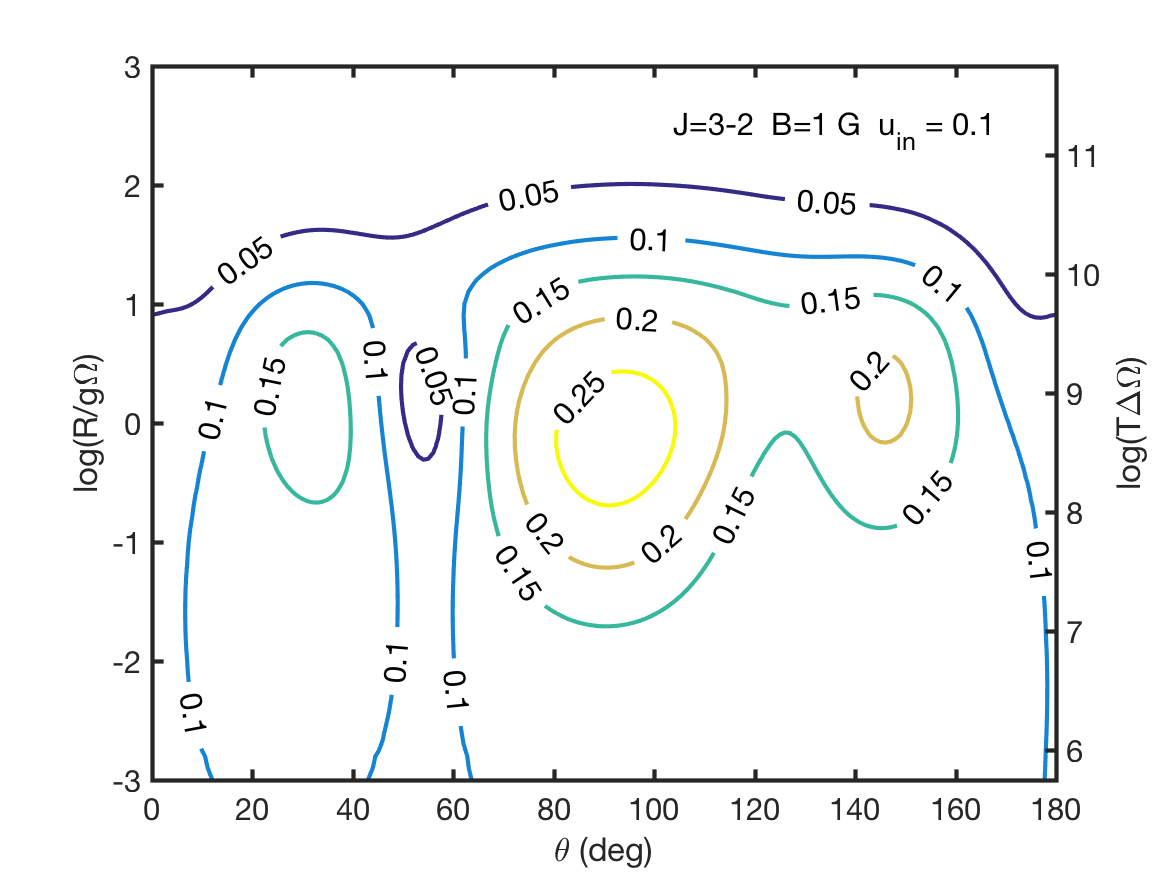}
       \caption{}
    \end{subfigure}
    ~
    \begin{subfigure}[b]{0.32\textwidth}
       \includegraphics[width=\textwidth]{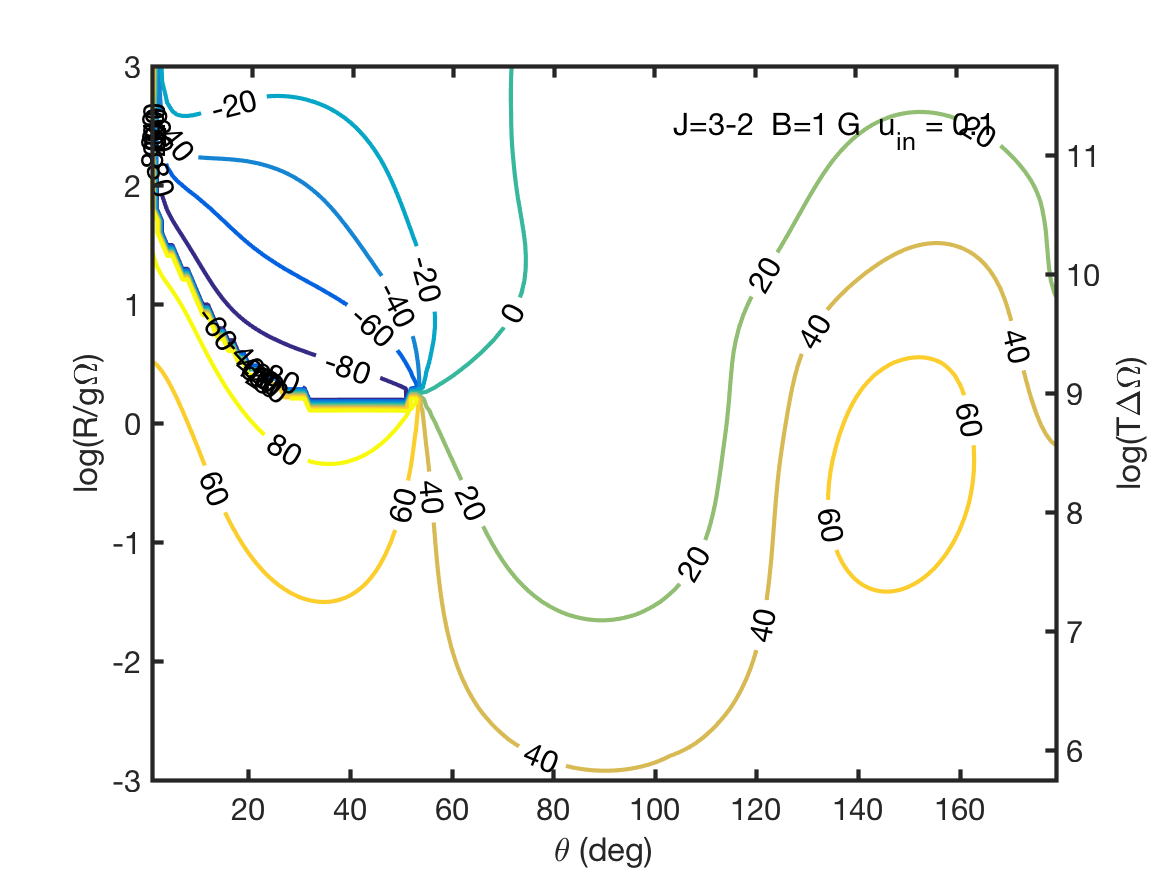}
       \caption{}
    \end{subfigure}
     ~
    \begin{subfigure}[b]{0.32\textwidth}
      \includegraphics[width=\textwidth]{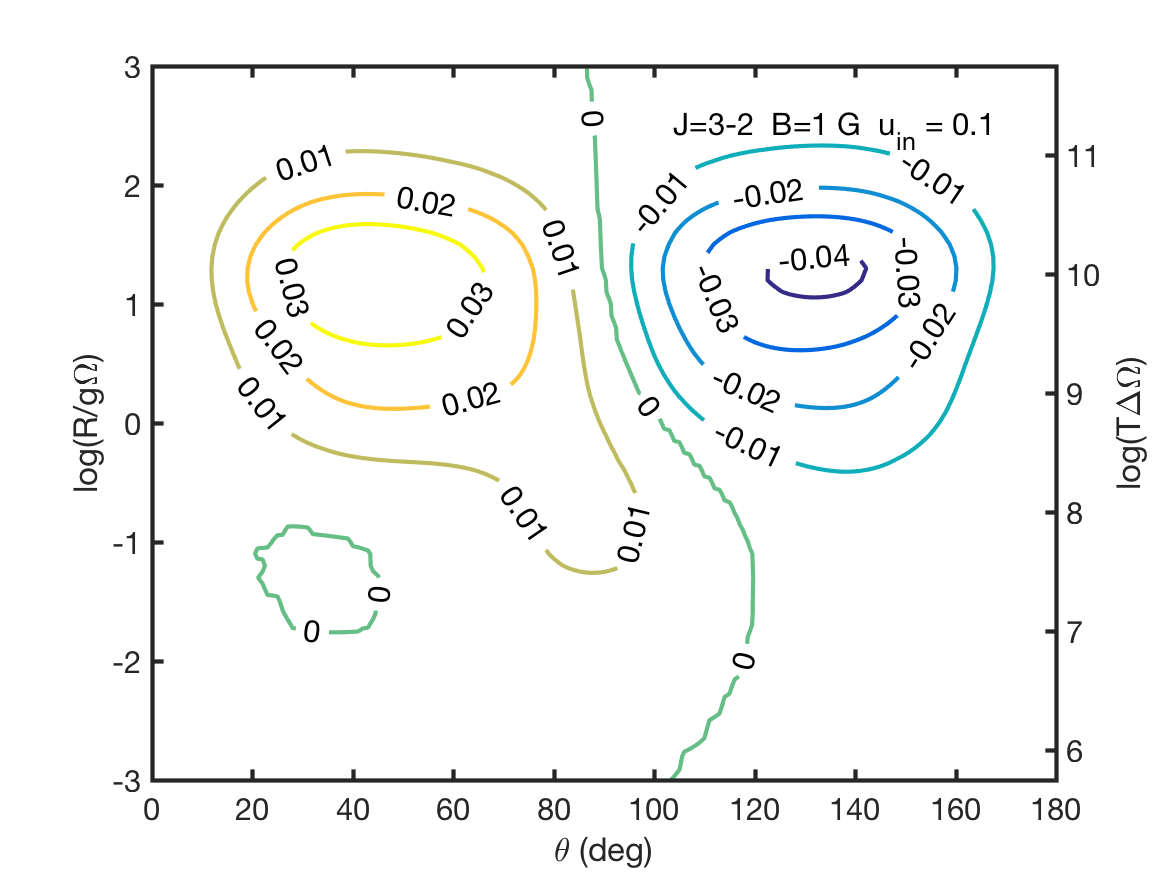}
      \caption{}
    \end{subfigure}
    ~
    \begin{subfigure}[b]{0.32\textwidth}
       \includegraphics[width=\textwidth]{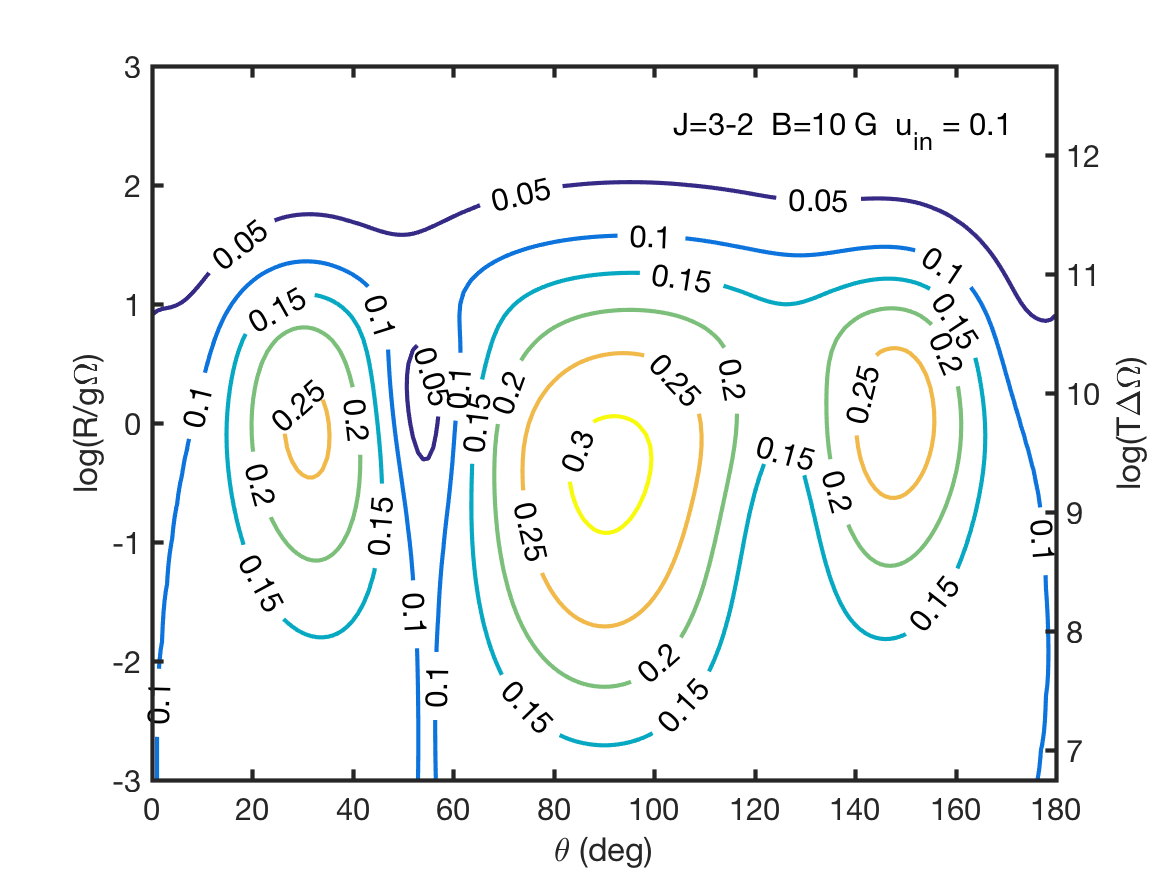}
       \caption{}
    \end{subfigure}
    ~
    \begin{subfigure}[b]{0.32\textwidth}
       \includegraphics[width=\textwidth]{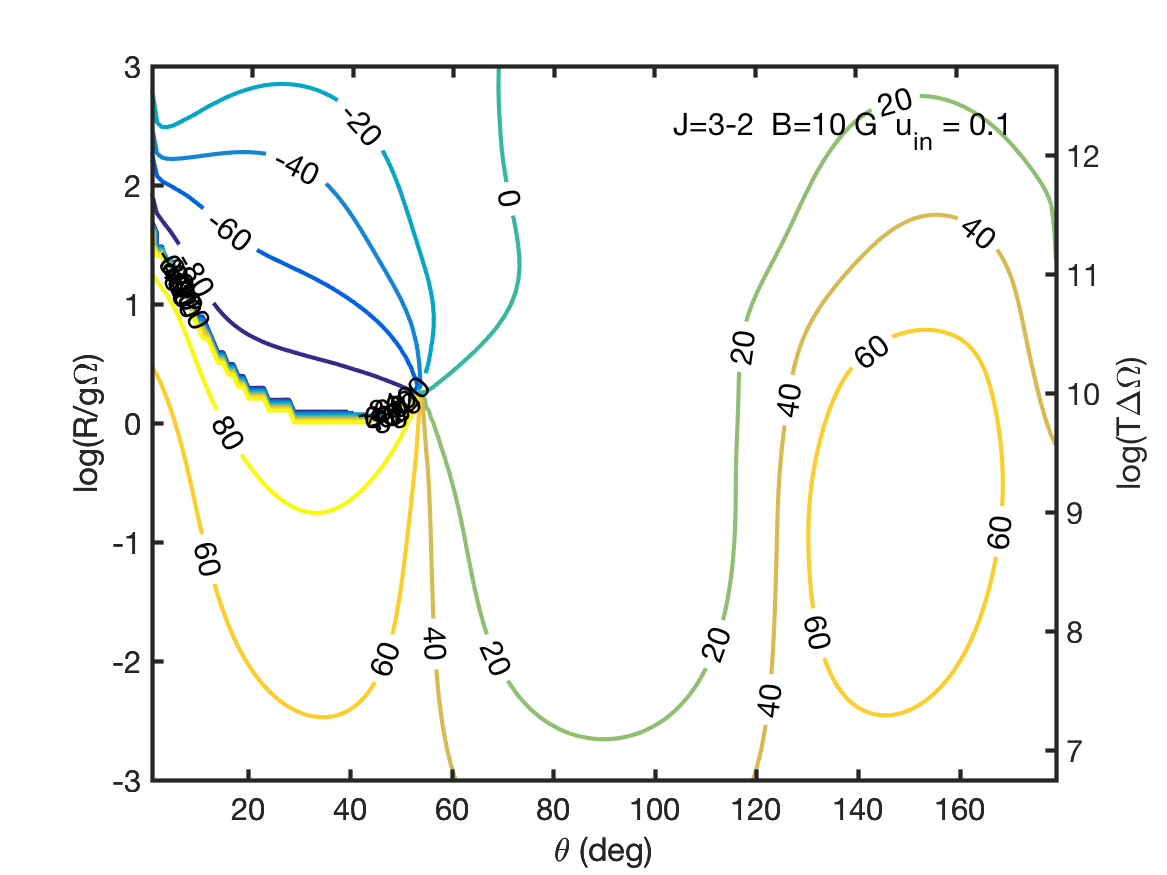}
       \caption{}
    \end{subfigure}
     ~
    \begin{subfigure}[b]{0.32\textwidth}
      \includegraphics[width=\textwidth]{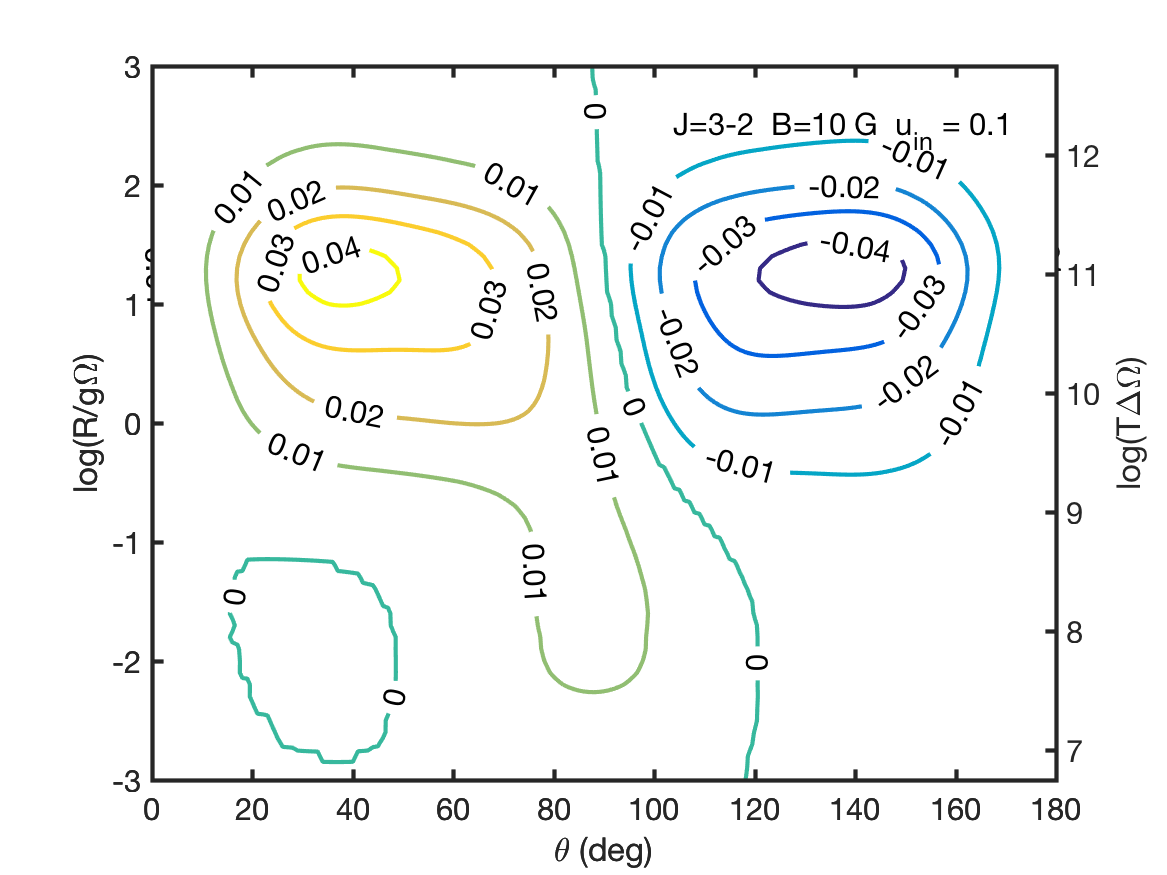}
      \caption{}
    \end{subfigure}
  \caption{Simulations of a SiO maser with $10\%$ polarized seed radiation. Linear polarization fraction (a,d,g) and angle (b,e,h) and circular polarization fraction (c,f,i). Magnetic field strength and transition angular momentum are denoted inside the figure.}
\end{figure*}

\begin{figure*}
    \centering
    \begin{subfigure}[b]{0.45\textwidth}
       \includegraphics[width=\textwidth]{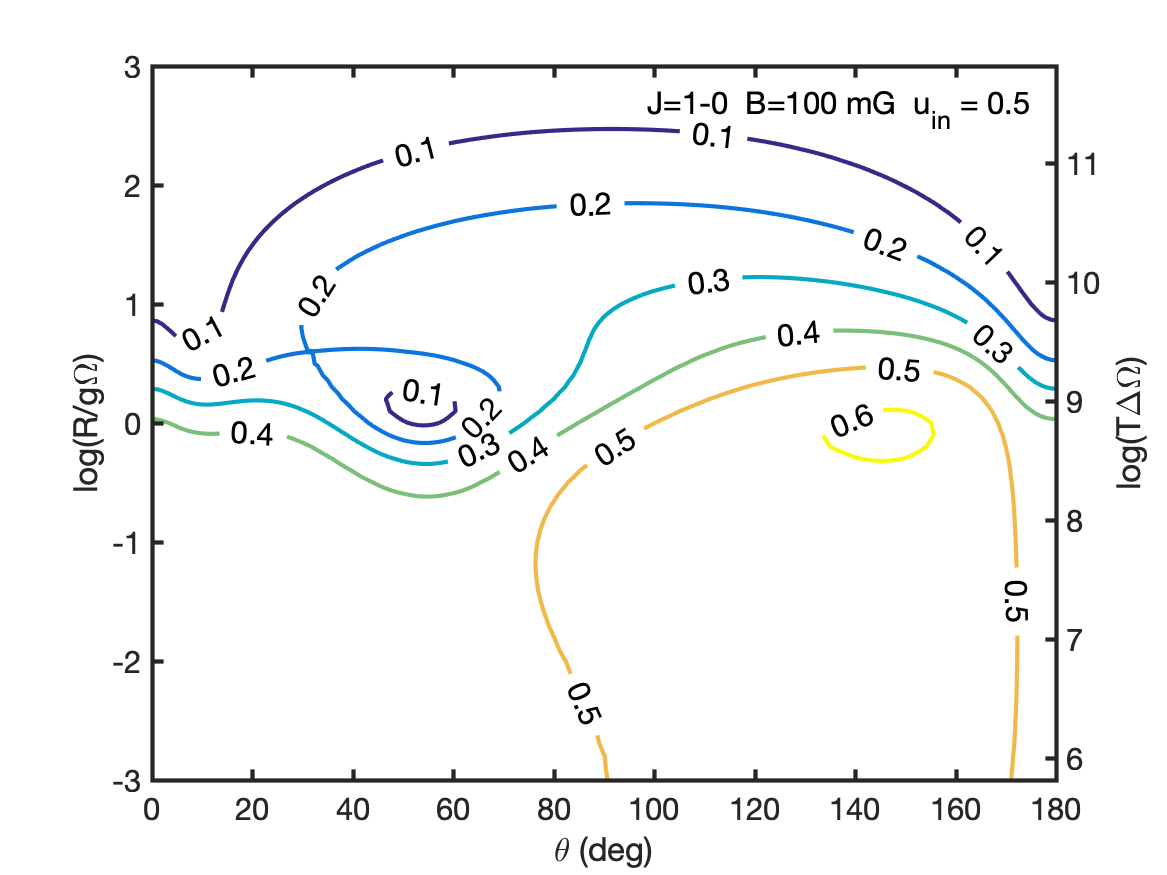} 
       \caption{}
    \end{subfigure}
    ~ 
    \begin{subfigure}[b]{0.45\textwidth}
       \includegraphics[width=\textwidth]{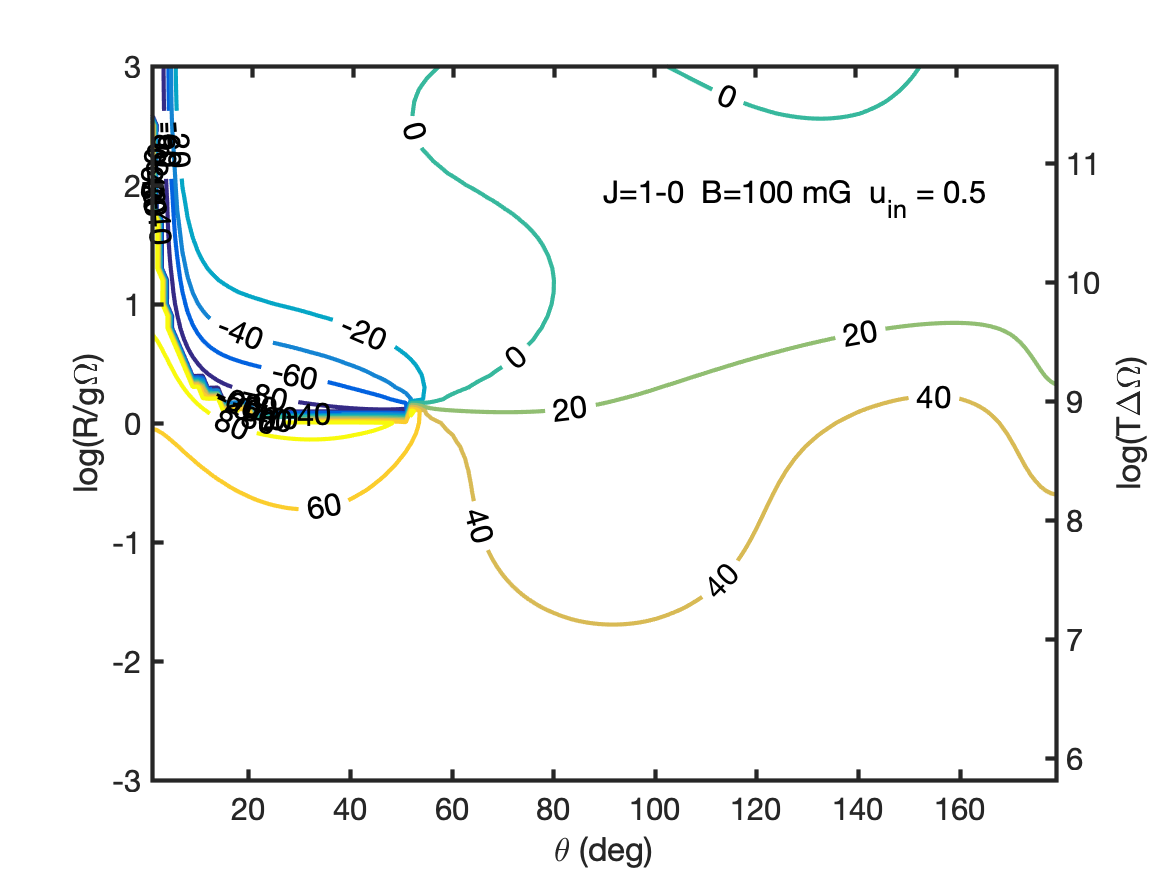} 
       \caption{}
    \end{subfigure}
     ~ 
    \begin{subfigure}[b]{0.45\textwidth}
      \includegraphics[width=\textwidth]{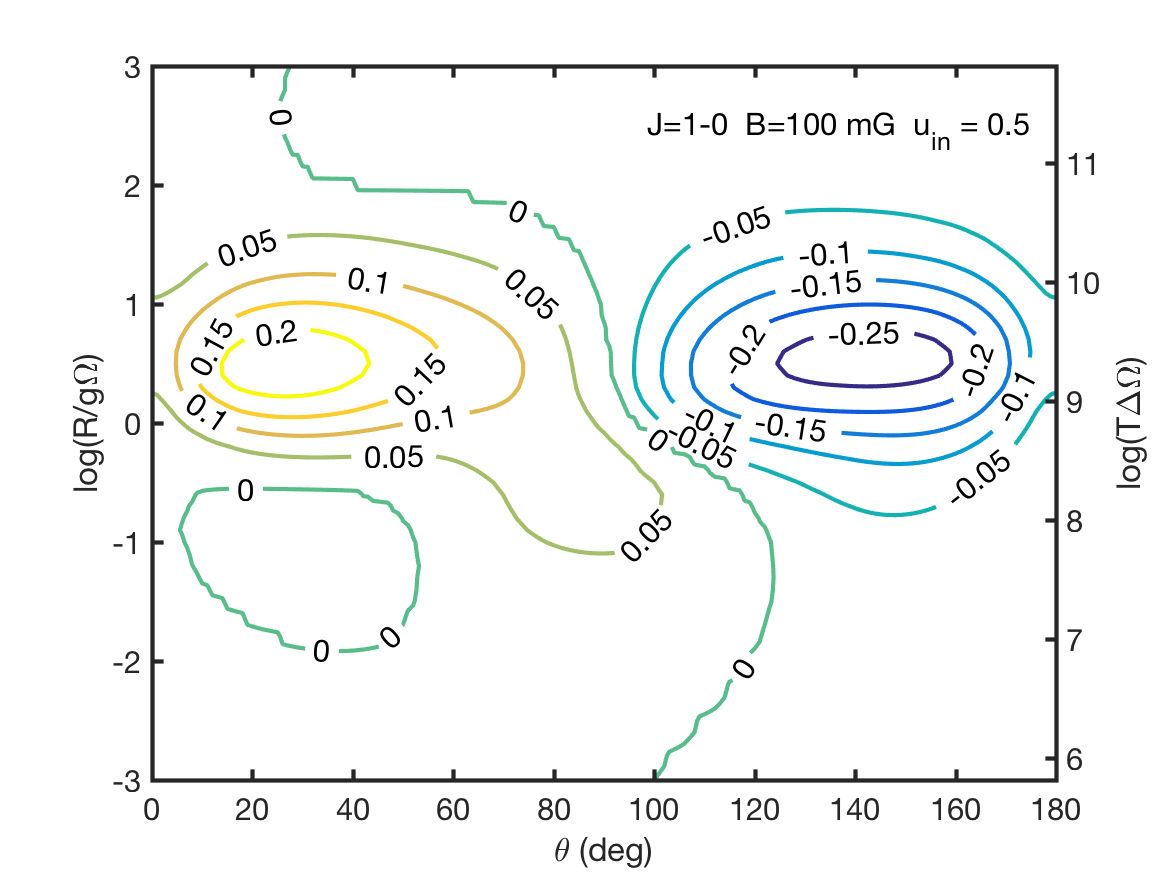}
      \caption{}
    \end{subfigure}
     ~
    \begin{subfigure}[b]{0.45\textwidth}
       \includegraphics[width=\textwidth]{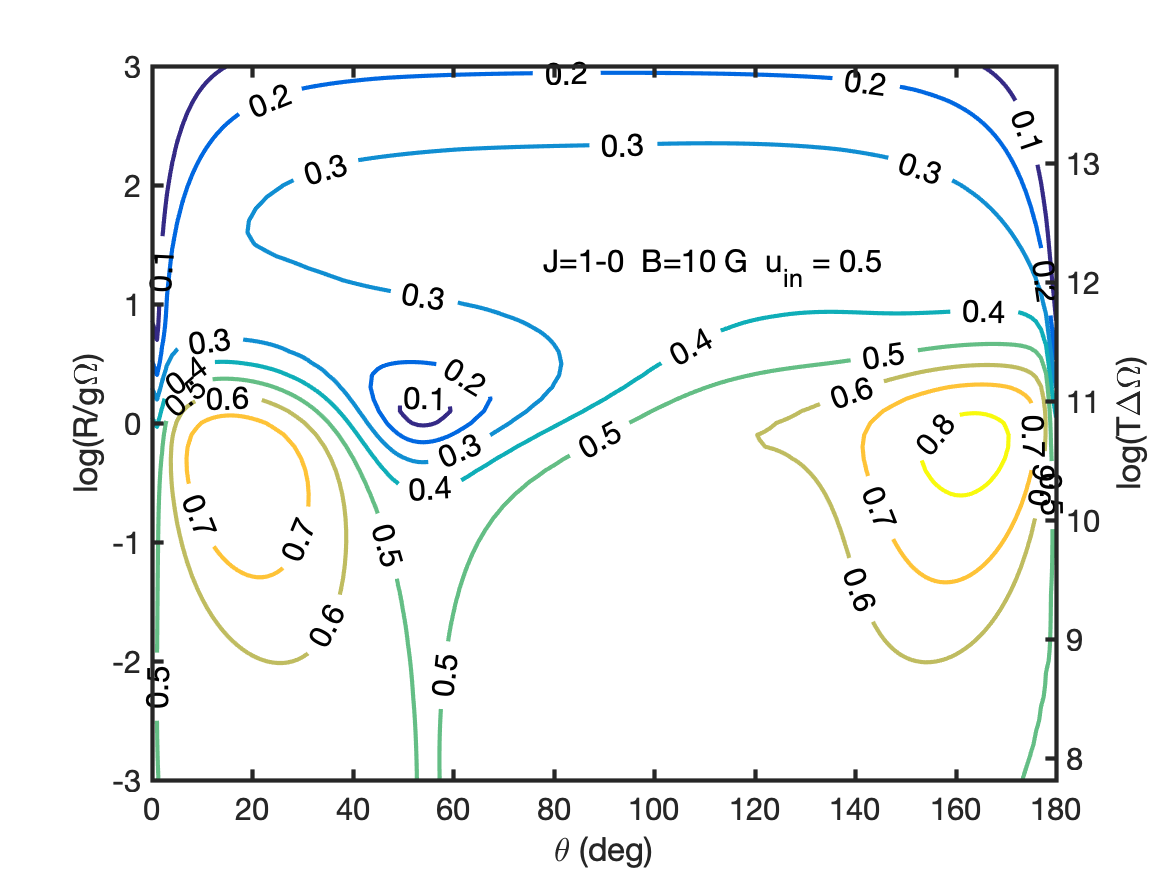} 
       \caption{}
    \end{subfigure}
    ~ 
    \begin{subfigure}[b]{0.45\textwidth}
       \includegraphics[width=\textwidth]{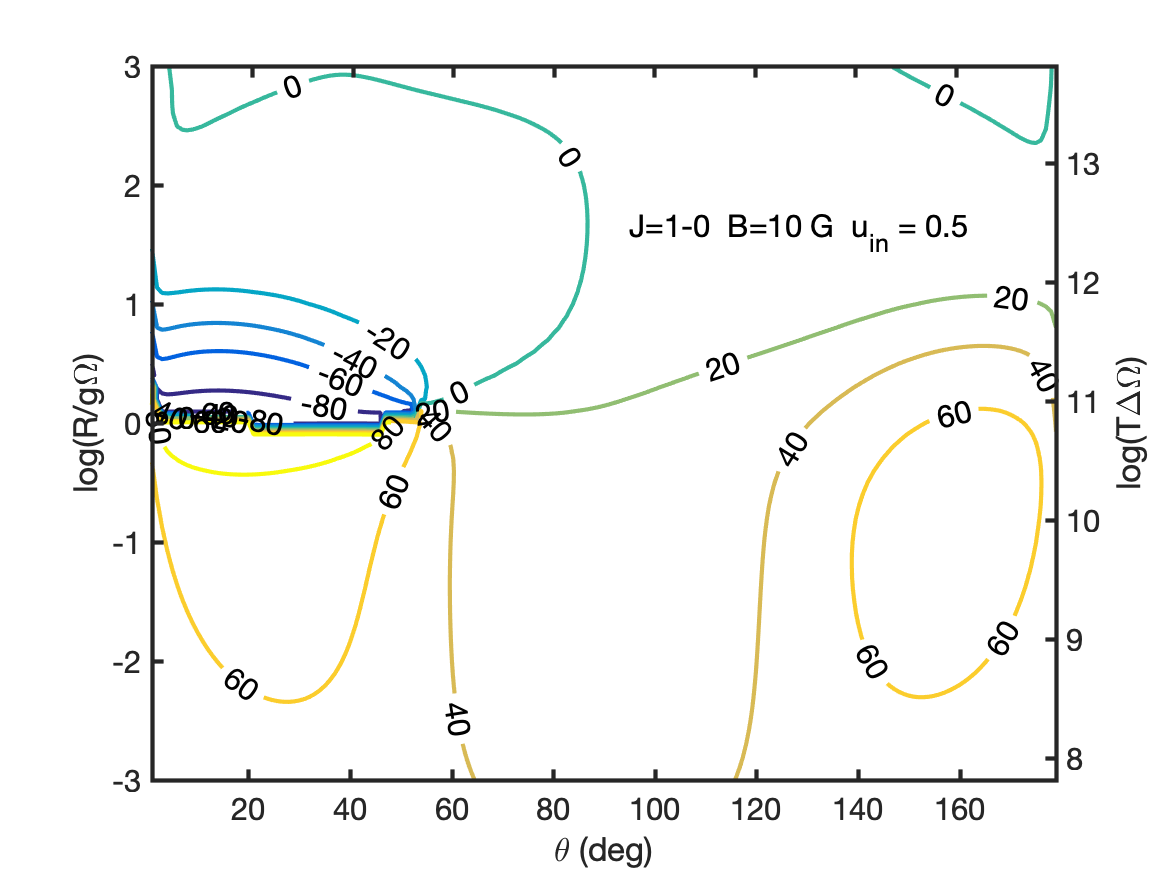} 
       \caption{}
    \end{subfigure}
     ~ 
    \begin{subfigure}[b]{0.45\textwidth}
      \includegraphics[width=\textwidth]{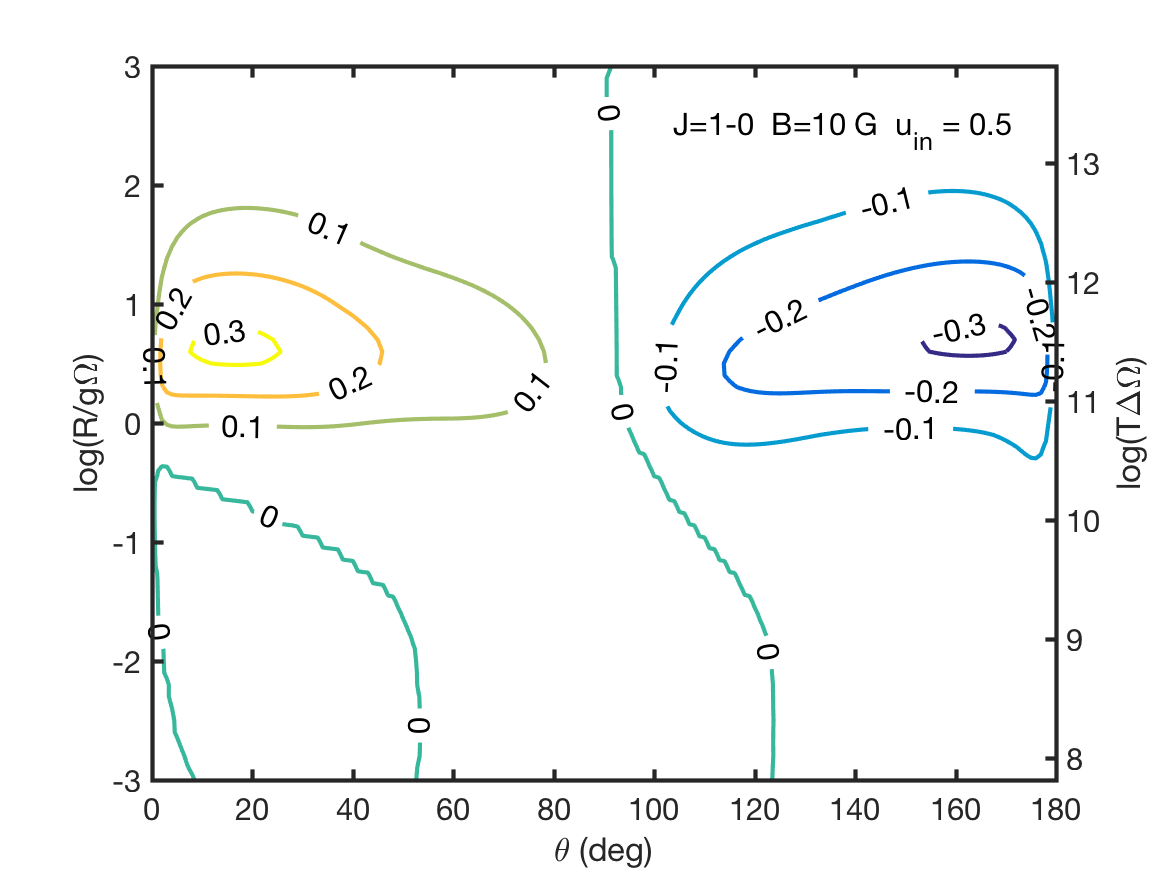}
      \caption{}
    \end{subfigure}
  \caption{Simulations of a SiO maser with $50\%$ polarized seed radiation. Linear polarization fraction (a,d) and angle (b,e) and circular polarization fraction (c,f). Magnetic field strength and transition angular momentum are denoted inside the figure.}
\end{figure*}

\begin{figure*}
    \centering
    \begin{subfigure}[b]{0.32\textwidth}
       \includegraphics[width=\textwidth]{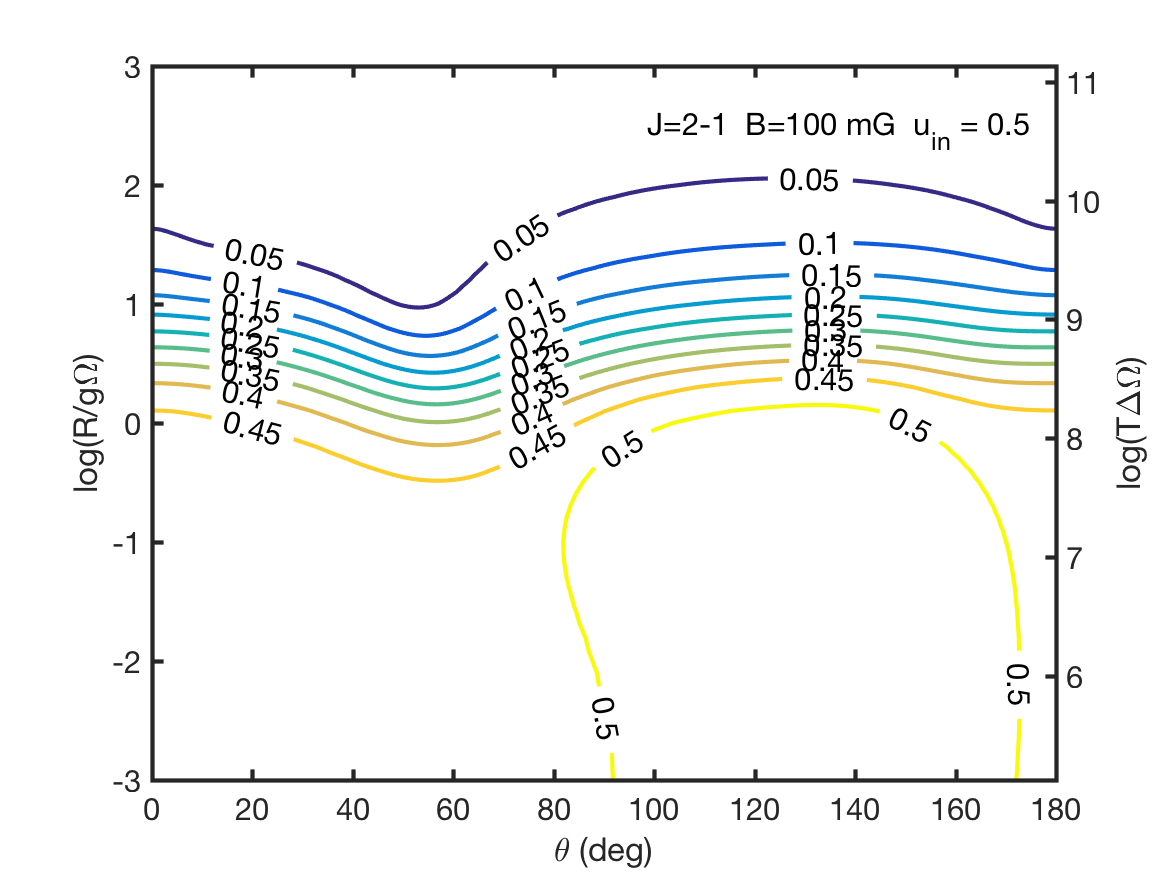}
       \caption{}
    \end{subfigure}
    ~
    \begin{subfigure}[b]{0.32\textwidth}
       \includegraphics[width=\textwidth]{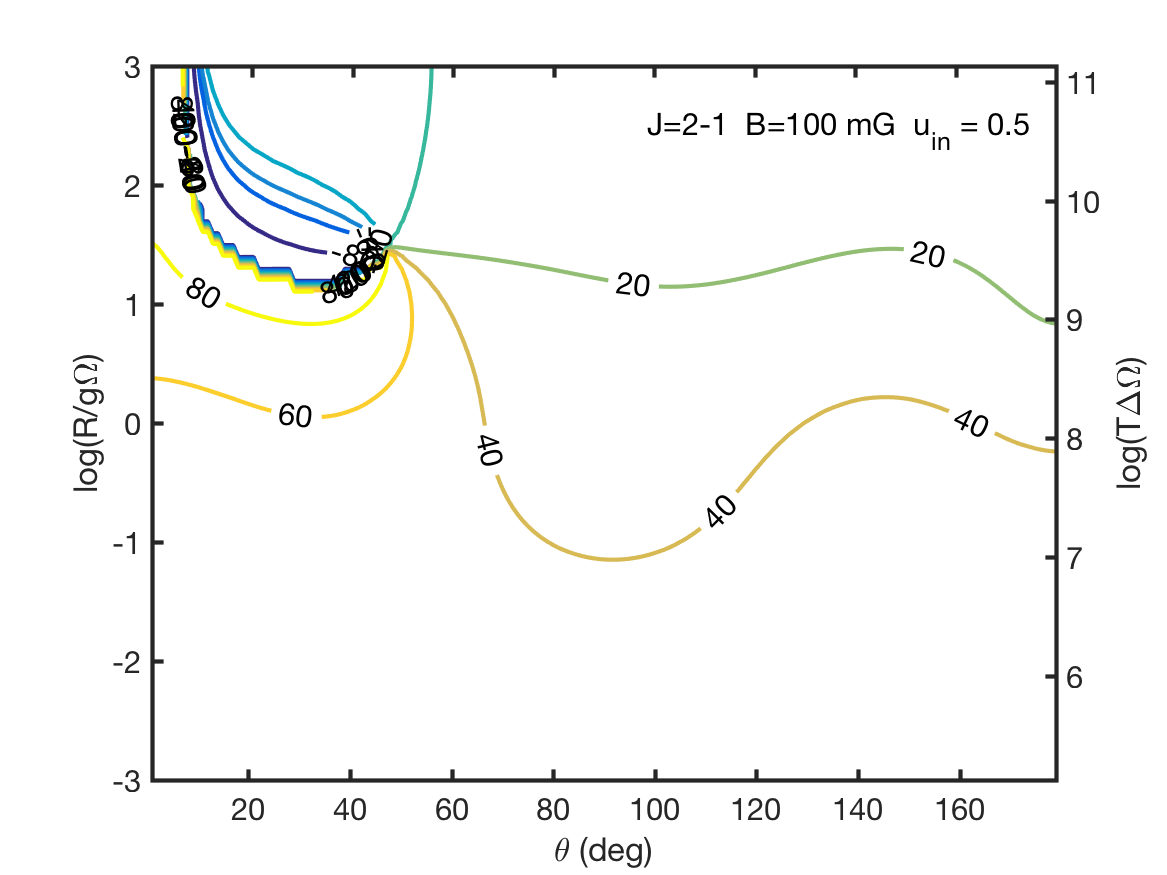}
       \caption{}
    \end{subfigure}
     ~
    \begin{subfigure}[b]{0.32\textwidth}
      \includegraphics[width=\textwidth]{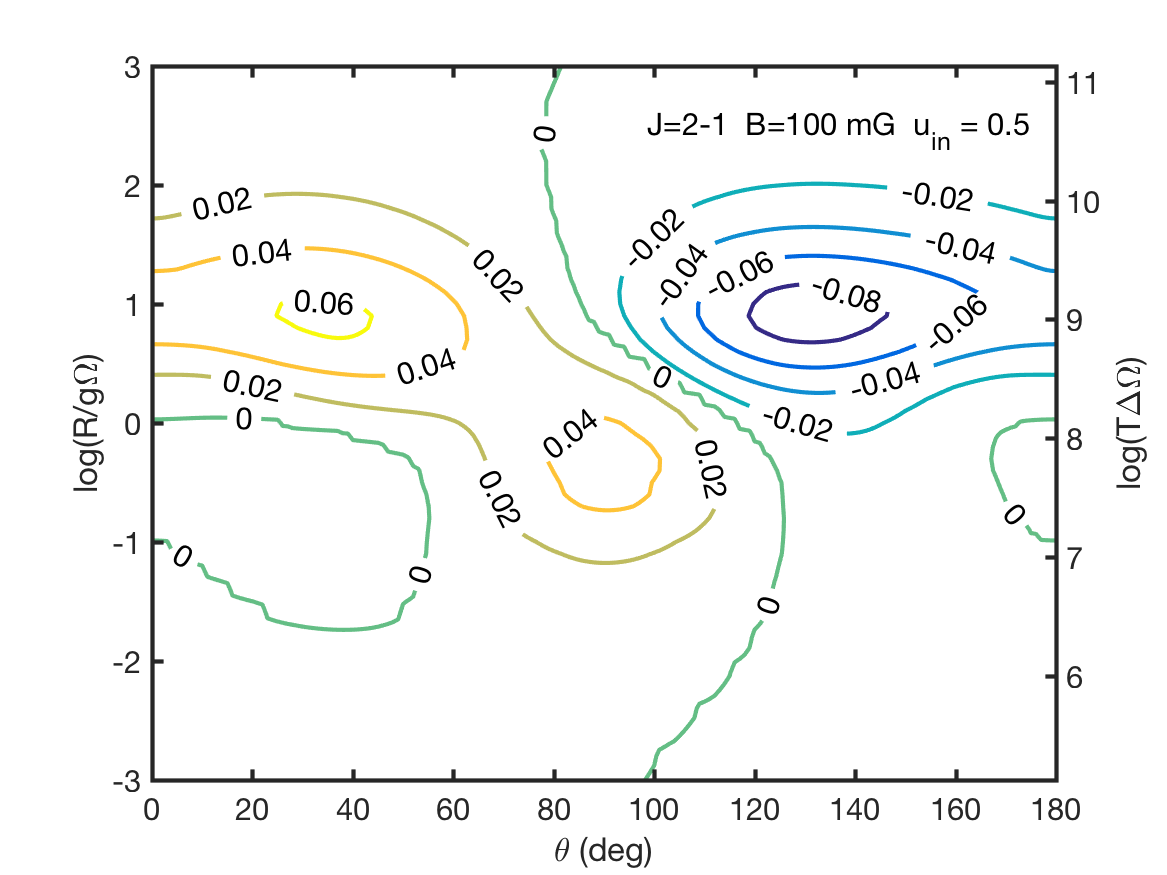}
      \caption{}
    \end{subfigure}
    ~
    \begin{subfigure}[b]{0.32\textwidth}
       \includegraphics[width=\textwidth]{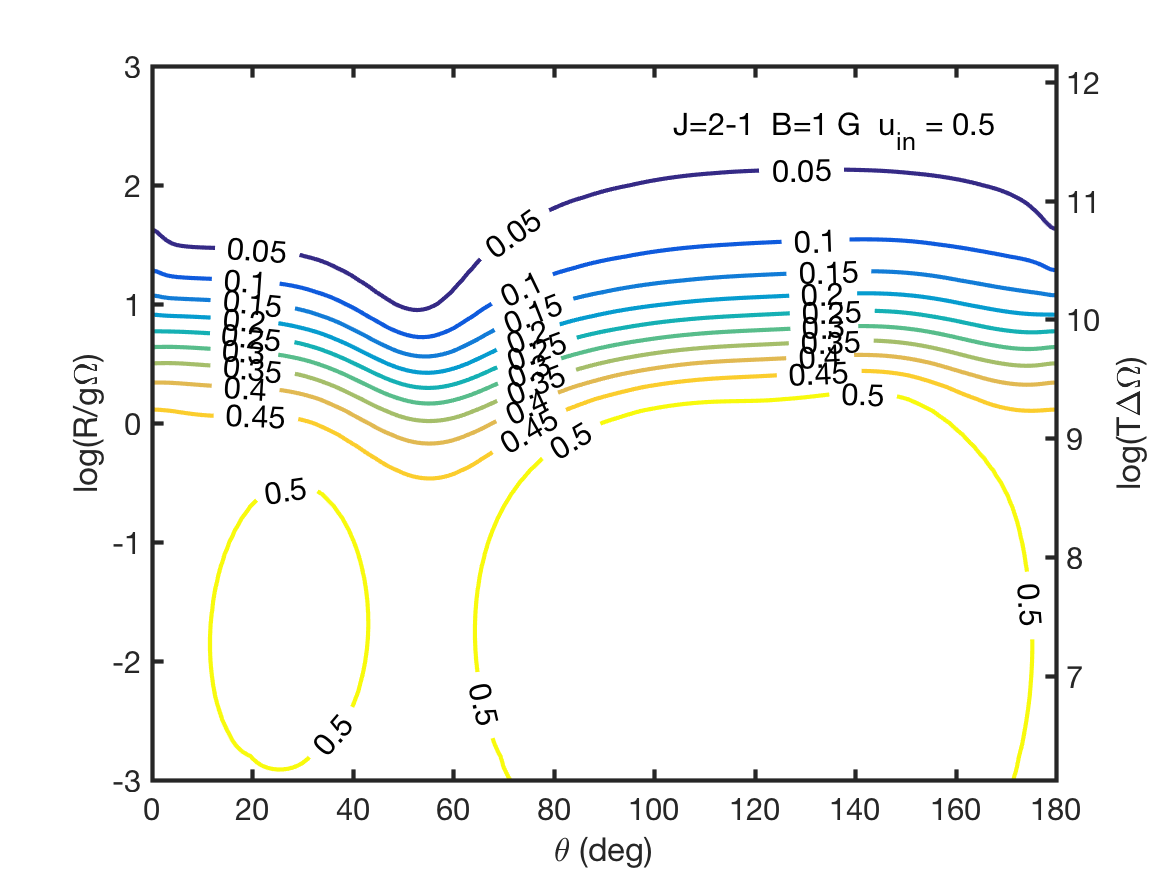}
       \caption{}
    \end{subfigure}
    ~
    \begin{subfigure}[b]{0.32\textwidth}
       \includegraphics[width=\textwidth]{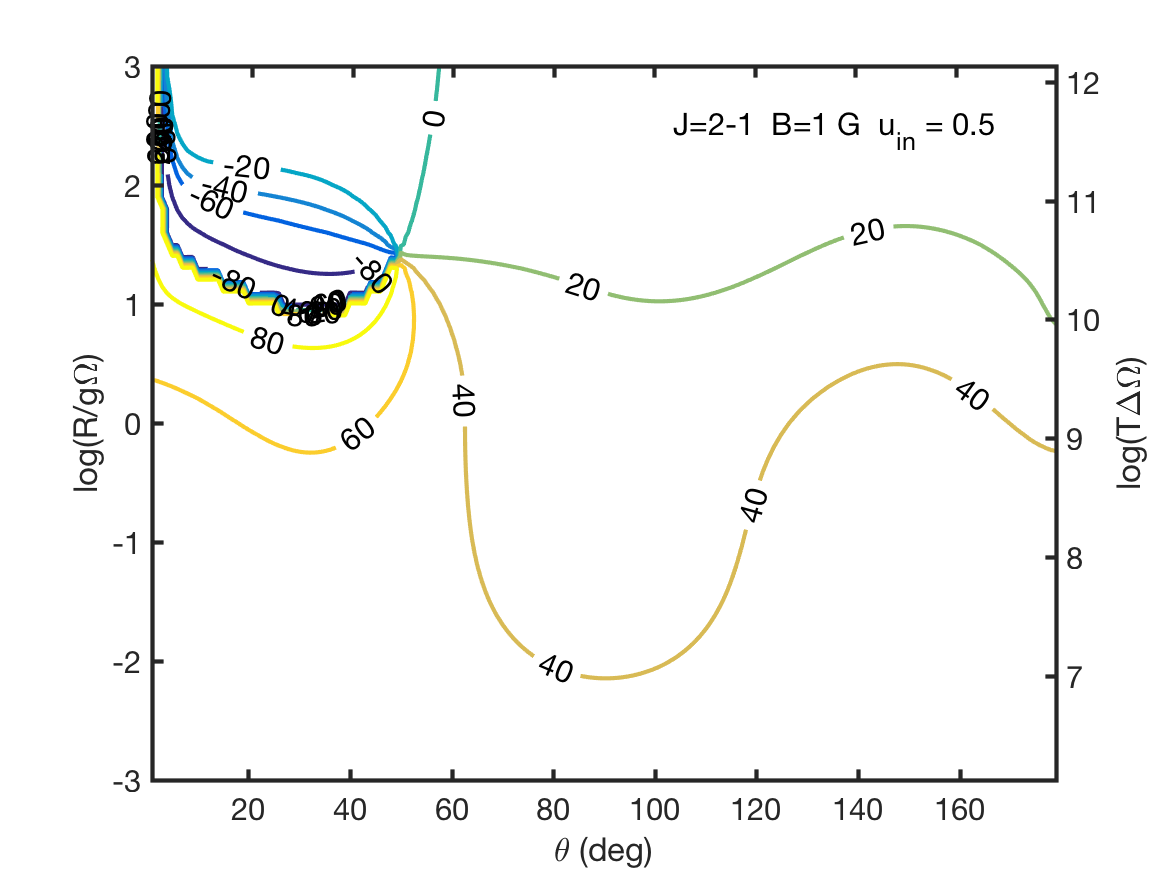}
       \caption{}
    \end{subfigure}
     ~
    \begin{subfigure}[b]{0.32\textwidth}
      \includegraphics[width=\textwidth]{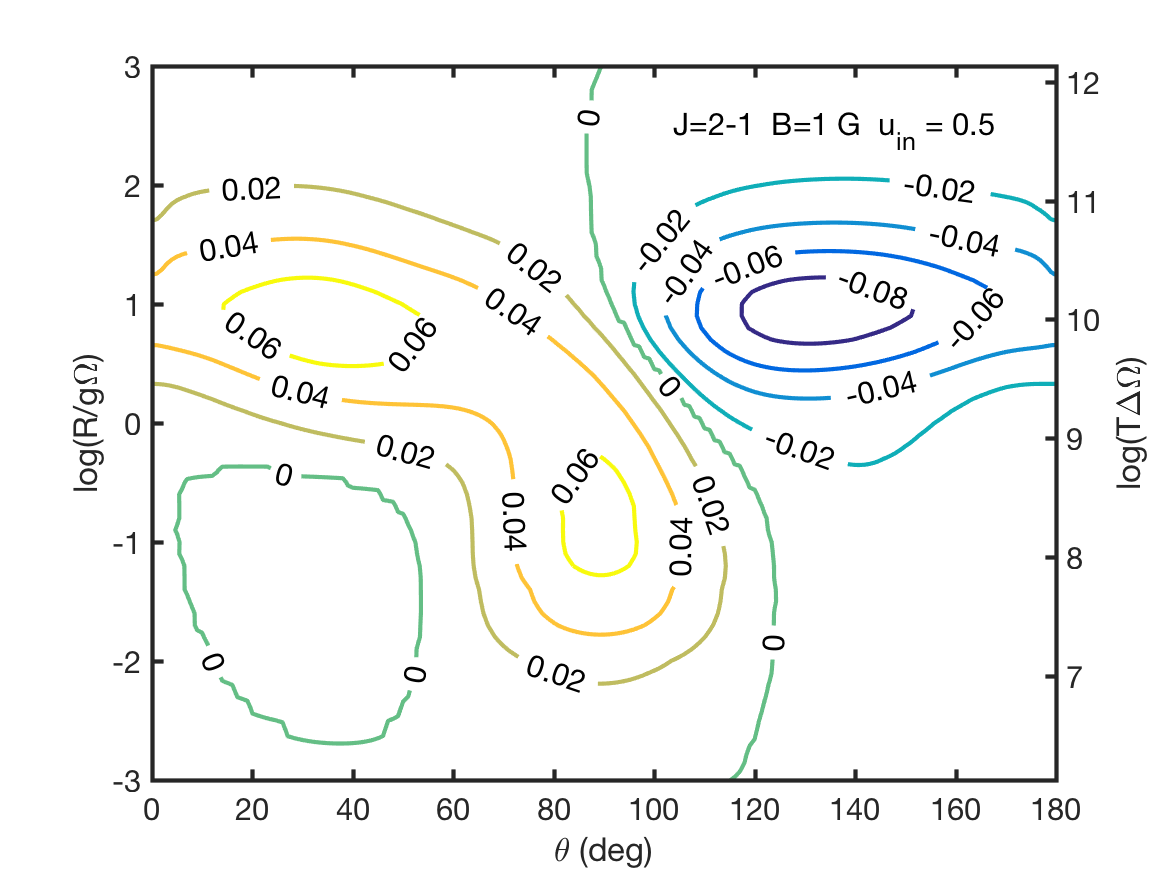}
      \caption{}
    \end{subfigure}
    ~
    \begin{subfigure}[b]{0.32\textwidth}
       \includegraphics[width=\textwidth]{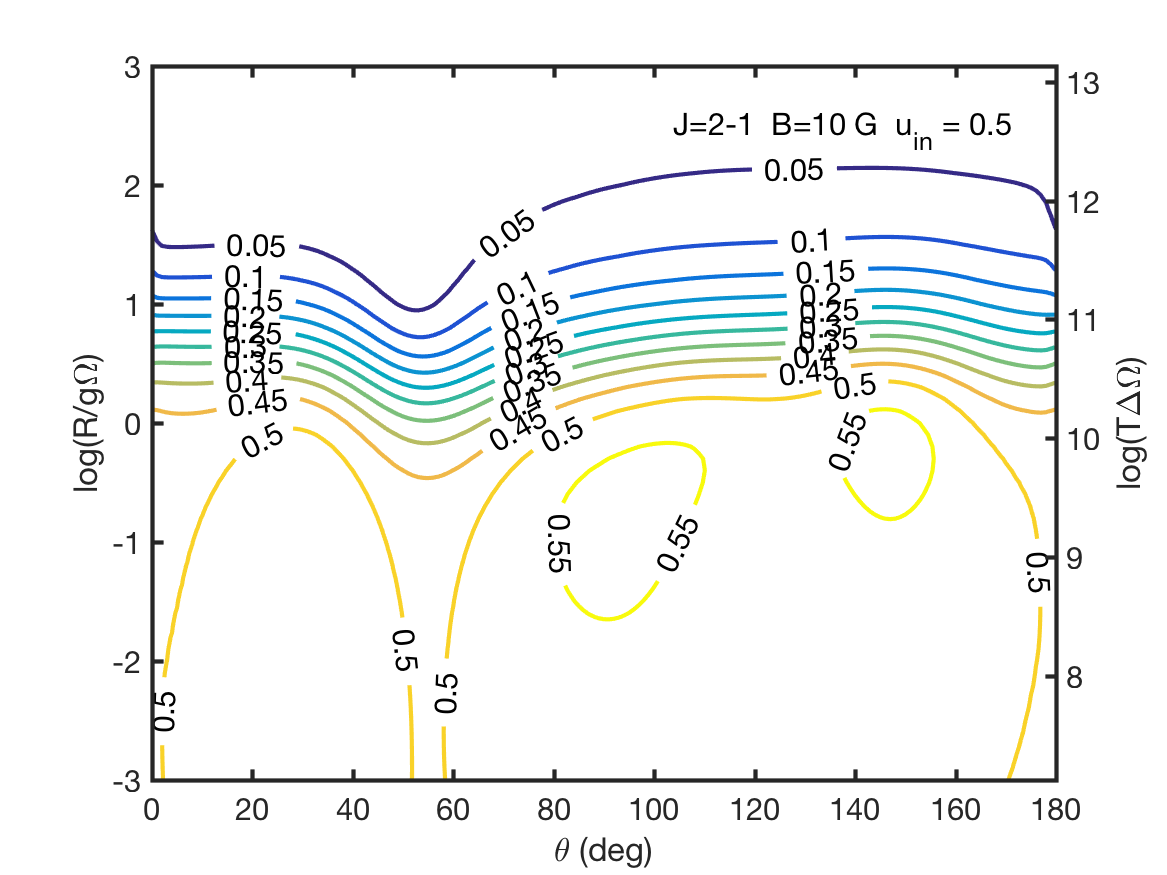}
       \caption{}
    \end{subfigure}
    ~
    \begin{subfigure}[b]{0.32\textwidth}
       \includegraphics[width=\textwidth]{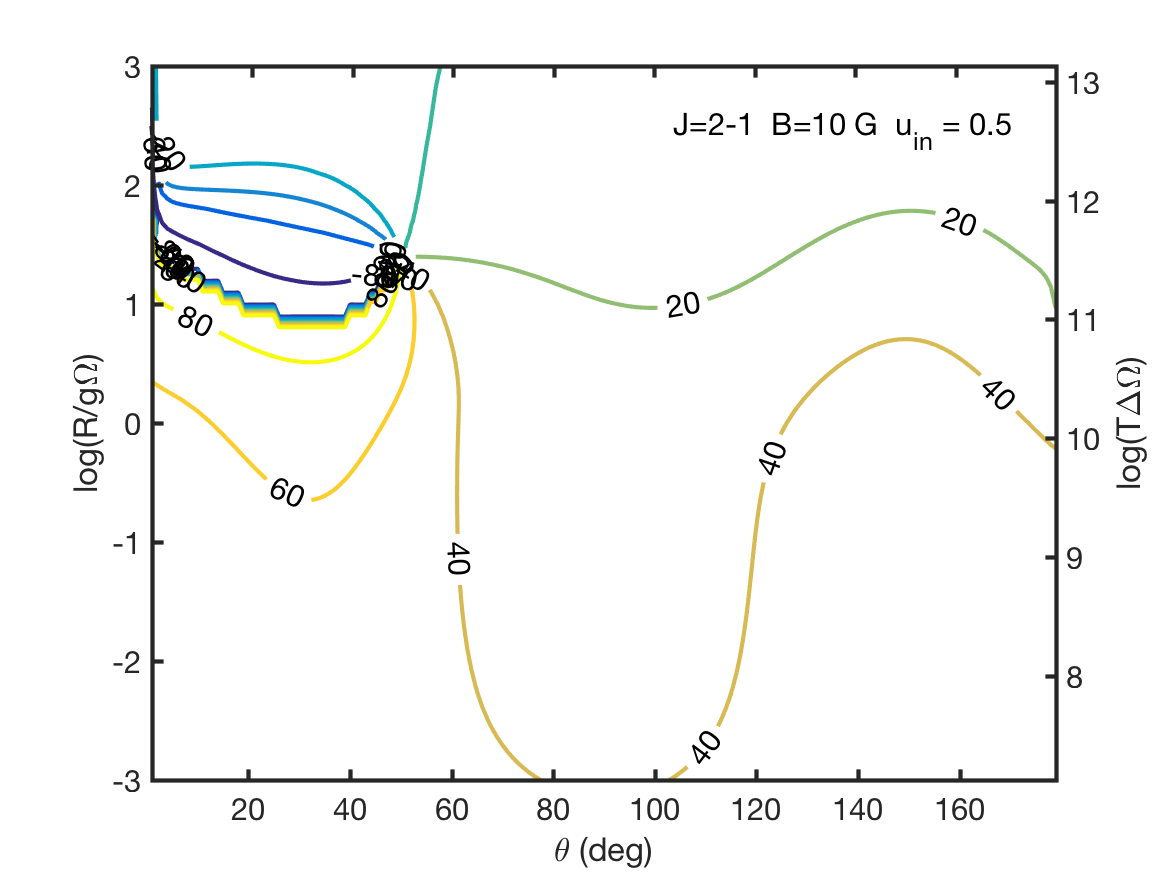}
       \caption{}
    \end{subfigure}
     ~
    \begin{subfigure}[b]{0.32\textwidth}
      \includegraphics[width=\textwidth]{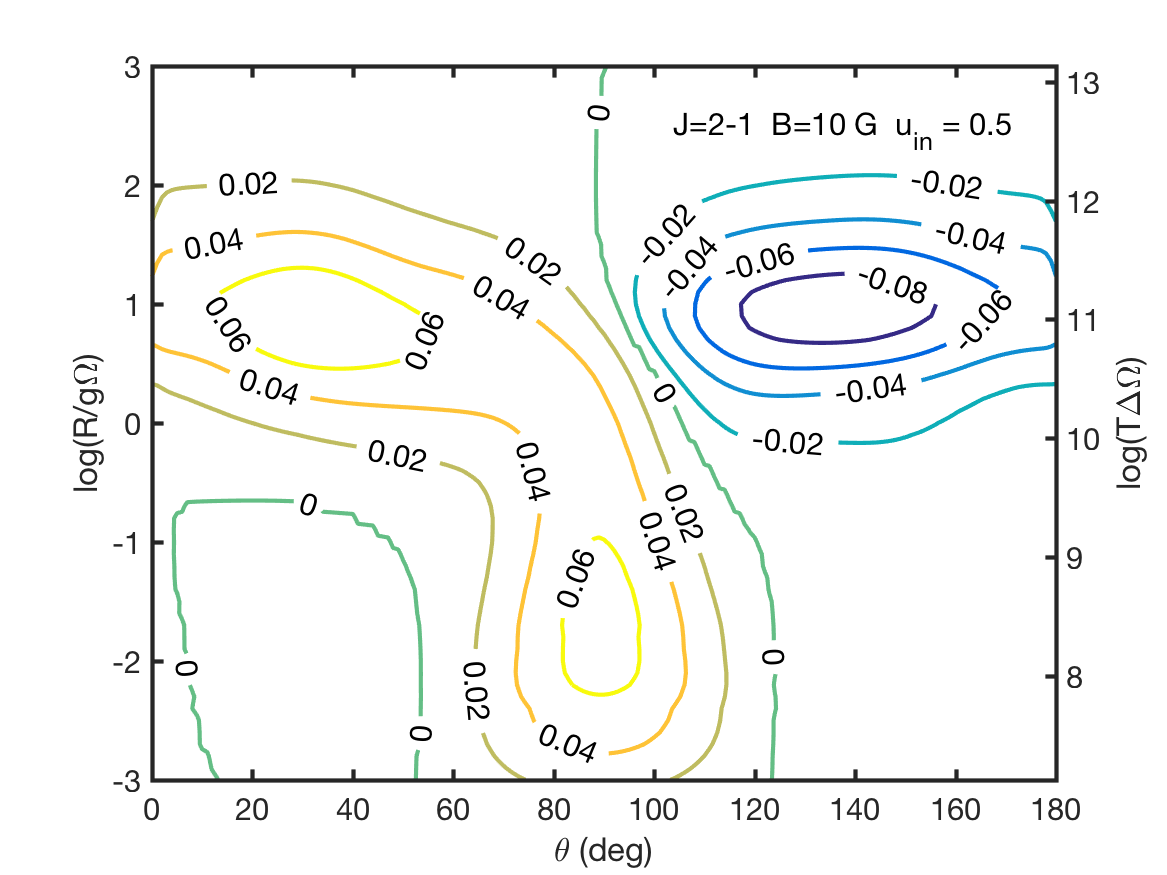}
      \caption{}
    \end{subfigure}
  \caption{Simulations of a SiO maser with $50\%$ polarized seed radiation. Linear polarization fraction (a,d,g) and angle (b,e,h) and circular polarization fraction (c,f,i). Magnetic field strength and transition angular momentum are denoted inside the figure.}
\end{figure*}

\begin{figure*}
    \centering
    \begin{subfigure}[b]{0.32\textwidth}
       \includegraphics[width=\textwidth]{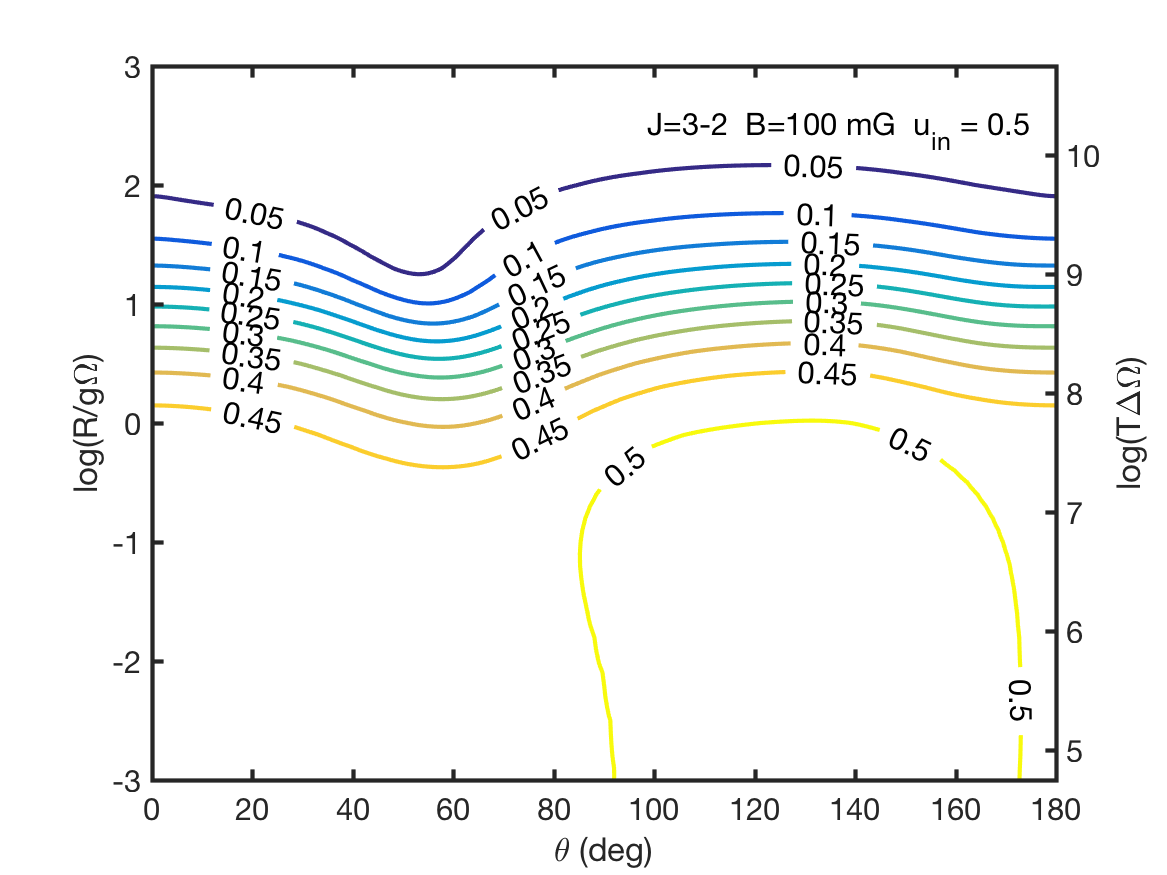}
       \caption{}
    \end{subfigure}
    ~
    \begin{subfigure}[b]{0.32\textwidth}
       \includegraphics[width=\textwidth]{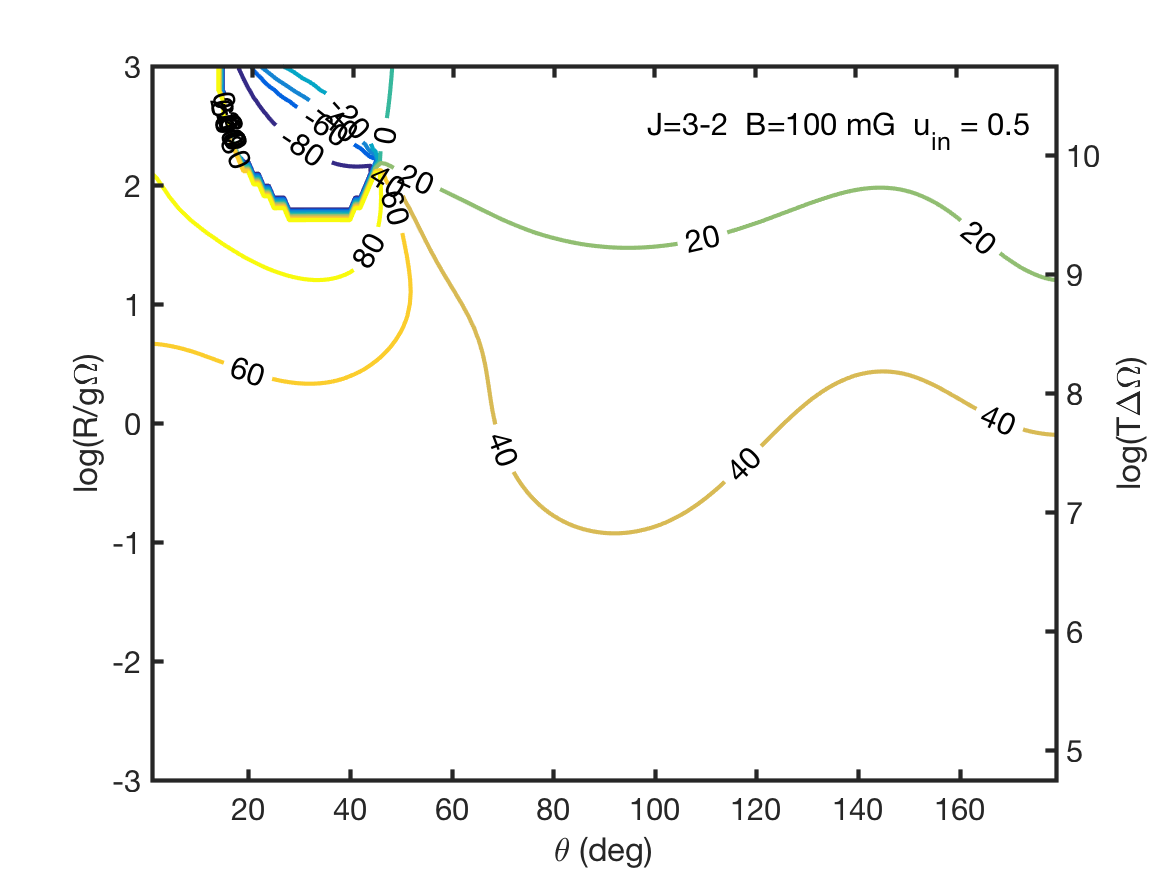}
       \caption{}
    \end{subfigure}
     ~
    \begin{subfigure}[b]{0.32\textwidth}
      \includegraphics[width=\textwidth]{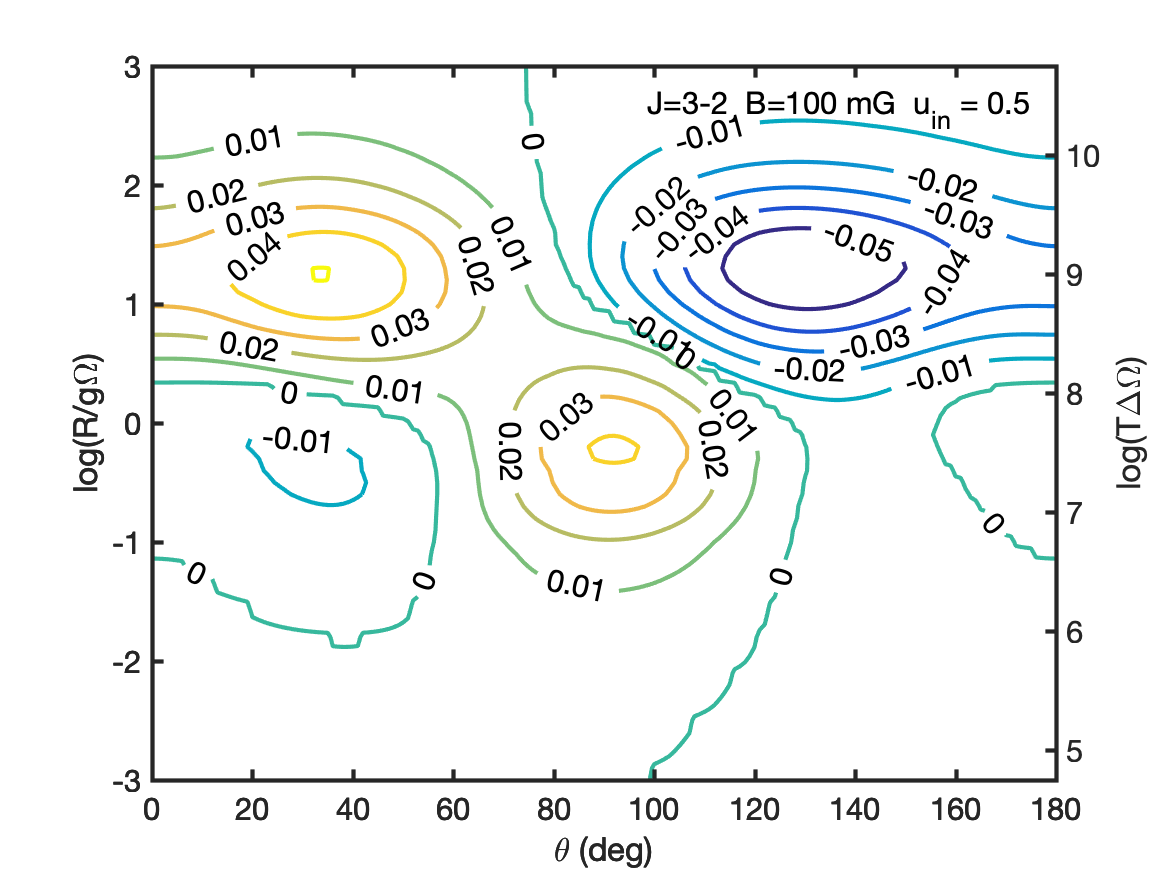}
      \caption{}
    \end{subfigure}
    ~
    \begin{subfigure}[b]{0.32\textwidth}
       \includegraphics[width=\textwidth]{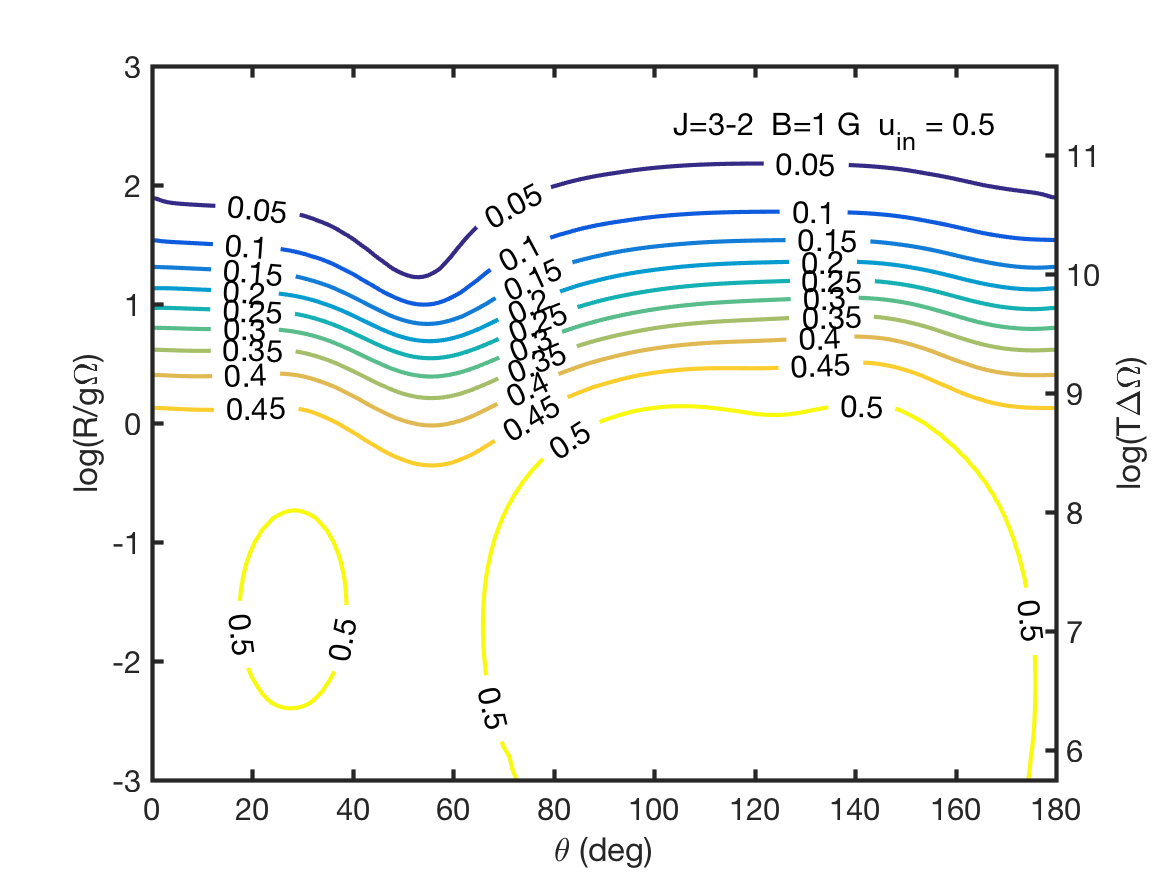}
       \caption{}
    \end{subfigure}
    ~
    \begin{subfigure}[b]{0.32\textwidth}
       \includegraphics[width=\textwidth]{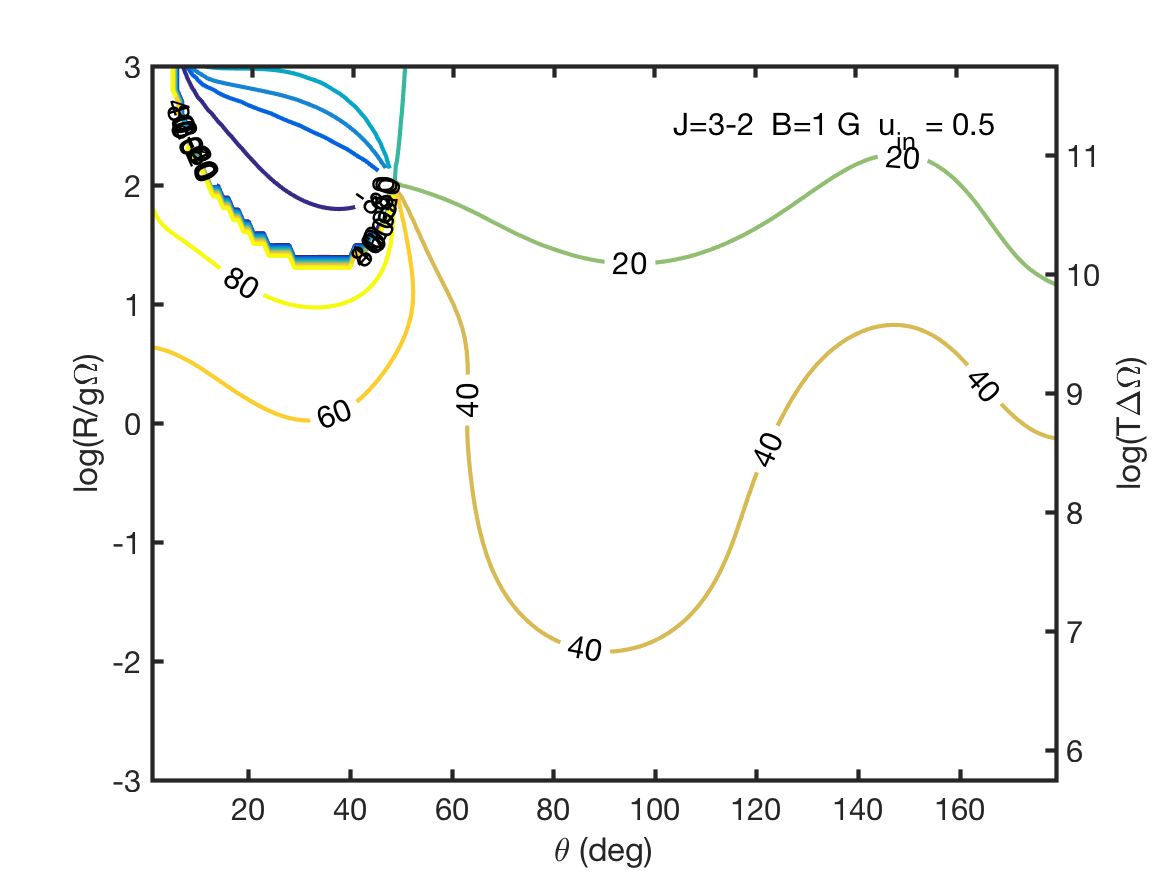}
       \caption{}
    \end{subfigure}
     ~
    \begin{subfigure}[b]{0.32\textwidth}
      \includegraphics[width=\textwidth]{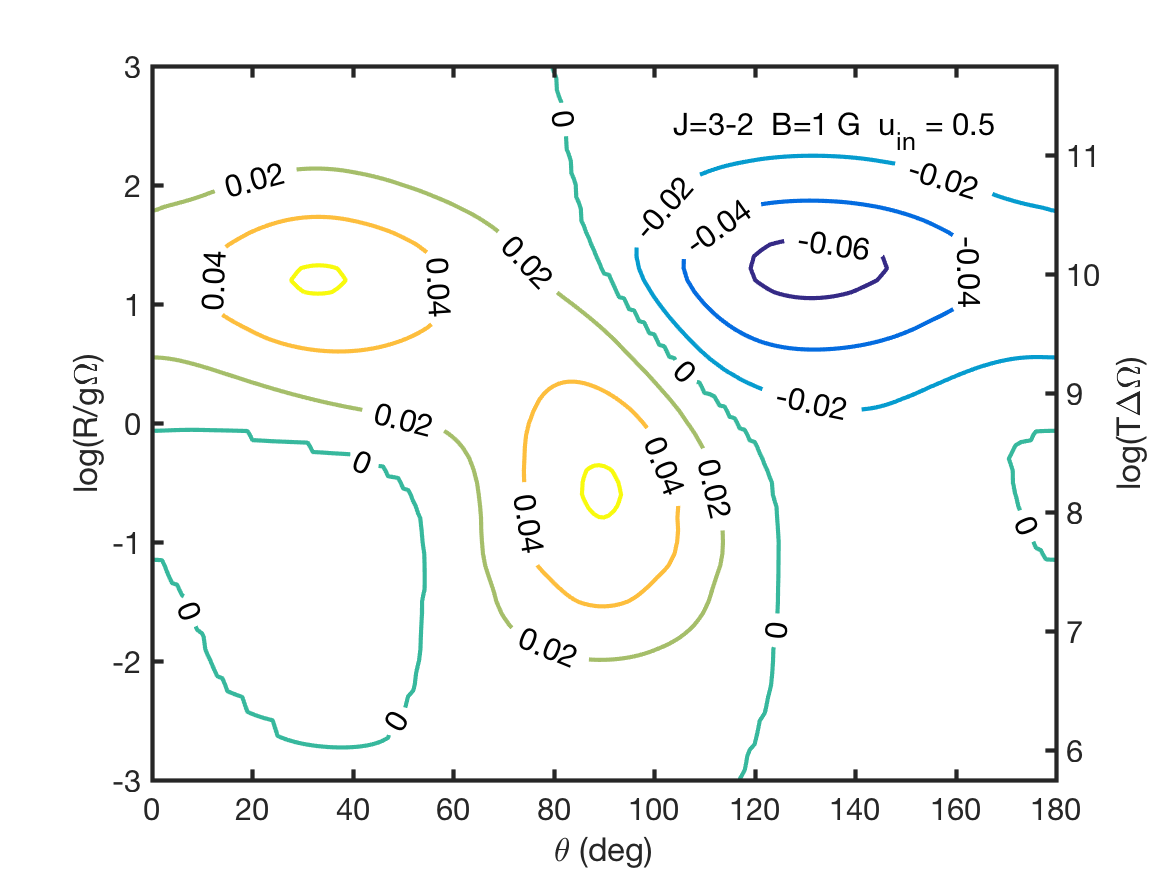}
      \caption{}
    \end{subfigure}
     ~
    \begin{subfigure}[b]{0.32\textwidth}
       \includegraphics[width=\textwidth]{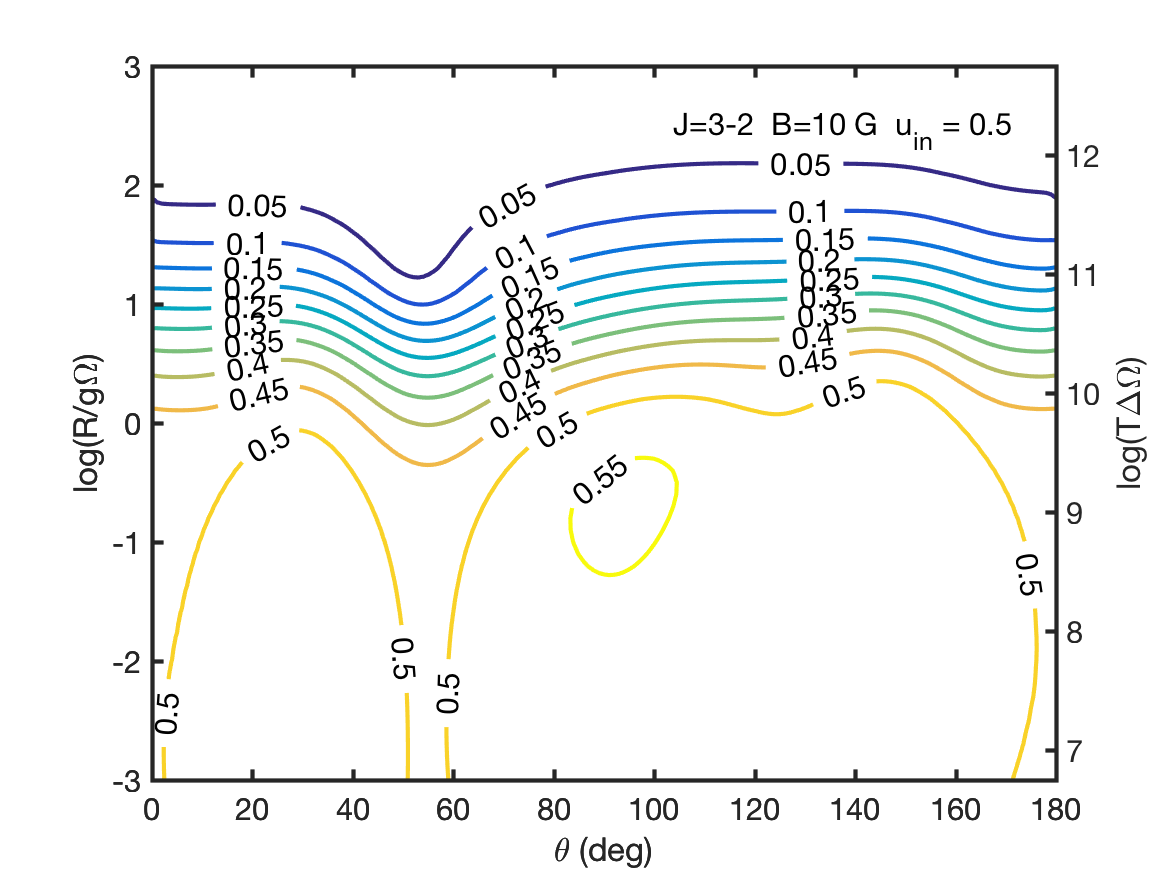}
       \caption{}
    \end{subfigure}
    ~
    \begin{subfigure}[b]{0.32\textwidth}
       \includegraphics[width=\textwidth]{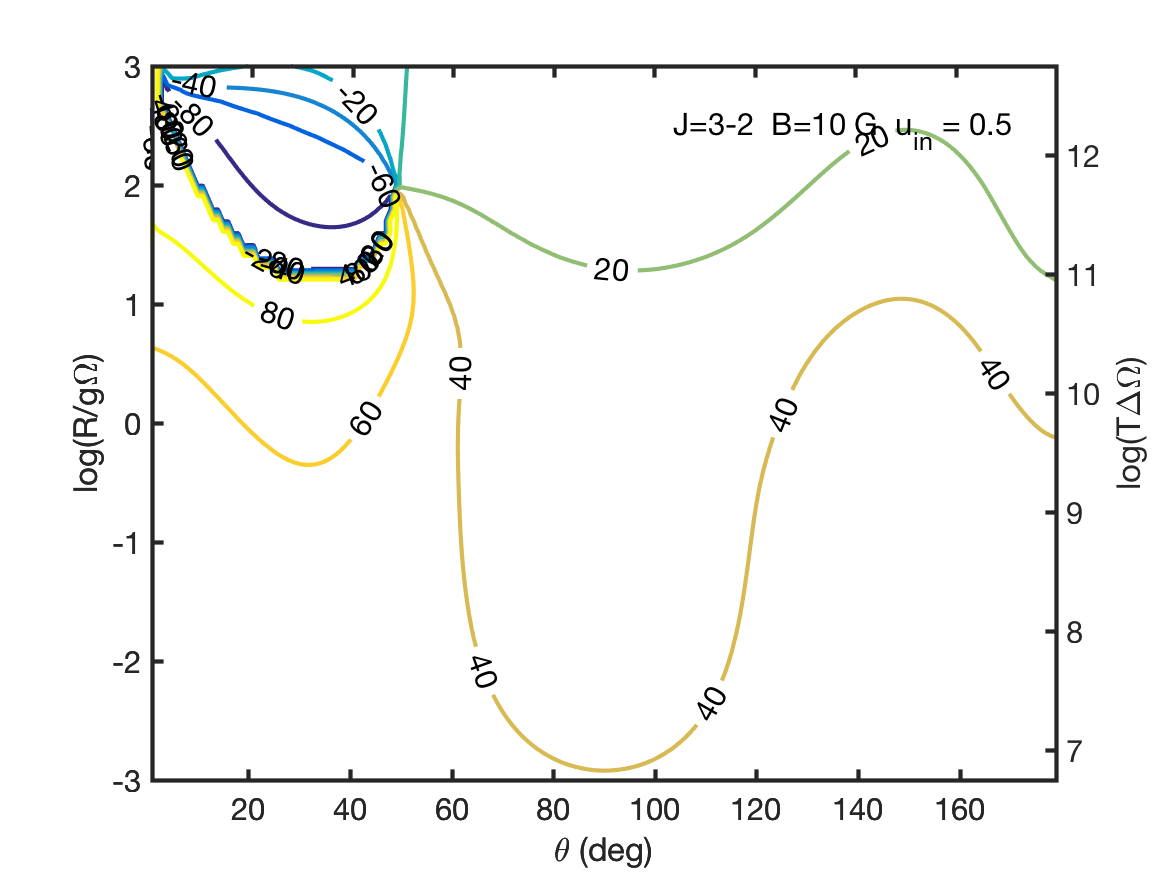}
       \caption{}
    \end{subfigure}
     ~
    \begin{subfigure}[b]{0.32\textwidth}
      \includegraphics[width=\textwidth]{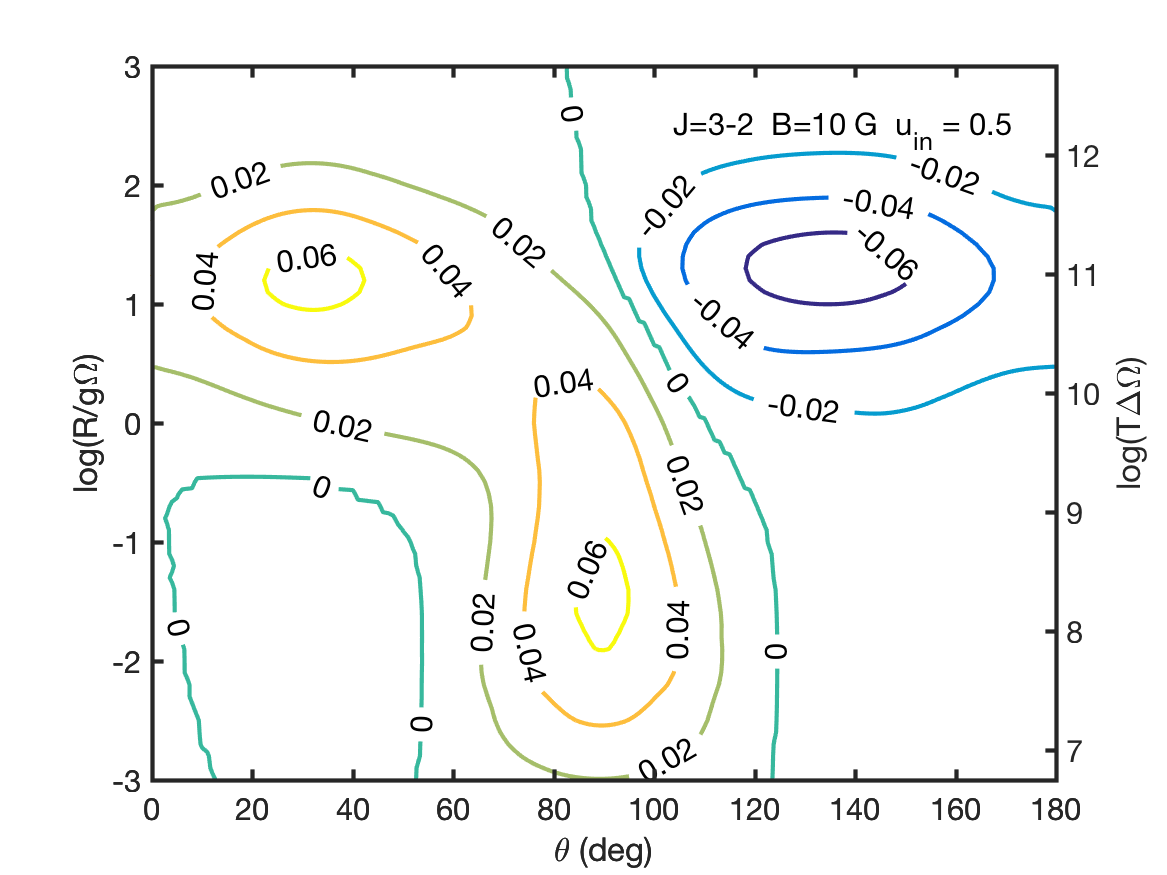}
      \caption{}
    \end{subfigure}
  \caption{Simulations of a SiO maser with $50\%$ polarized seed radiation. Linear polarization fraction (a,d,g) and angle (b,e,h) and circular polarization fraction (c,f,i). Magnetic field strength and transition angular momentum are denoted inside the figure.}
\end{figure*}

\begin{figure*}
    \centering
    \begin{subfigure}[b]{0.45\textwidth}
       \includegraphics[width=\textwidth]{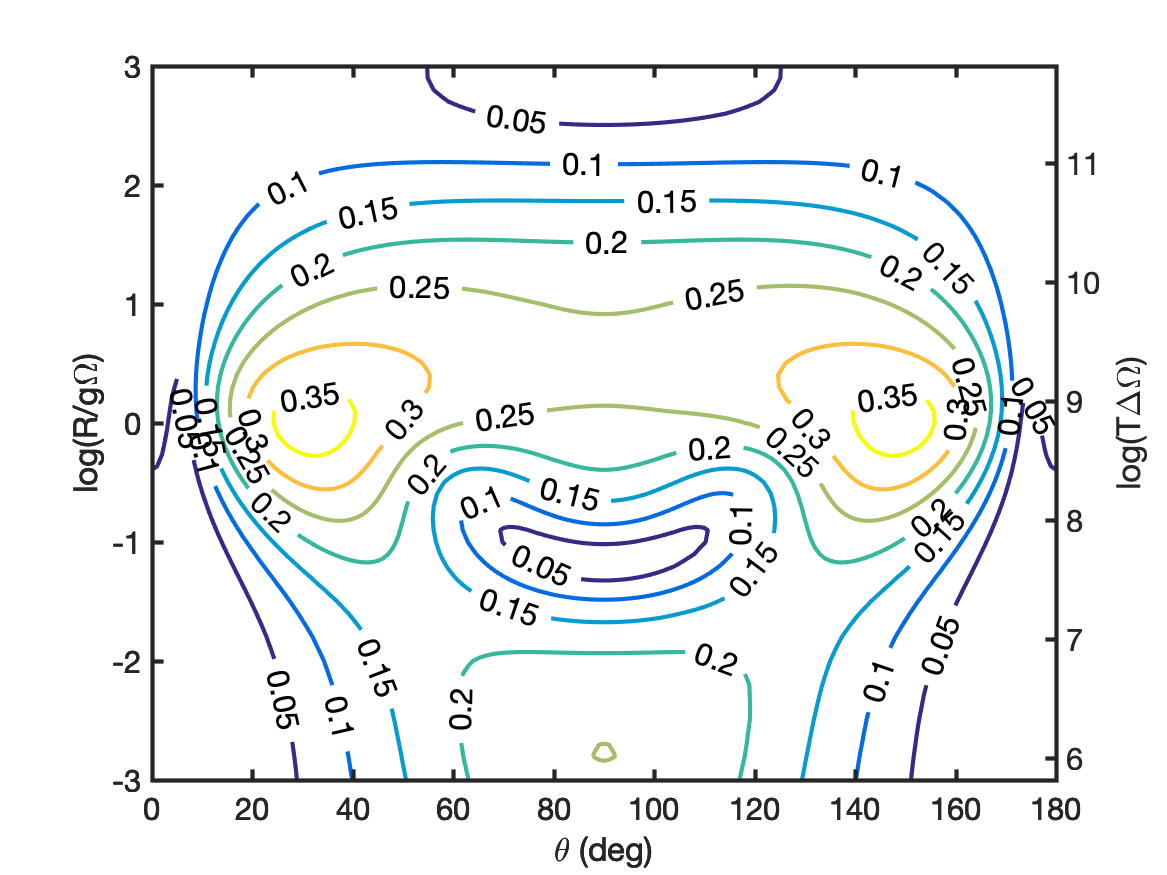} 
       \caption{}
    \end{subfigure}
    ~ 
    \begin{subfigure}[b]{0.45\textwidth}
       \includegraphics[width=\textwidth]{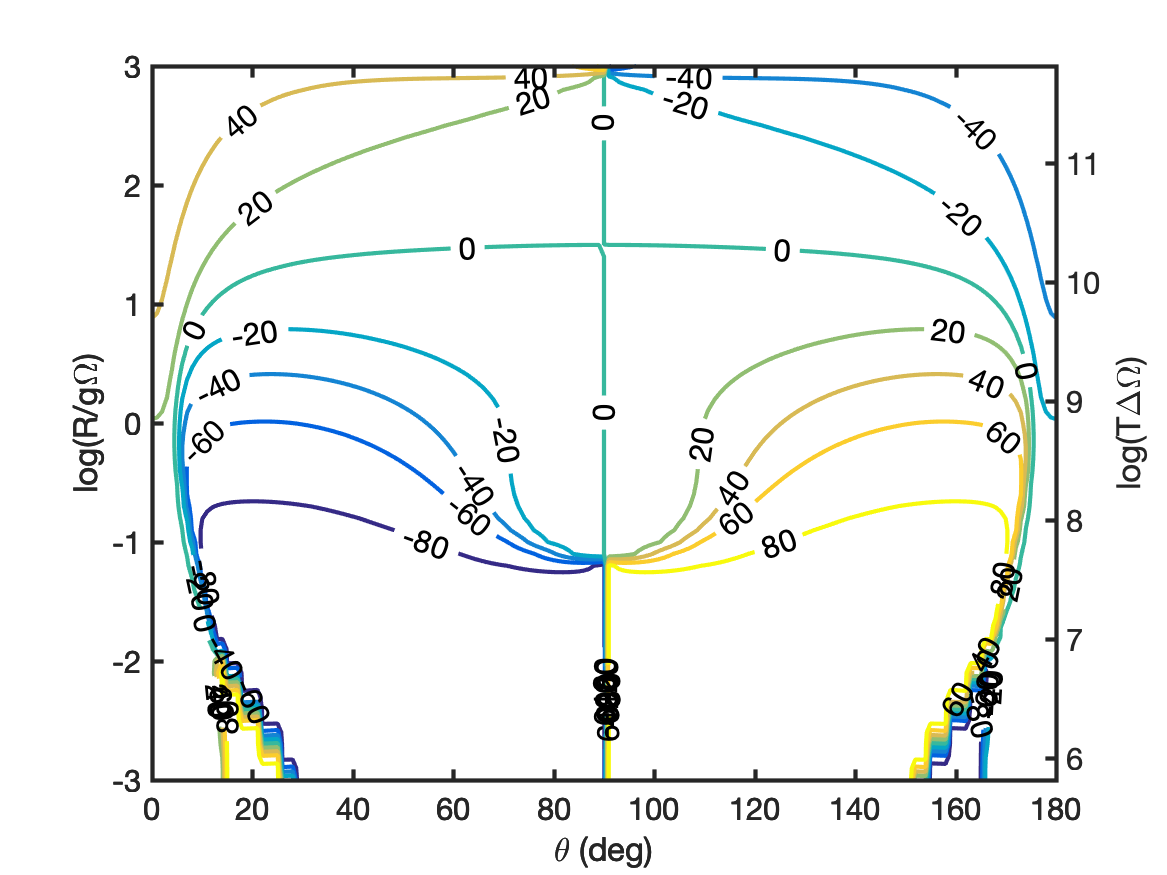} 
       \caption{}
    \end{subfigure}
     ~ 
    \begin{subfigure}[b]{0.45\textwidth}
      \includegraphics[width=\textwidth]{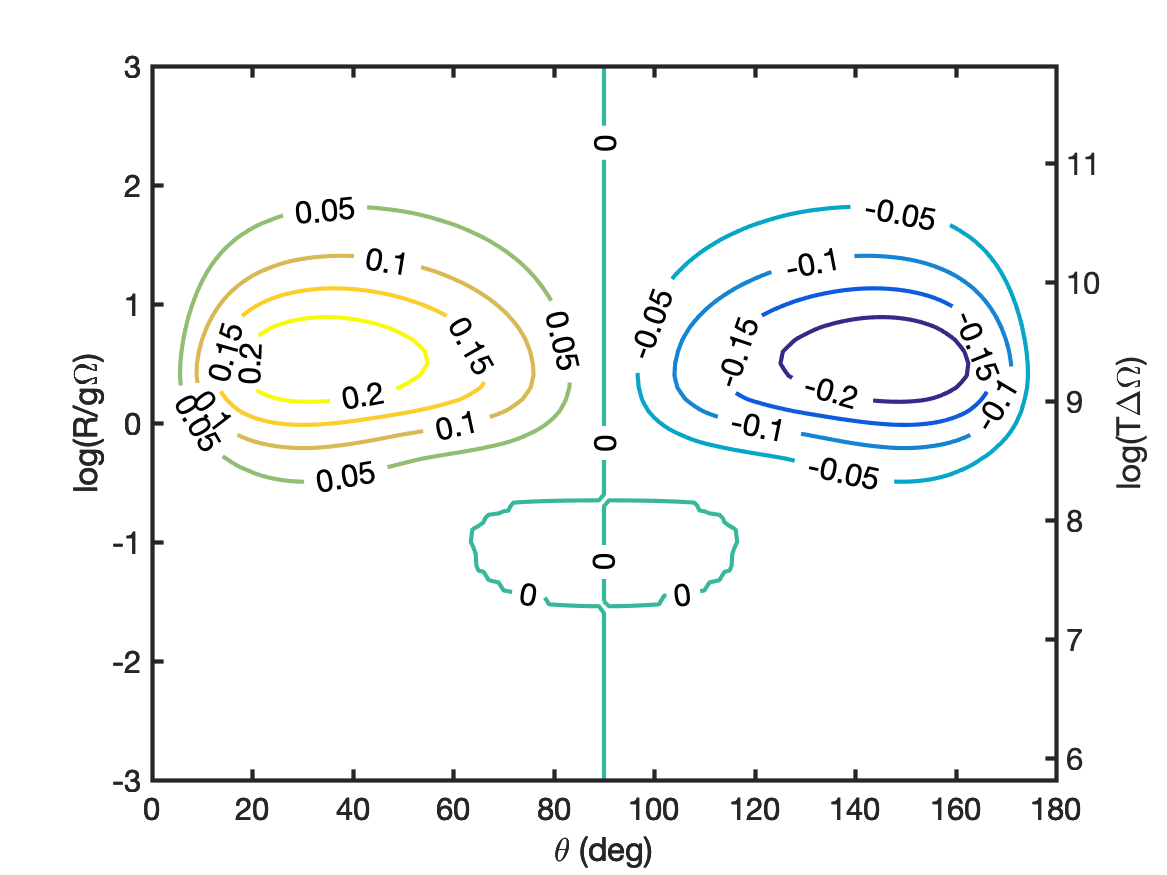}
      \caption{}
    \end{subfigure}
     ~
    \begin{subfigure}[b]{0.45\textwidth}
       \includegraphics[width=\textwidth]{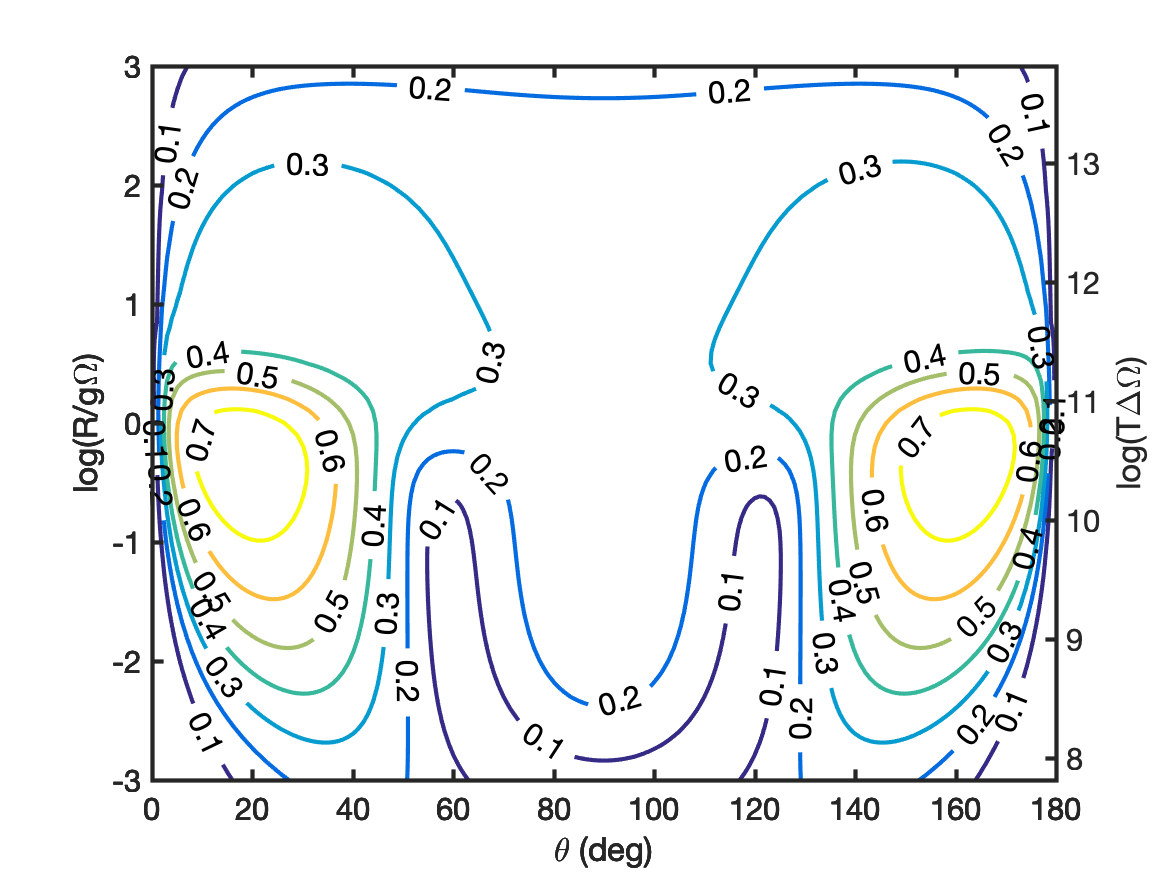} 
       \caption{}
    \end{subfigure}
    ~ 
    \begin{subfigure}[b]{0.45\textwidth}
       \includegraphics[width=\textwidth]{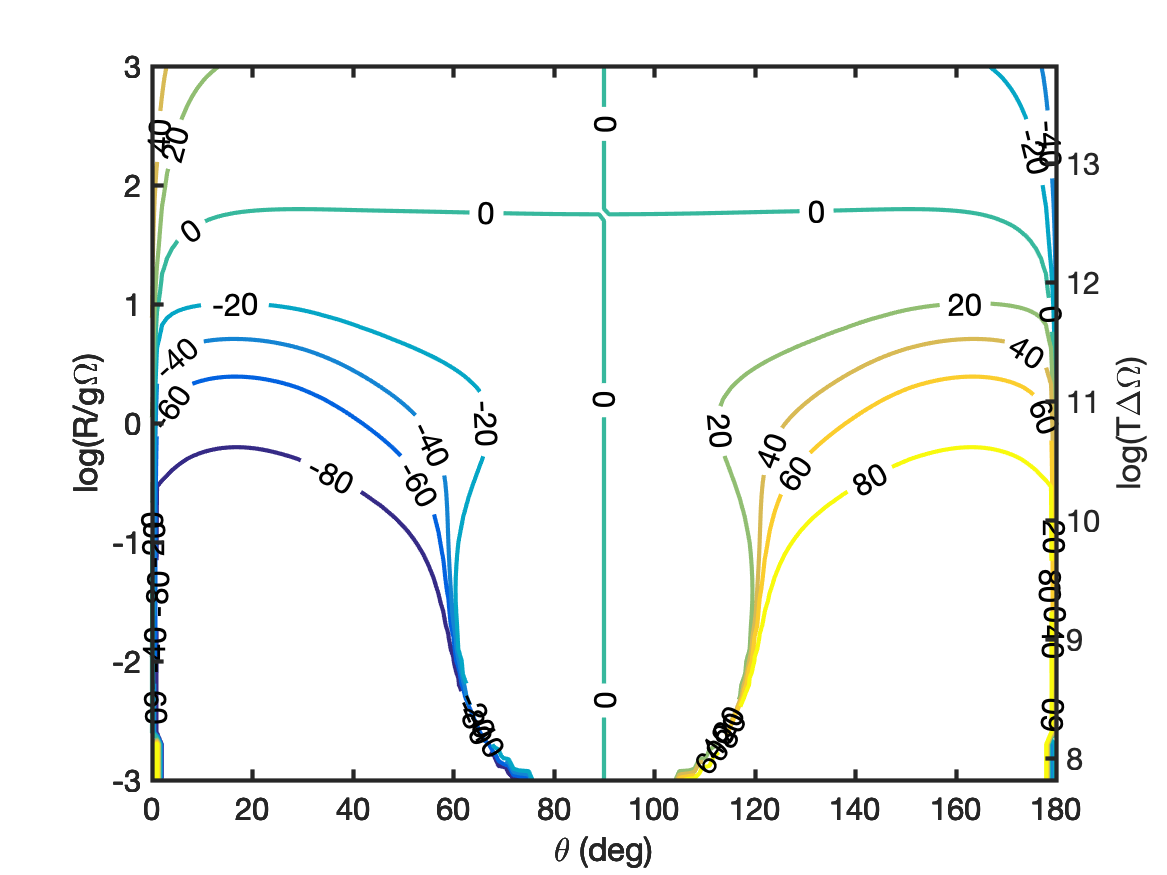} 
       \caption{}
    \end{subfigure}
     ~ 
    \begin{subfigure}[b]{0.45\textwidth}
      \includegraphics[width=\textwidth]{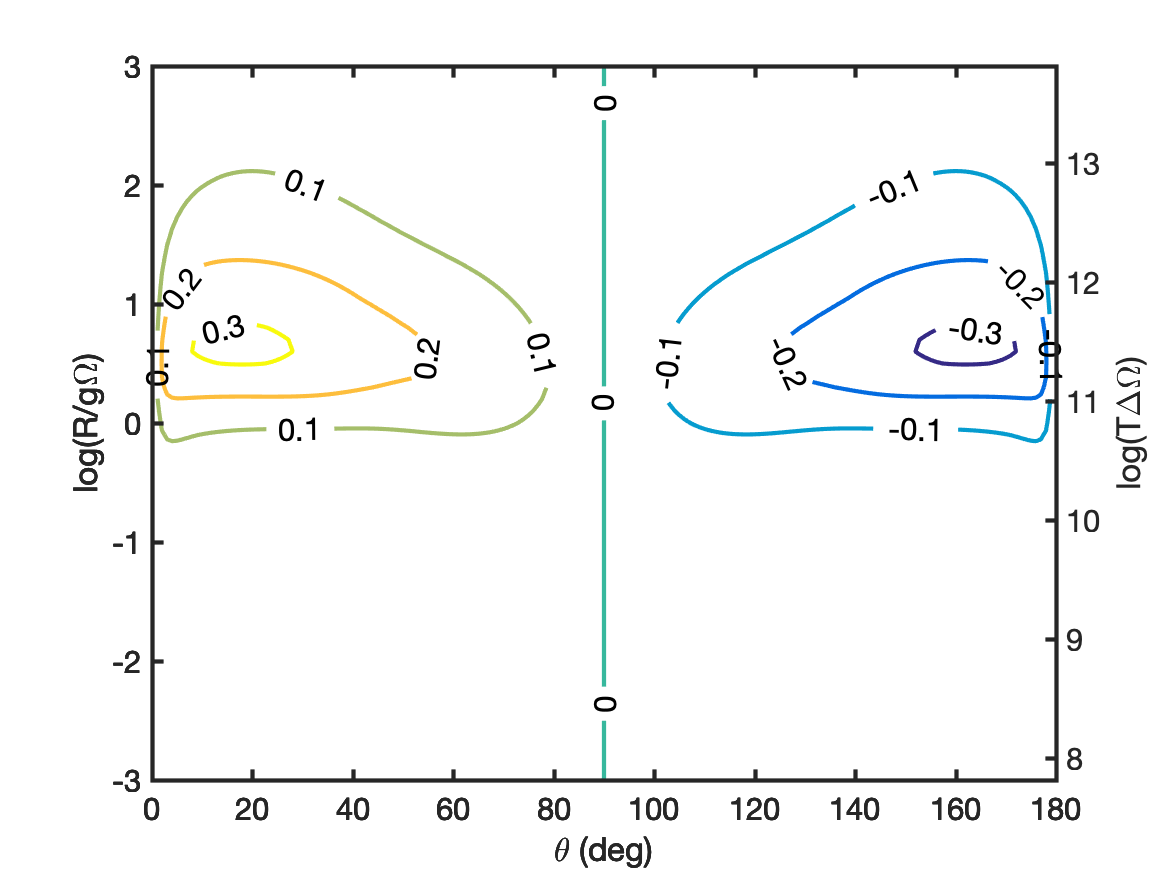}
      \caption{}
    \end{subfigure}
  \caption{Simulations of $J=1-0$ SiO masers with anisotropic pumping direction parallel to the magnetic field. Linear polarization fraction (a,d) and angle (b,e) and circular polarization fraction (c,f). Magnetic field strengths are $B=100$ mG for (a,b,c) and $B=10$ G for (d,e,f).}
\end{figure*}

\begin{figure*}
    \centering
    \begin{subfigure}[b]{0.32\textwidth}
       \includegraphics[width=\textwidth]{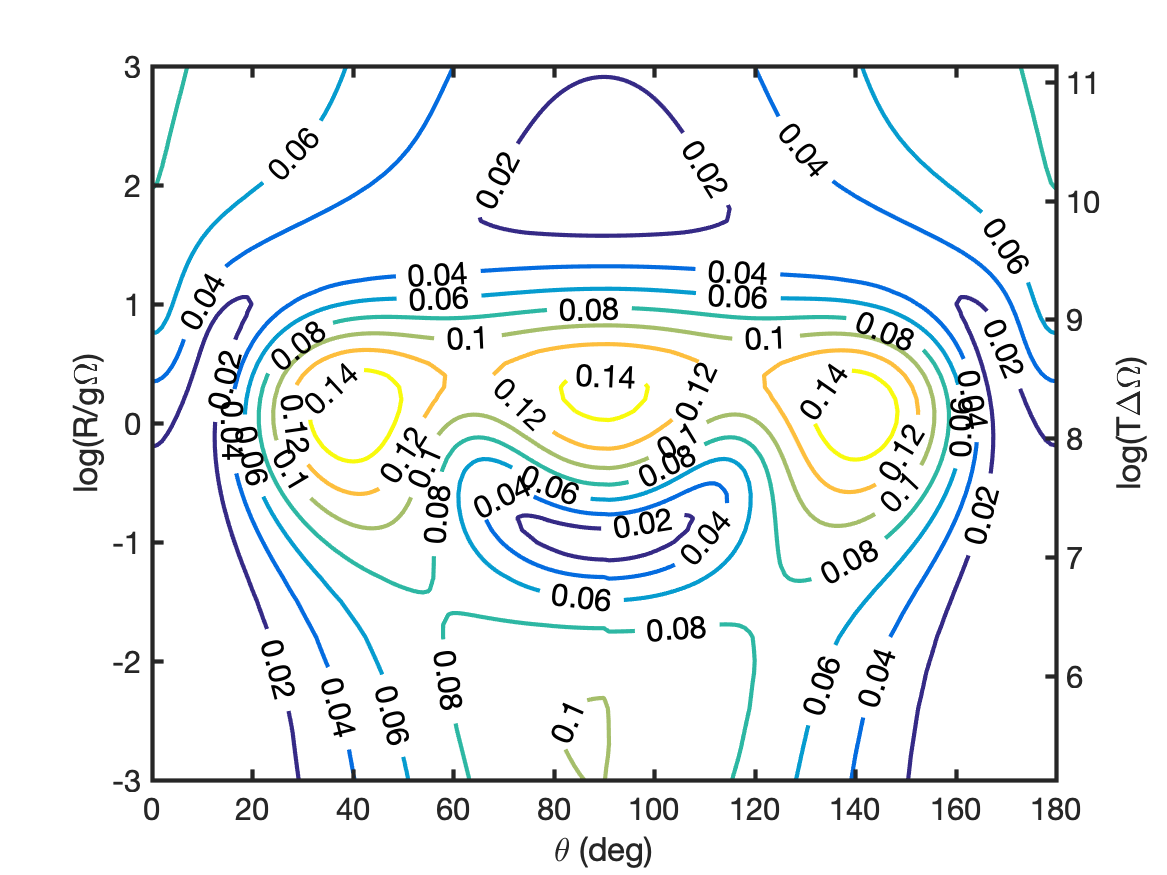}
       \caption{}
    \end{subfigure}
    ~
    \begin{subfigure}[b]{0.32\textwidth}
       \includegraphics[width=\textwidth]{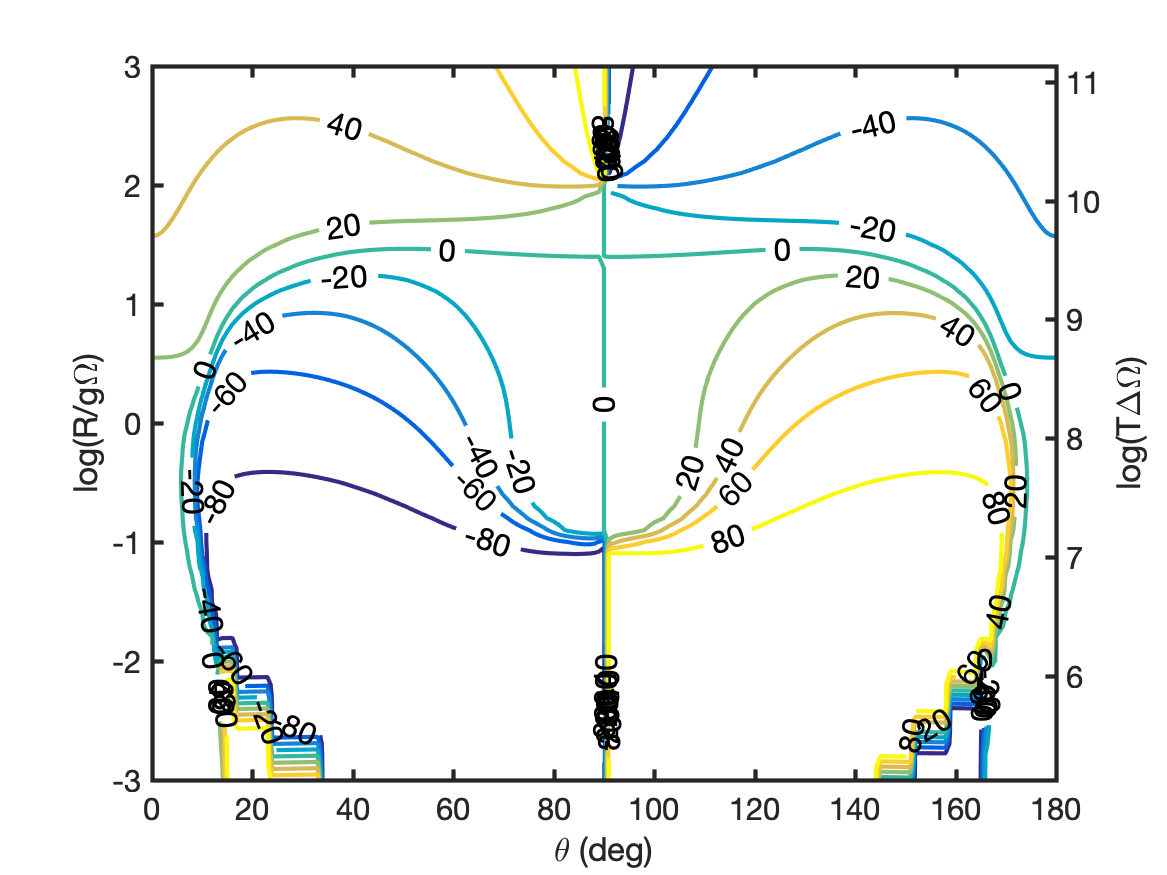}
       \caption{}
    \end{subfigure}
     ~
    \begin{subfigure}[b]{0.32\textwidth}
      \includegraphics[width=\textwidth]{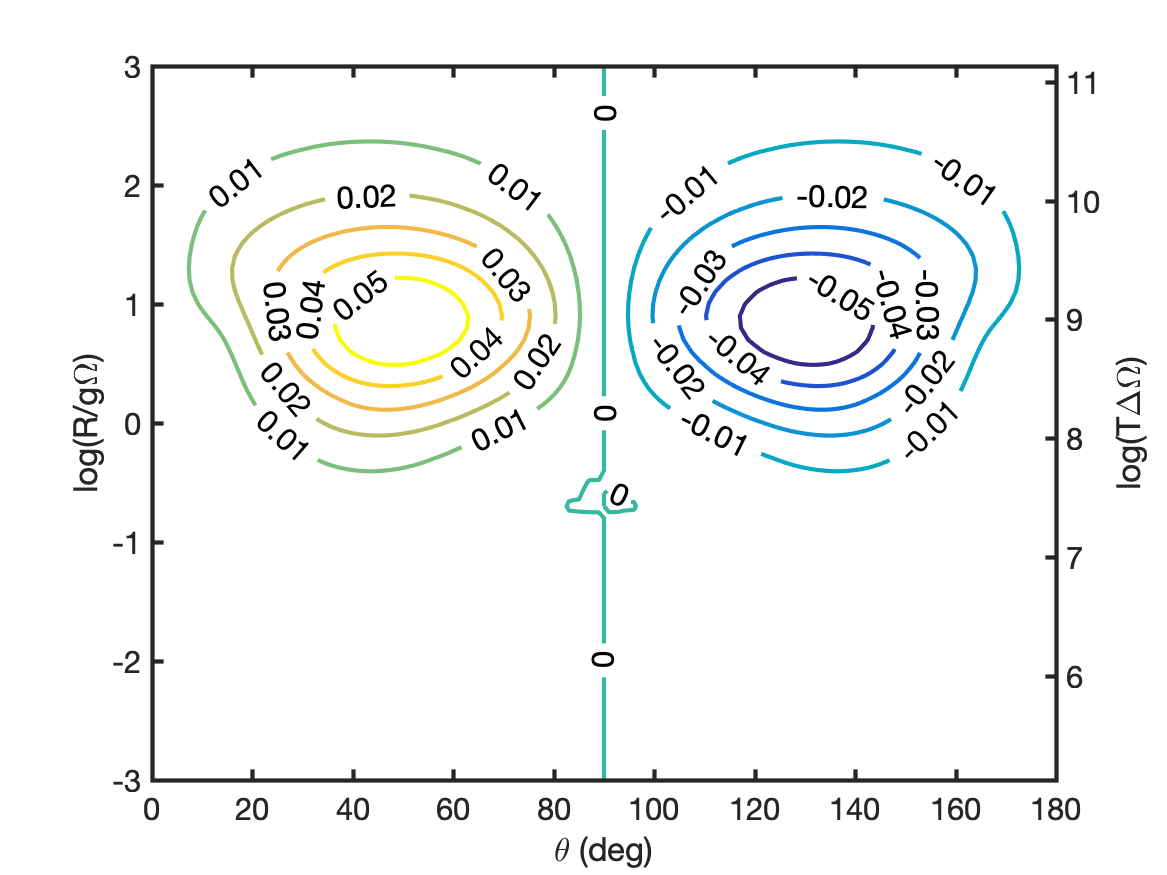}
      \caption{}
    \end{subfigure}
    ~
    \begin{subfigure}[b]{0.32\textwidth}
       \includegraphics[width=\textwidth]{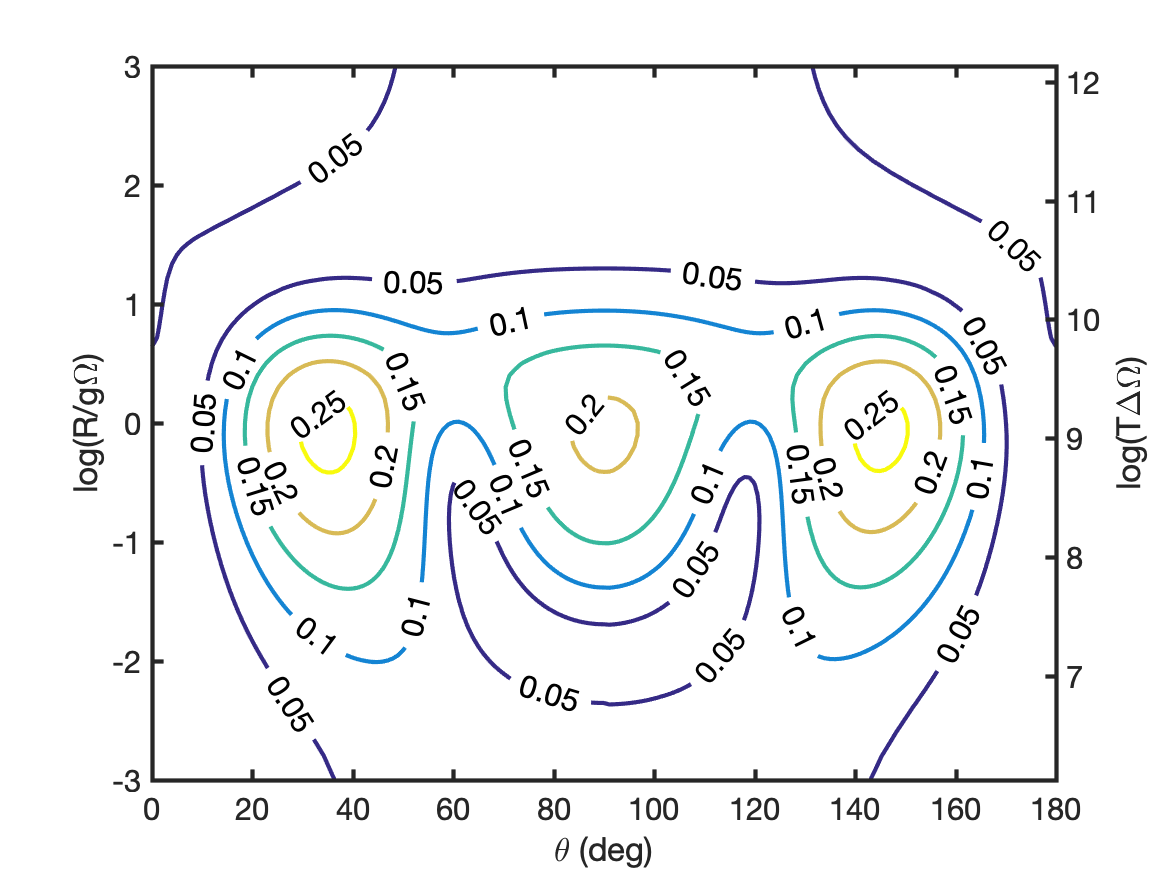}
       \caption{}
    \end{subfigure}
    ~
    \begin{subfigure}[b]{0.32\textwidth}
       \includegraphics[width=\textwidth]{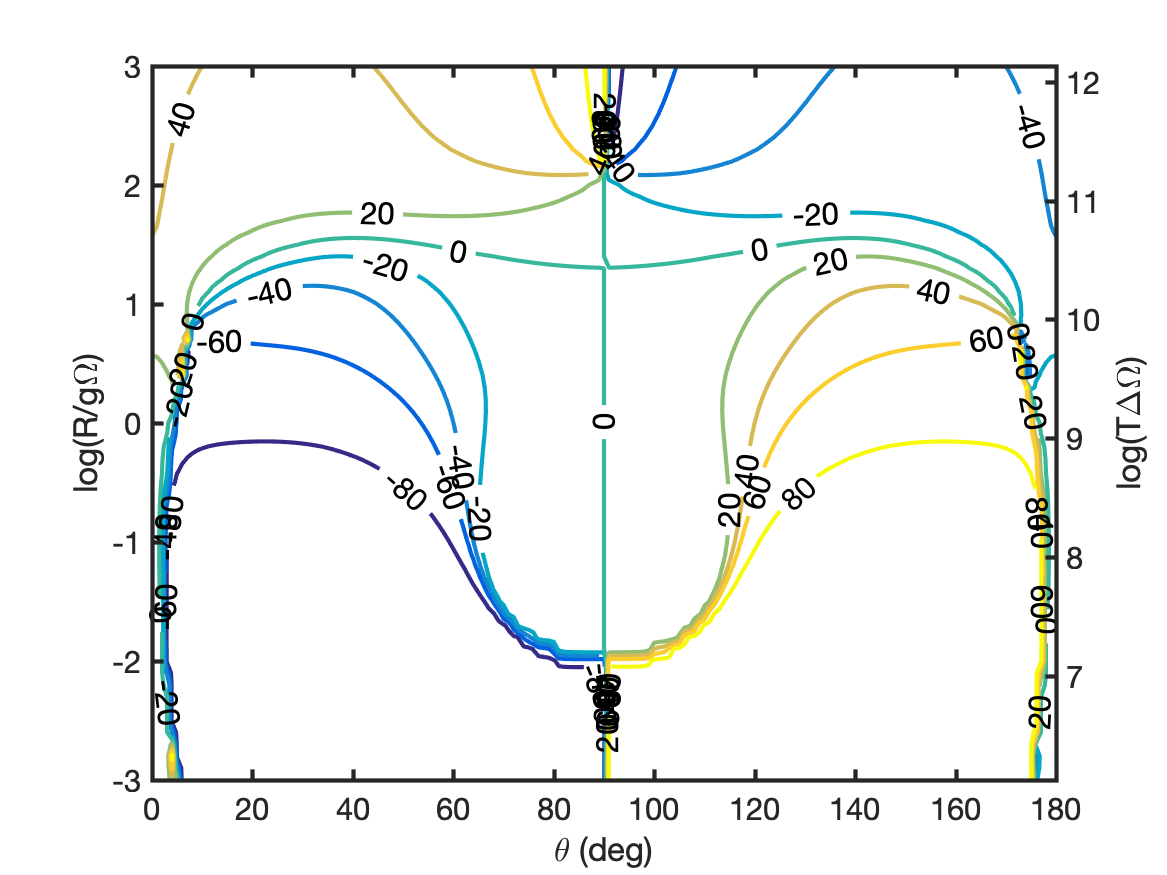}
       \caption{}
    \end{subfigure}
     ~
    \begin{subfigure}[b]{0.32\textwidth}
      \includegraphics[width=\textwidth]{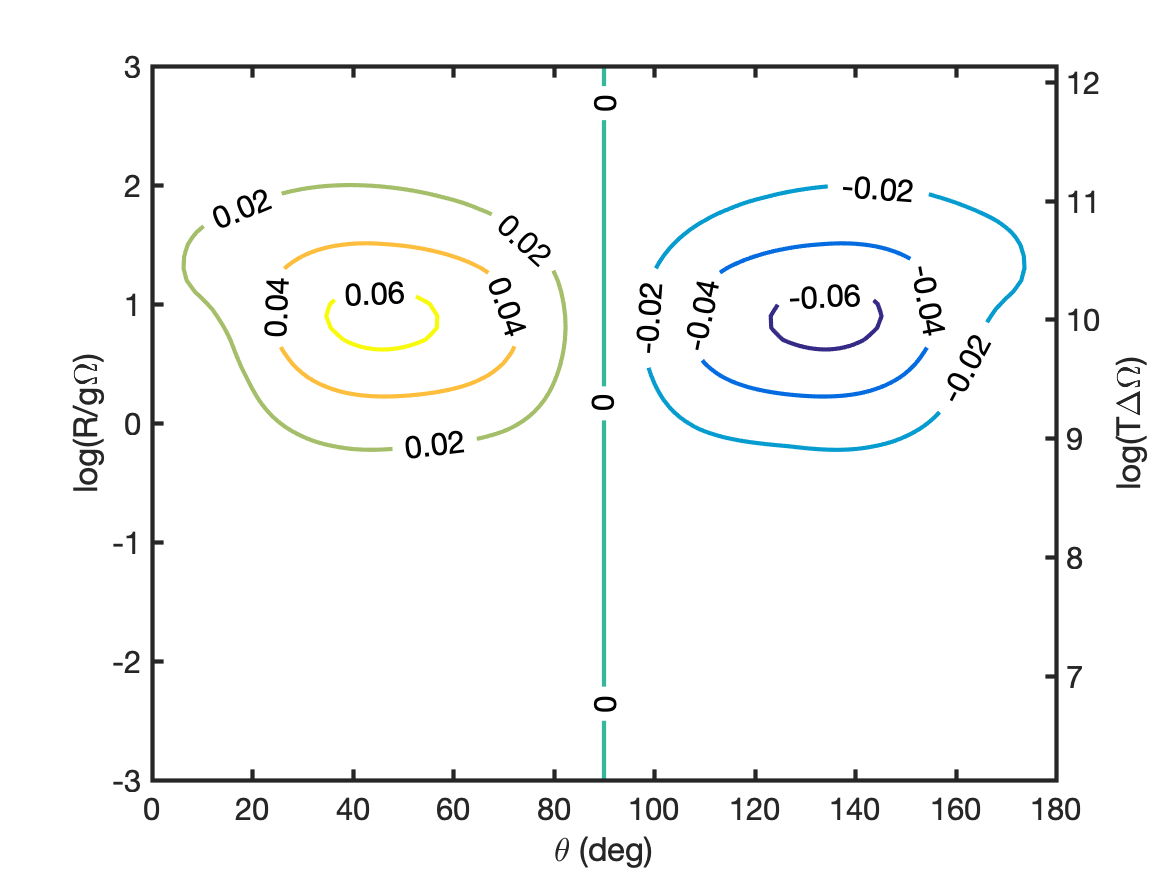}
      \caption{}
    \end{subfigure}
     ~
    \begin{subfigure}[b]{0.32\textwidth}
       \includegraphics[width=\textwidth]{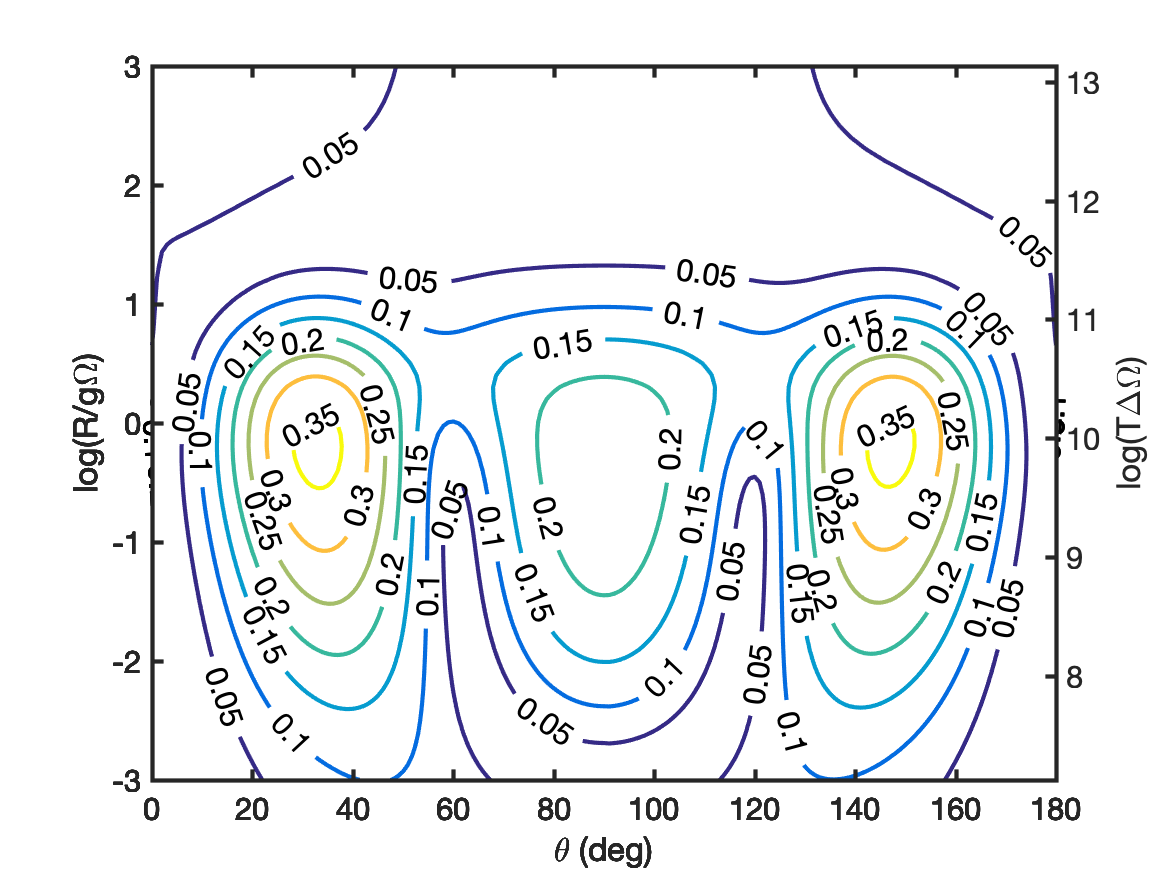}
       \caption{}
    \end{subfigure}
    ~
    \begin{subfigure}[b]{0.32\textwidth}
       \includegraphics[width=\textwidth]{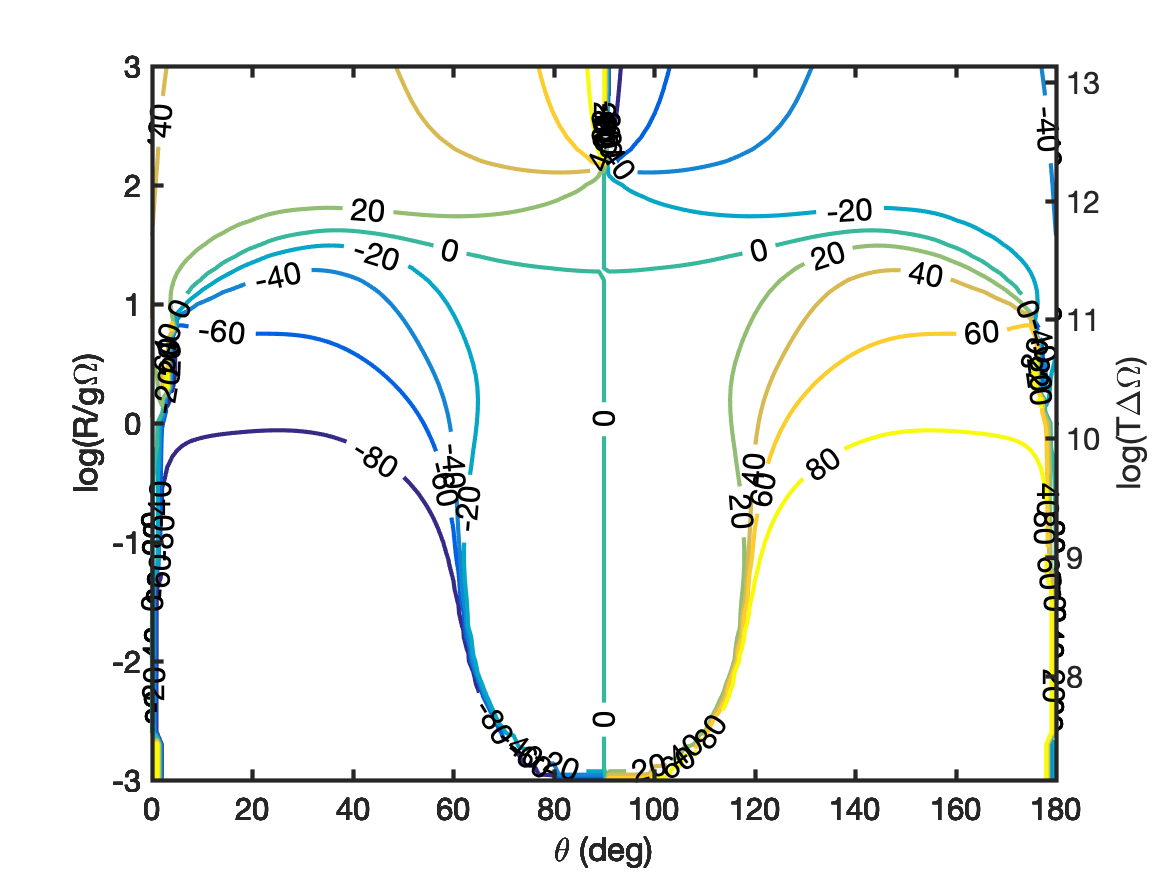}
       \caption{}
    \end{subfigure}
     ~
    \begin{subfigure}[b]{0.32\textwidth}
      \includegraphics[width=\textwidth]{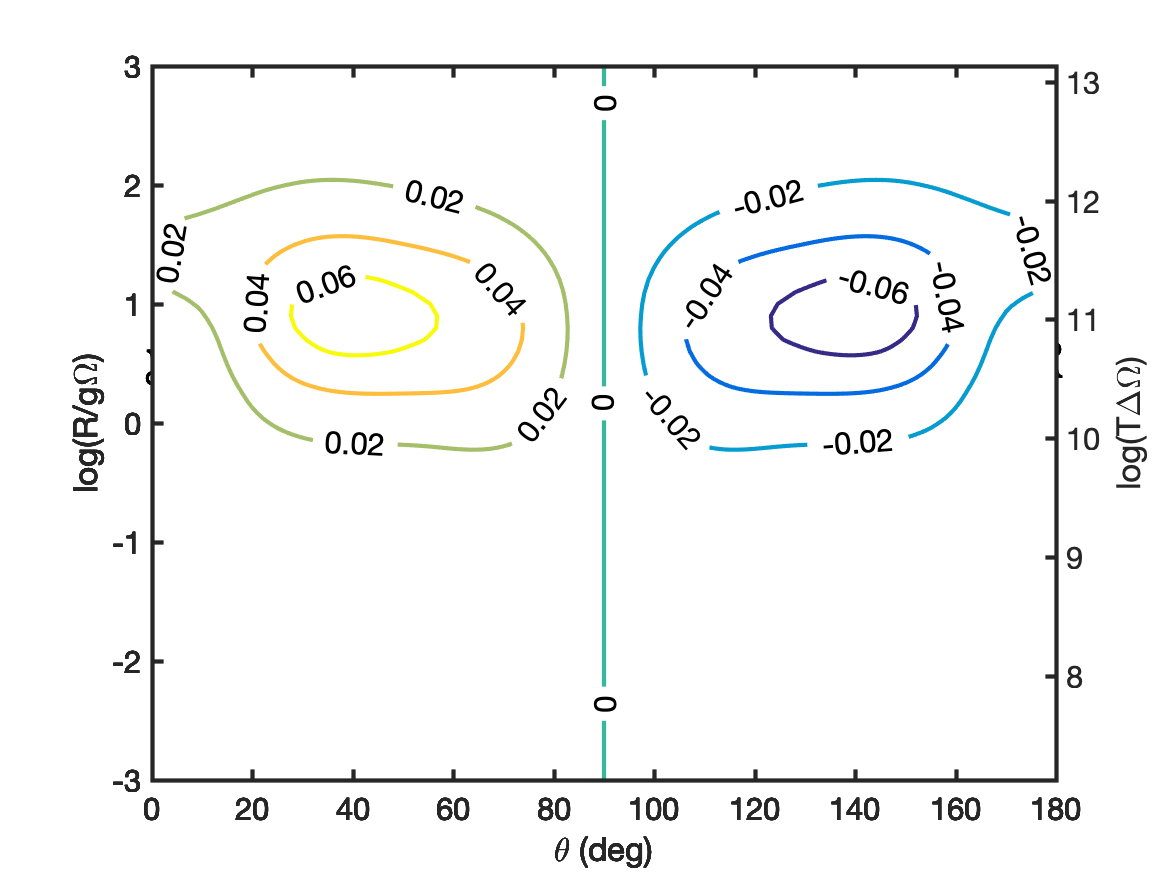}
      \caption{}
    \end{subfigure}
  \caption{Simulations of $J=2-1$ SiO masers with anisotropic pumping direction parallel to the magnetic field. Linear polarization fraction (a,d,f) and angle (b,e,h) and circular polarization fraction (c,f,g). Magnetic field strengths are $B=100$ mG for (a,b,c), $B=1$ G for (d,e,f) and $B=10$ G for (g,h,i).}
\end{figure*}

\begin{figure*}
    \centering
    \begin{subfigure}[b]{0.45\textwidth}
       \includegraphics[width=\textwidth]{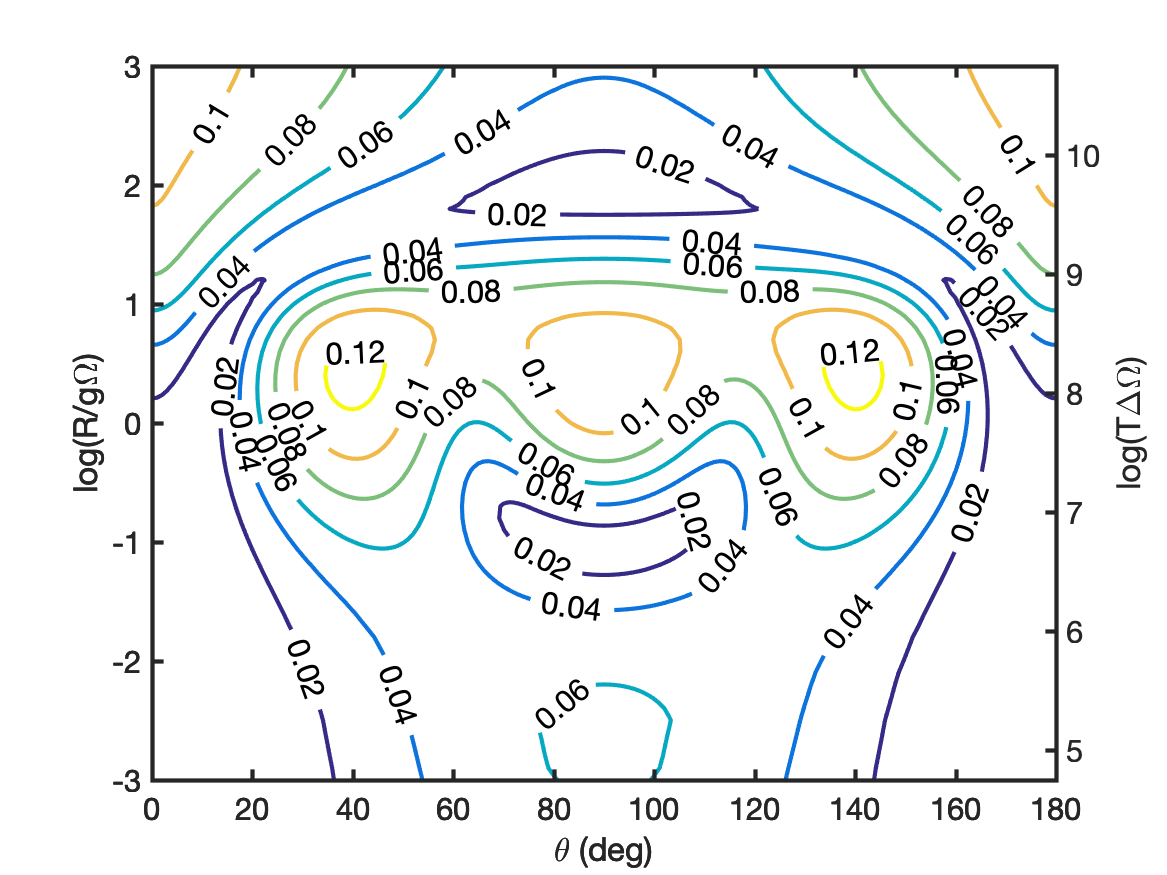}
       \caption{}
    \end{subfigure}
    ~
    \begin{subfigure}[b]{0.45\textwidth}
       \includegraphics[width=\textwidth]{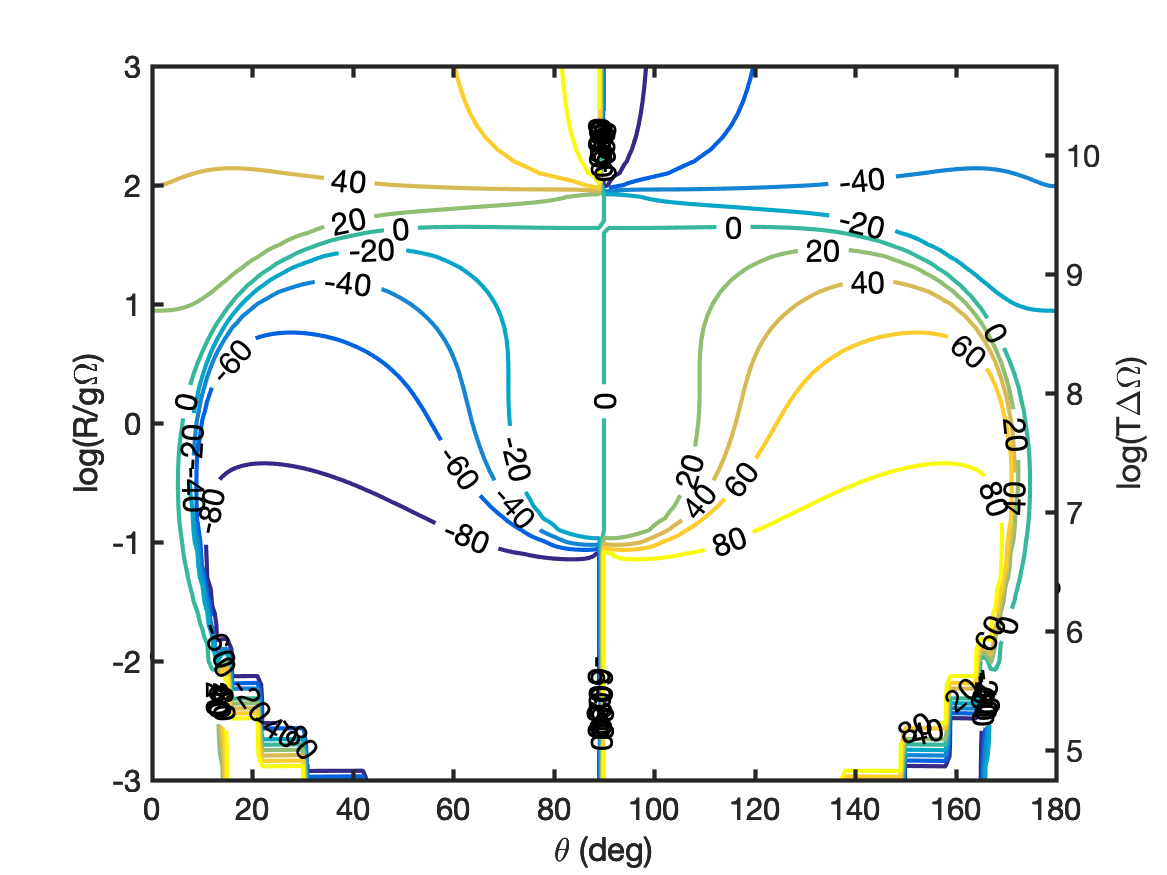}
       \caption{}
    \end{subfigure}
     ~
    \begin{subfigure}[b]{0.45\textwidth}
      \includegraphics[width=\textwidth]{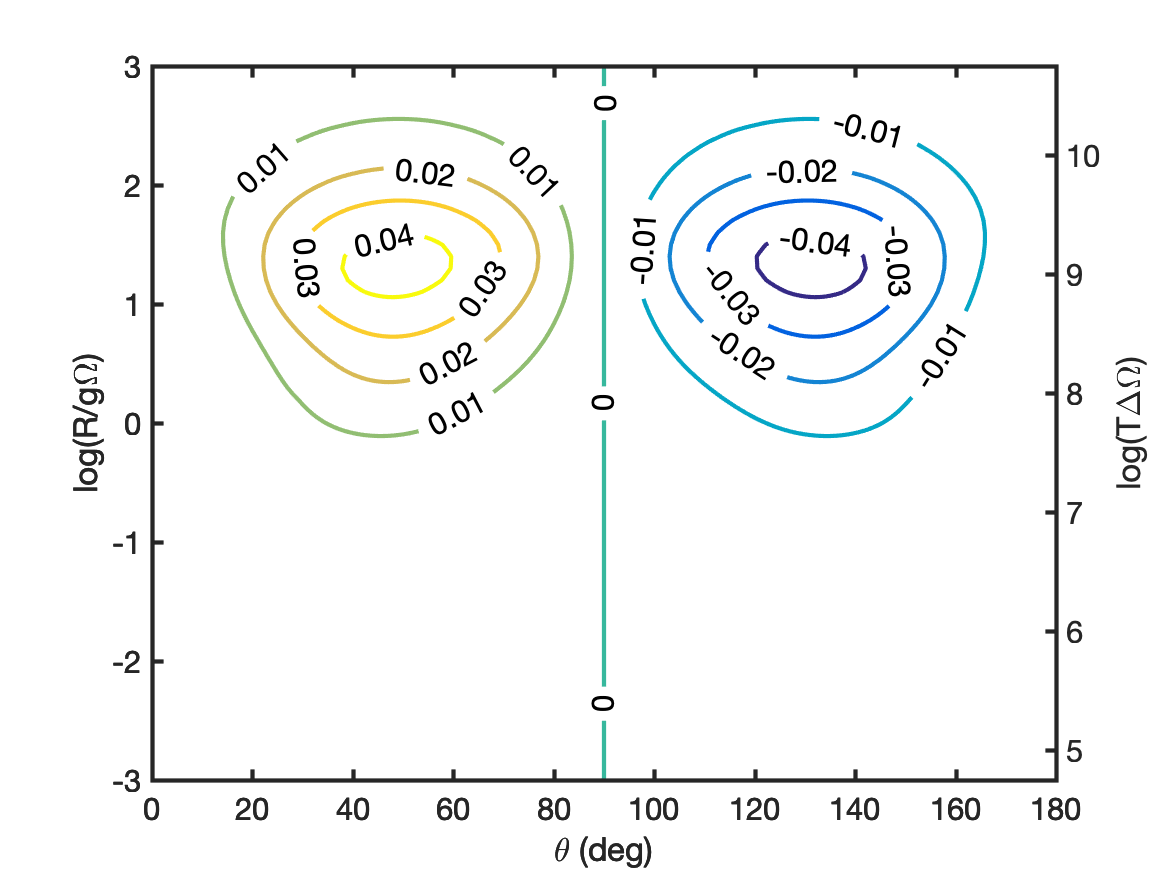}
      \caption{}
    \end{subfigure}
     ~
    \begin{subfigure}[b]{0.45\textwidth}
       \includegraphics[width=\textwidth]{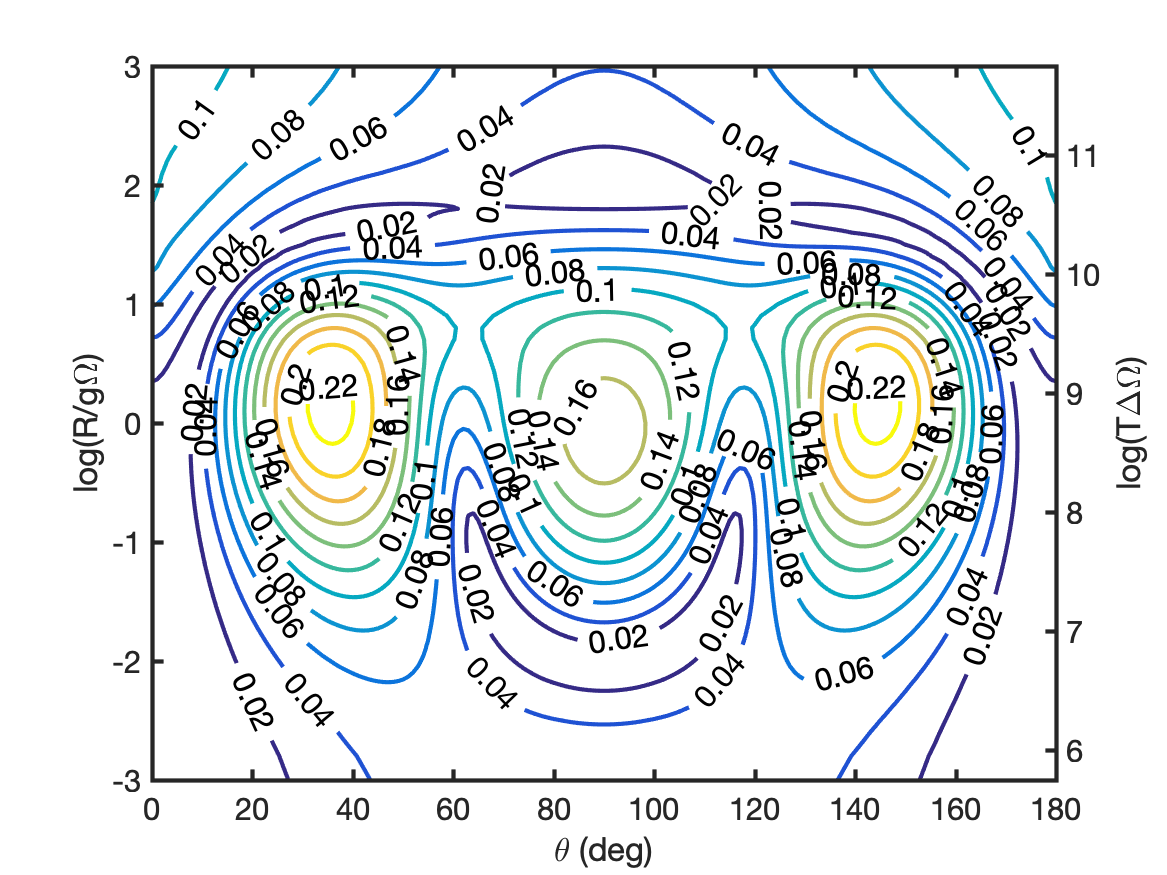}
       \caption{}
    \end{subfigure}
    ~
    \begin{subfigure}[b]{0.45\textwidth}
       \includegraphics[width=\textwidth]{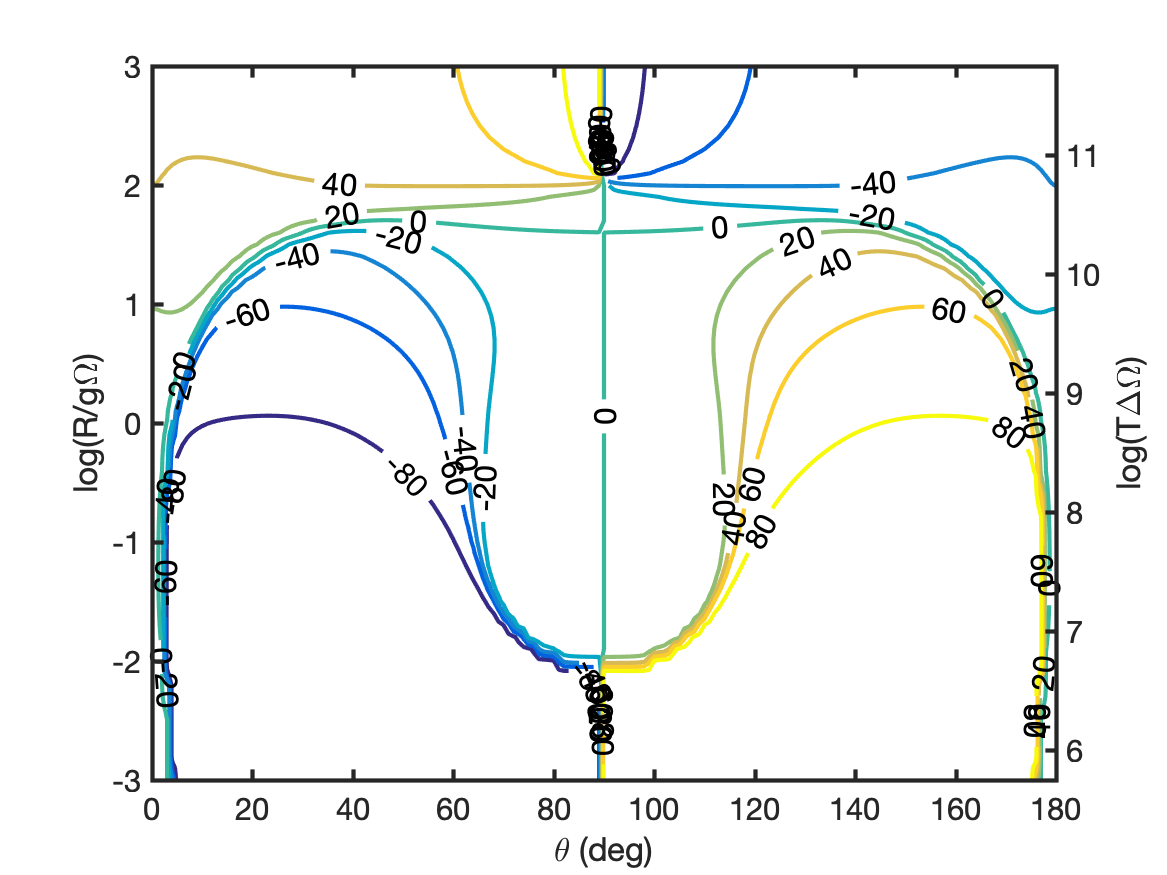}
       \caption{}
    \end{subfigure}
     ~
    \begin{subfigure}[b]{0.45\textwidth}
      \includegraphics[width=\textwidth]{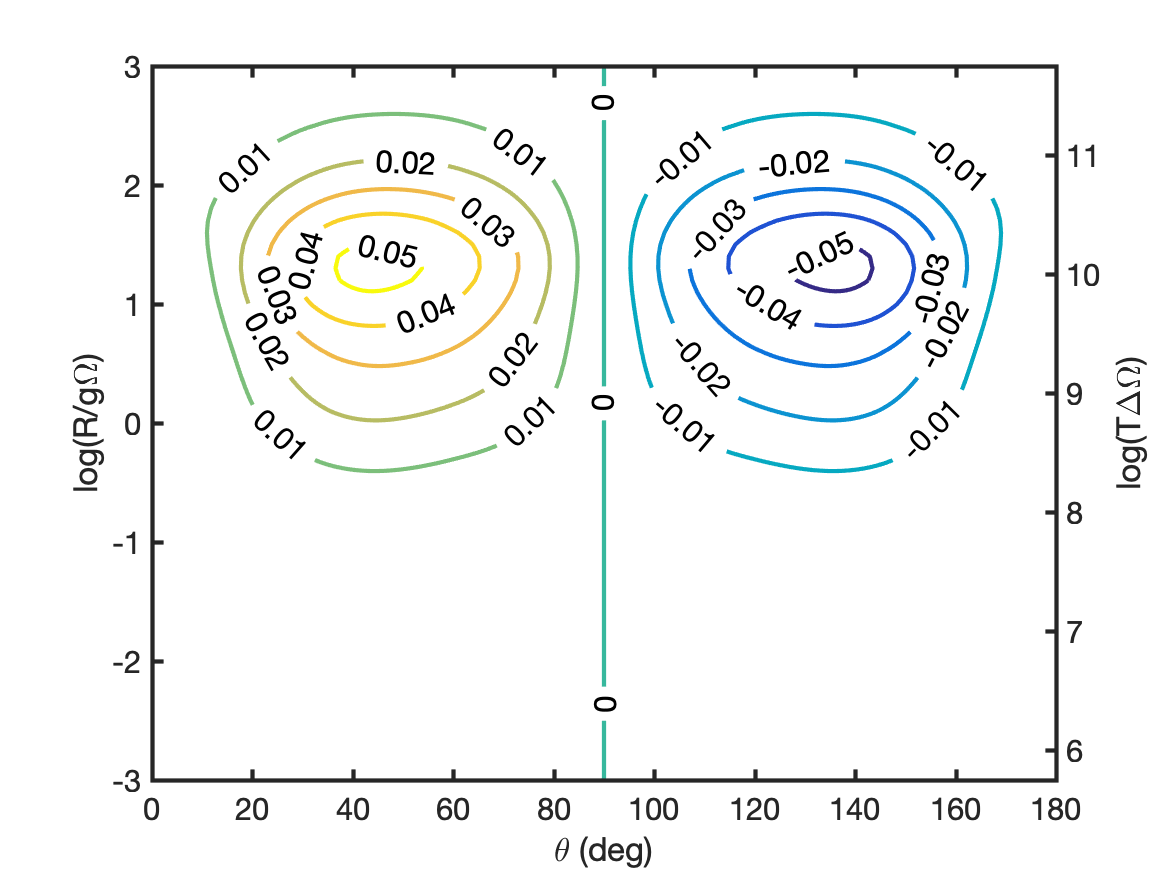}
      \caption{}
    \end{subfigure}
  \caption{Simulations of $J=3-2$ SiO masers with anisotropic pumping direction parallel to the magnetic field. Linear polarization fraction (a,d) and angle (b,e) and circular polarization fraction (c,f). Magnetic field strengths are $B=100$ mG for (a,b,c) and $B=1$ G for (d,e,f).}
\end{figure*}
    
\begin{figure*}
    \centering
    \begin{subfigure}[b]{0.45\textwidth}
       \includegraphics[width=\textwidth]{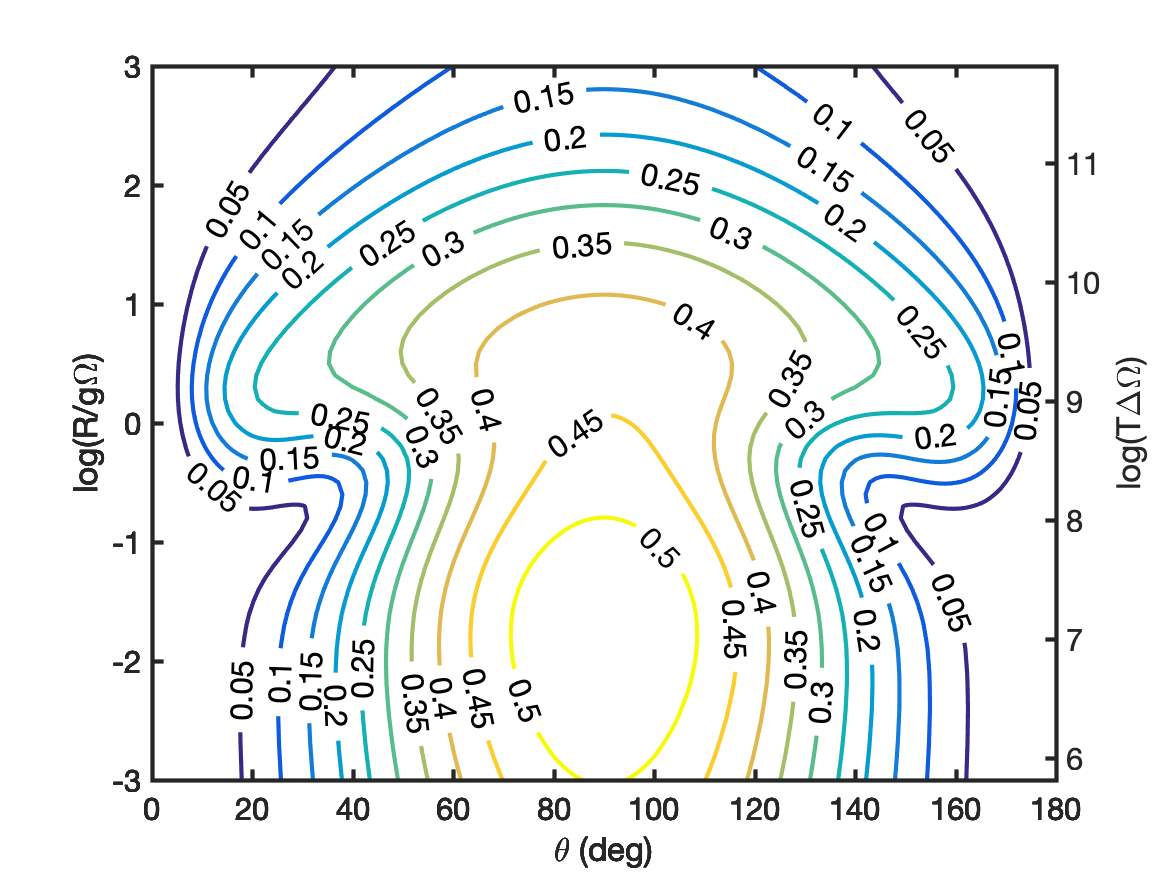} 
       \caption{}
    \end{subfigure}
    ~ 
    \begin{subfigure}[b]{0.45\textwidth}
       \includegraphics[width=\textwidth]{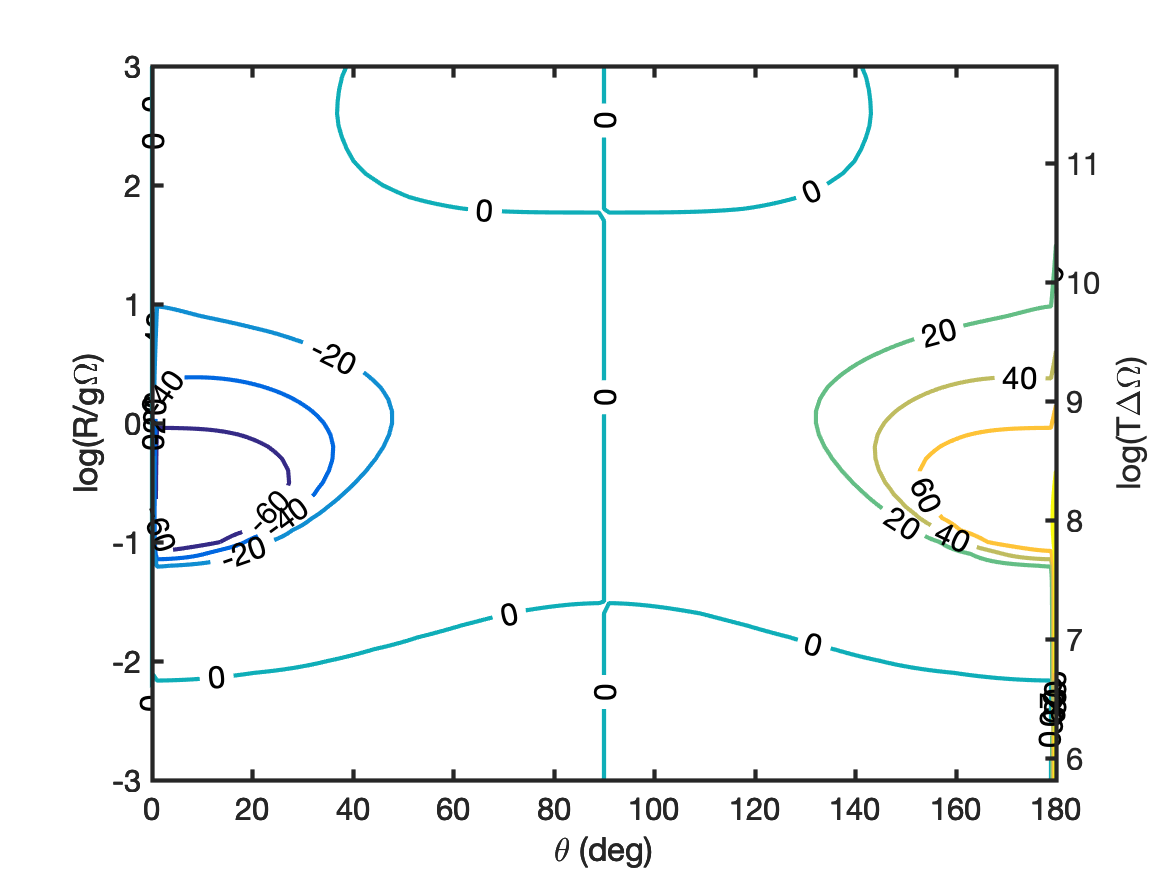} 
       \caption{}
    \end{subfigure}
     ~ 
    \begin{subfigure}[b]{0.45\textwidth}
      \includegraphics[width=\textwidth]{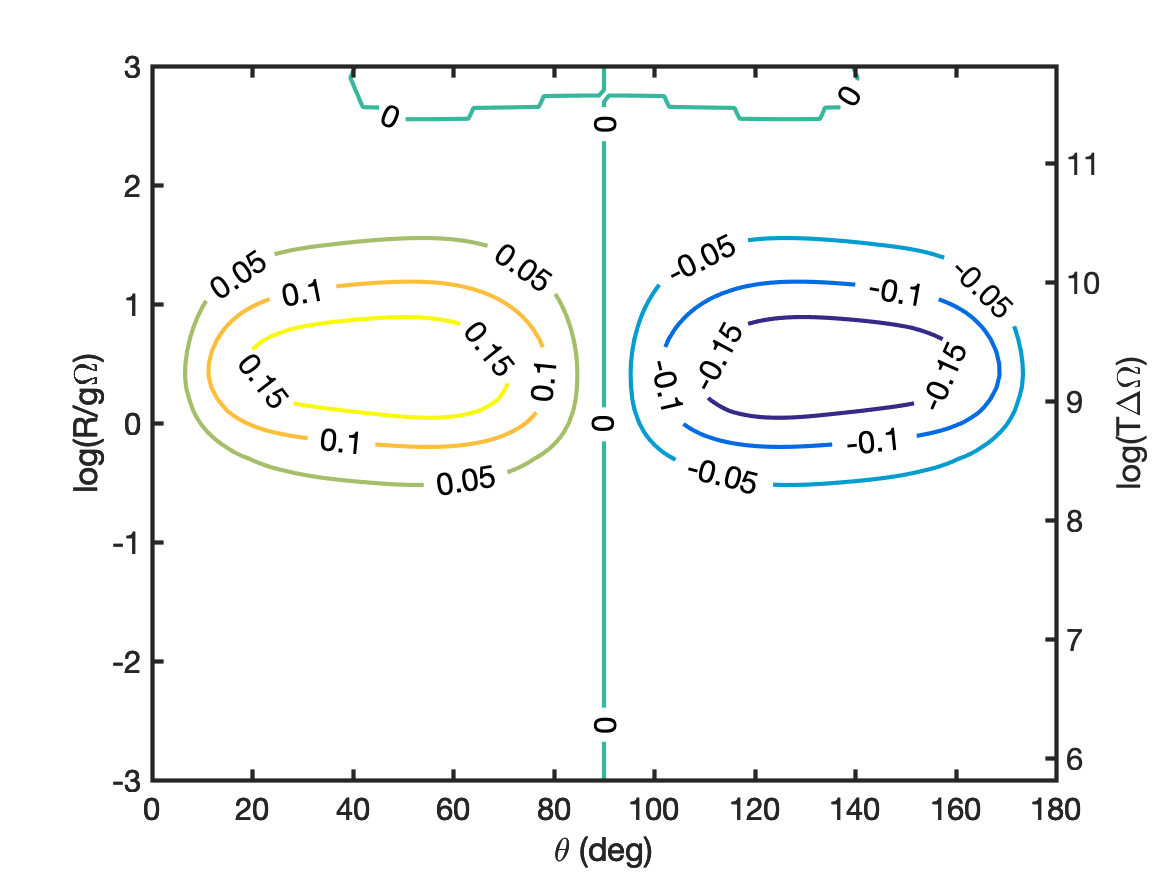}
      \caption{}
    \end{subfigure}
    ~
    \begin{subfigure}[b]{0.45\textwidth}
       \includegraphics[width=\textwidth]{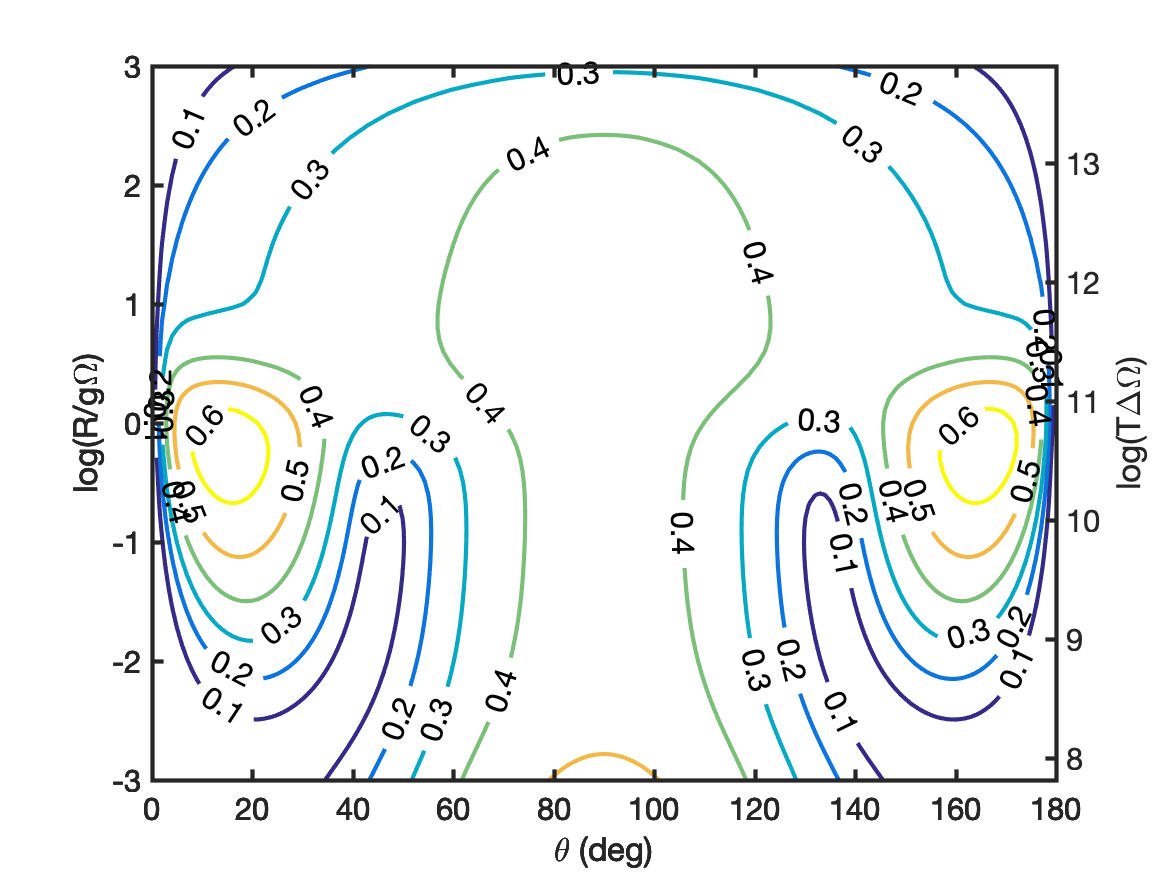} 
       \caption{}
    \end{subfigure}
    ~ 
    \begin{subfigure}[b]{0.45\textwidth}
       \includegraphics[width=\textwidth]{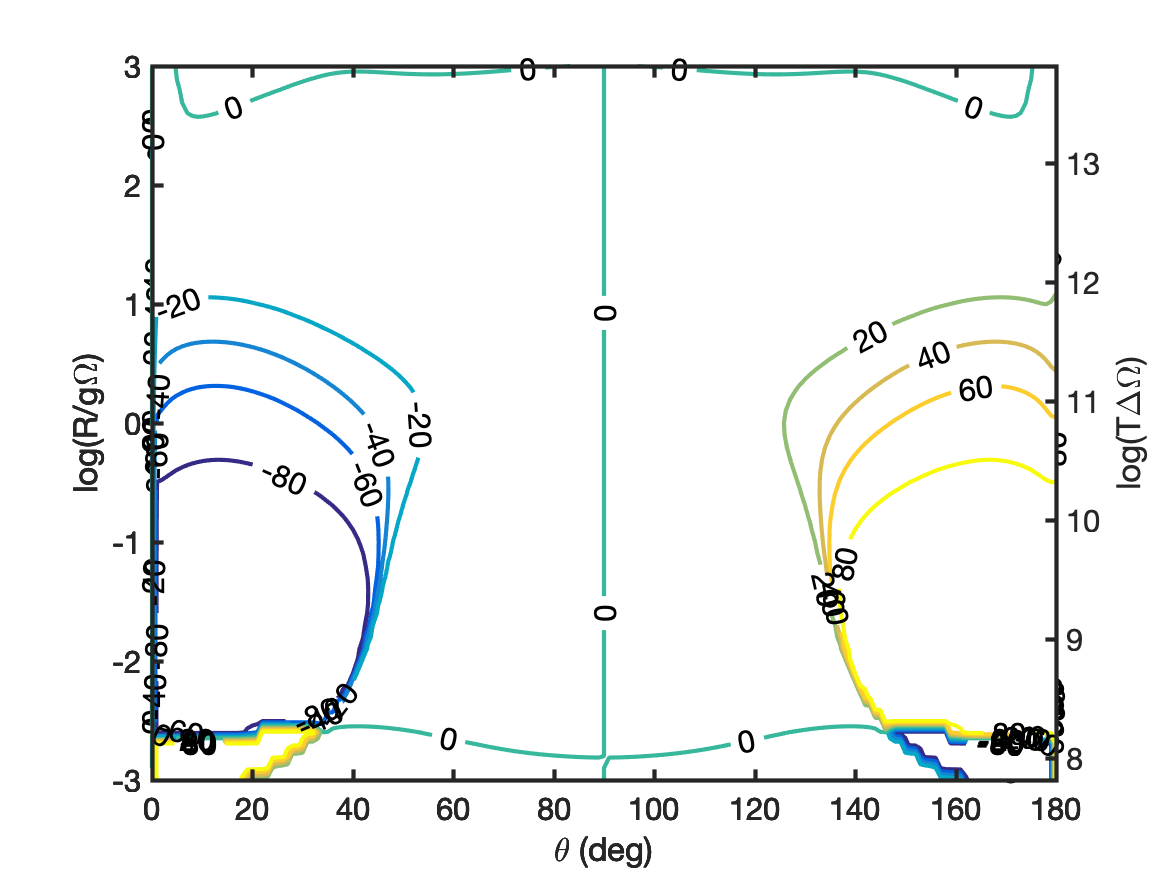} 
       \caption{}
    \end{subfigure}
     ~ 
    \begin{subfigure}[b]{0.45\textwidth}
      \includegraphics[width=\textwidth]{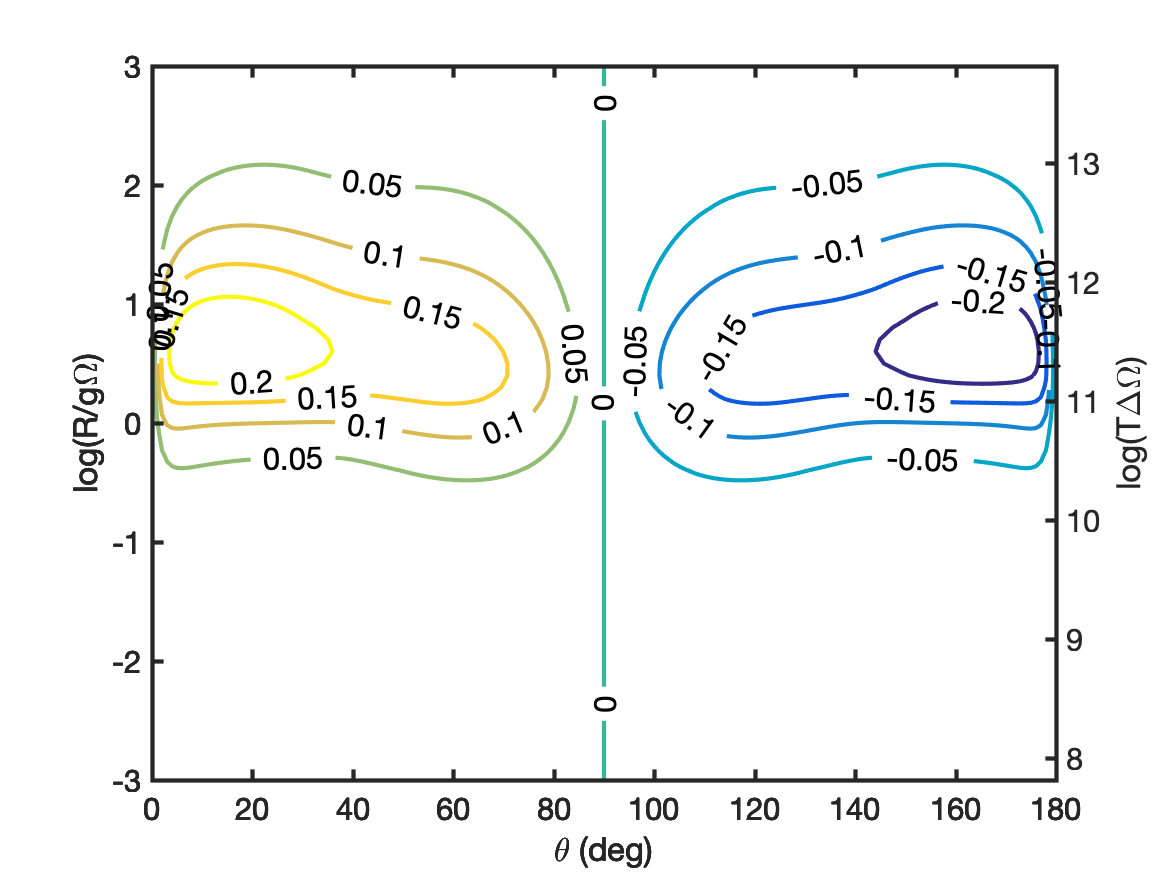}
      \caption{}
    \end{subfigure}
  \caption{Simulations of $J=1-0$ SiO masers with anisotropic pumping direction perpendicular to the magnetic field and propagation direction. Linear polarization fraction (a,d) and angle (b,e) and circular polarization fraction (c,f). Magnetic field strengths are $B=100$ mG for (a,b,c) and $B=10$ G for (d,e,f).}
\end{figure*}

\begin{figure*}
    \centering
    \begin{subfigure}[b]{0.45\textwidth}
       \includegraphics[width=\textwidth]{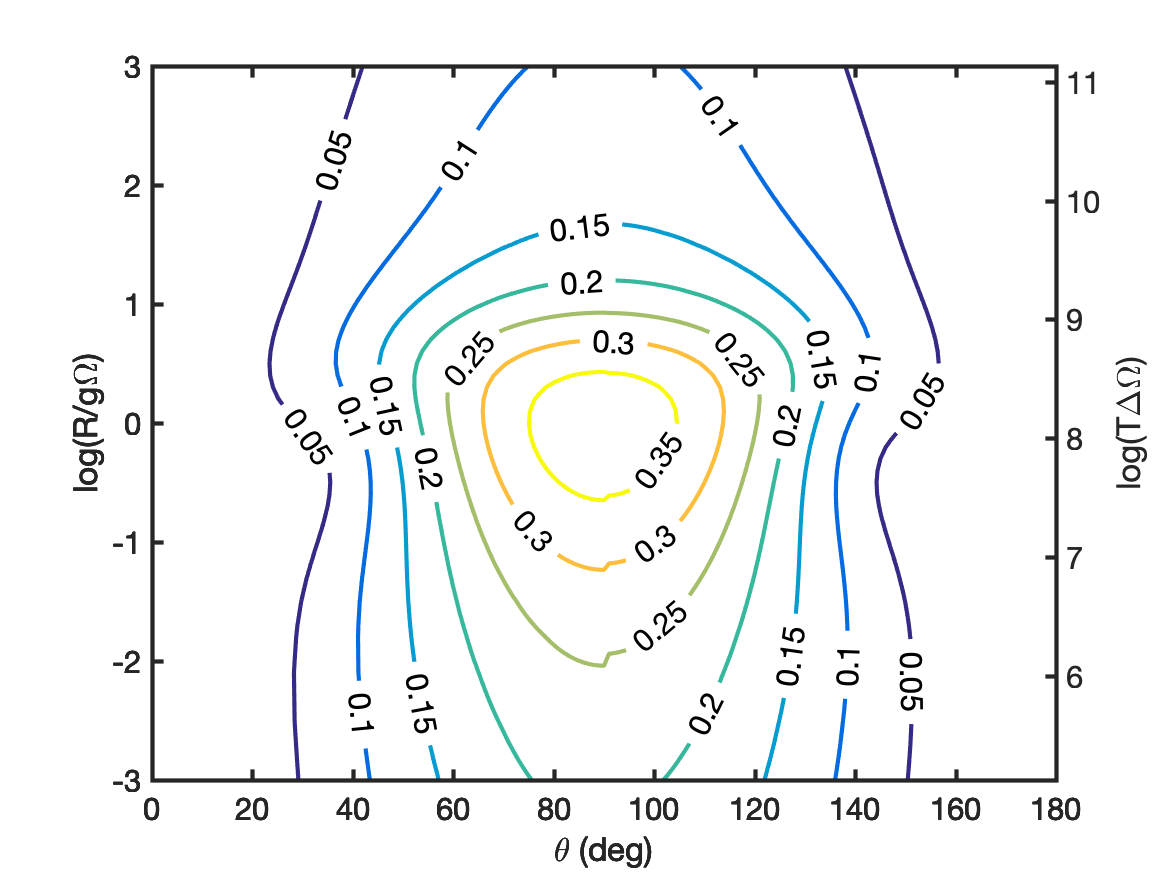}
       \caption{}
    \end{subfigure}
    ~
    \begin{subfigure}[b]{0.45\textwidth}
       \includegraphics[width=\textwidth]{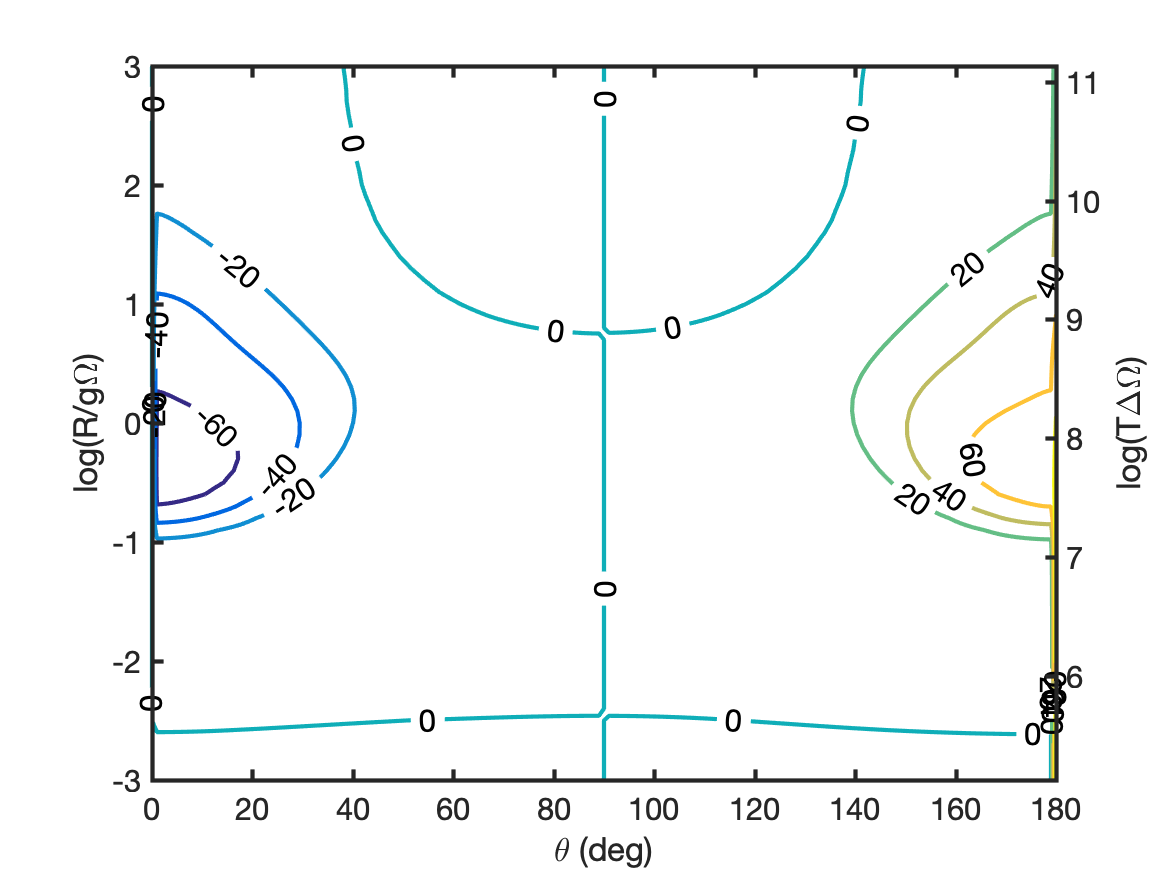}
       \caption{}
    \end{subfigure}
     ~
    \begin{subfigure}[b]{0.45\textwidth}
      \includegraphics[width=\textwidth]{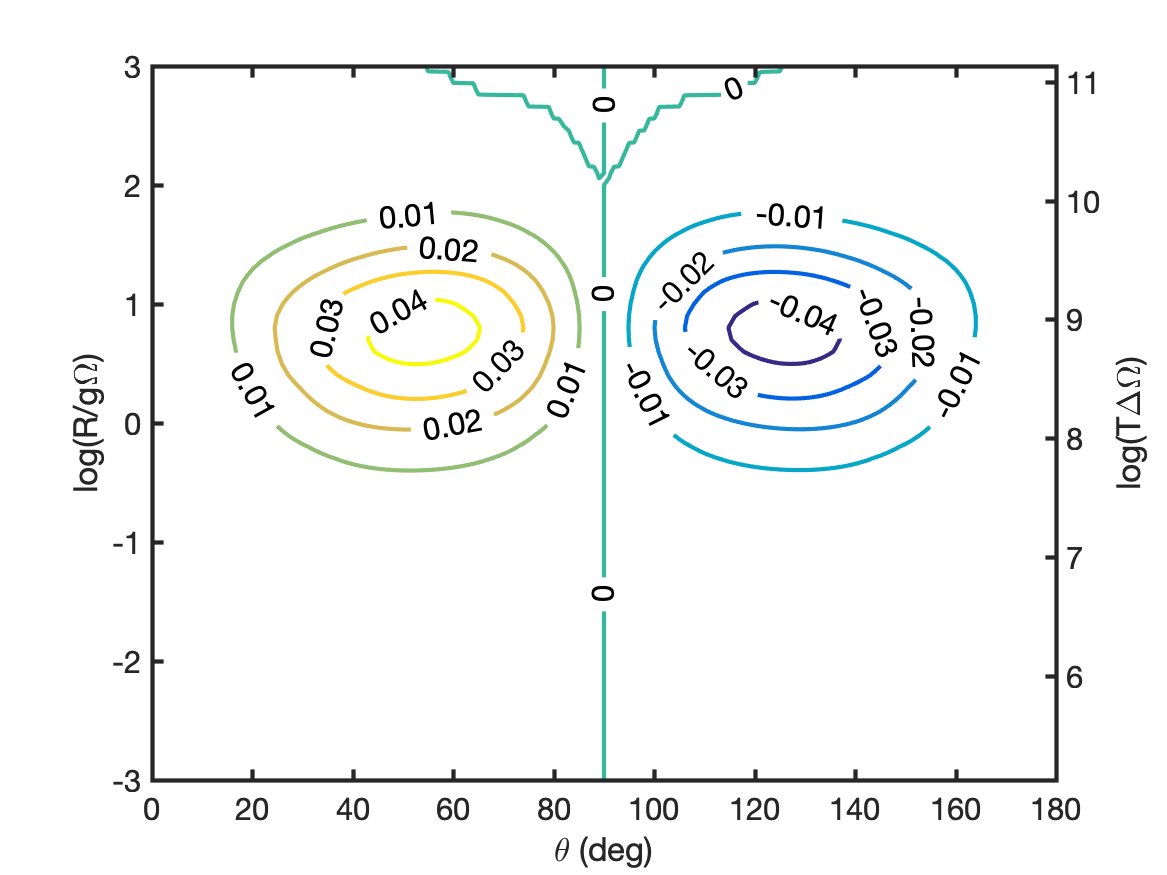}
      \caption{}
    \end{subfigure}
    ~
    \begin{subfigure}[b]{0.45\textwidth}
       \includegraphics[width=\textwidth]{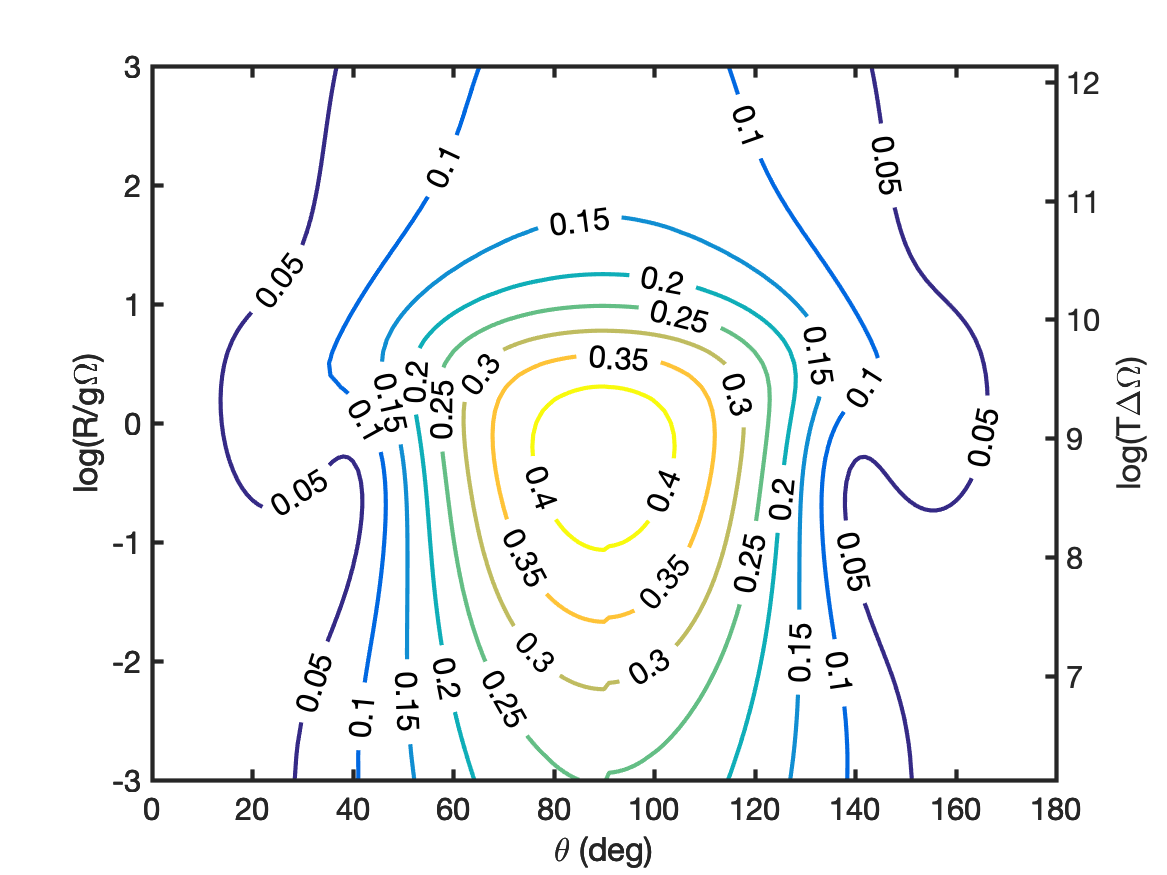}
       \caption{}
    \end{subfigure}
    ~
    \begin{subfigure}[b]{0.45\textwidth}
       \includegraphics[width=\textwidth]{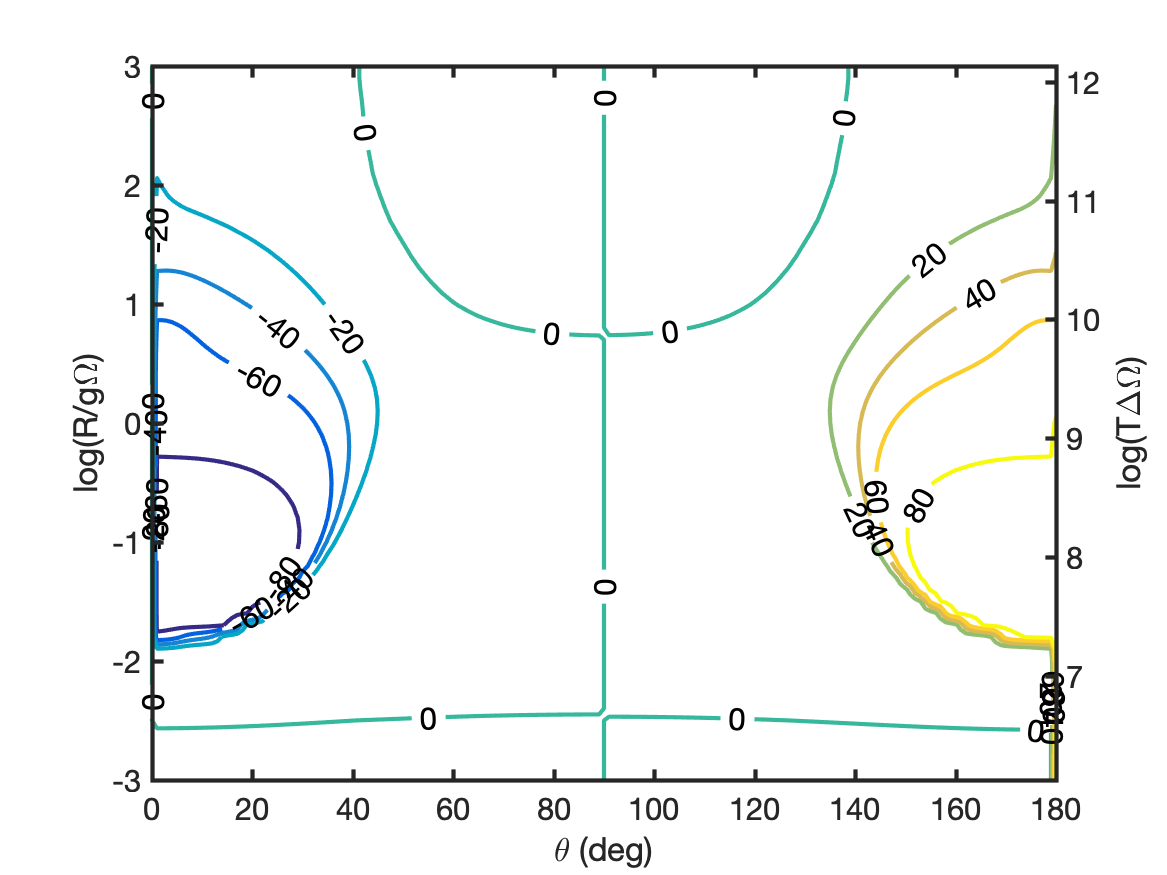}
       \caption{}
    \end{subfigure}
     ~
    \begin{subfigure}[b]{0.45\textwidth}
      \includegraphics[width=\textwidth]{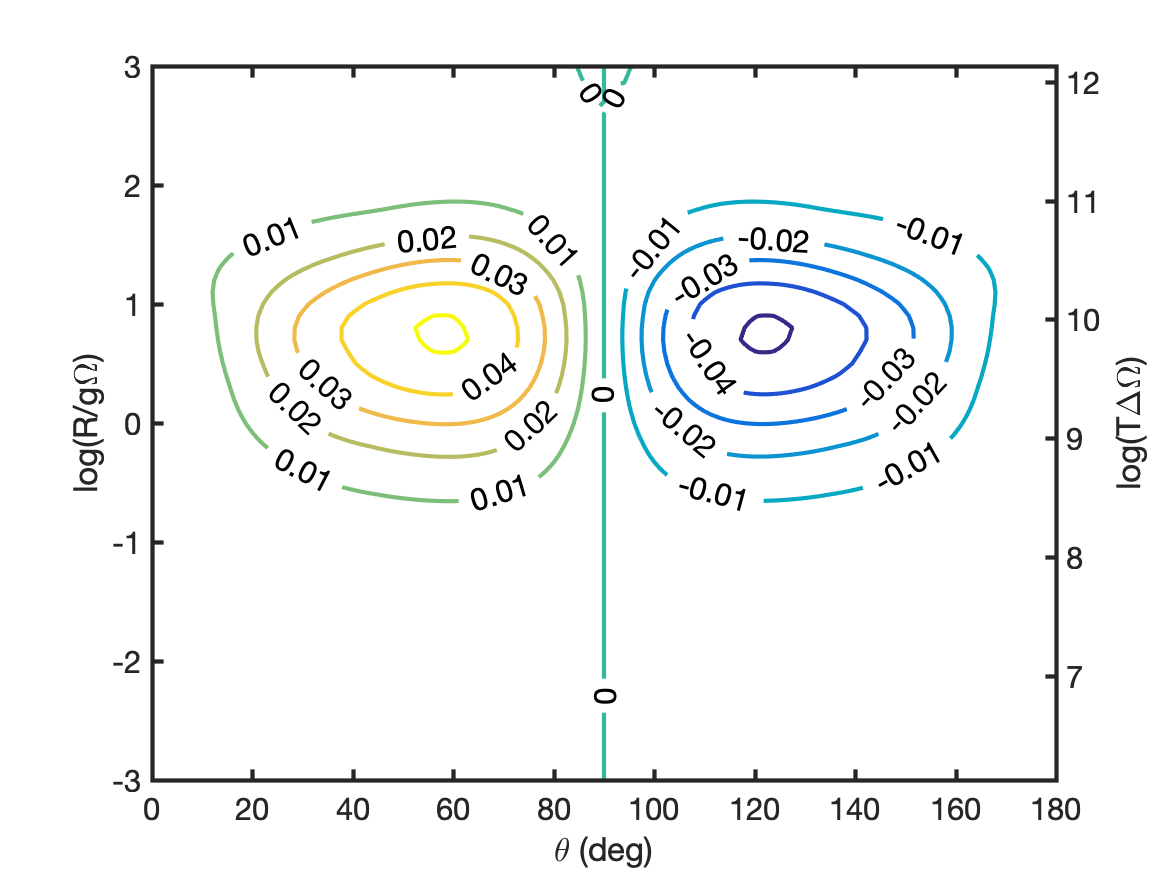}
      \caption{}
    \end{subfigure}
  \caption{Simulations of $J=2-1$ SiO masers with anisotropic pumping direction perpendicular to the magnetic field and propagation direction. Linear polarization fraction (a,d) and angle (b,e) and circular polarization fraction (c,f). Magnetic field strengths are $B=100$ mG for (a,b,c) and $B=1$ G for (d,e,f).}
\end{figure*}

\begin{figure*}
    \centering
    \begin{subfigure}[b]{0.45\textwidth}
       \includegraphics[width=\textwidth]{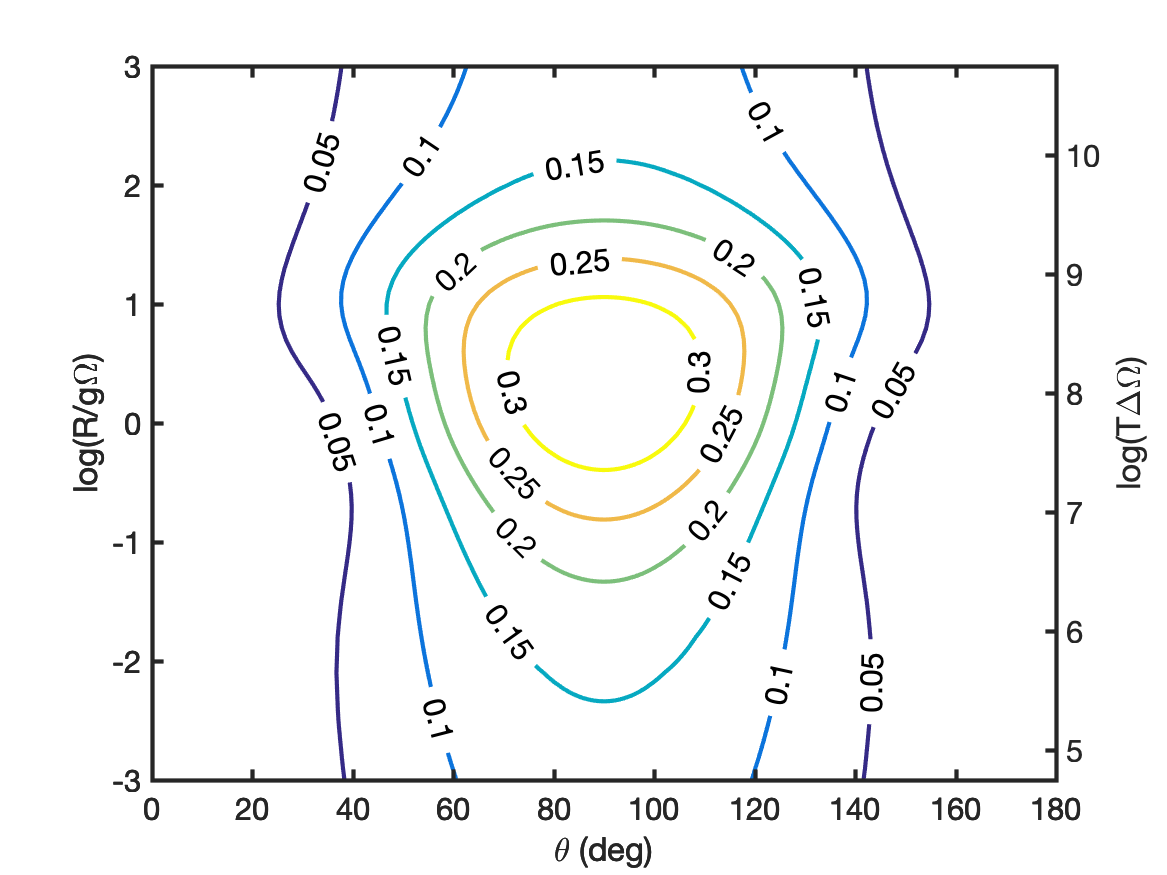}
       \caption{}
    \end{subfigure}
    ~
    \begin{subfigure}[b]{0.45\textwidth}
       \includegraphics[width=\textwidth]{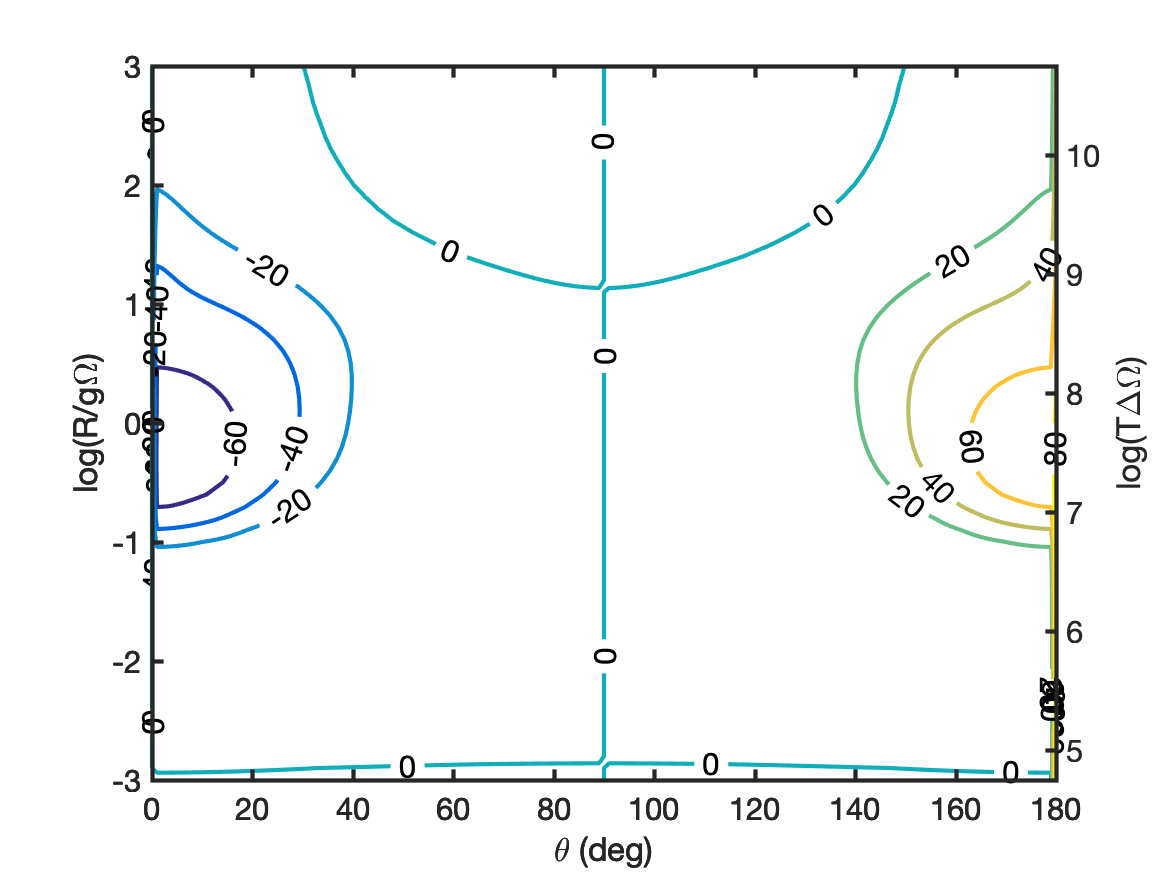}
       \caption{}
    \end{subfigure}
     ~
    \begin{subfigure}[b]{0.45\textwidth}
      \includegraphics[width=\textwidth]{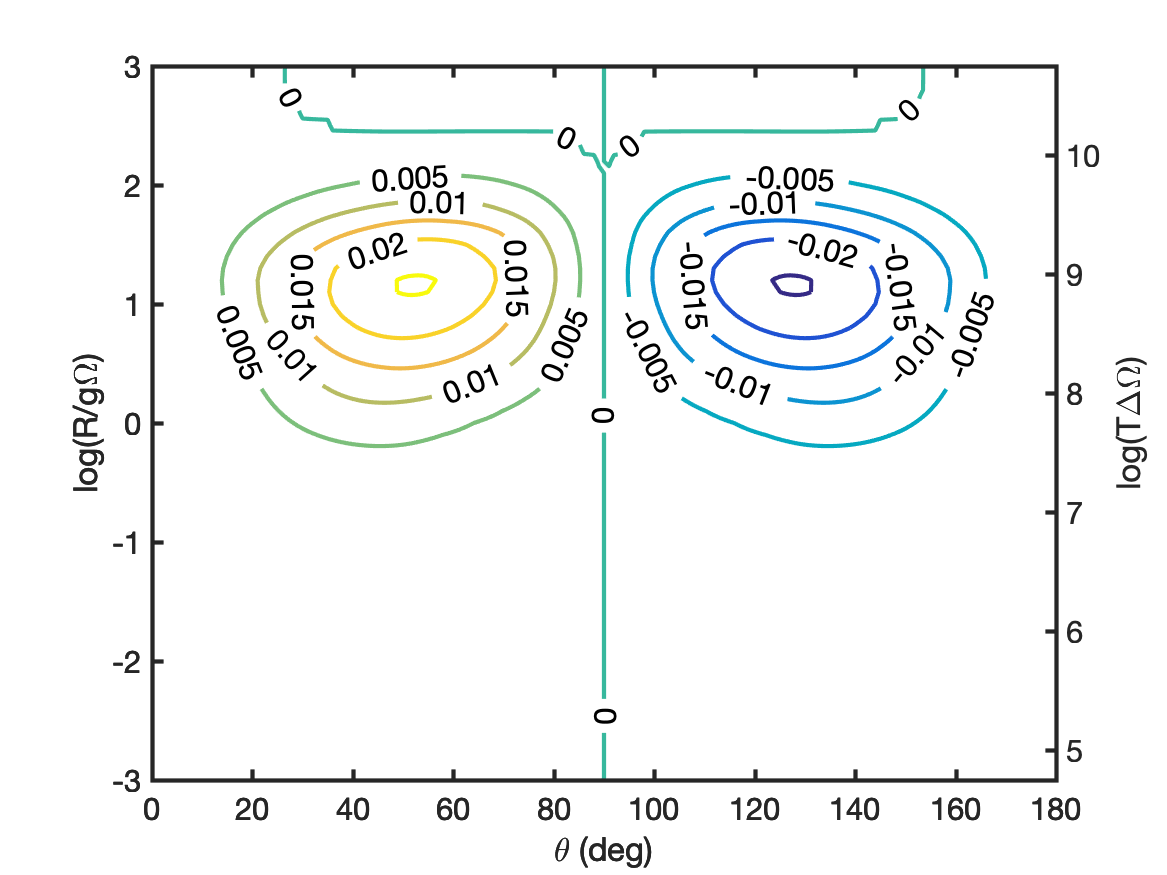}
      \caption{}
    \end{subfigure}
     ~
    \begin{subfigure}[b]{0.45\textwidth}
       \includegraphics[width=\textwidth]{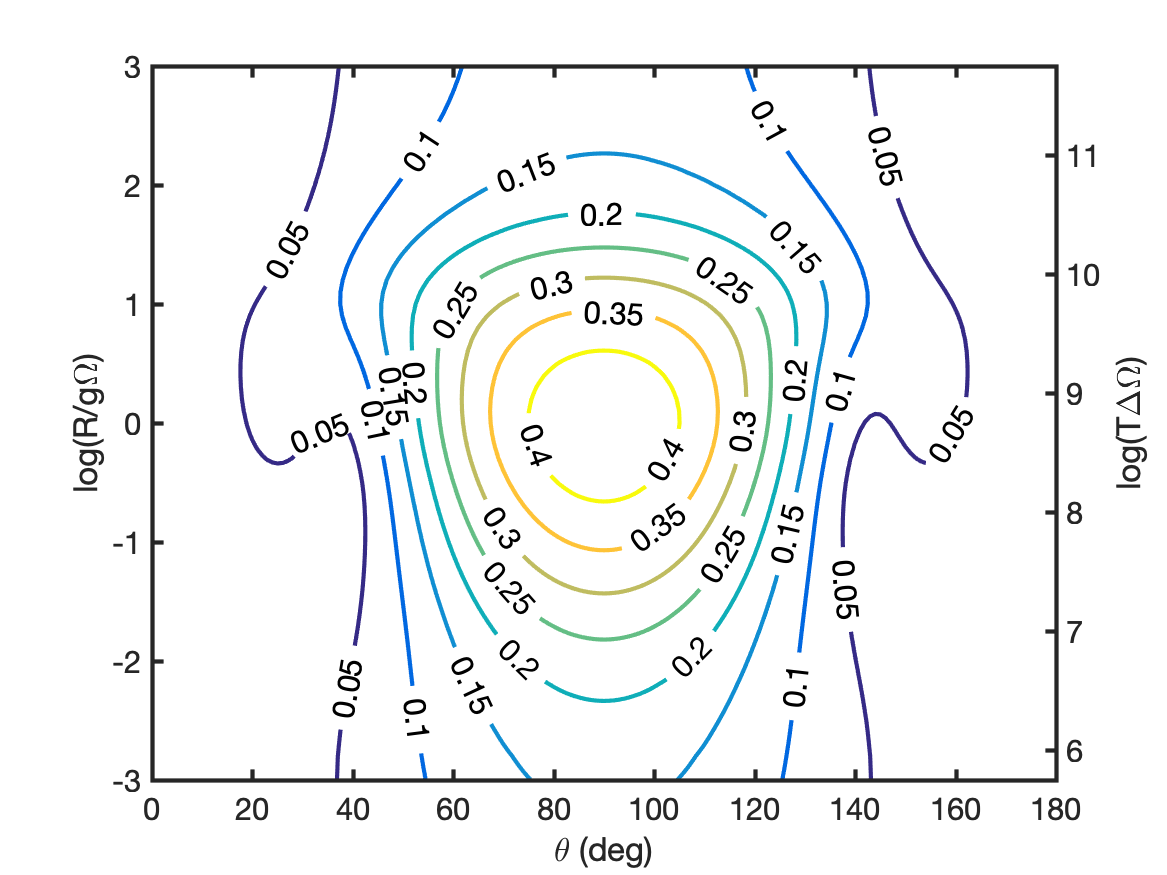}
       \caption{}
    \end{subfigure}
    ~
    \begin{subfigure}[b]{0.45\textwidth}
       \includegraphics[width=\textwidth]{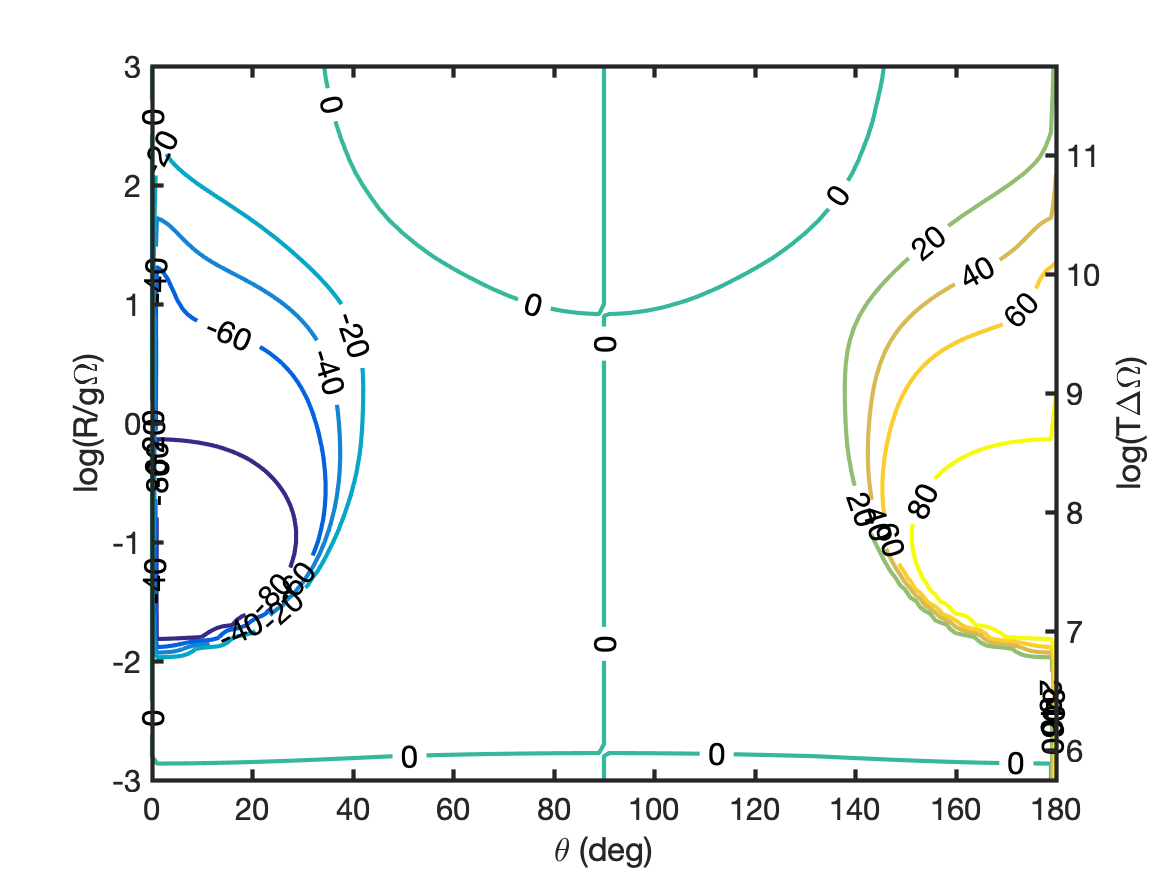}
       \caption{}
    \end{subfigure}
     ~
    \begin{subfigure}[b]{0.45\textwidth}
      \includegraphics[width=\textwidth]{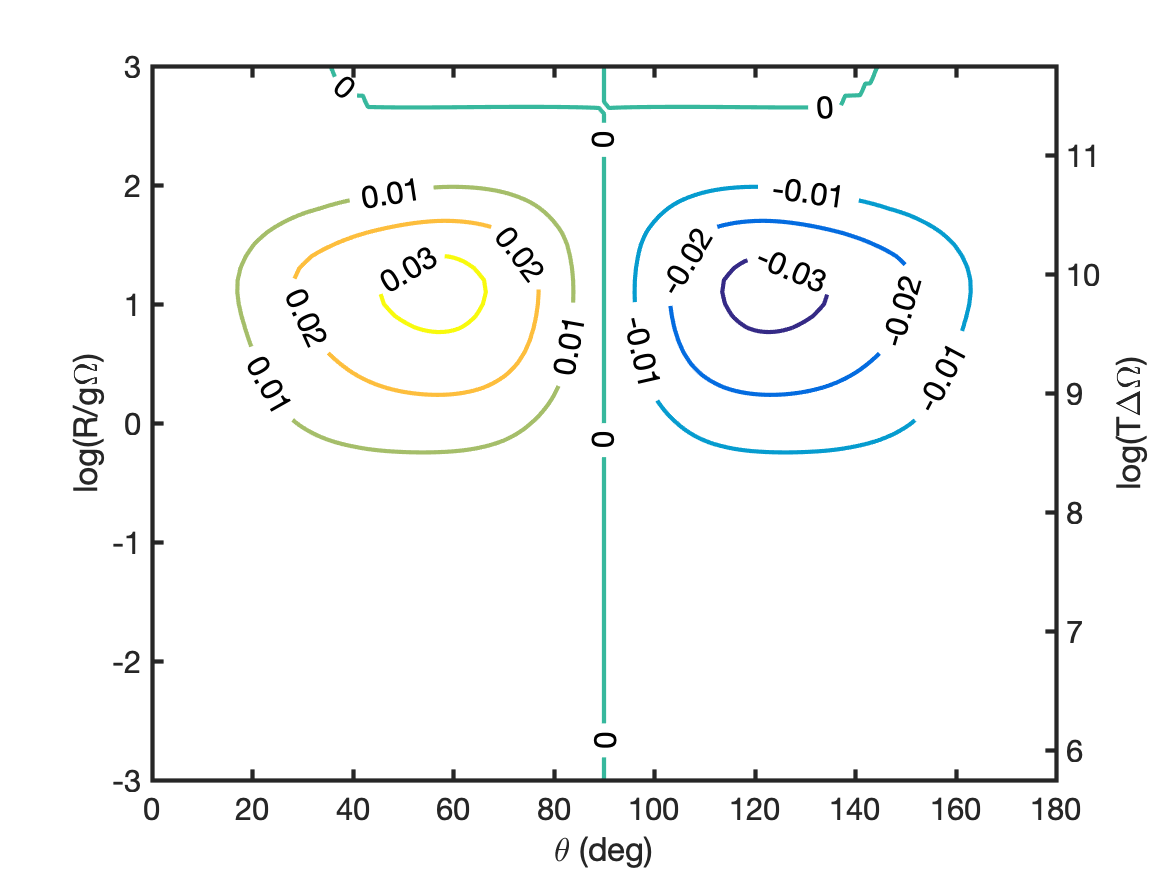}
      \caption{}
    \end{subfigure}
  \caption{Simulations of $J=3-2$ SiO masers with anisotropic pumping direction perpendicular to the magnetic field and propagation direction. Linear polarization fraction (a,d) and angle (b,e) and circular polarization fraction (c,f). Magnetic field strengths are $B=100$ mG for (a,b,c) and $B=1$ G for (d,e,f).}
\end{figure*}

\begin{figure*}
    \centering
    \begin{subfigure}[b]{0.45\textwidth}
       \includegraphics[width=\textwidth]{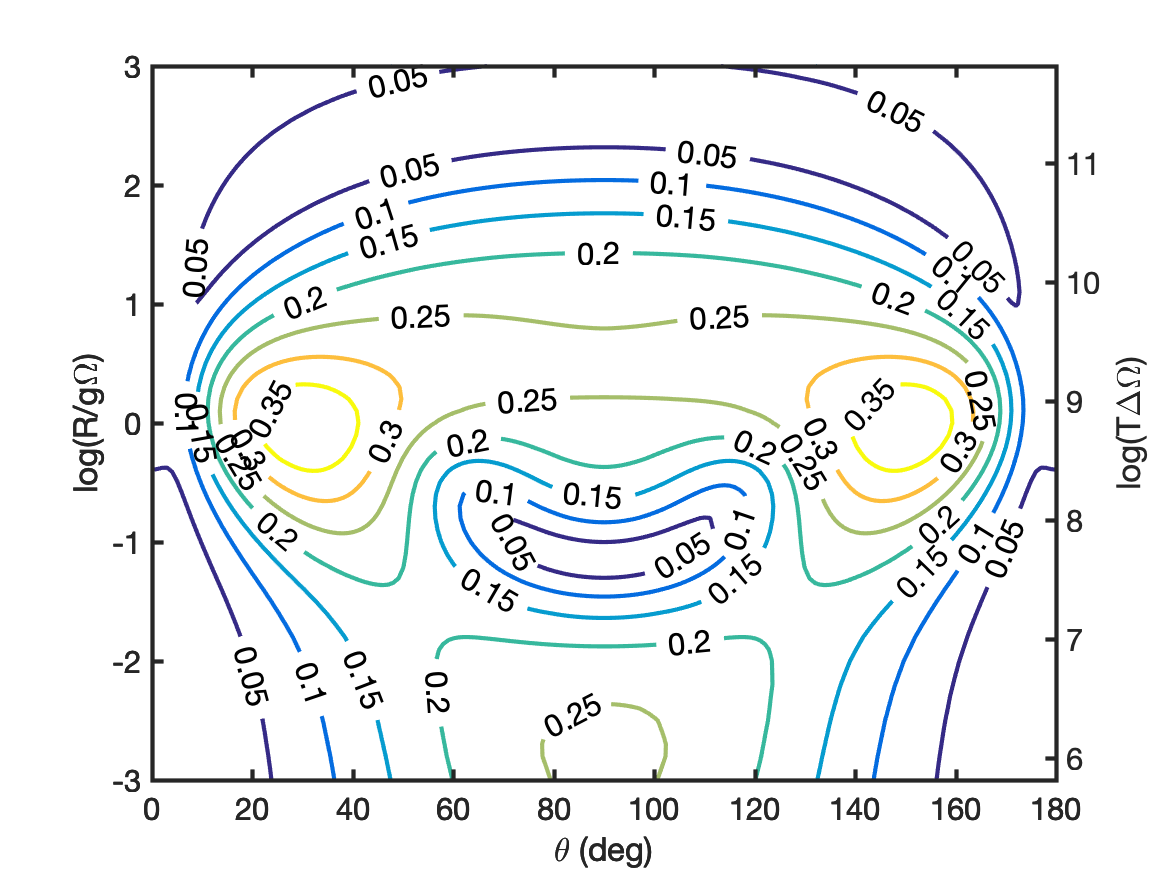} 
       \caption{}
    \end{subfigure}
    ~ 
    \begin{subfigure}[b]{0.45\textwidth}
       \includegraphics[width=\textwidth]{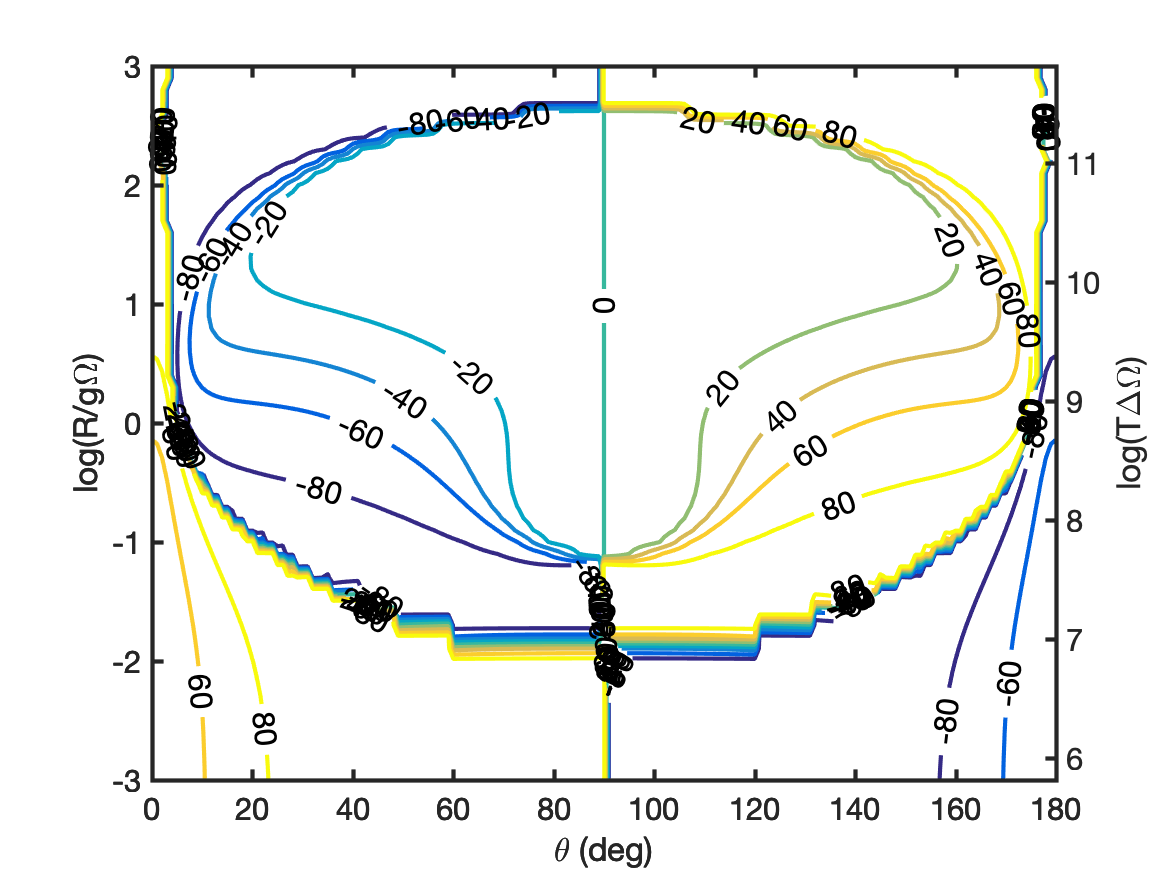} 
       \caption{}
    \end{subfigure}
     ~ 
    \begin{subfigure}[b]{0.45\textwidth}
      \includegraphics[width=\textwidth]{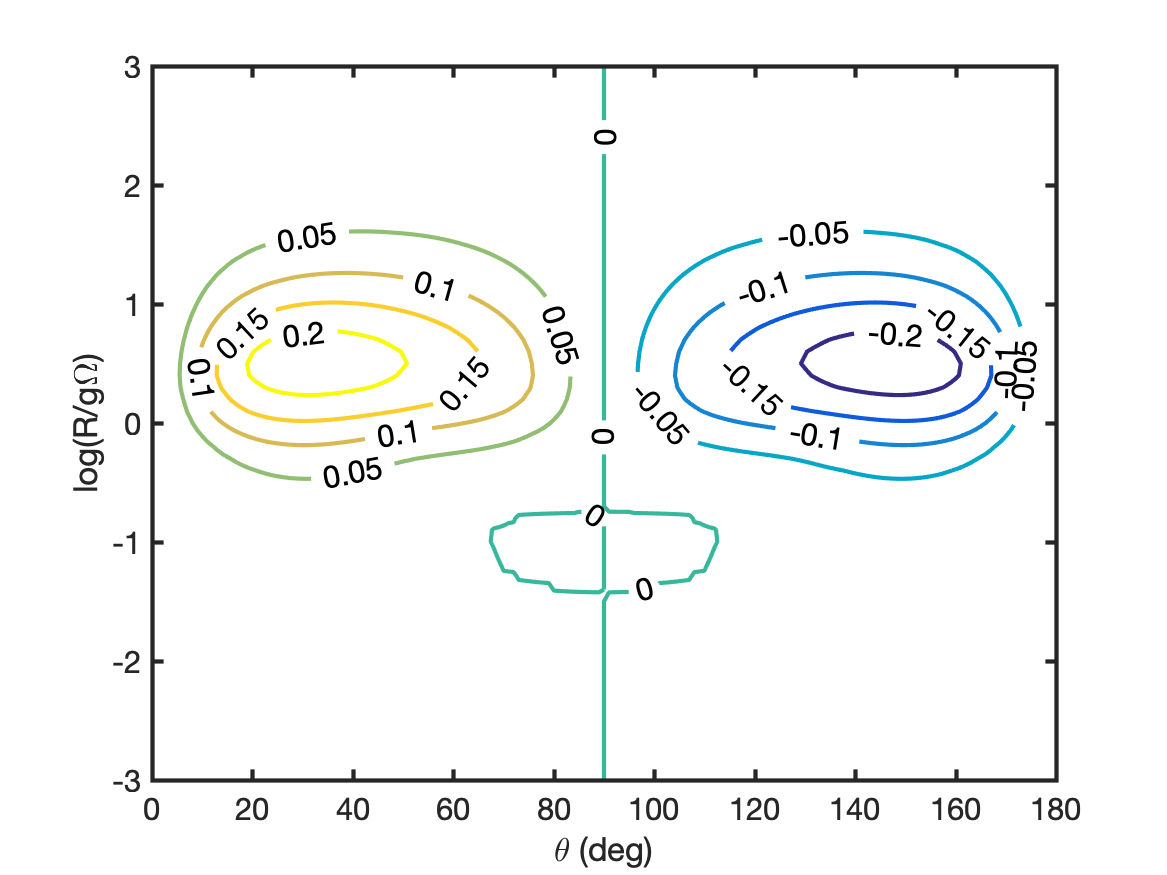}
      \caption{}
    \end{subfigure}
     ~
    \begin{subfigure}[b]{0.45\textwidth}
       \includegraphics[width=\textwidth]{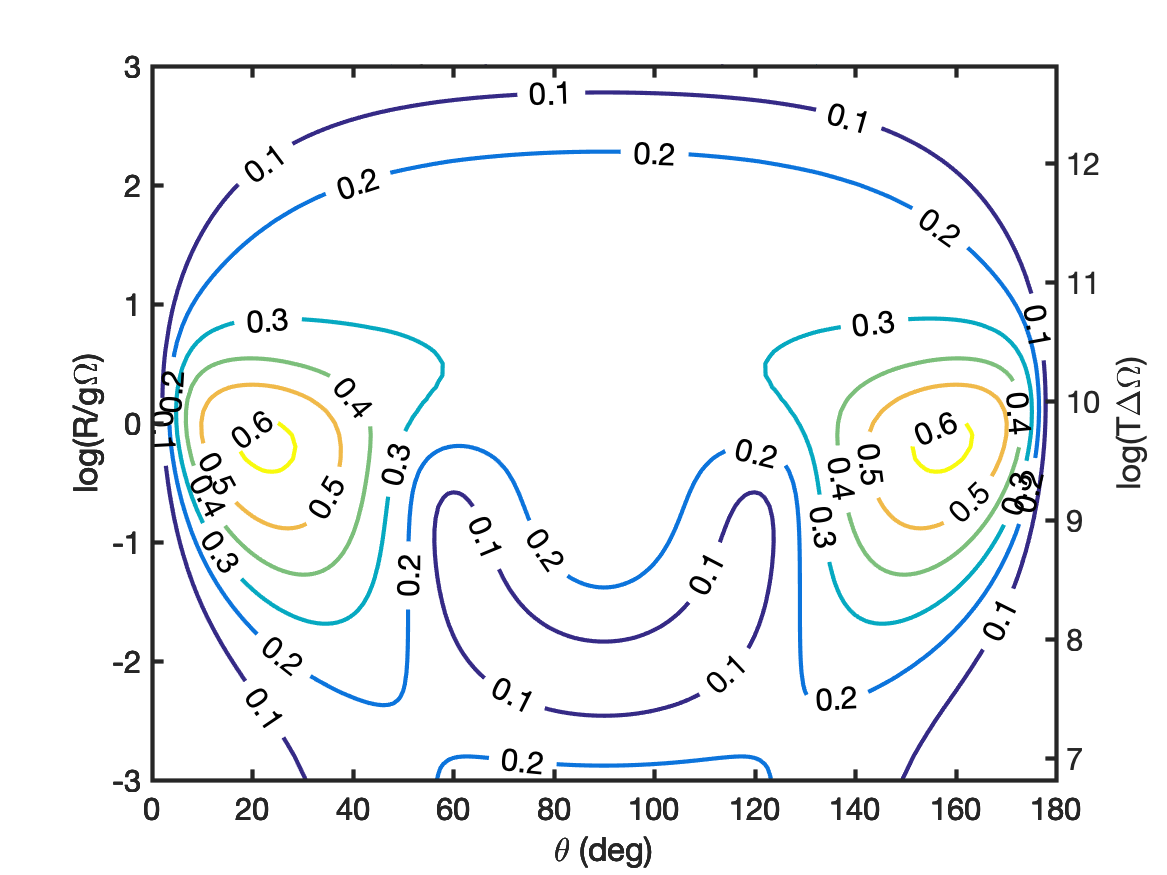} 
       \caption{}
    \end{subfigure}
    ~ 
    \begin{subfigure}[b]{0.45\textwidth}
       \includegraphics[width=\textwidth]{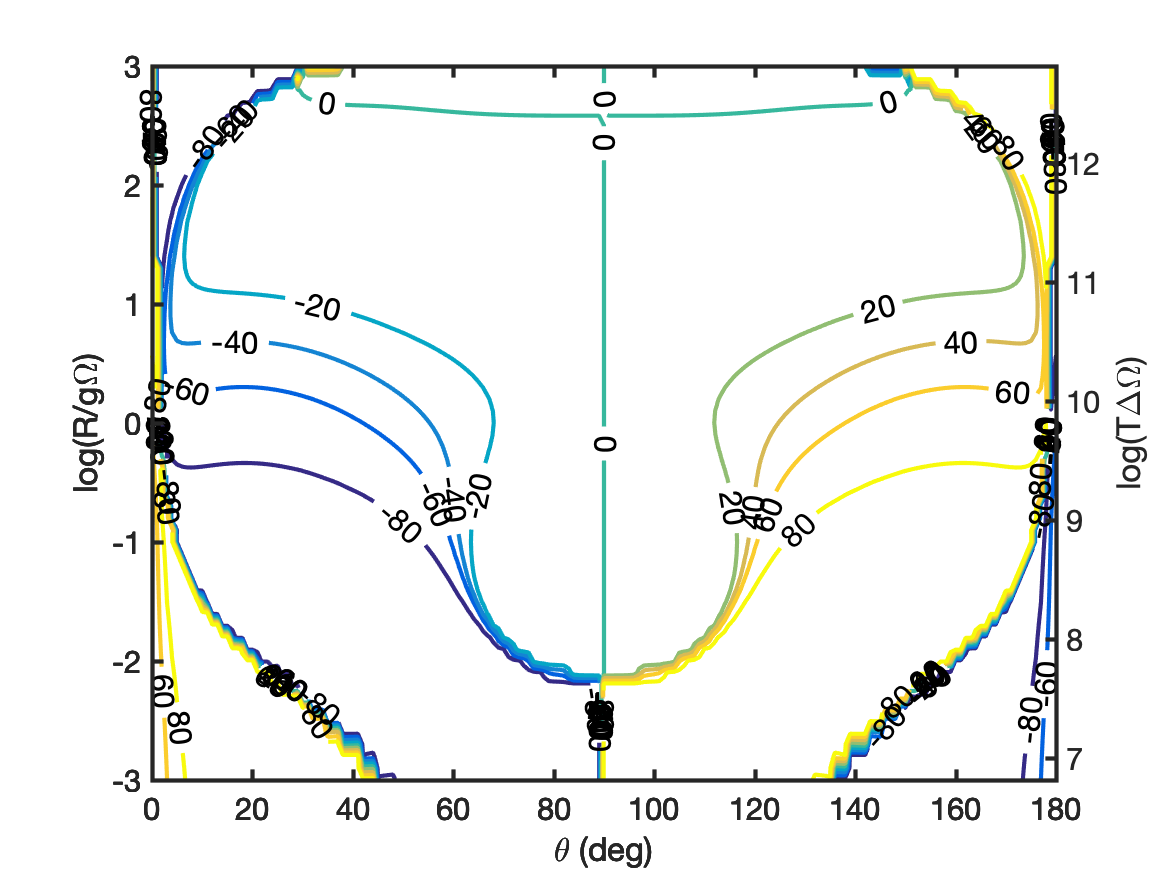} 
       \caption{}
    \end{subfigure}
     ~ 
    \begin{subfigure}[b]{0.45\textwidth}
      \includegraphics[width=\textwidth]{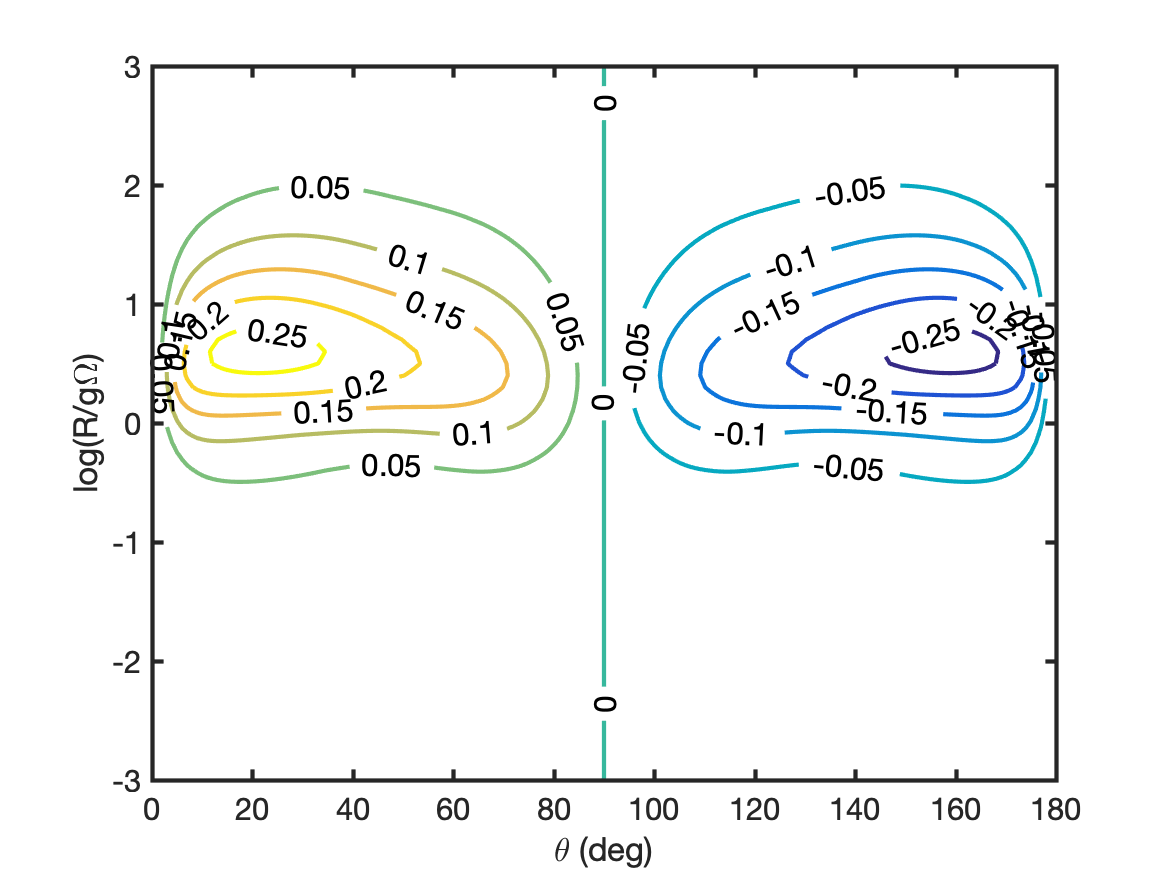}
      \caption{}
    \end{subfigure}
  \caption{Simulations of $J=1-0$ SiO masers with anisotropic pumping direction at $45^o$ from the magnetic field in the plane perpendicular to the propagation direction. Linear polarization fraction (a,d) and angle (b,e) and circular polarization fraction (c,f). Magnetic field strengths are $B=100$ mG for (a,b,c) and $B=1$ G for (d,e,f).}
\end{figure*}

\begin{figure*}
    \centering
    \begin{subfigure}[b]{0.45\textwidth}
       \includegraphics[width=\textwidth]{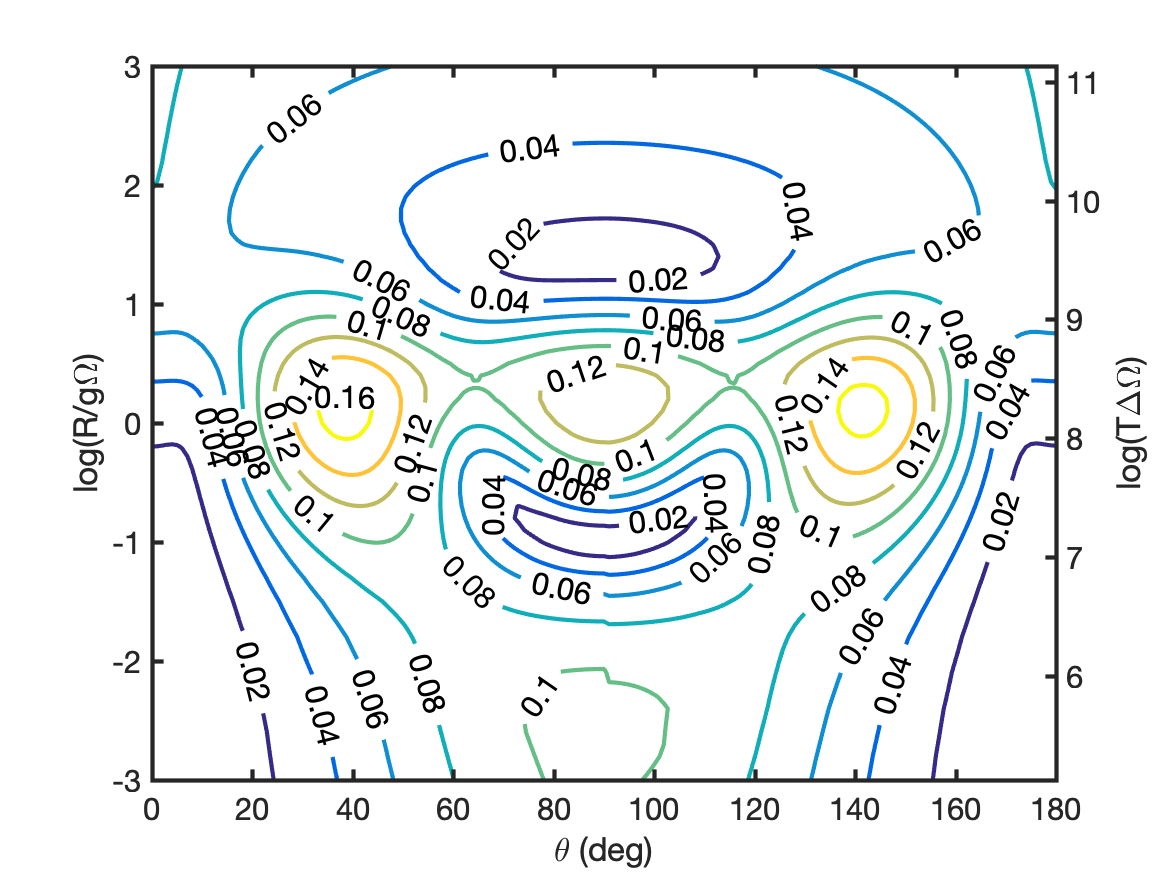}
       \caption{}
    \end{subfigure}
    ~
    \begin{subfigure}[b]{0.45\textwidth}
       \includegraphics[width=\textwidth]{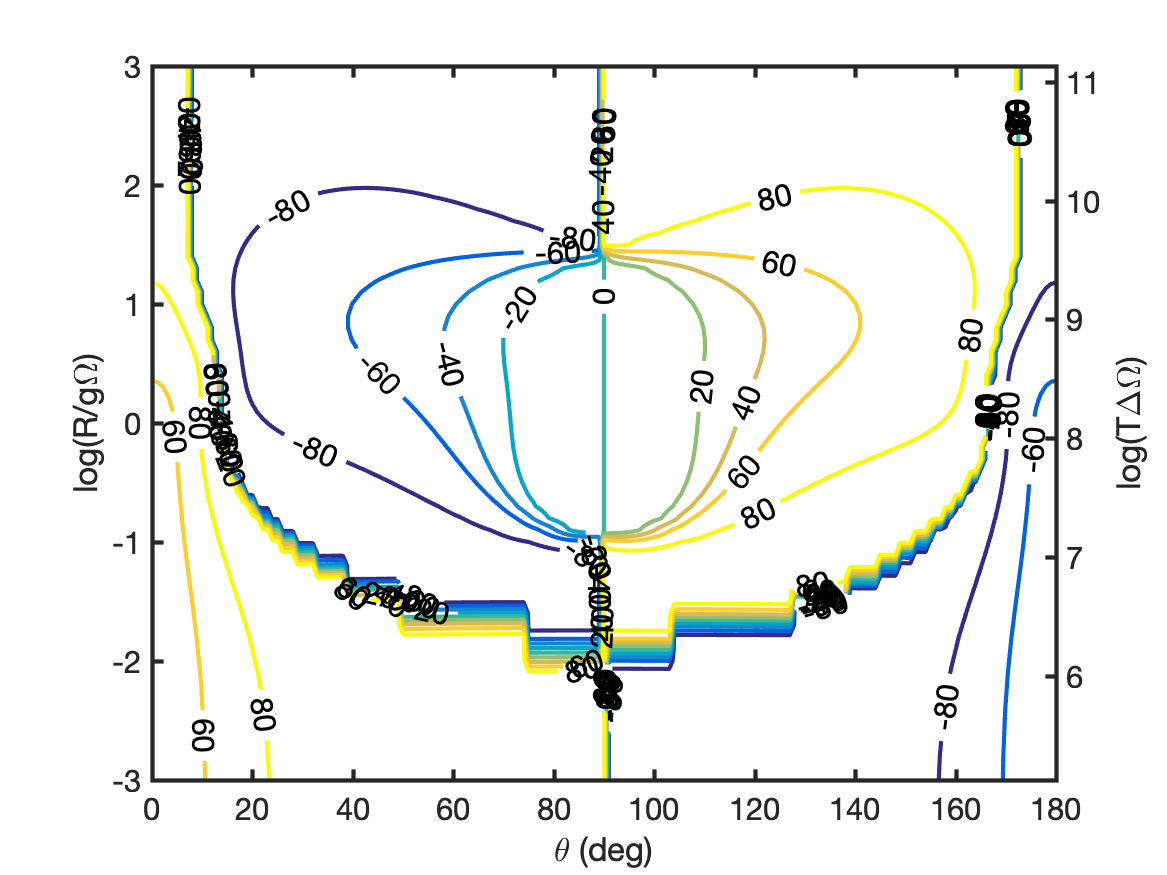}
       \caption{}
    \end{subfigure}
     ~
    \begin{subfigure}[b]{0.45\textwidth}
      \includegraphics[width=\textwidth]{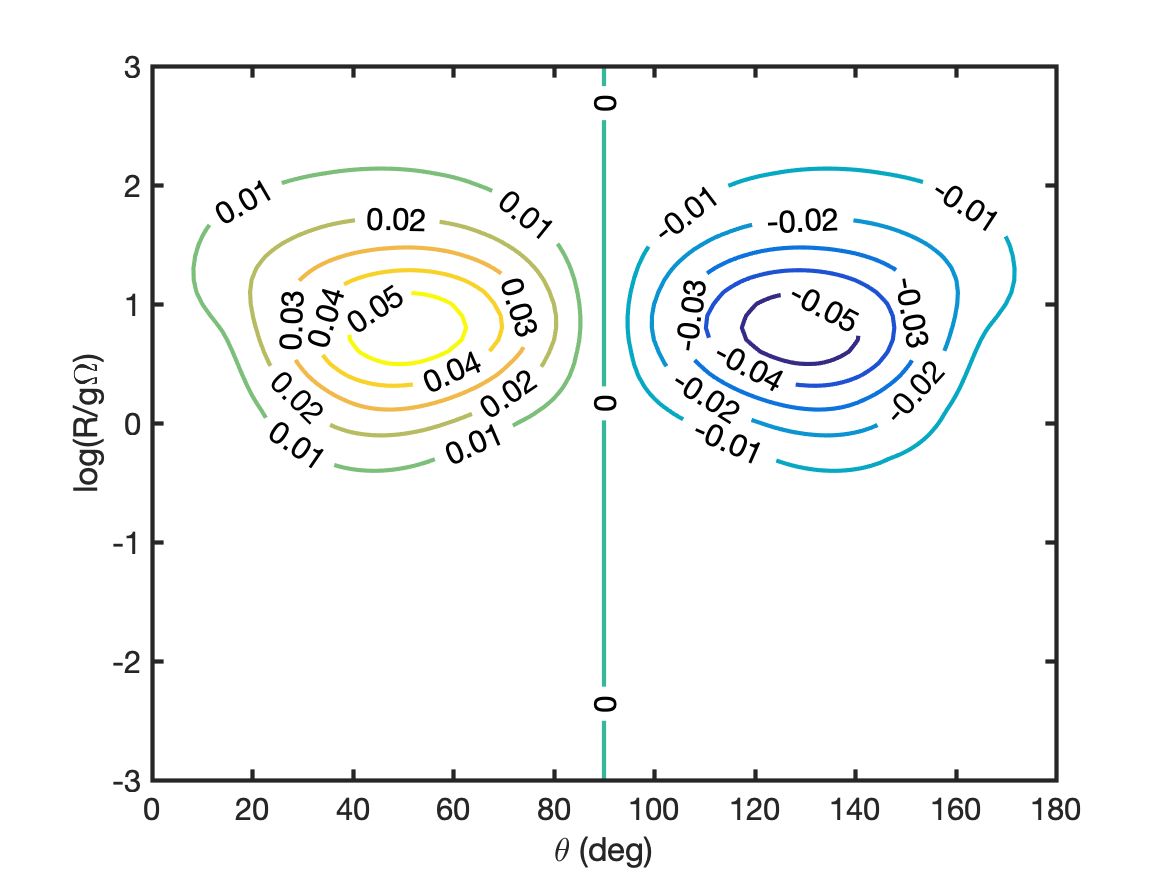}
      \caption{}
    \end{subfigure}
    ~
    \begin{subfigure}[b]{0.45\textwidth}
       \includegraphics[width=\textwidth]{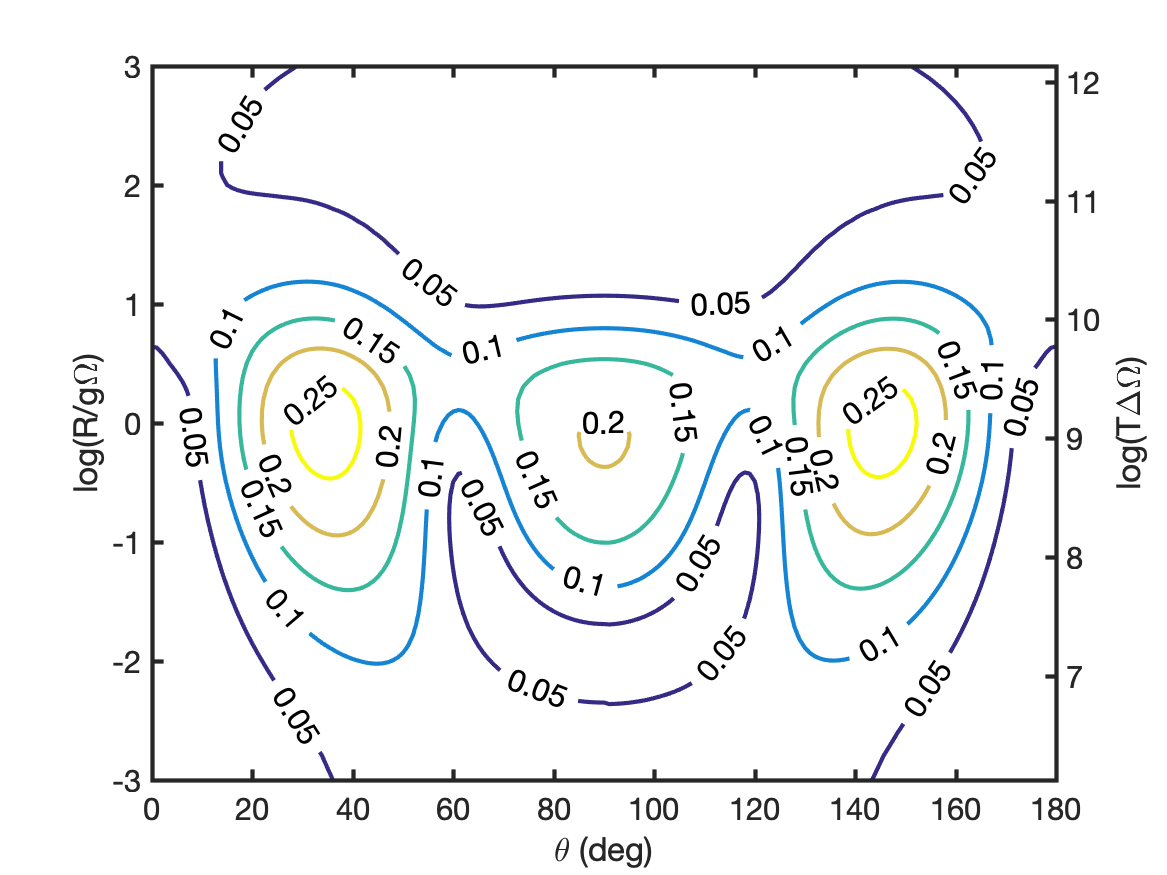}
       \caption{}
    \end{subfigure}
    ~
    \begin{subfigure}[b]{0.45\textwidth}
       \includegraphics[width=\textwidth]{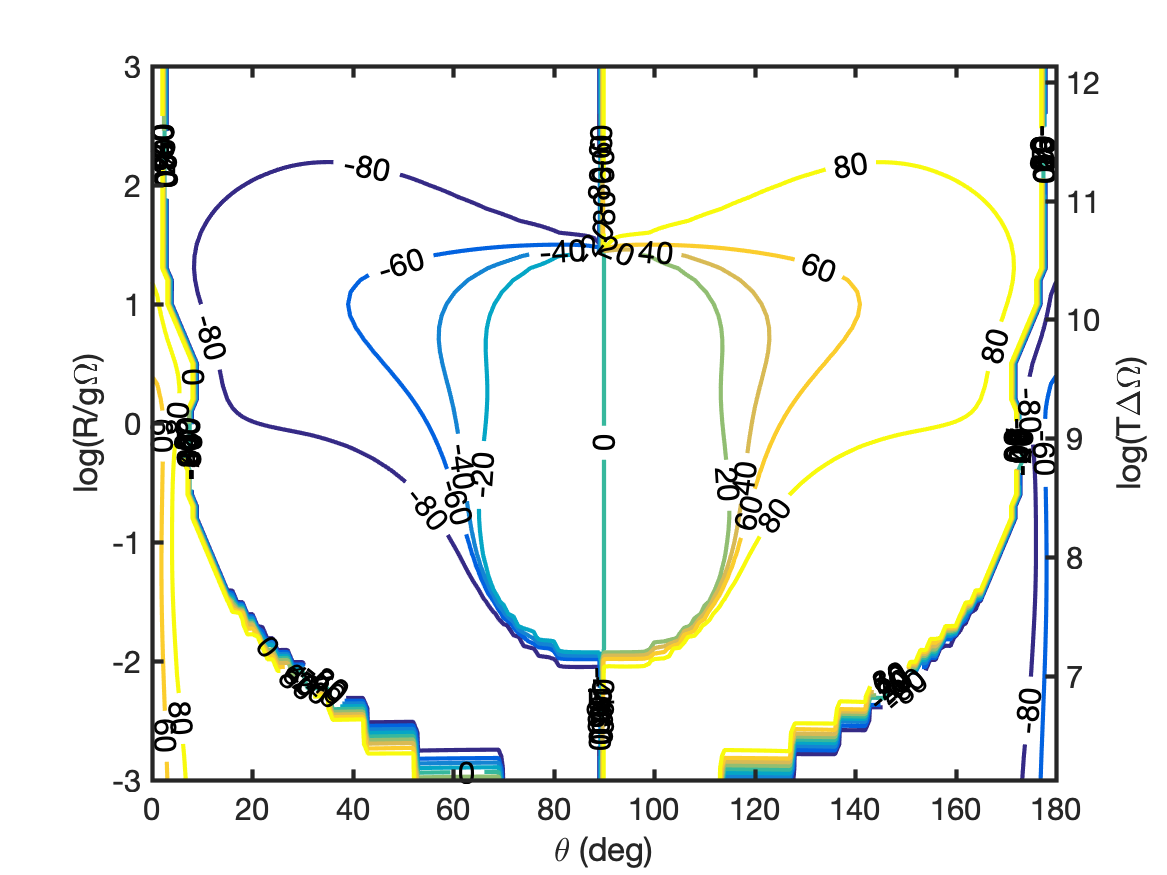}
       \caption{}
    \end{subfigure}
     ~
    \begin{subfigure}[b]{0.45\textwidth}
      \includegraphics[width=\textwidth]{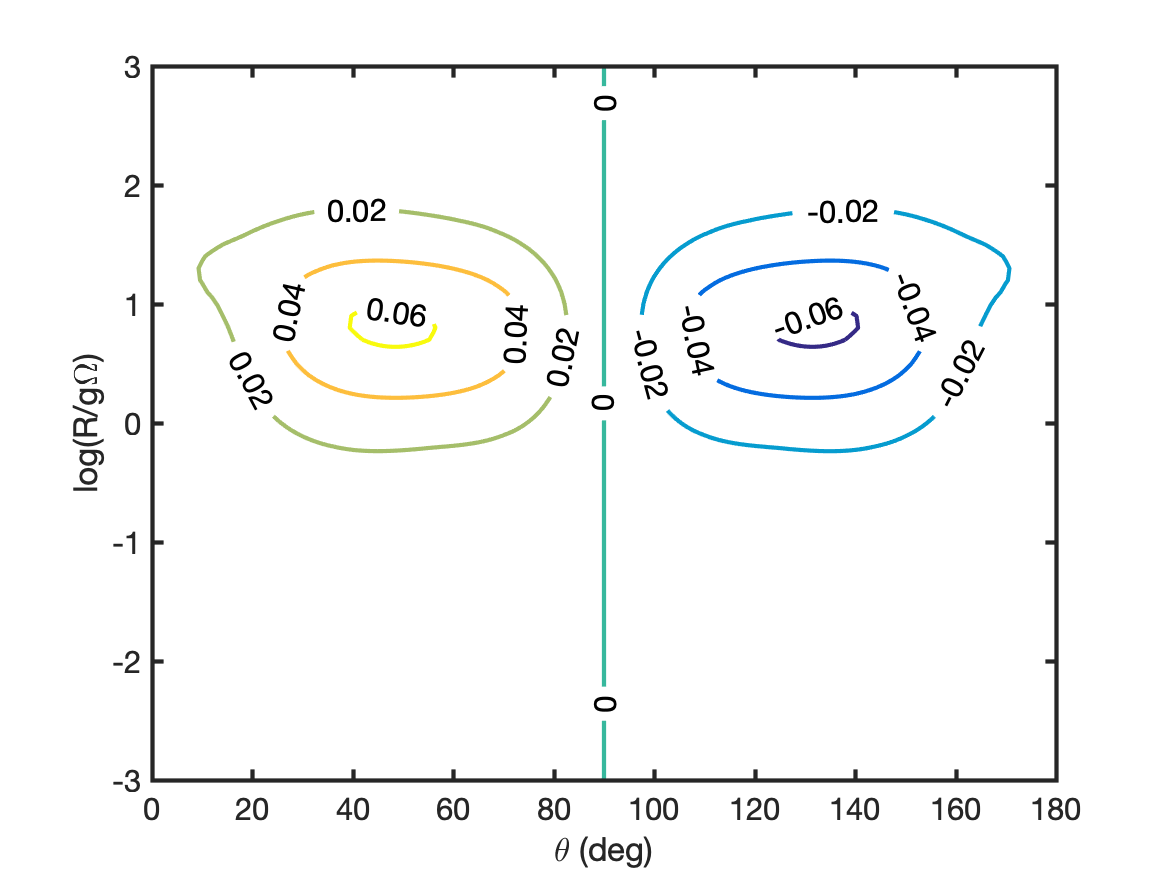}
      \caption{}
    \end{subfigure}
  \caption{Simulations of $J=2-1$ SiO masers with anisotropic pumping direction at $45^o$ from the magnetic field in the plane perpendicular to the propagation direction. Linear polarization fraction (a,d) and angle (b,e) and circular polarization fraction (c,f). Magnetic field strengths are $B=100$ mG for (a,b,c) and $B=1$ G for (d,e,f).}
\end{figure*}

\begin{figure*}
    \centering    
    \begin{subfigure}[b]{0.45\textwidth}
       \includegraphics[width=\textwidth]{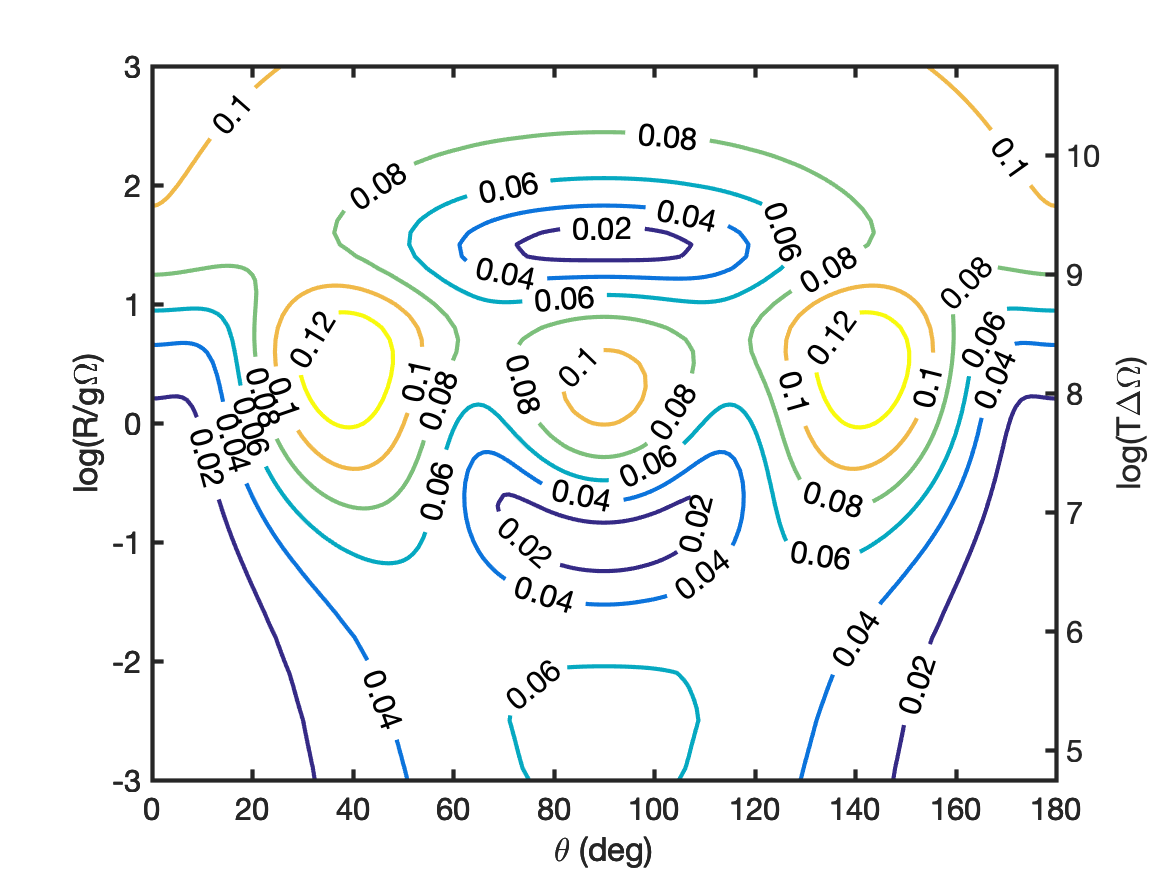}
       \caption{}
    \end{subfigure}
    ~
    \begin{subfigure}[b]{0.45\textwidth}
       \includegraphics[width=\textwidth]{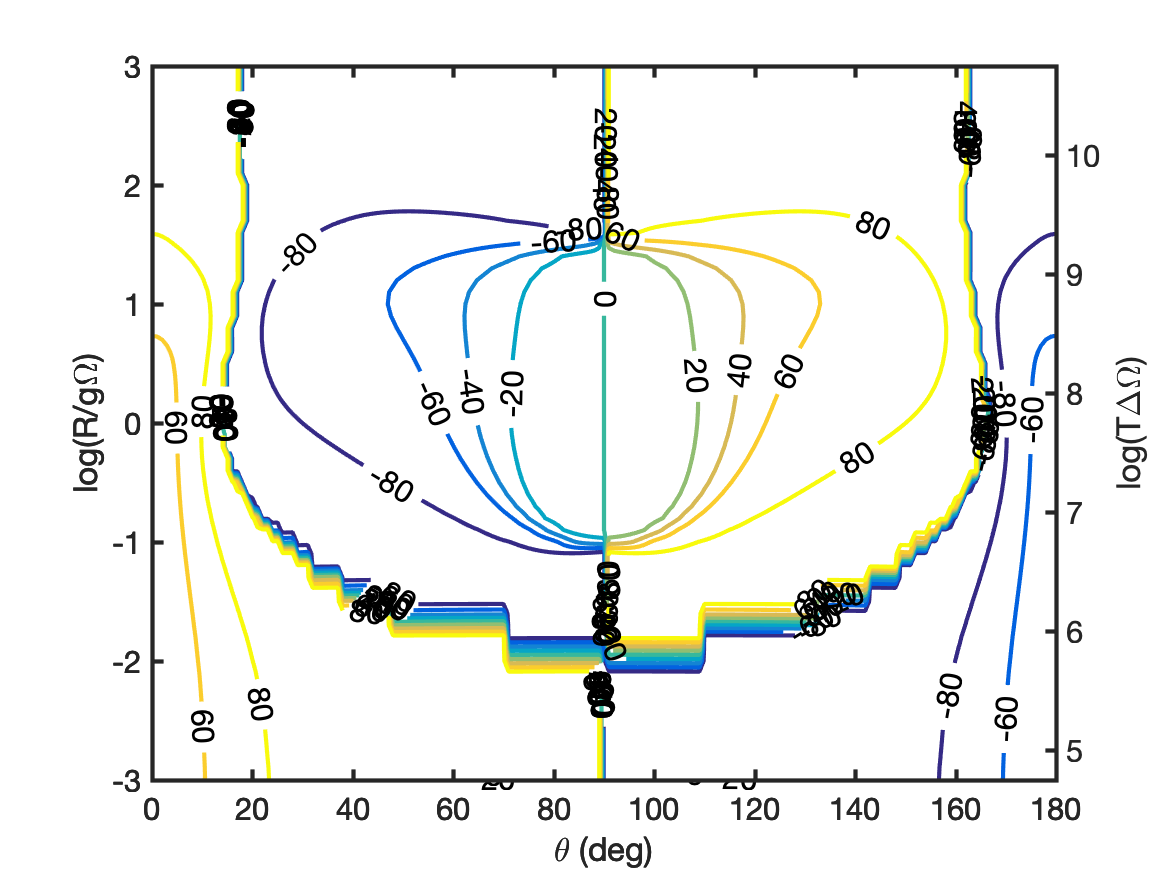}
       \caption{}
    \end{subfigure}
     ~
    \begin{subfigure}[b]{0.45\textwidth}
      \includegraphics[width=\textwidth]{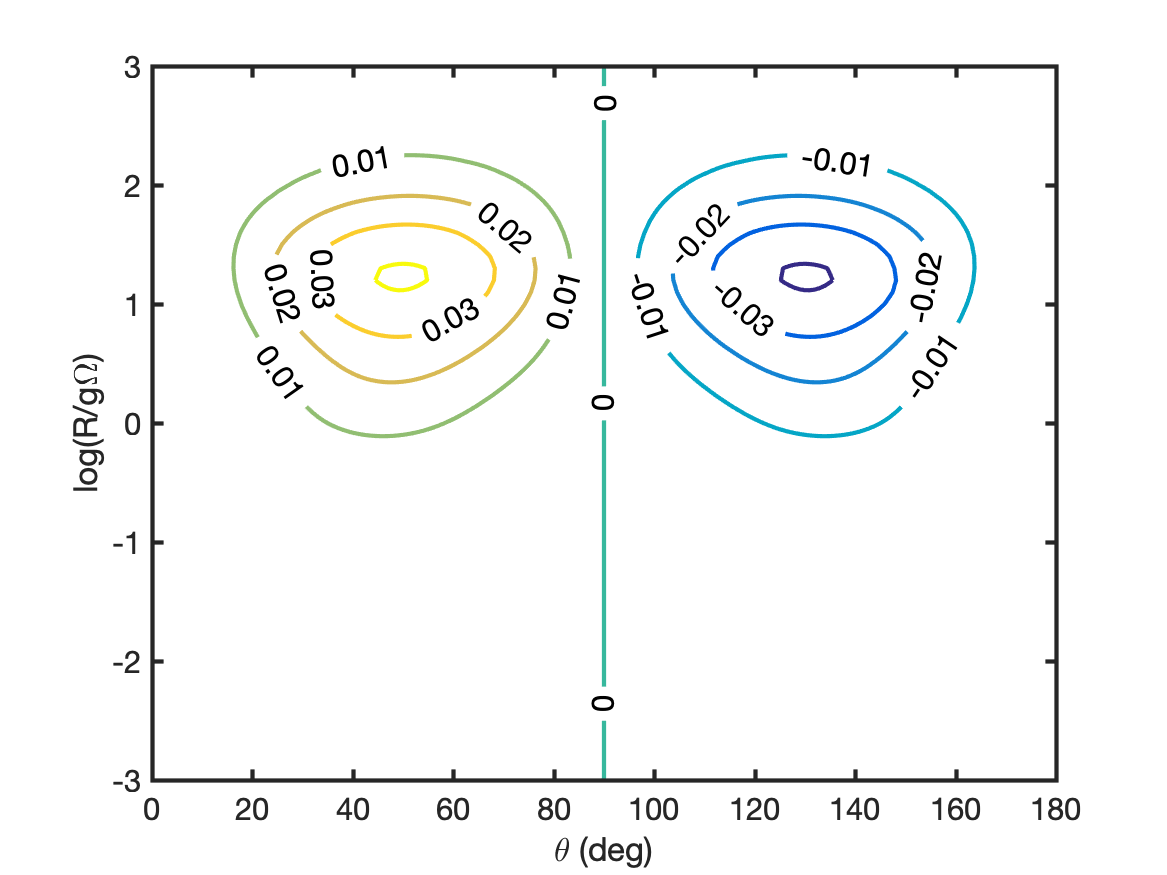}
      \caption{}
    \end{subfigure}
    ~
    \begin{subfigure}[b]{0.45\textwidth}
       \includegraphics[width=\textwidth]{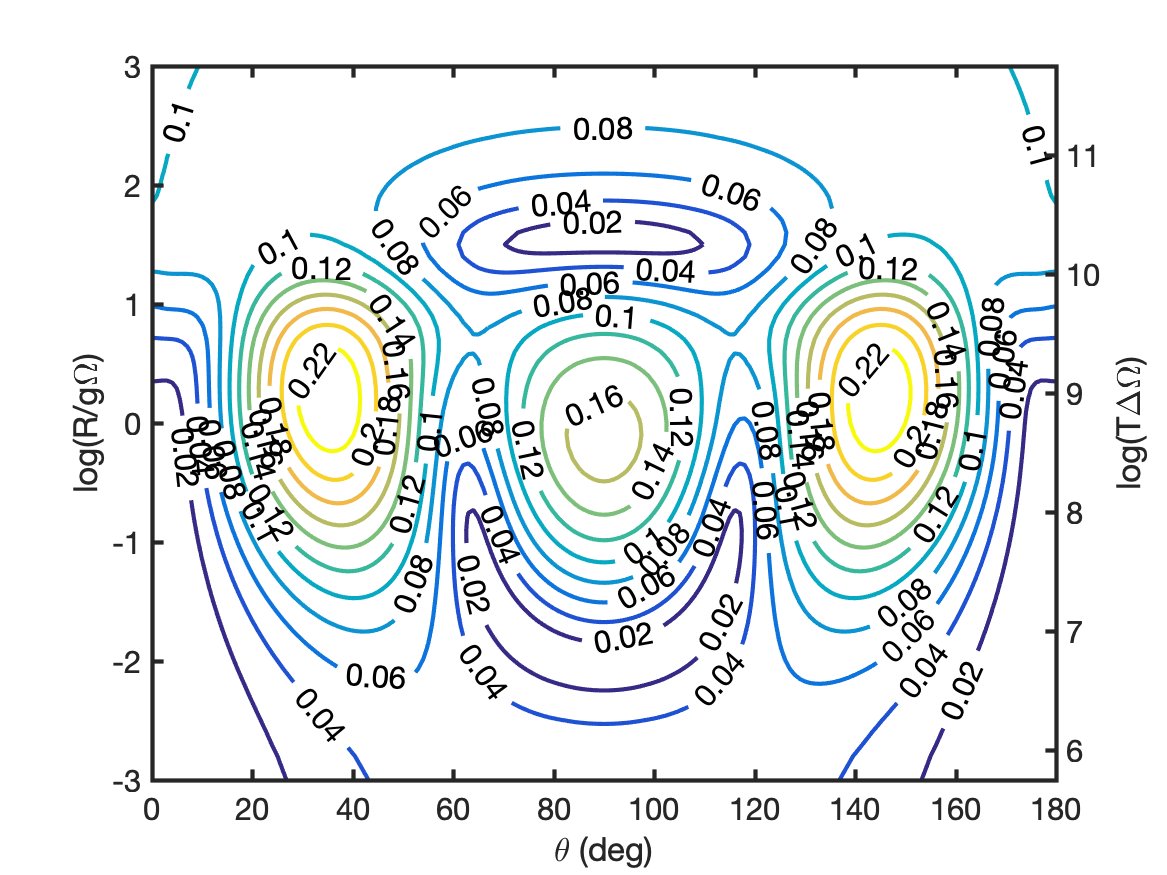}
       \caption{}
    \end{subfigure}
    ~
    \begin{subfigure}[b]{0.45\textwidth}
       \includegraphics[width=\textwidth]{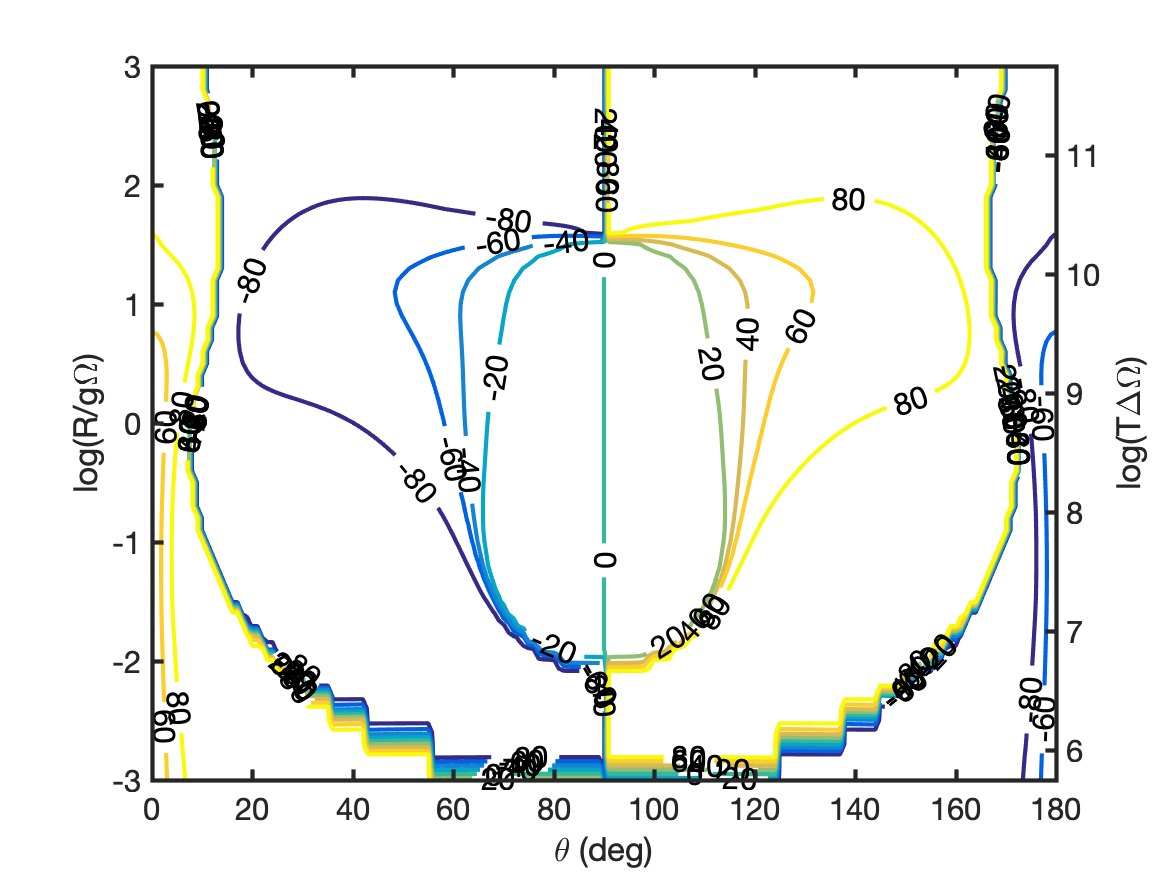}
       \caption{}
    \end{subfigure}
     ~
    \begin{subfigure}[b]{0.45\textwidth}
      \includegraphics[width=\textwidth]{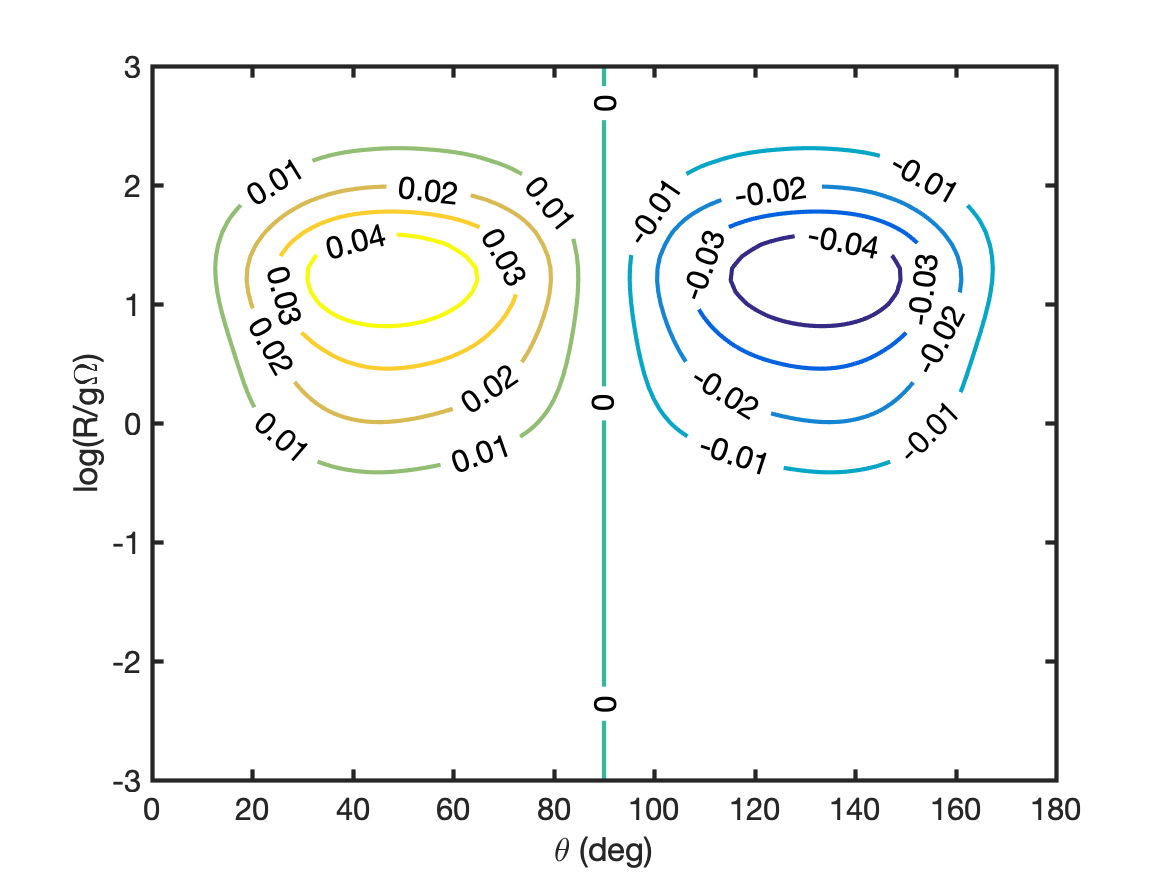}
      \caption{}
    \end{subfigure}
  \caption{Simulations of $J=3-2$ SiO masers with anisotropic pumping direction at $45^o$ from the magnetic field in the plane perpendicular to the propagation direction. Linear polarization fraction (a,d) and angle (b,e) and circular polarization fraction (c,f). Magnetic field strengths are $B=100$ mG for (a,b,c) and $B=1$ G for (d,e,f).}
\end{figure*}

\begin{figure*}
    \centering
    \begin{subfigure}[b]{0.32\textwidth}
       \includegraphics[width=\textwidth]{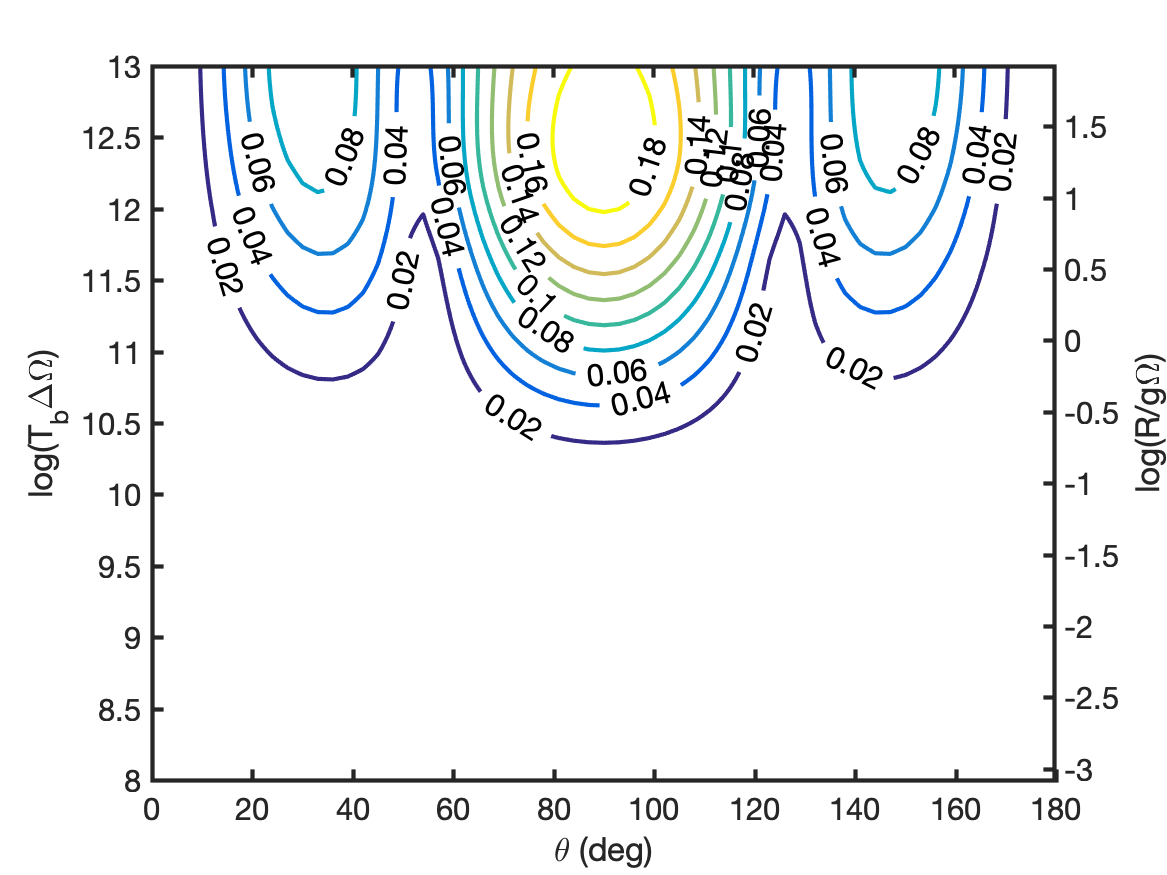}
       \caption{}
    \end{subfigure}
    ~
    \begin{subfigure}[b]{0.32\textwidth}
       \includegraphics[width=\textwidth]{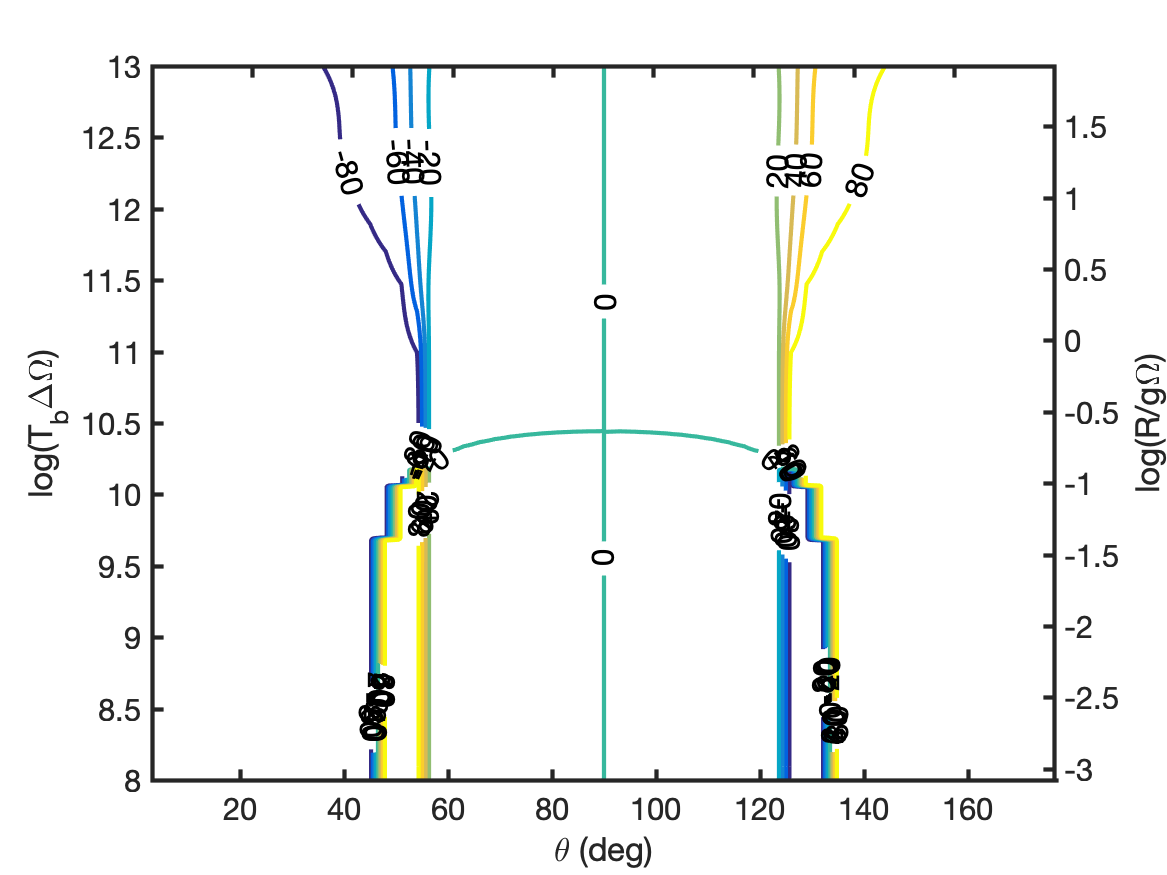}
       \caption{}
    \end{subfigure}
     ~
    \begin{subfigure}[b]{0.32\textwidth}
       \includegraphics[width=\textwidth]{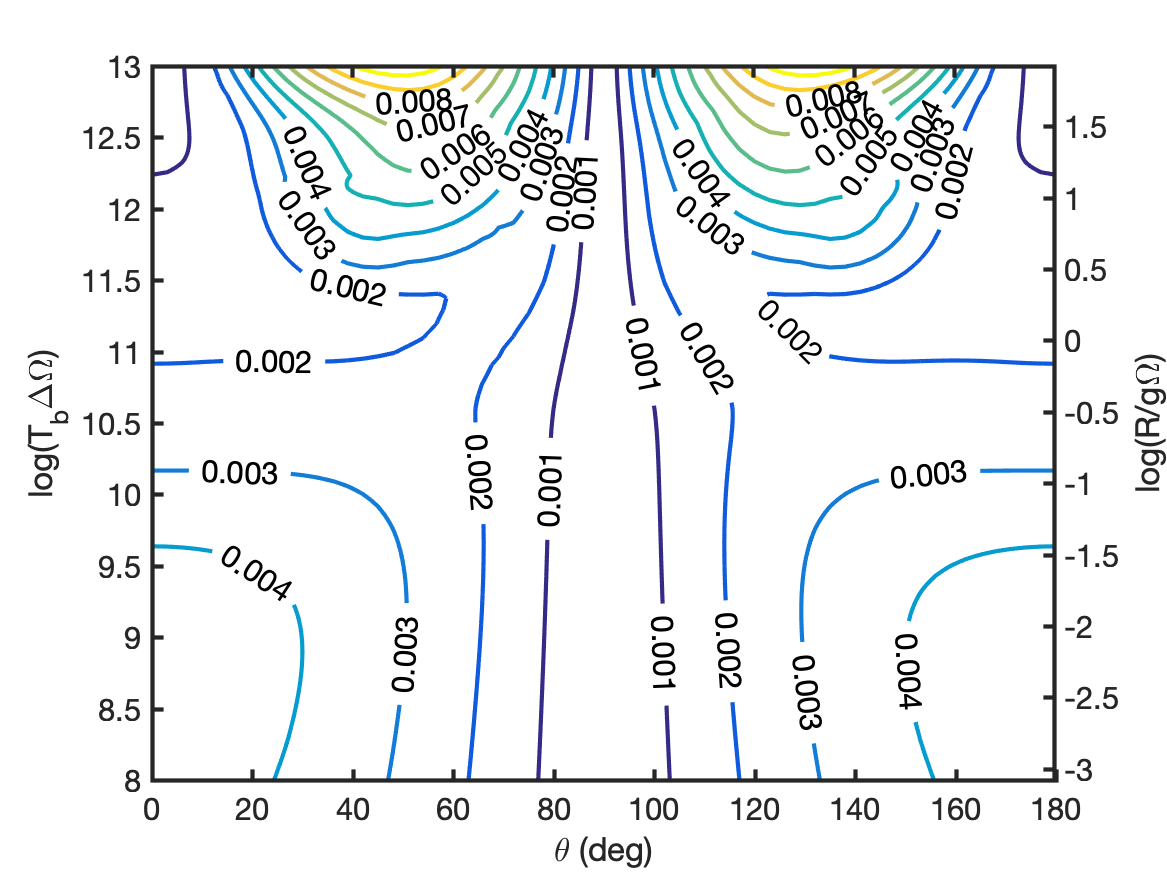}
      \caption{}
    \end{subfigure}
     ~
    \begin{subfigure}[b]{0.32\textwidth}
       \includegraphics[width=\textwidth]{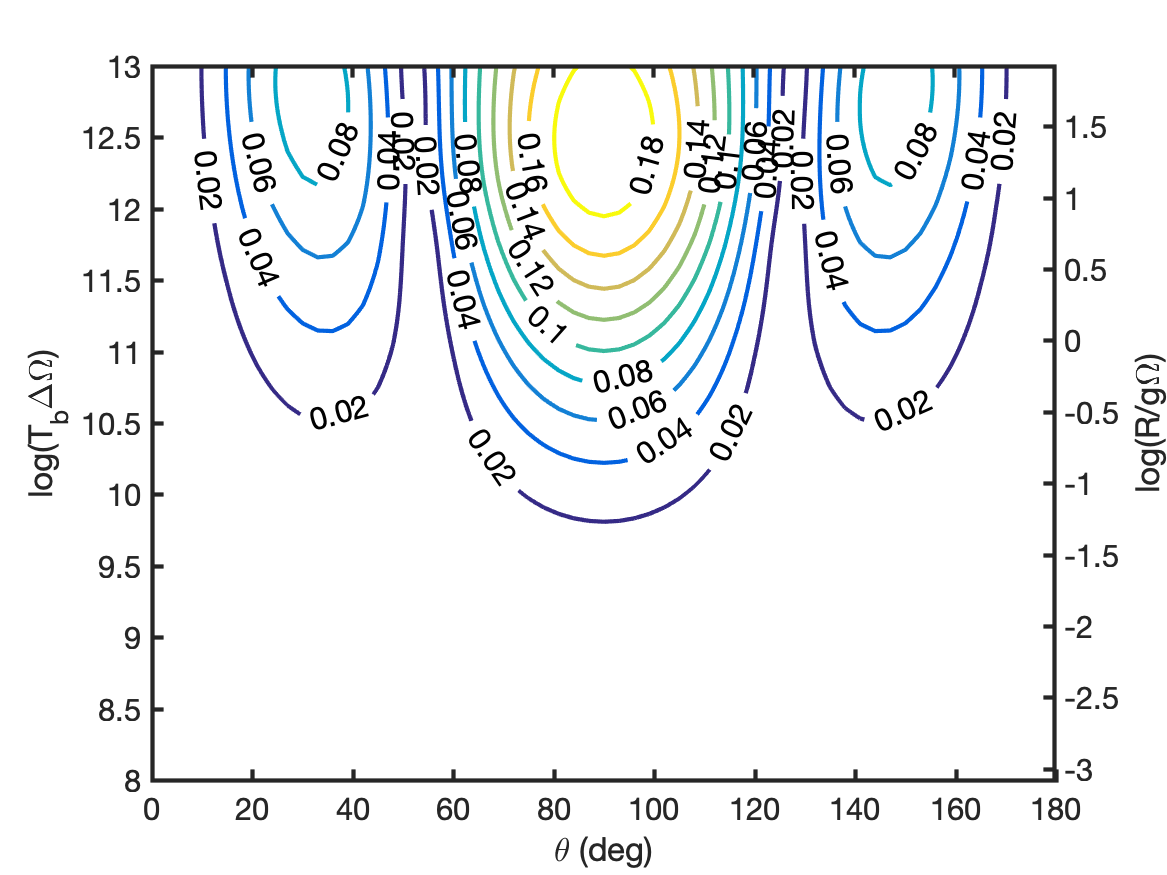}
       \caption{}
    \end{subfigure}
    ~
    \begin{subfigure}[b]{0.32\textwidth}
       \includegraphics[width=\textwidth]{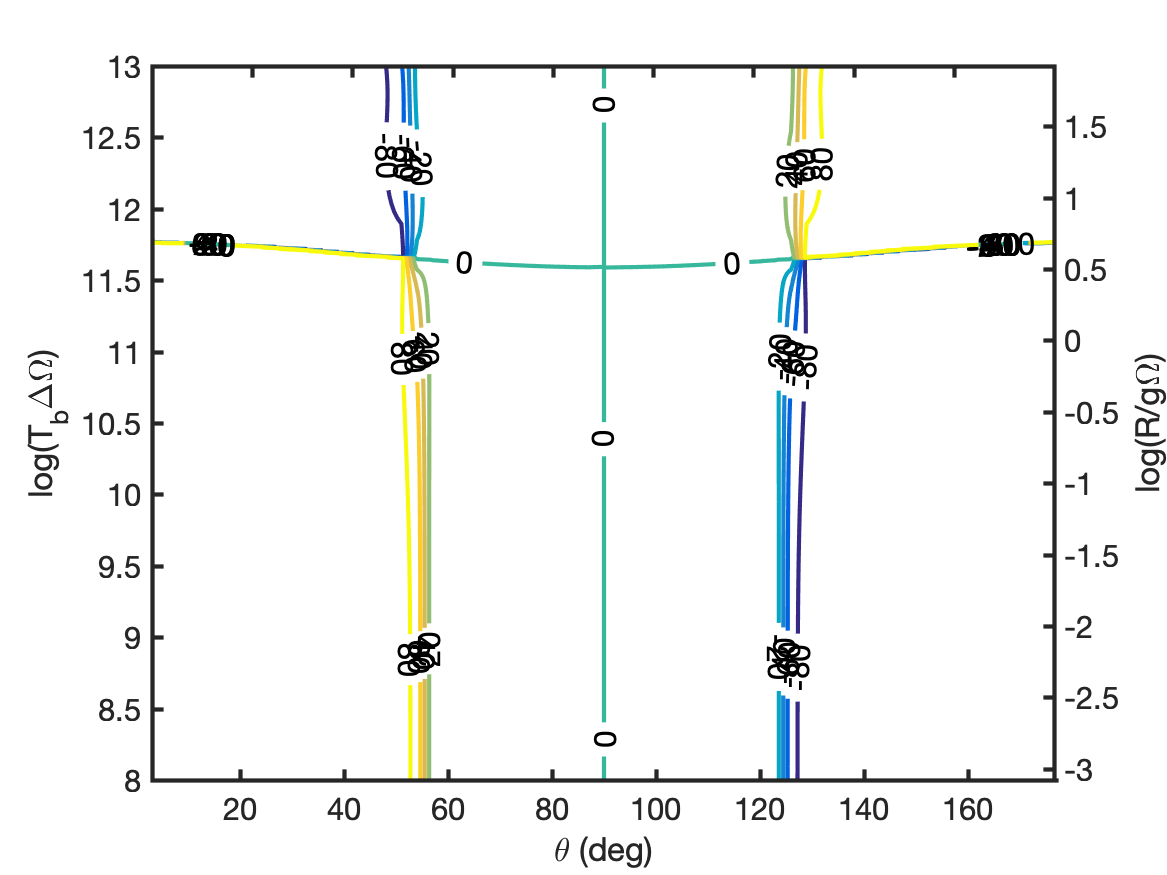}
       \caption{}
    \end{subfigure}
     ~
    \begin{subfigure}[b]{0.32\textwidth}
       \includegraphics[width=\textwidth]{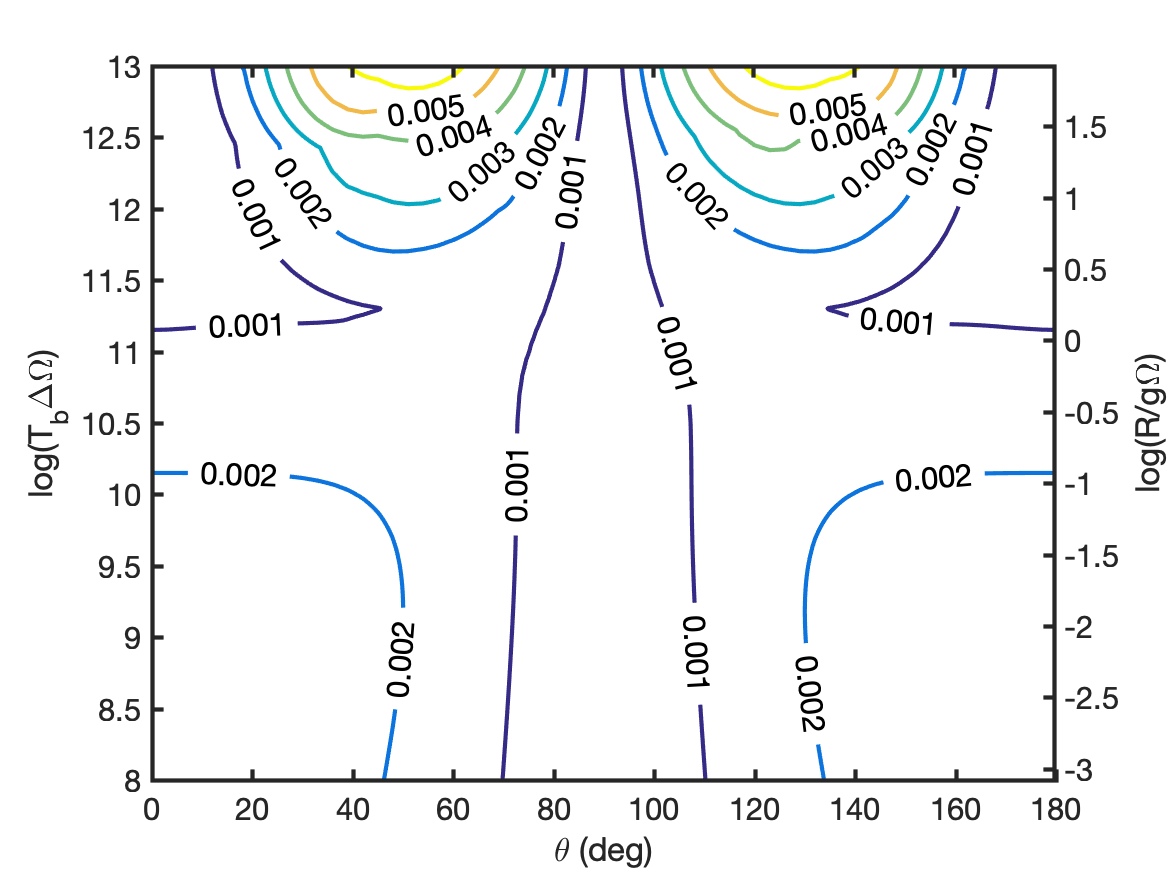}
      \caption{}
    \end{subfigure}
     ~
    \begin{subfigure}[b]{0.32\textwidth}
       \includegraphics[width=\textwidth]{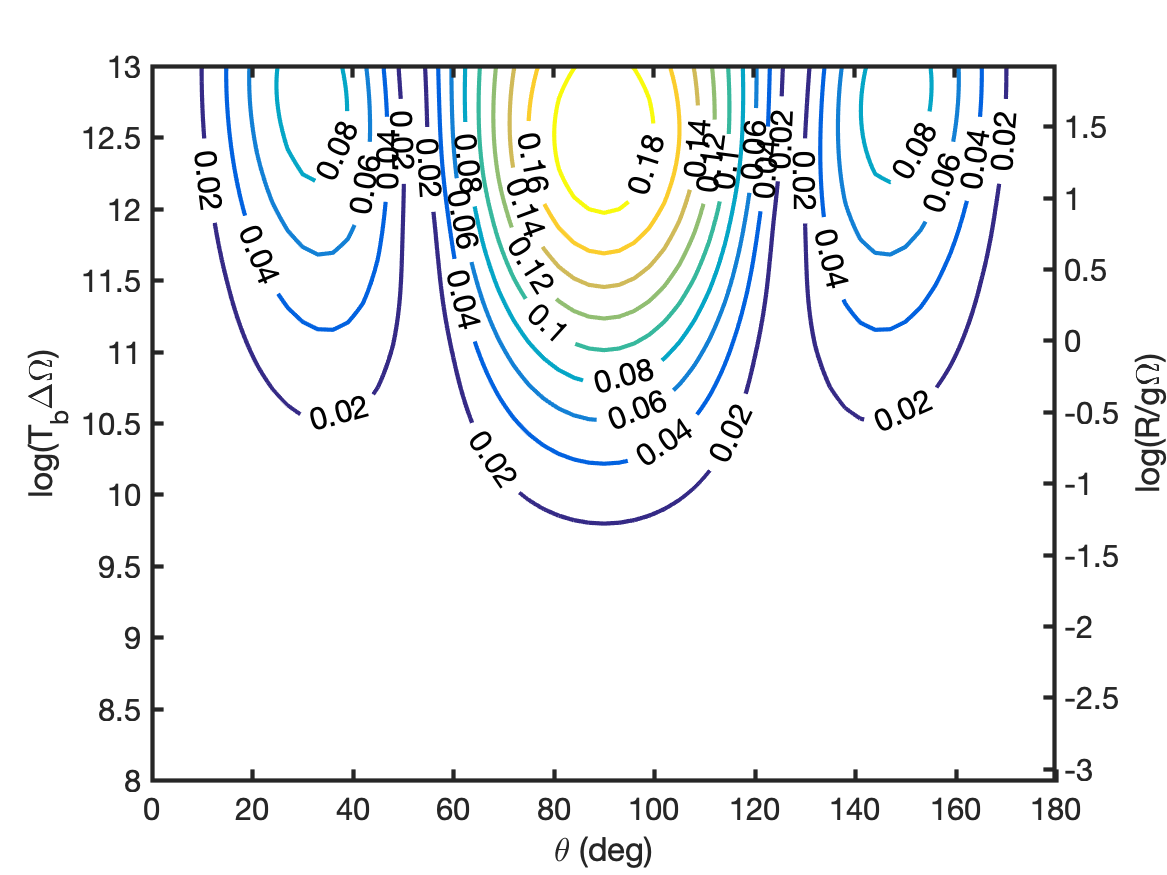}
       \caption{}
    \end{subfigure}
    ~
    \begin{subfigure}[b]{0.32\textwidth}
       \includegraphics[width=\textwidth]{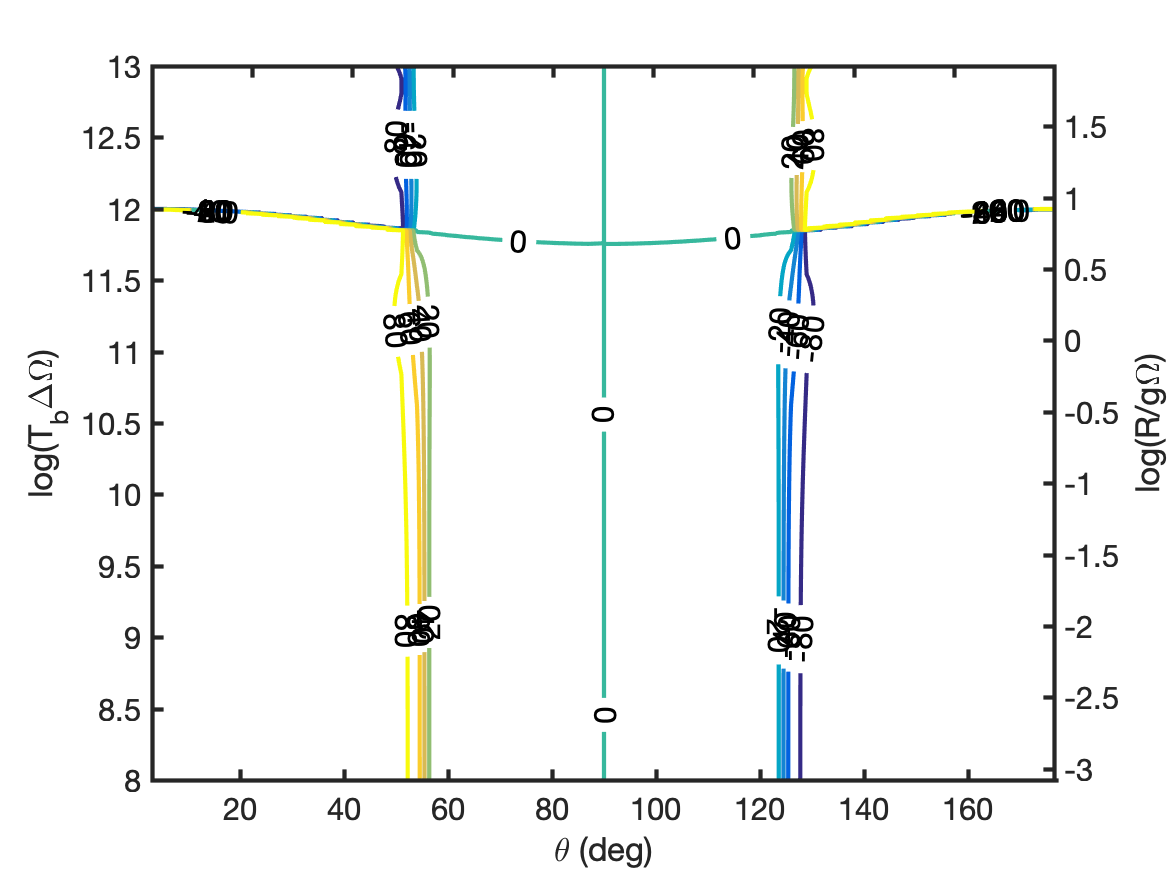}
       \caption{}
    \end{subfigure}
     ~
    \begin{subfigure}[b]{0.32\textwidth}
       \includegraphics[width=\textwidth]{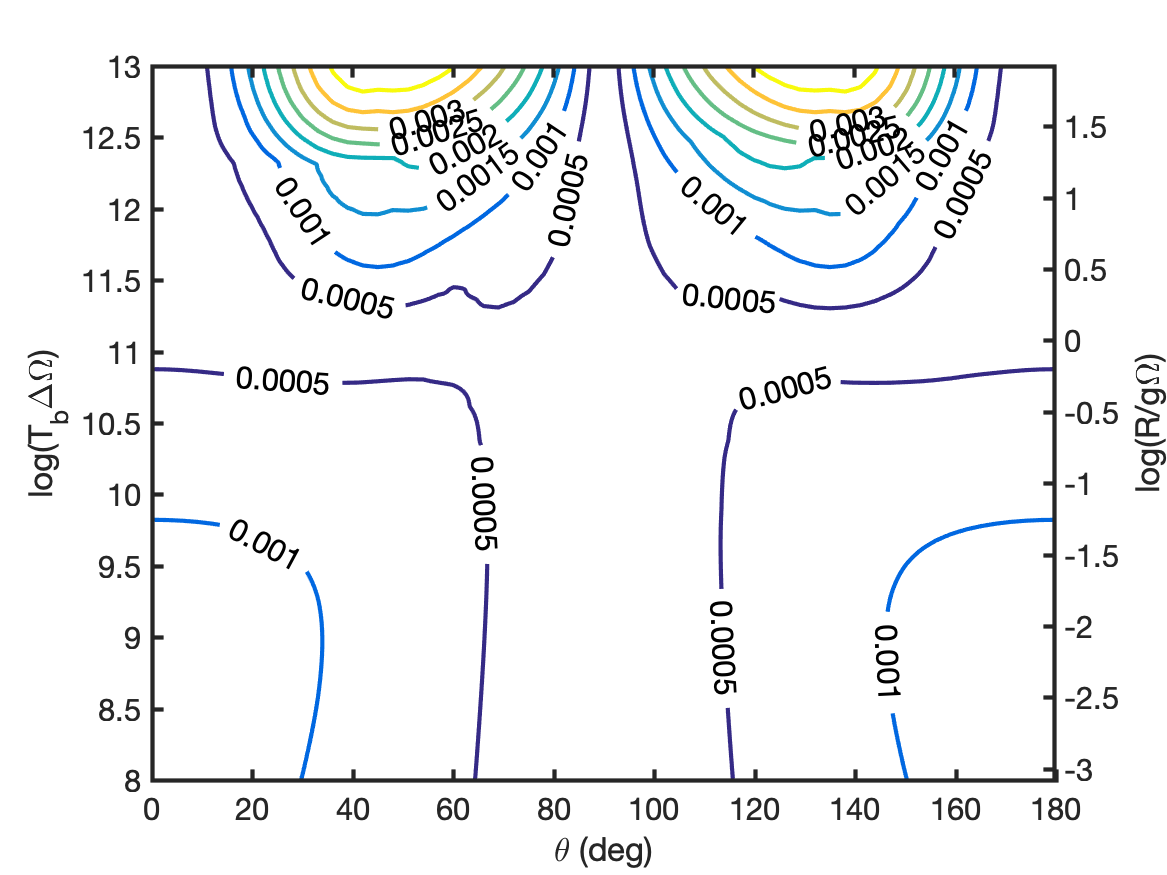}
      \caption{}
    \end{subfigure}
  \caption{Polarization of a water maser isotropically pumped at $B=20\ \mathrm{mG}$. Linear polarization fraction (a,d,g), angle (b,e,h) and circular polarization fraction (c,f,i). Thermal width used $v_{th} = 0.6$ km/s (a,b,c), $1$ km/s (d,e,f) and $2$ km/s (g,h,i).}
\end{figure*}

\begin{figure*}
    \centering
    \begin{subfigure}[b]{0.32\textwidth}
       \includegraphics[width=\textwidth]{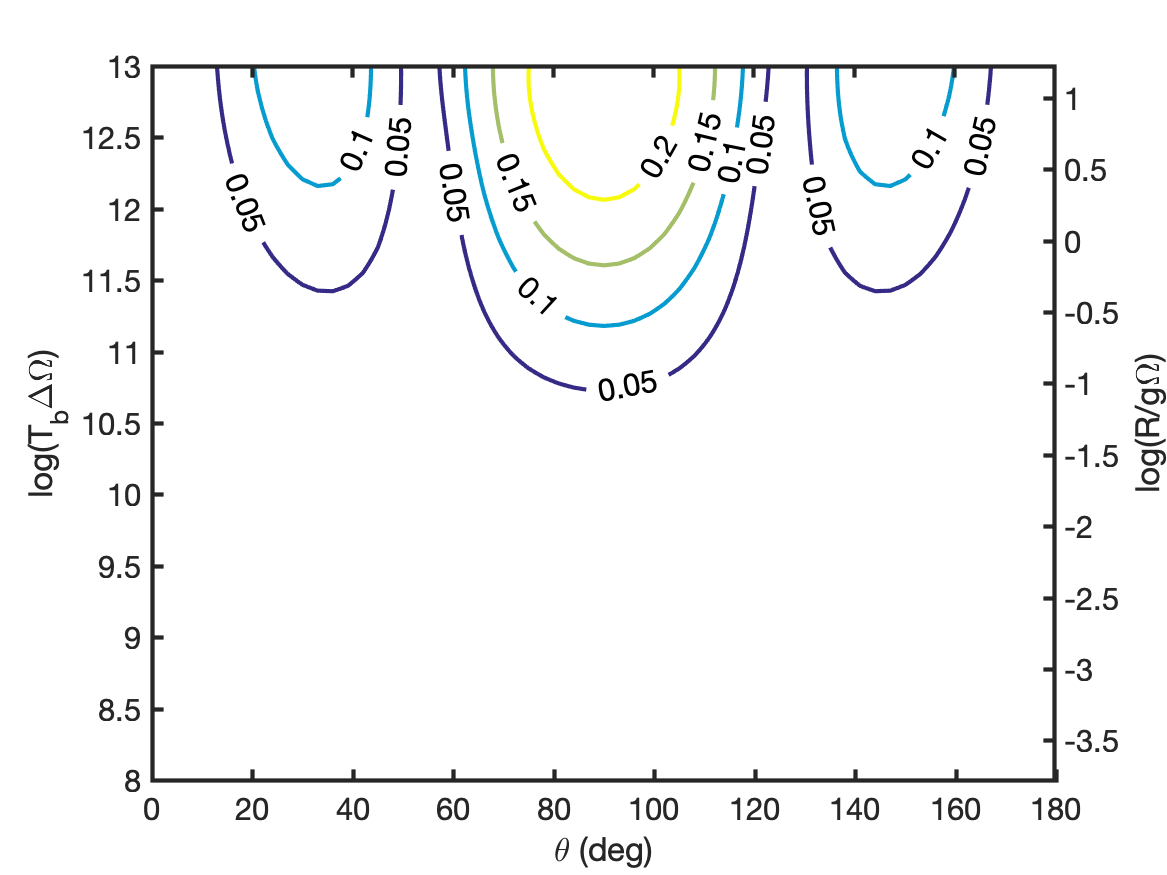}
       \caption{}
    \end{subfigure}
    ~
    \begin{subfigure}[b]{0.32\textwidth}
       \includegraphics[width=\textwidth]{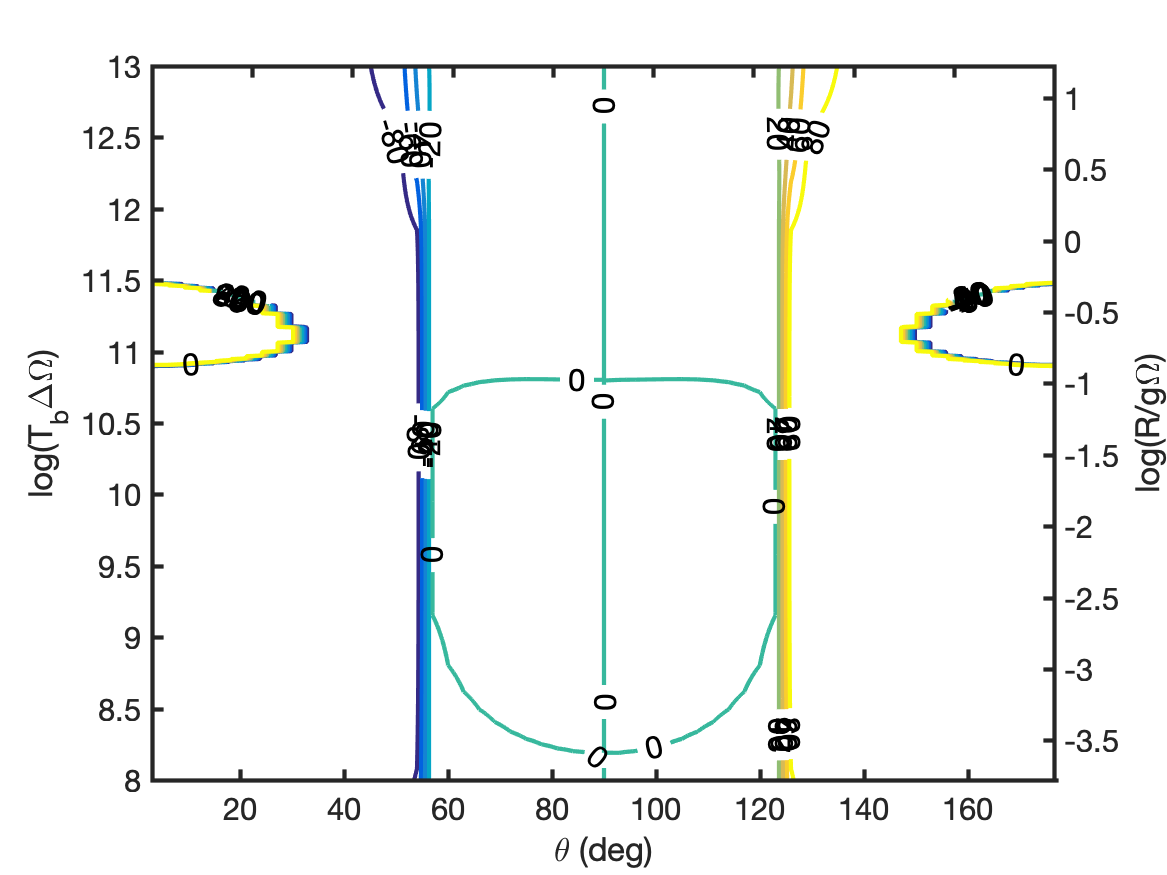}
       \caption{}
    \end{subfigure}
     ~
    \begin{subfigure}[b]{0.32\textwidth}
       \includegraphics[width=\textwidth]{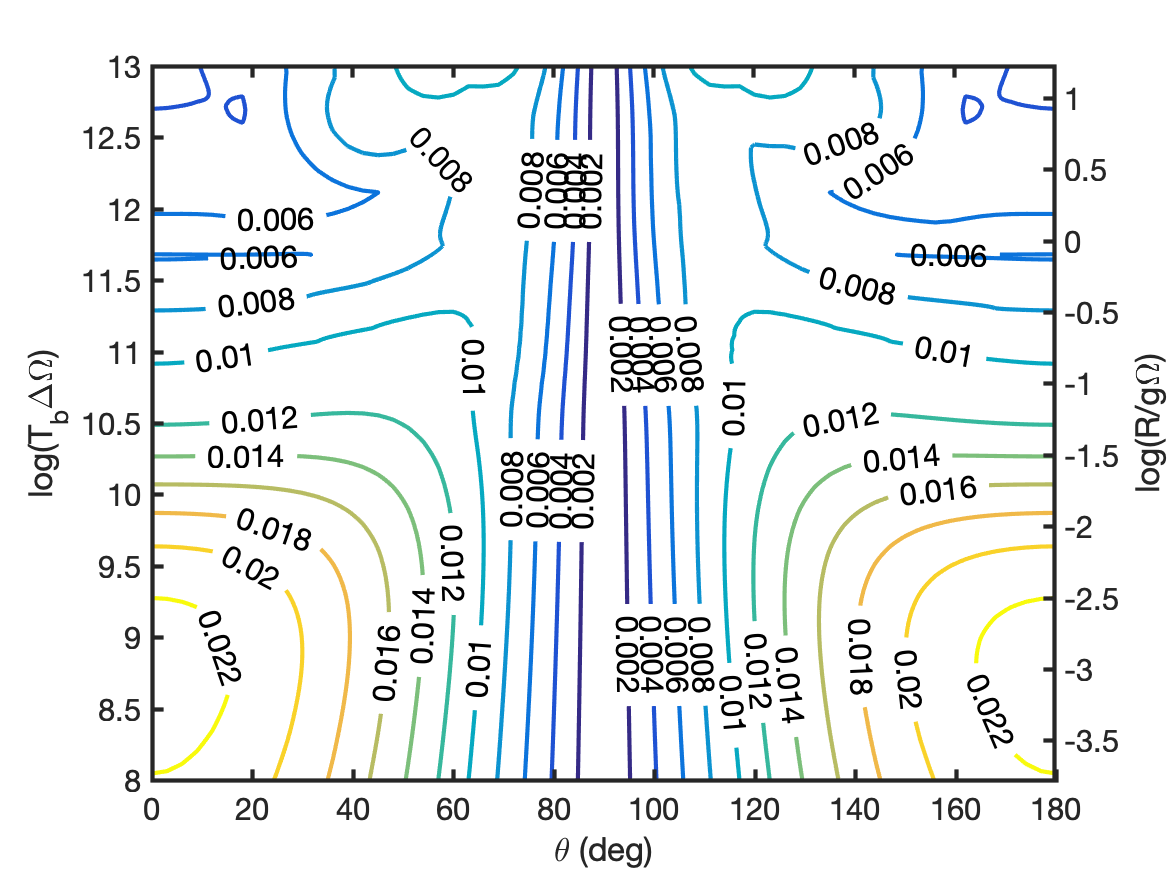}
      \caption{}
    \end{subfigure}
    ~
    \begin{subfigure}[b]{0.32\textwidth}
       \includegraphics[width=\textwidth]{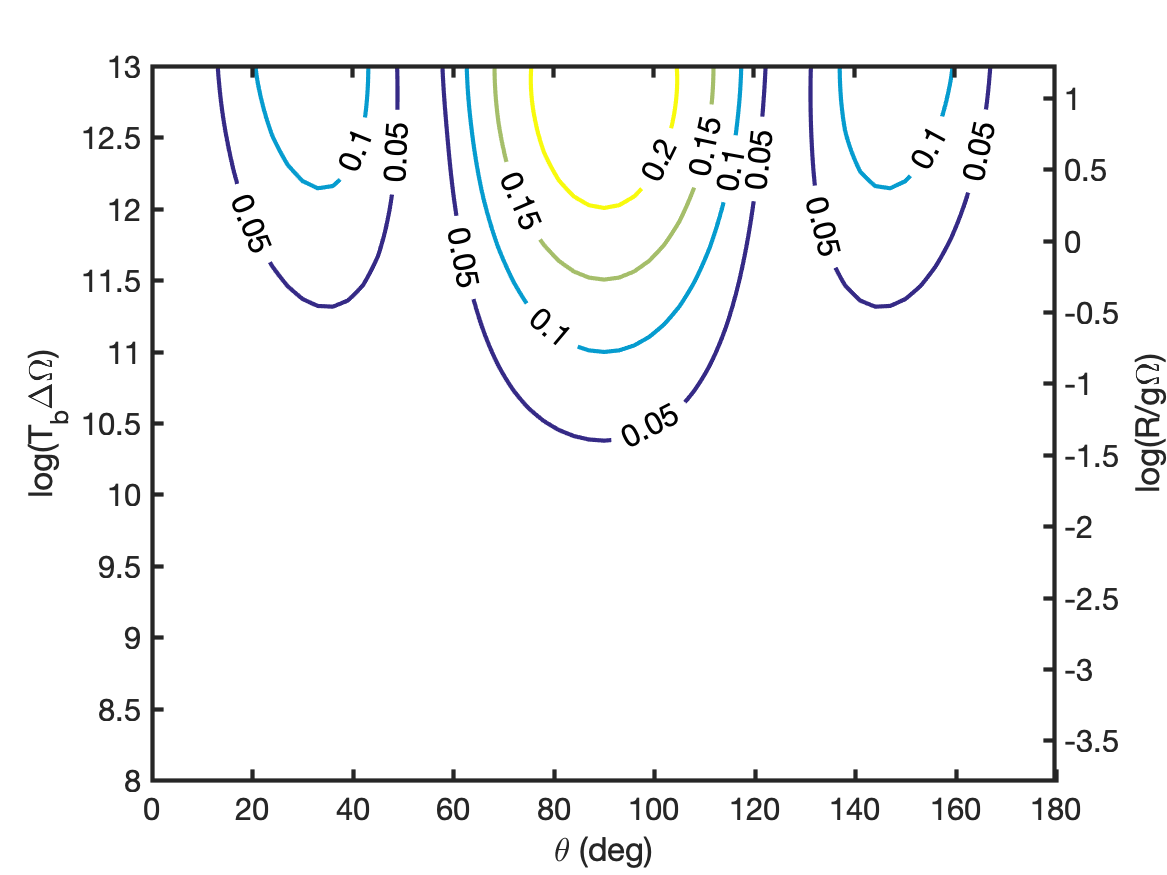}
       \caption{}
    \end{subfigure}
    ~
    \begin{subfigure}[b]{0.32\textwidth}
       \includegraphics[width=\textwidth]{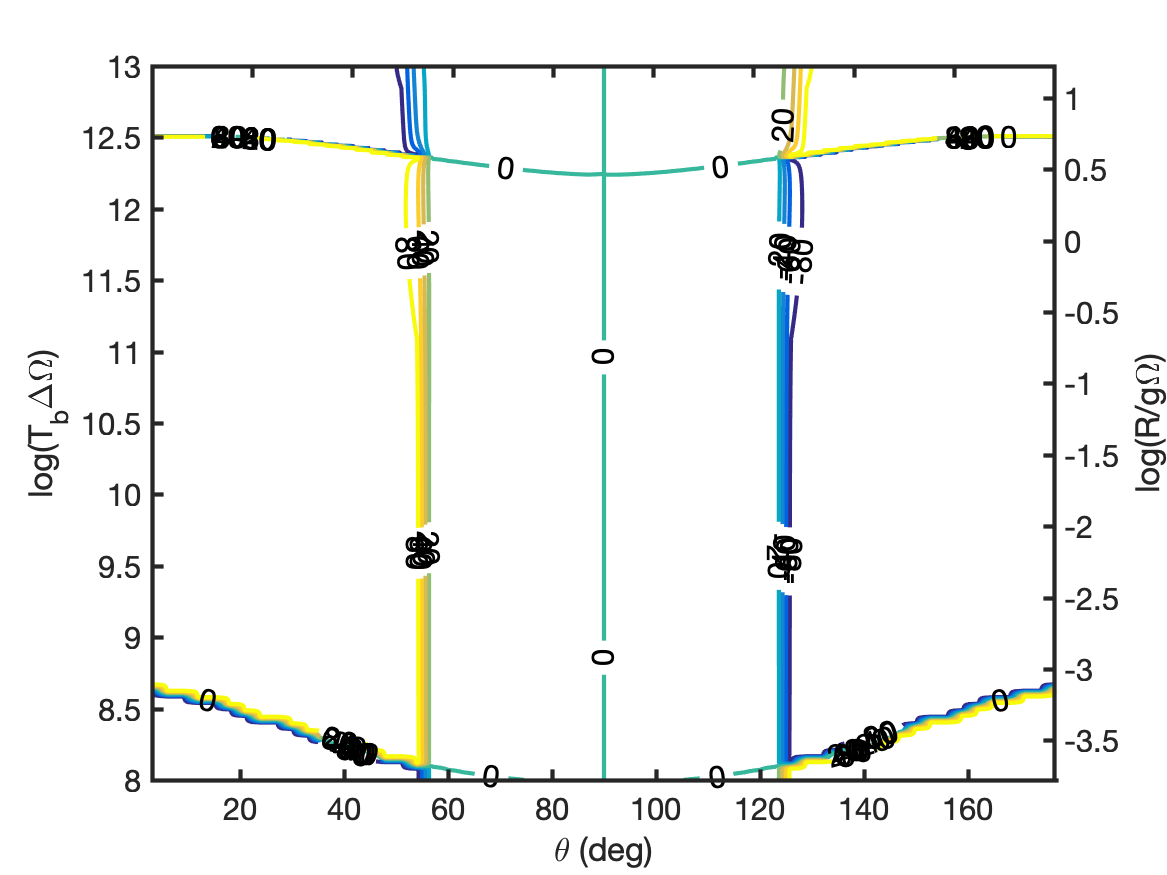}
       \caption{}
    \end{subfigure}
     ~
    \begin{subfigure}[b]{0.32\textwidth}
       \includegraphics[width=\textwidth]{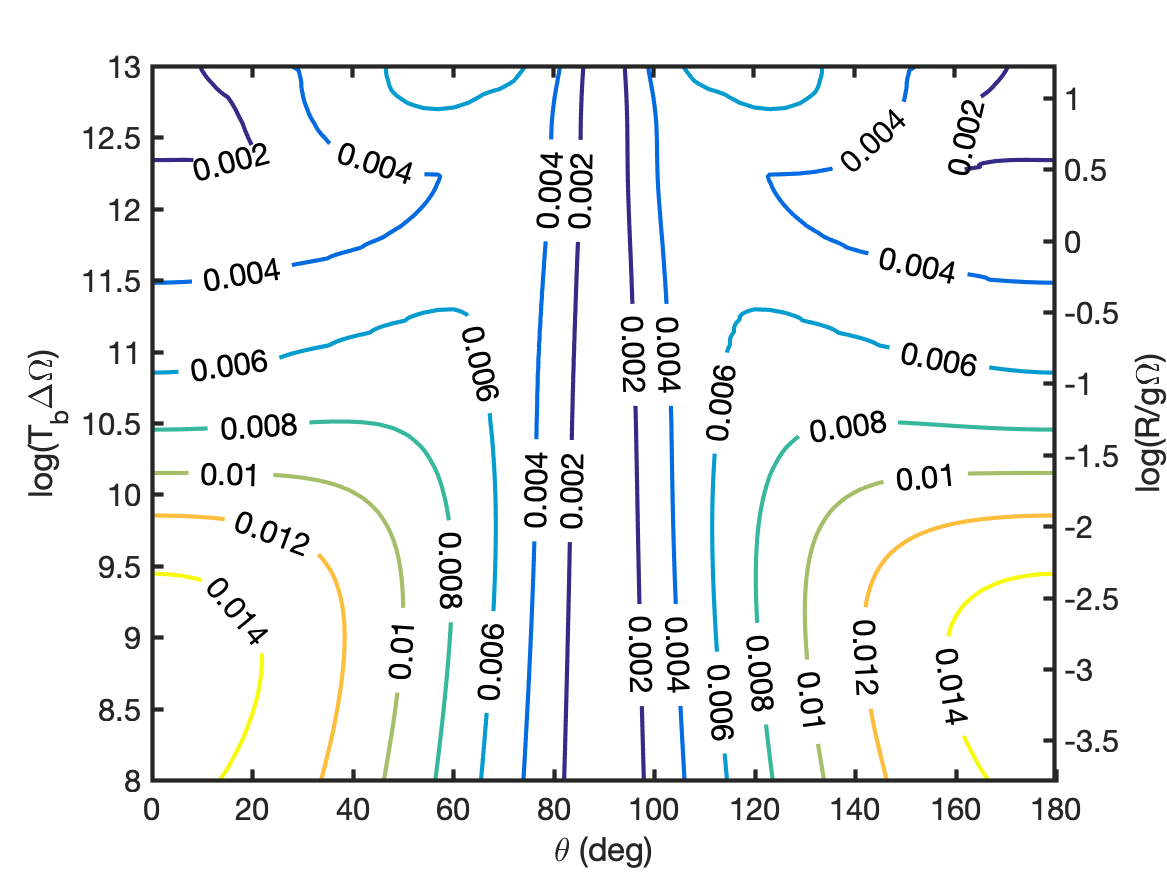}
      \caption{}
    \end{subfigure}
     ~
    \begin{subfigure}[b]{0.32\textwidth}
       \includegraphics[width=\textwidth]{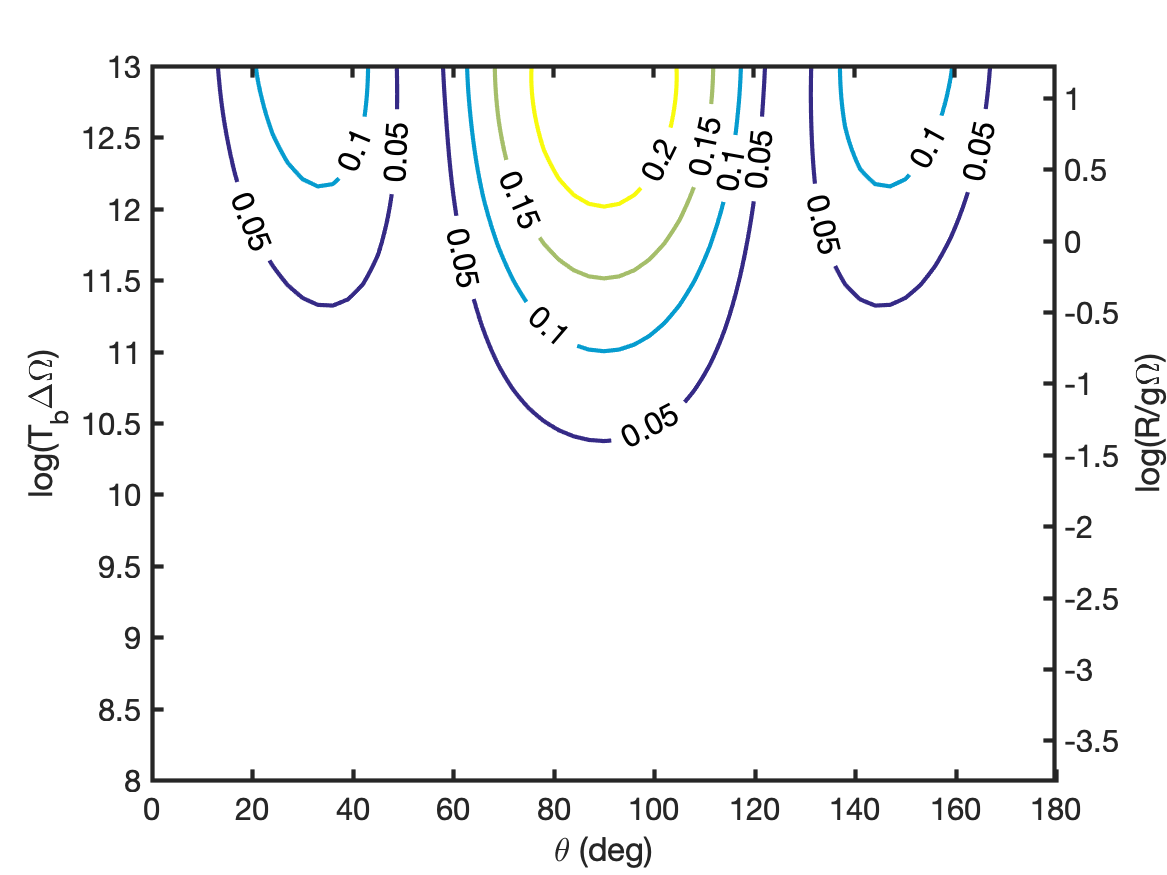}
       \caption{}
    \end{subfigure}
    ~
    \begin{subfigure}[b]{0.32\textwidth}
       \includegraphics[width=\textwidth]{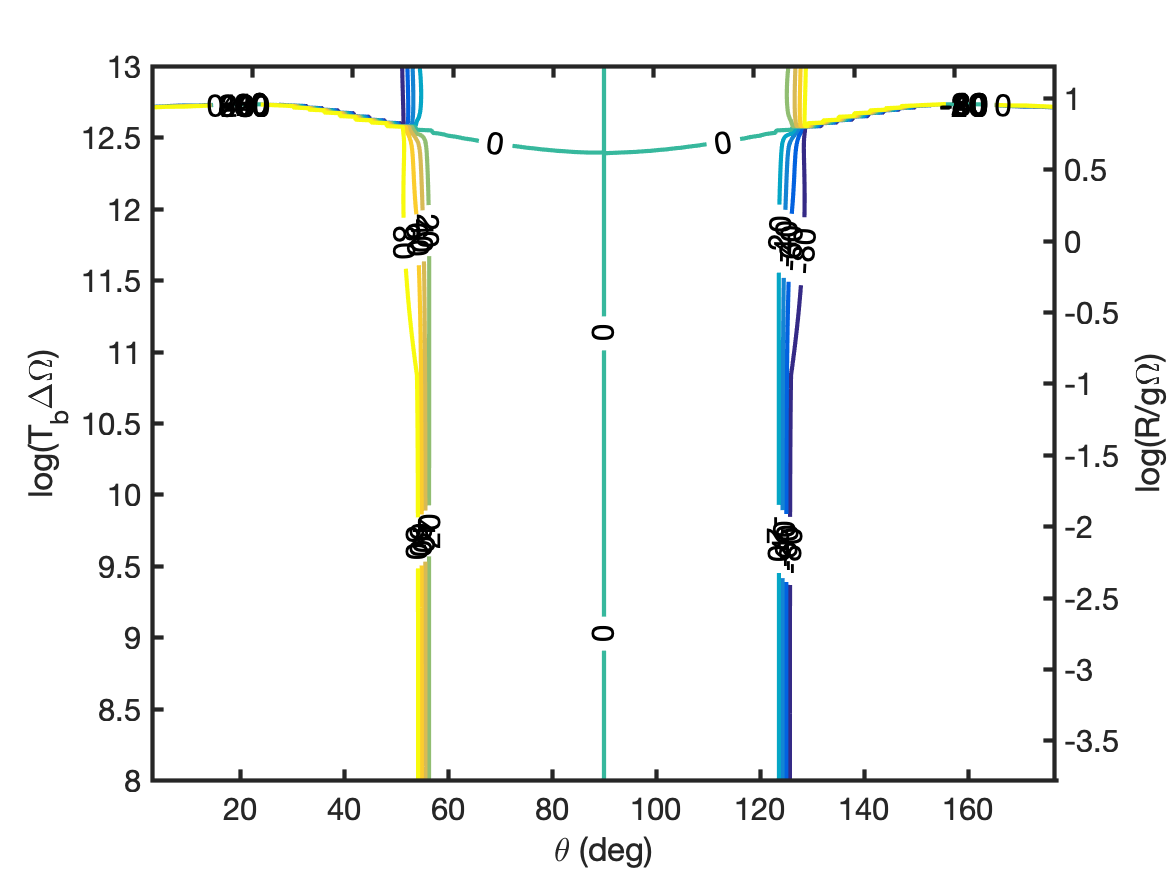}
       \caption{}
    \end{subfigure}
     ~
    \begin{subfigure}[b]{0.32\textwidth}
       \includegraphics[width=\textwidth]{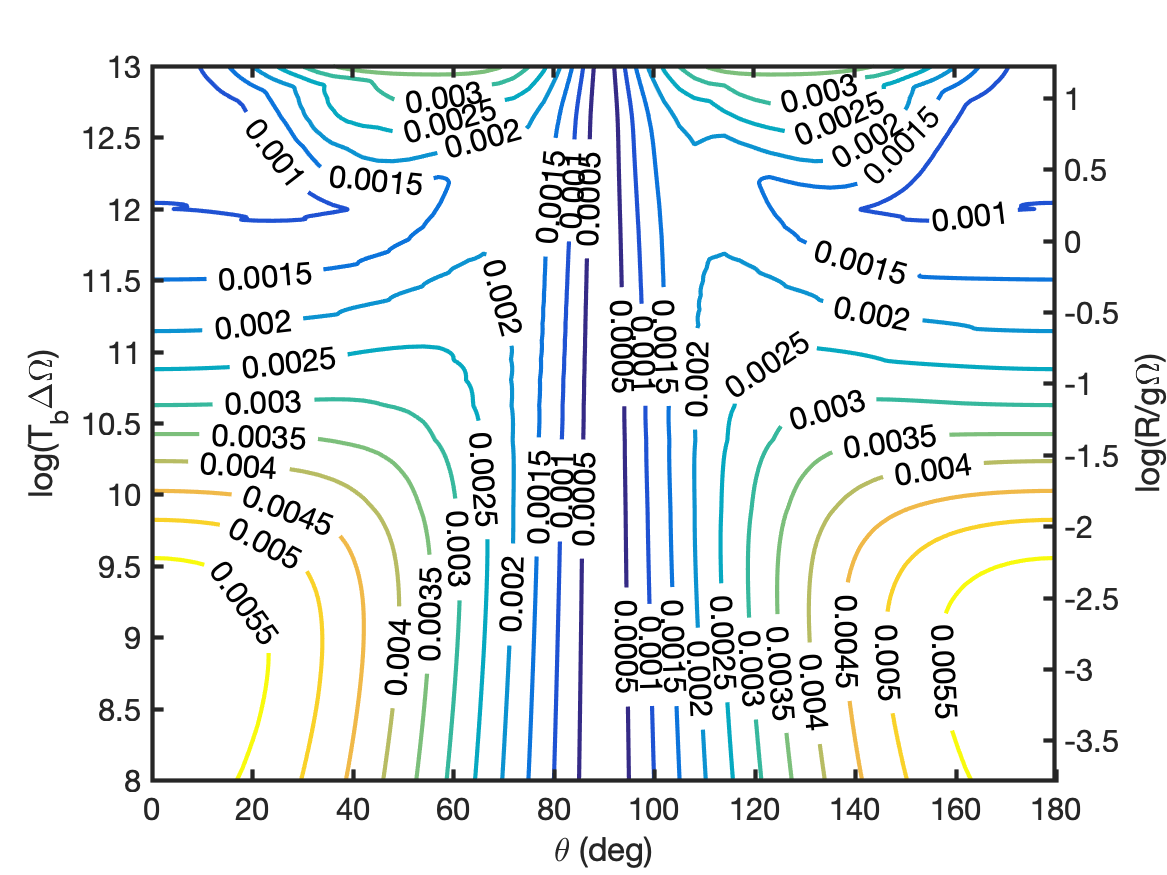}
      \caption{}
    \end{subfigure}
  \caption{Polarization of a water maser isotropically pumped at $B=100\ \mathrm{mG}$. Linear polarization fraction (a,d,g), angle (b,e,h) and circular polarization fraction (c,f,i). Thermal width used $v_{th} = 0.6$ km/s (a,b,c), $1$ km/s (d,e,f) and $2$ km/s (g,h,i).}
\end{figure*}

\begin{figure*}
    \centering
    \begin{subfigure}[b]{0.32\textwidth}
       \includegraphics[width=\textwidth]{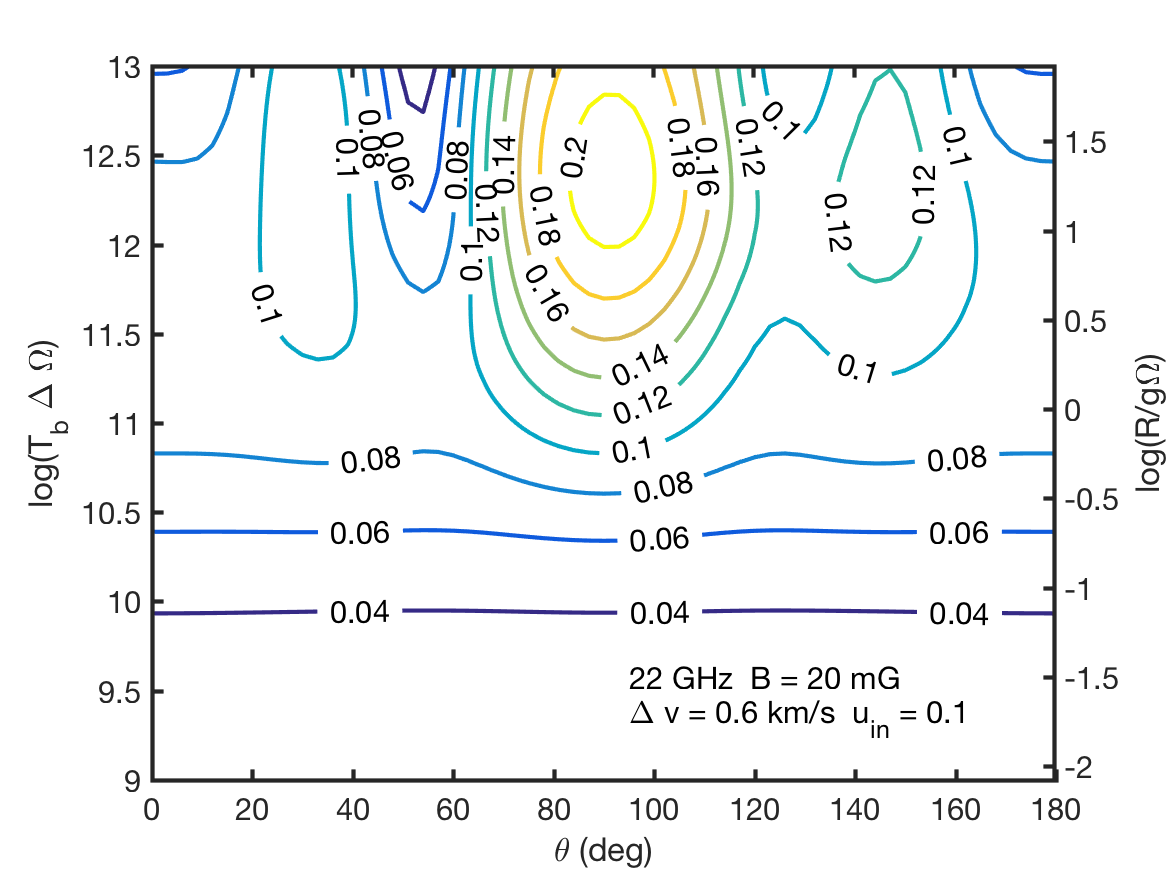}
       \caption{}
    \end{subfigure}
    ~
    \begin{subfigure}[b]{0.32\textwidth}
       \includegraphics[width=\textwidth]{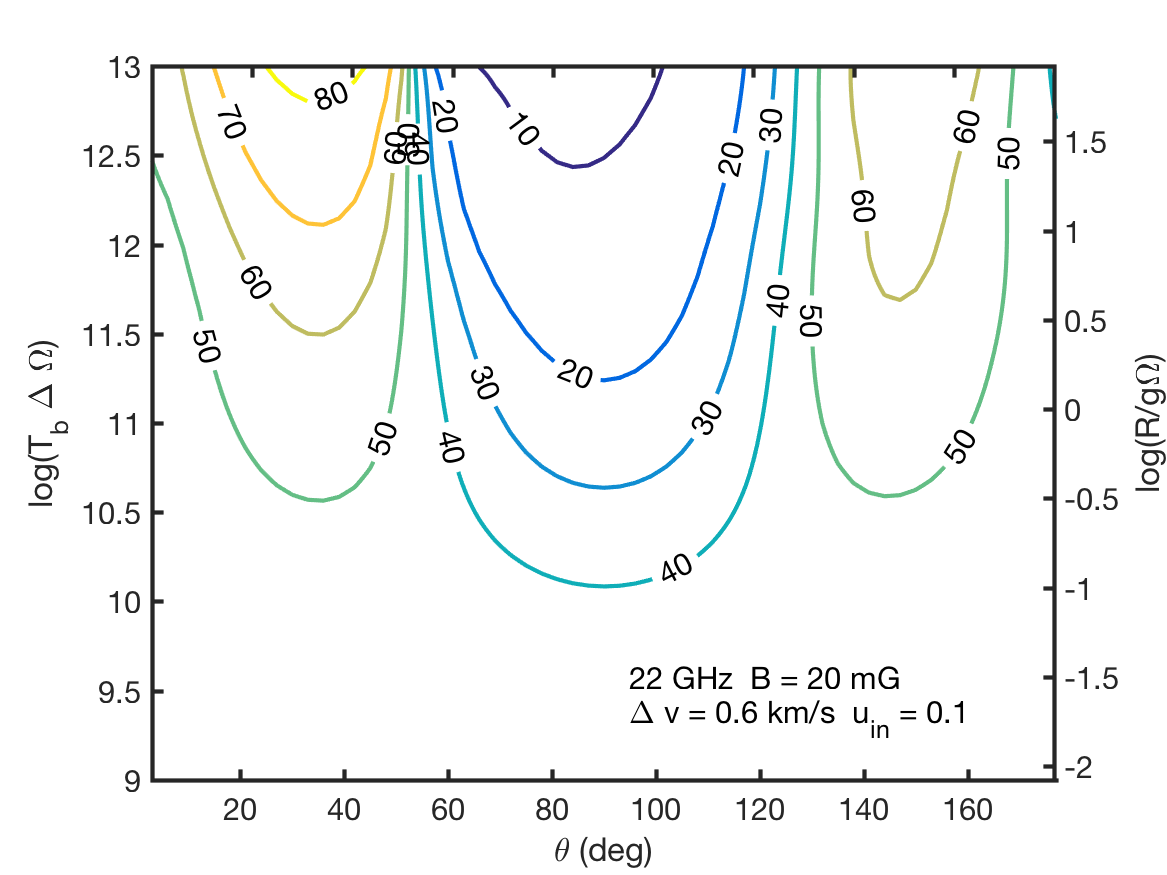}
       \caption{}
    \end{subfigure}
     ~
    \begin{subfigure}[b]{0.32\textwidth}
       \includegraphics[width=\textwidth]{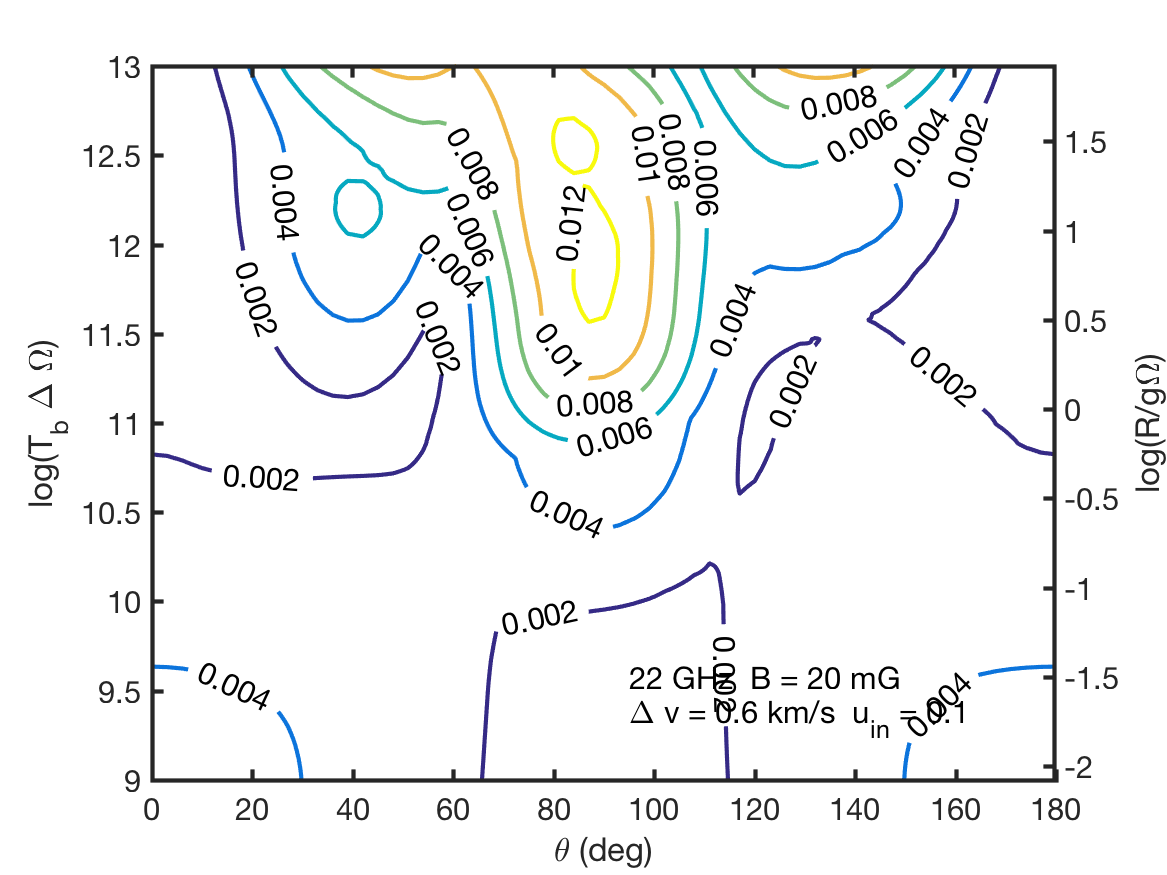}
      \caption{}
    \end{subfigure}
     ~
    \begin{subfigure}[b]{0.32\textwidth}
       \includegraphics[width=\textwidth]{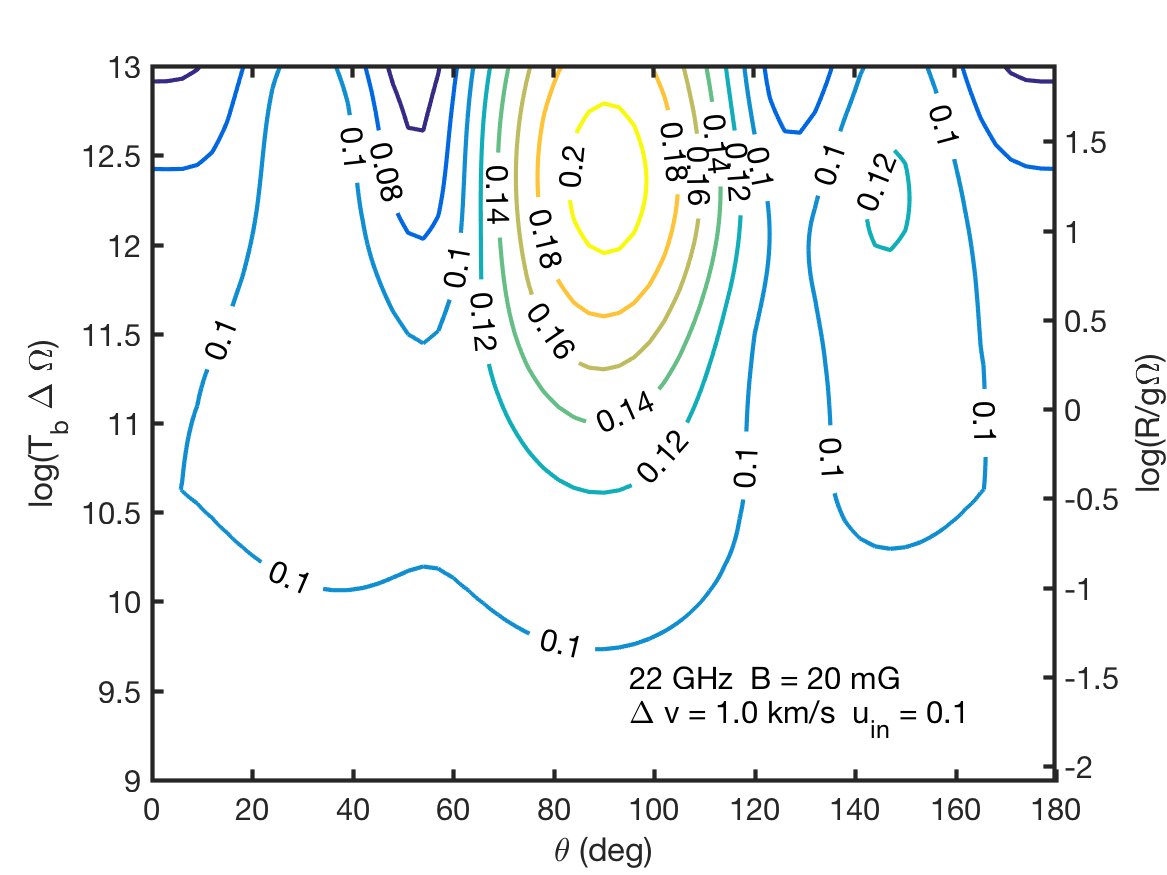}
       \caption{}
    \end{subfigure}
    ~
    \begin{subfigure}[b]{0.32\textwidth}
       \includegraphics[width=\textwidth]{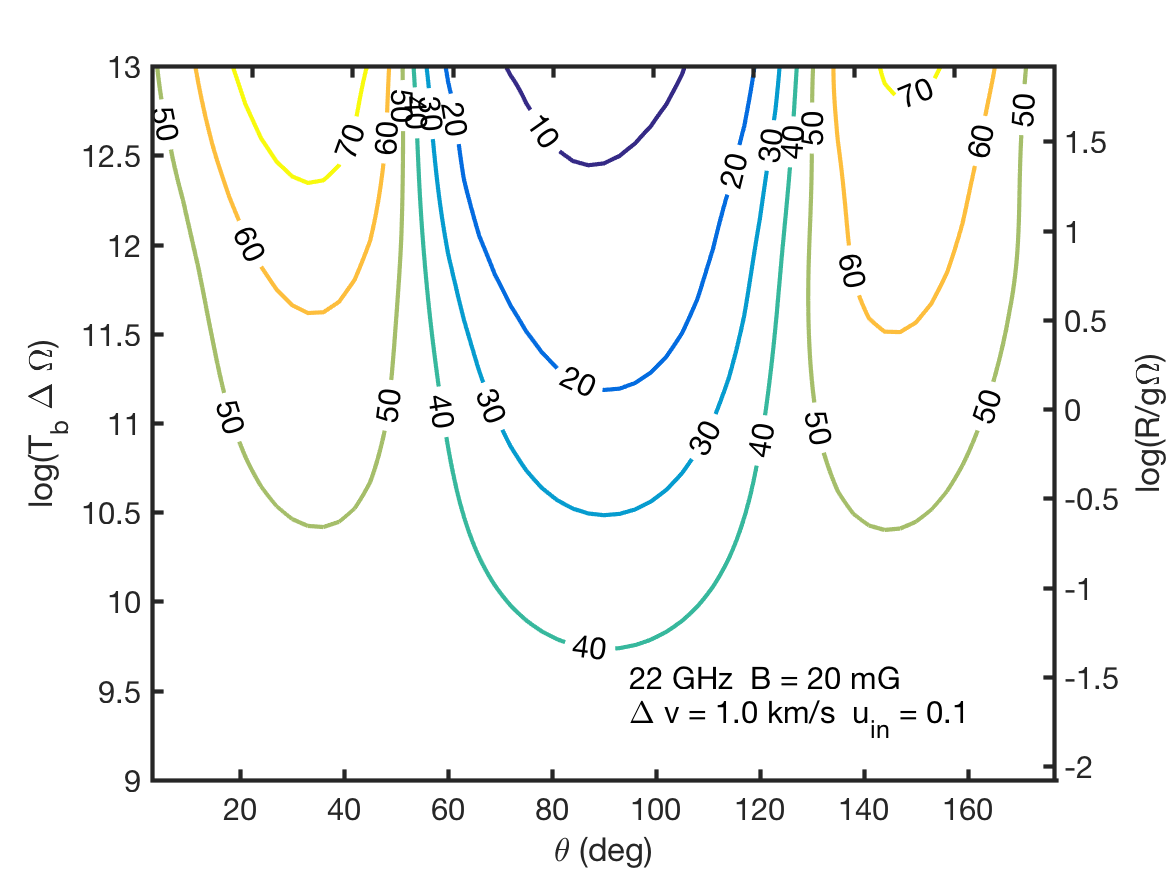}
       \caption{}
    \end{subfigure}
     ~
    \begin{subfigure}[b]{0.32\textwidth}      
       \includegraphics[width=\textwidth]{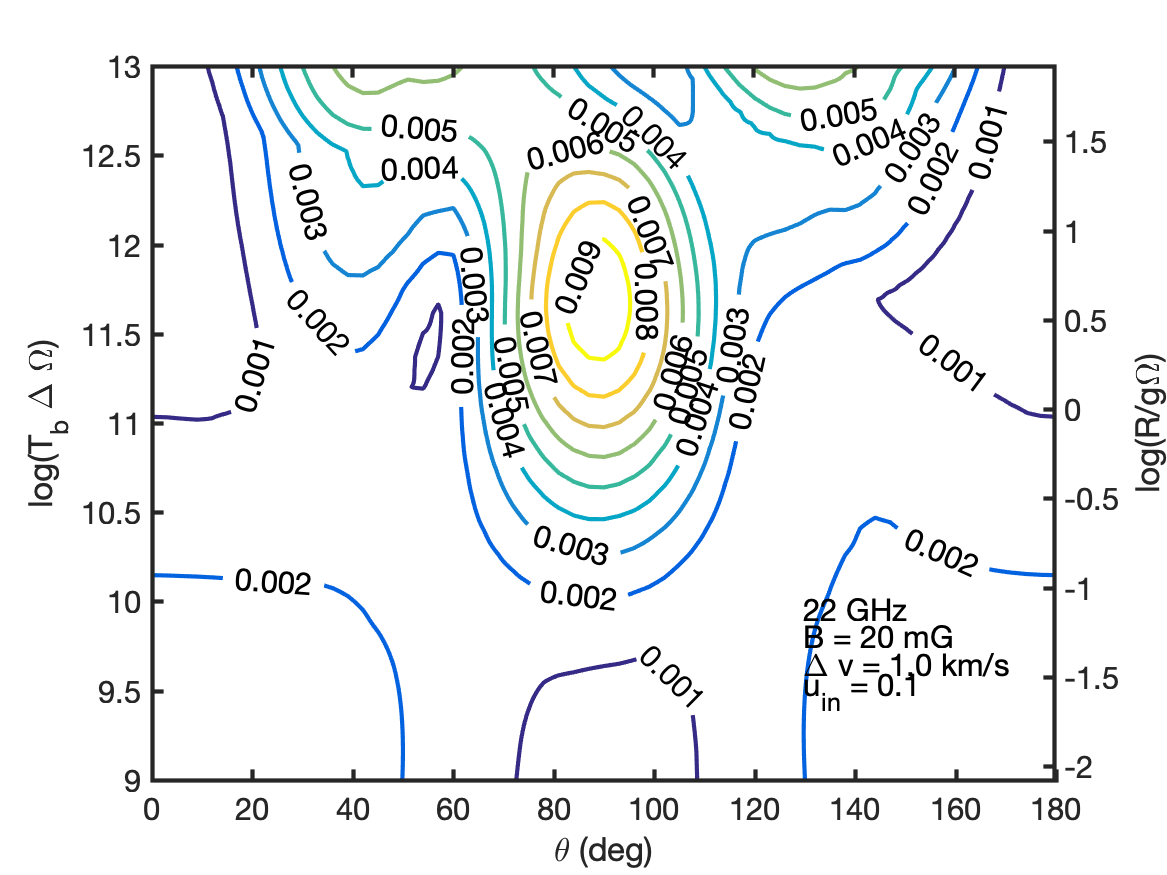}
      \caption{}
    \end{subfigure}
    ~
    \begin{subfigure}[b]{0.32\textwidth}
       \includegraphics[width=\textwidth]{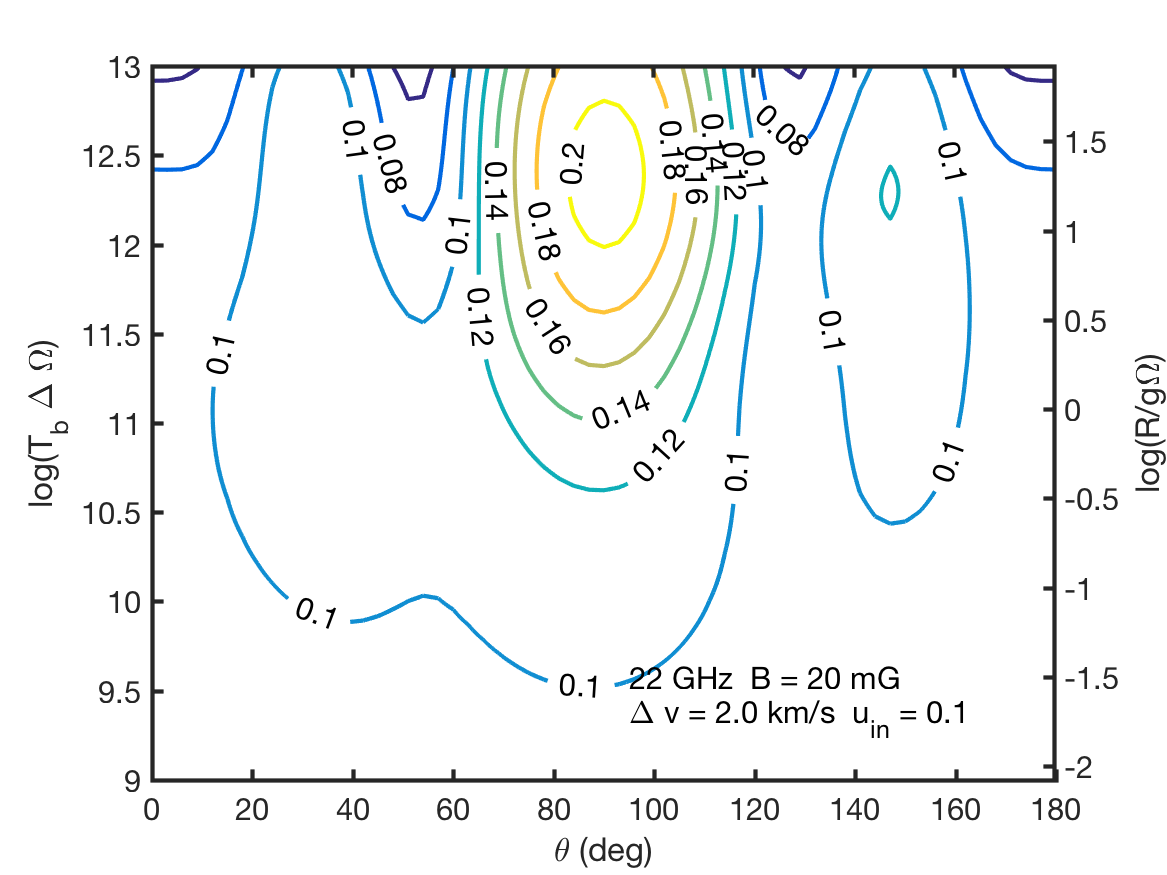}
       \caption{}
    \end{subfigure}
    ~
    \begin{subfigure}[b]{0.32\textwidth}
       \includegraphics[width=\textwidth]{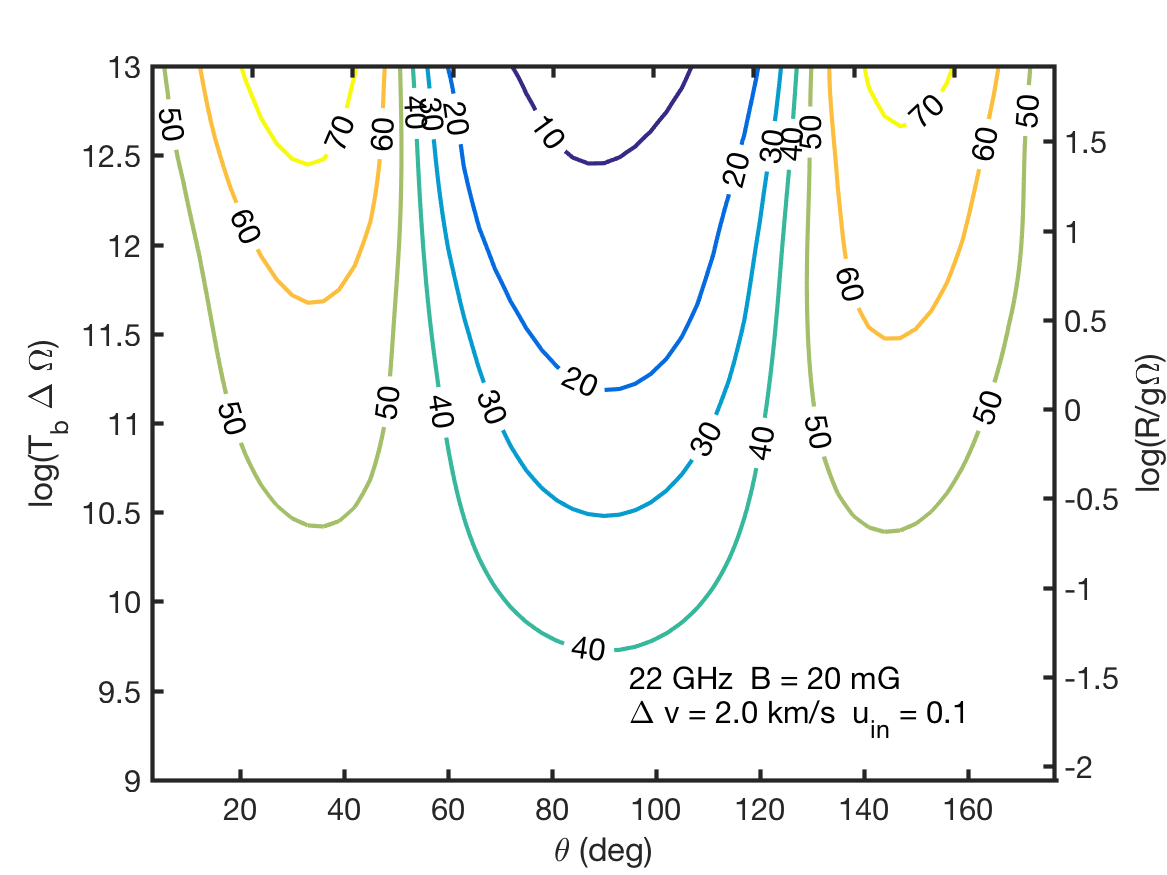}
       \caption{}
    \end{subfigure}
     ~
    \begin{subfigure}[b]{0.32\textwidth}      
       \includegraphics[width=\textwidth]{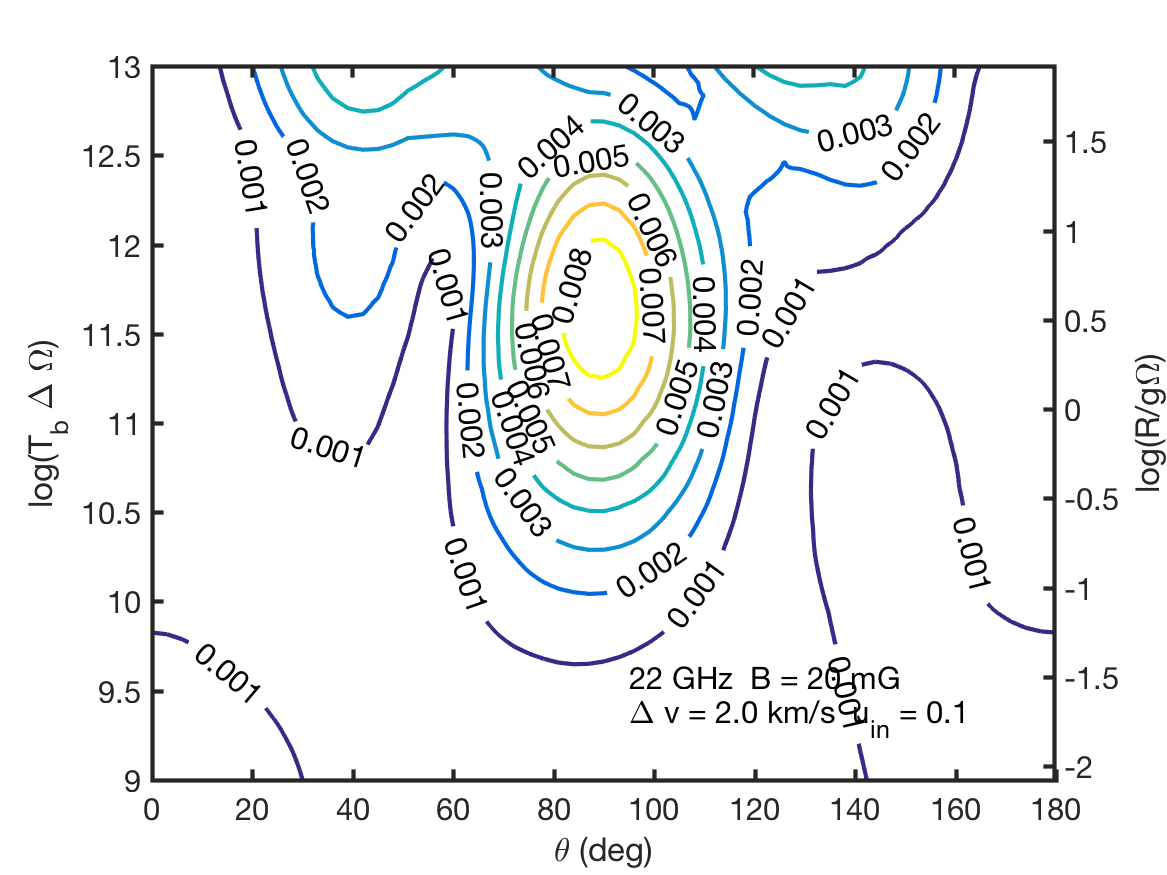}
      \caption{}
    \end{subfigure}
  \caption{Polarization of a water maser with $10\%$ polarized seed radiation at $B=20\ \mathrm{mG}$. Linear polarization fraction (a,d,g), angle (b,e,h) and circular polarization fraction (c,f,i). Thermal width used $v_{th} = 0.6$ km/s (a,b,c), $1$ km/s (d,e,f) and $2$ km/s (g,h,i).}
\end{figure*}

\begin{figure*}
    \centering
    \begin{subfigure}[b]{0.32\textwidth}
       \includegraphics[width=\textwidth]{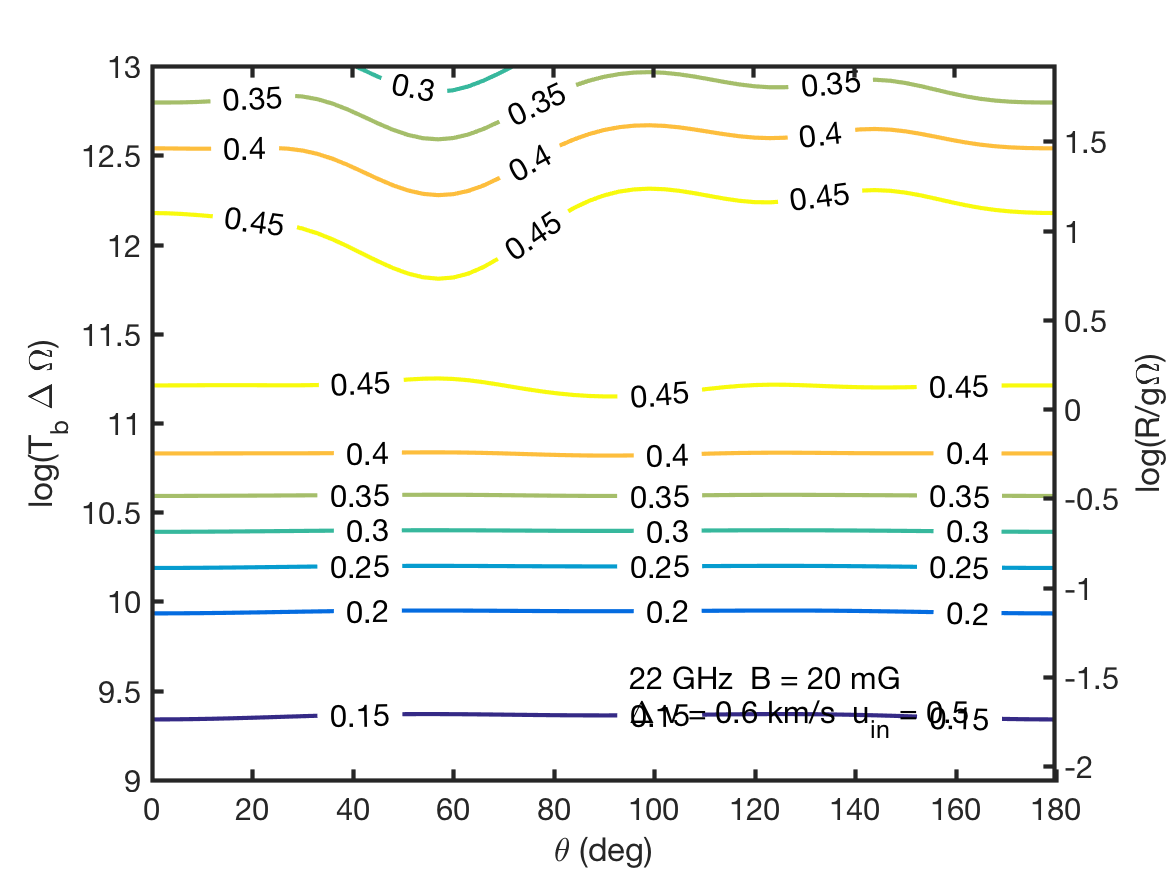}
       \caption{}
    \end{subfigure}
    ~
    \begin{subfigure}[b]{0.32\textwidth}
       \includegraphics[width=\textwidth]{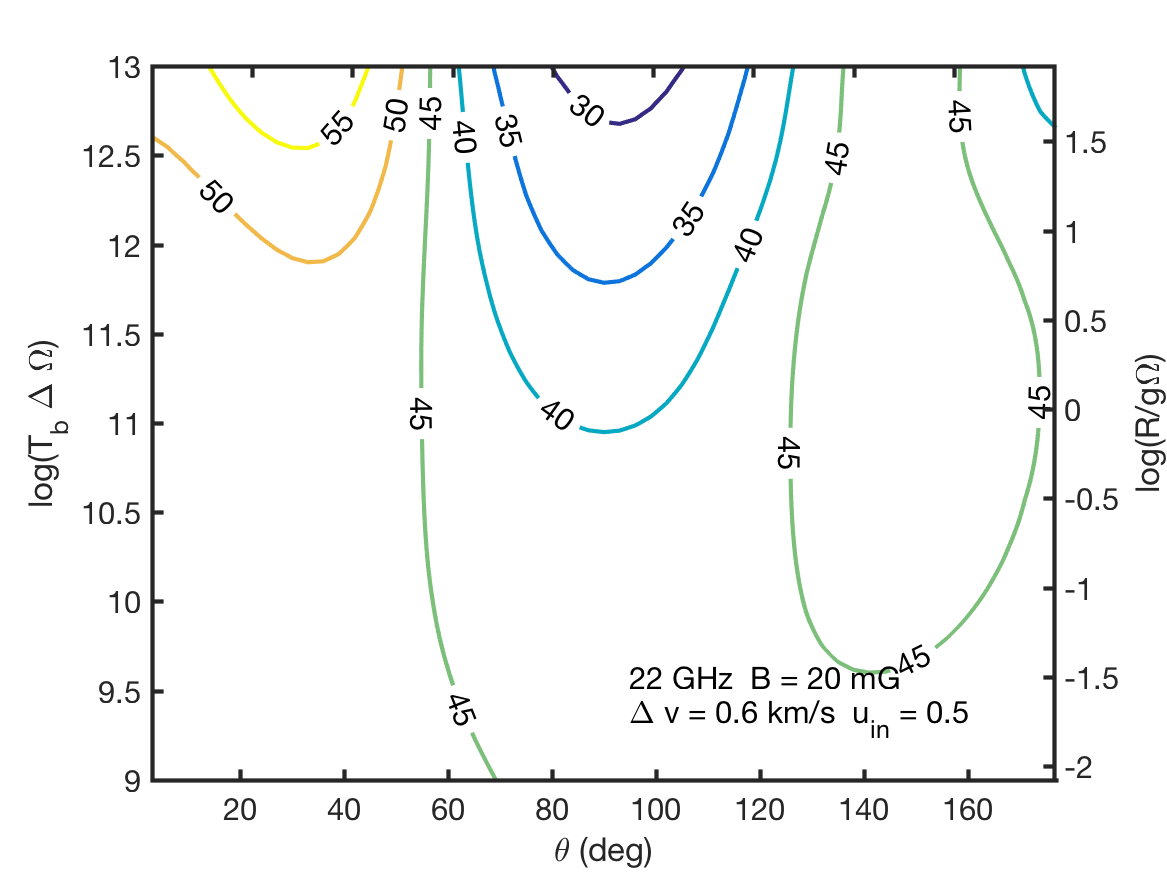}
       \caption{}
    \end{subfigure}
     ~
    \begin{subfigure}[b]{0.32\textwidth}      
       \includegraphics[width=\textwidth]{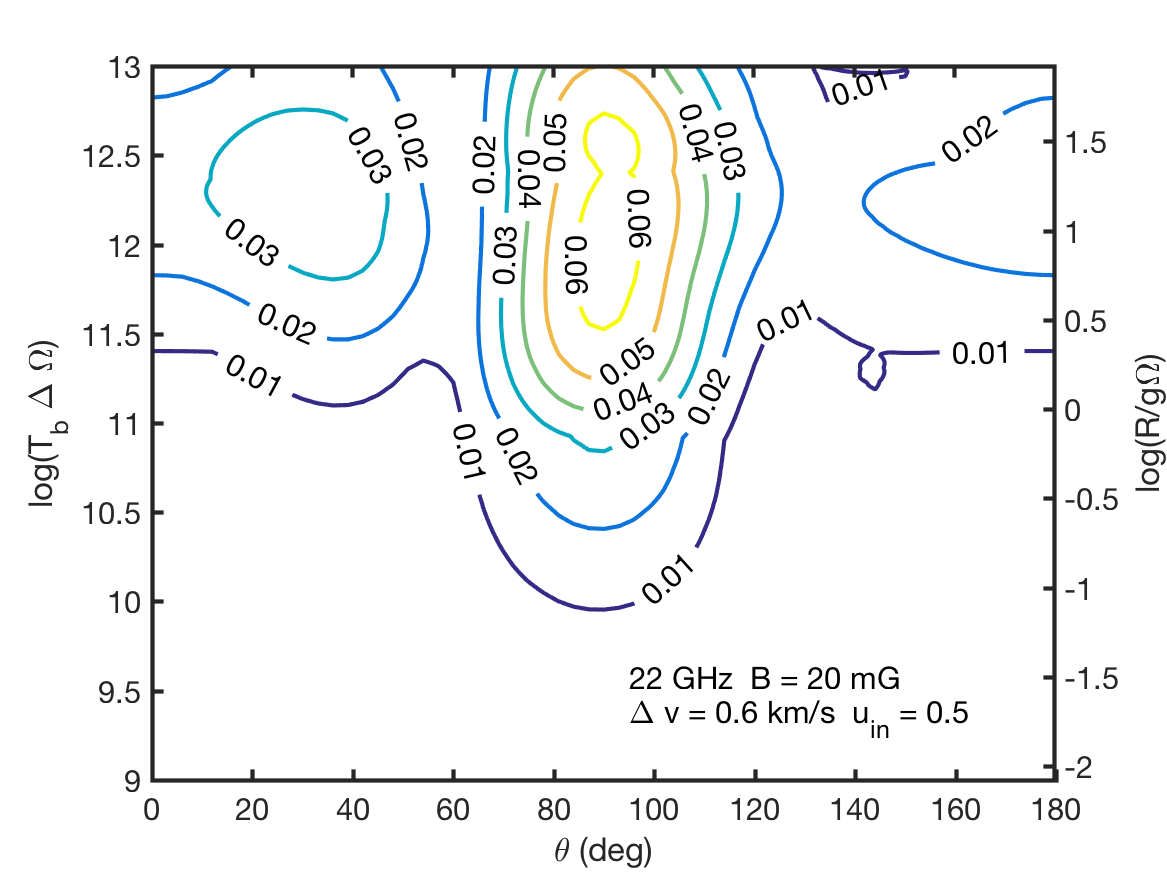}
      \caption{}
    \end{subfigure}
     ~
    \begin{subfigure}[b]{0.32\textwidth}
       \includegraphics[width=\textwidth]{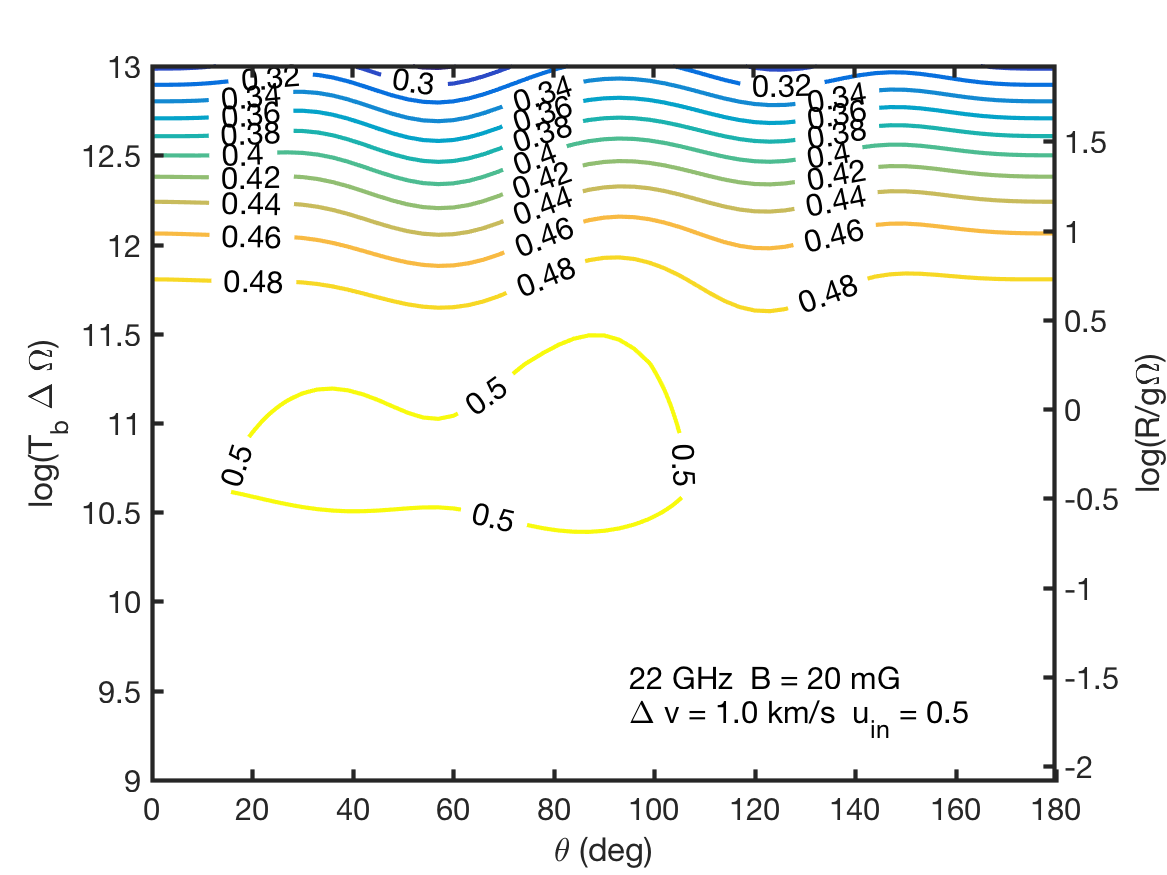}
       \caption{} 
    \end{subfigure}
    ~
    \begin{subfigure}[b]{0.32\textwidth}
       \includegraphics[width=\textwidth]{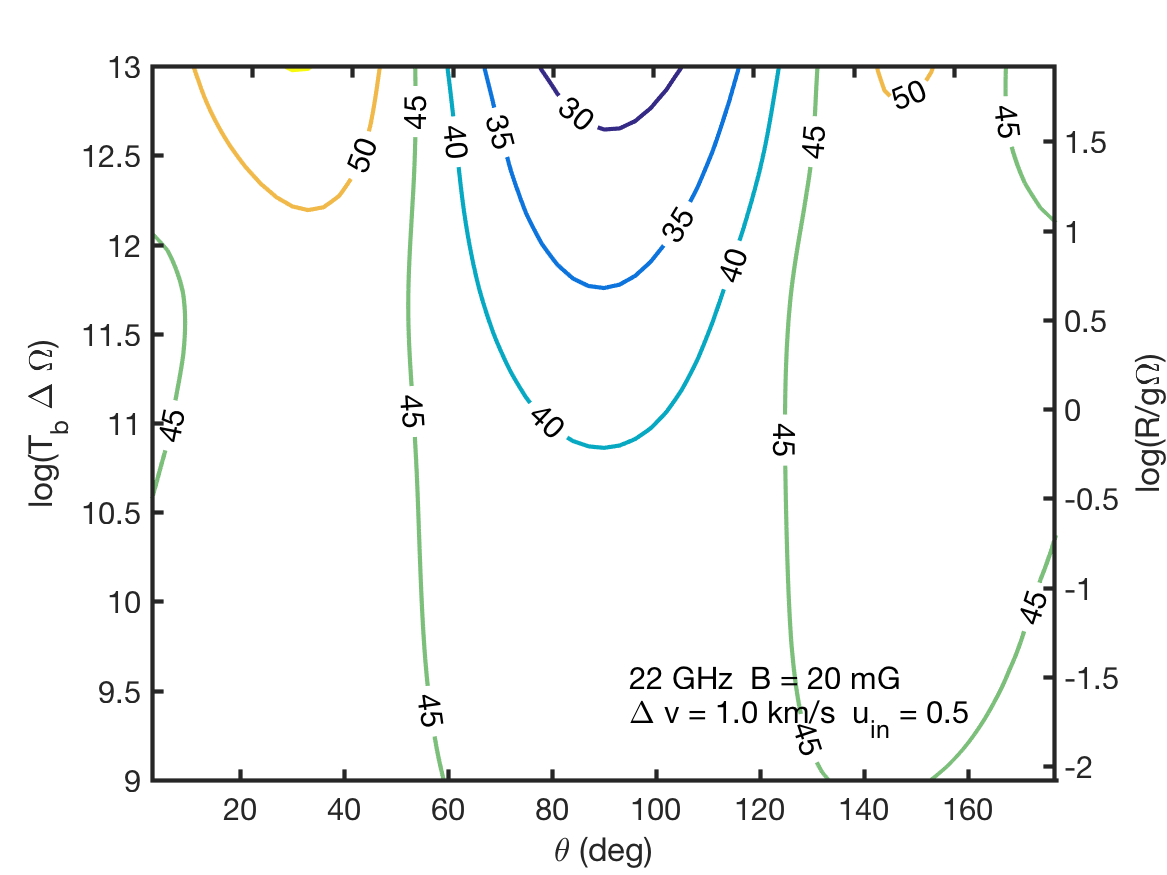}
       \caption{}
    \end{subfigure}
     ~
    \begin{subfigure}[b]{0.32\textwidth}      
       \includegraphics[width=\textwidth]{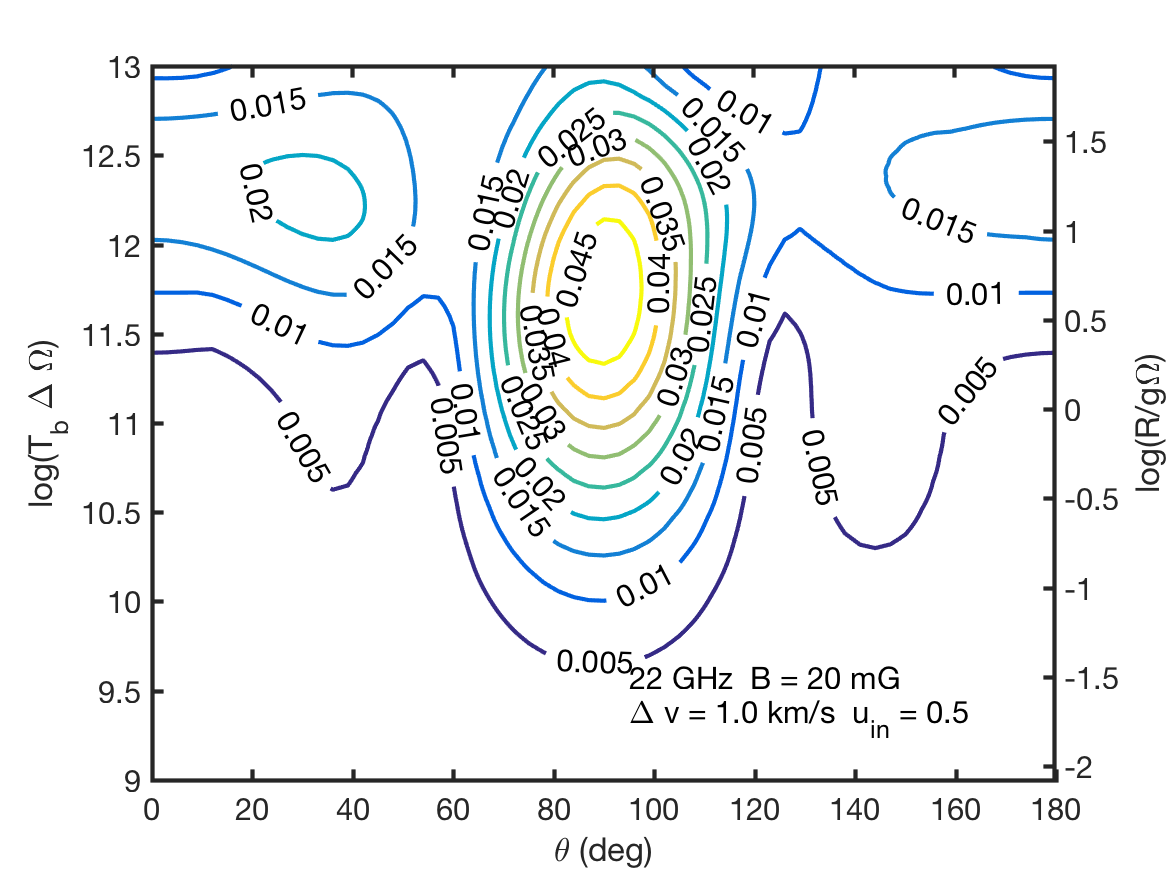}
      \caption{}
    \end{subfigure}
     ~
    \begin{subfigure}[b]{0.32\textwidth}
       \includegraphics[width=\textwidth]{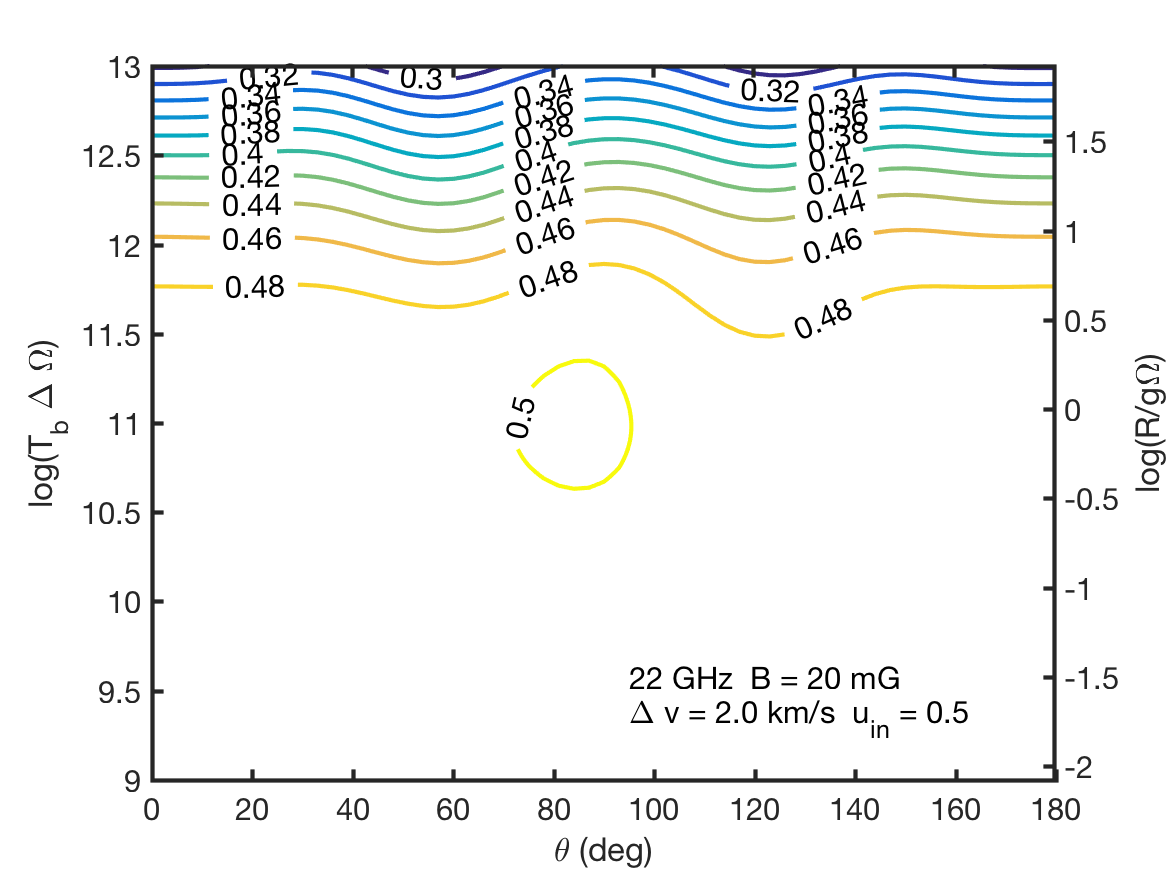}
       \caption{}
    \end{subfigure}
    ~
    \begin{subfigure}[b]{0.32\textwidth}
       \includegraphics[width=\textwidth]{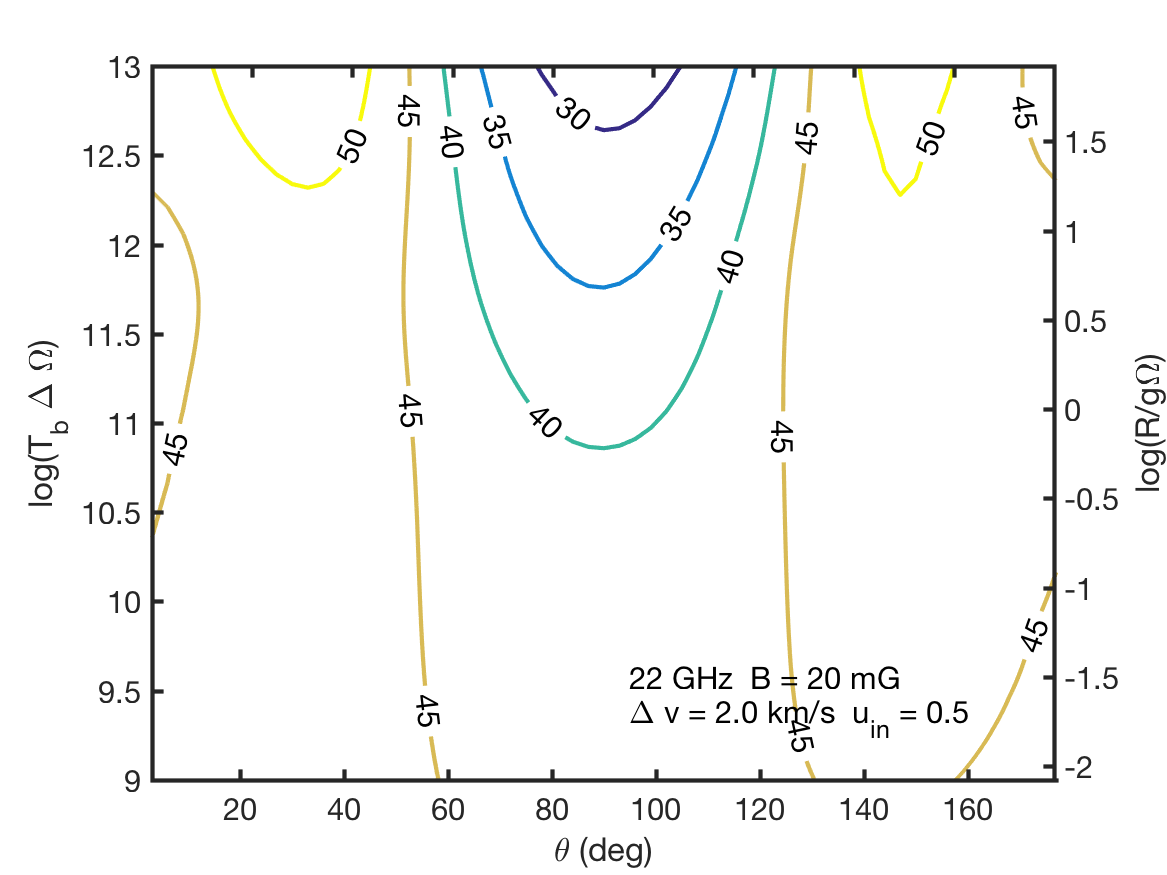}
       \caption{}
    \end{subfigure}
     ~
    \begin{subfigure}[b]{0.32\textwidth}      
       \includegraphics[width=\textwidth]{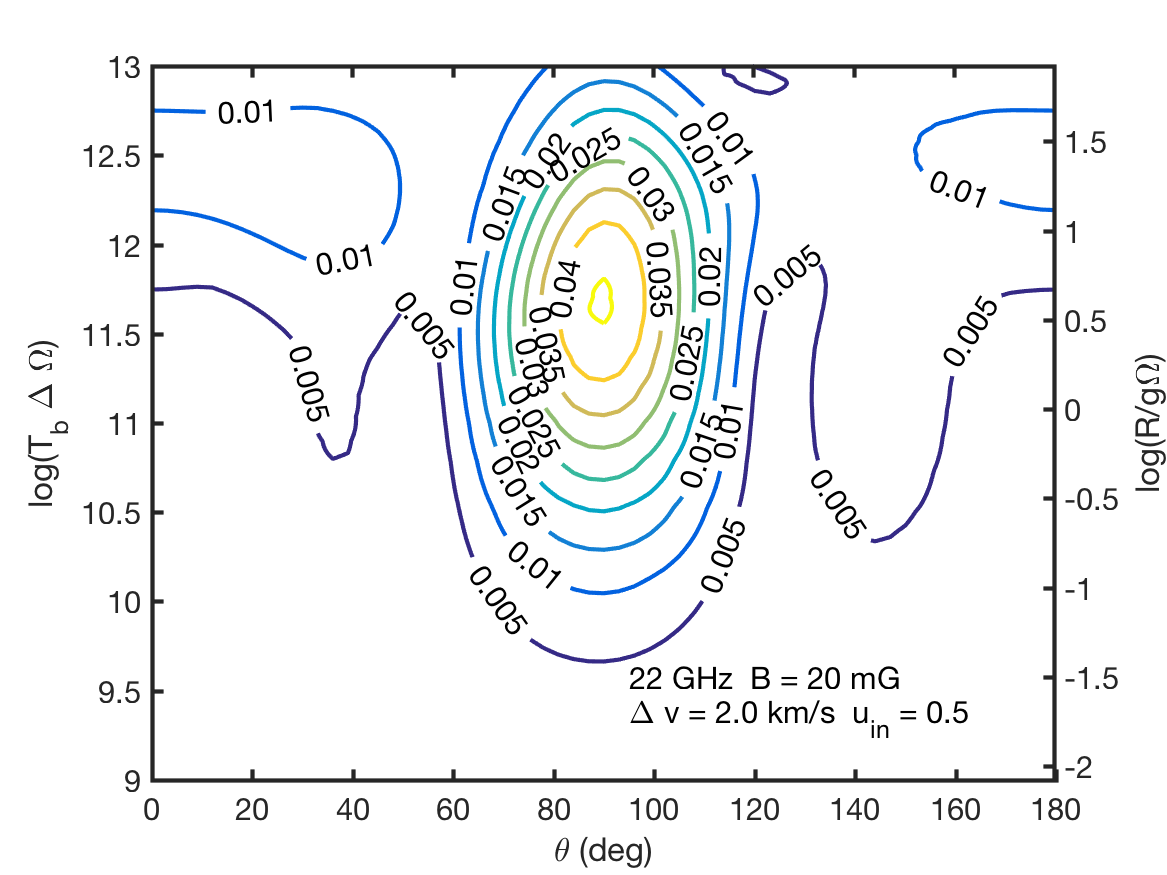}
      \caption{}
    \end{subfigure}
  \caption{Polarization of a water maser with $50\%$ polarized seed radiation at $B=20\ \mathrm{mG}$. Linear polarization fraction (a,d,g), angle (b,e,h) and circular polarization fraction (c,f,i). Thermal width used $v_{th} = 0.6$ km/s (a,b,c), $1$ km/s (d,e,f) and $2$ km/s (g,h,i).}
\end{figure*}
\end{appendix}

\end{document}